\documentclass[lettersize,journal]{IEEEtran}

\usepackage{amsmath,amssymb,amsfonts}
\usepackage{booktabs}
\usepackage{array}
\usepackage{graphicx}
\usepackage{xcolor}
\usepackage{colortbl}
\usepackage{microtype}
\usepackage[hidelinks]{hyperref}
\usepackage{cleveref}
\crefname{figure}{Fig.}{Figs.}\Crefname{figure}{Fig.}{Figs.}
\crefname{table}{Table}{Tables}\Crefname{table}{Table}{Tables}
\usepackage{multirow}
\definecolor{grey}{HTML}{EFEEEA}
\definecolor{greytint}{HTML}{F6F5F2}
\definecolor{ink}{HTML}{222222}
\definecolor{pblue}{HTML}{CAD5F1}
\definecolor{pamber}{HTML}{F6E2BF}
\definecolor{ppink}{HTML}{F0B8B8}
\definecolor{pteal}{HTML}{C0DDE3}
\definecolor{pviolet}{HTML}{D0C2EA}
\definecolor{rk1}{HTML}{B8E2FC}
\definecolor{rk2}{HTML}{DCF0FE}
\definecolor{rk3}{HTML}{F0F8FF}
\definecolor{hdr}{HTML}{EFEEEA}
\definecolor{pool}{HTML}{F6F5F2}
\definecolor{phase1bg}{HTML}{FBF2E2}
\definecolor{phase2bg}{HTML}{F9E6E8}

\graphicspath{{./}}

\newcommand{\oic}{\textsc{OIC}}
\newcommand{\V}[1]{\ifcsname val@#1\endcsname\csname val@#1\endcsname\else\textcolor{red}{\textbf{[#1]}}\fi}
\IfFileExists{numbers/numbers.tex}{
\expandafter\def\csname val@c1.dir.body.ci\endcsname{[87.7, 99.6]}
\expandafter\def\csname val@c1.dir.body.frac\endcsname{41/42}
\expandafter\def\csname val@c1.dir.body.full\endcsname{41/42 (97.6\%)}
\expandafter\def\csname val@c1.dir.body.pct\endcsname{97.6\%}
\expandafter\def\csname val@c1.dir.copied.ci\endcsname{[0.0, 8.4]}
\expandafter\def\csname val@c1.dir.copied.frac\endcsname{0/42}
\expandafter\def\csname val@c1.dir.copied.full\endcsname{0/42 (0.0\%)}
\expandafter\def\csname val@c1.dir.copied.pct\endcsname{0.0\%}
\expandafter\def\csname val@c1.dir.materialized.ci\endcsname{[0.0, 8.4]}
\expandafter\def\csname val@c1.dir.materialized.frac\endcsname{0/42}
\expandafter\def\csname val@c1.dir.materialized.full\endcsname{0/42 (0.0\%)}
\expandafter\def\csname val@c1.dir.materialized.pct\endcsname{0.0\%}
\expandafter\def\csname val@c1.dir.overwrite.ci\endcsname{[0.0, 8.4]}
\expandafter\def\csname val@c1.dir.overwrite.frac\endcsname{0/42}
\expandafter\def\csname val@c1.dir.overwrite.full\endcsname{0/42 (0.0\%)}
\expandafter\def\csname val@c1.dir.overwrite.pct\endcsname{0.0\%}
\expandafter\def\csname val@c1.dir.reported.ci\endcsname{[46.8, 75.0]}
\expandafter\def\csname val@c1.dir.reported.frac\endcsname{26/42}
\expandafter\def\csname val@c1.dir.reported.full\endcsname{26/42 (61.9\%)}
\expandafter\def\csname val@c1.dir.reported.pct\endcsname{61.9\%}
\expandafter\def\csname val@c1.tar.body.ci\endcsname{[87.7, 99.6]}
\expandafter\def\csname val@c1.tar.body.frac\endcsname{41/42}
\expandafter\def\csname val@c1.tar.body.full\endcsname{41/42 (97.6\%)}
\expandafter\def\csname val@c1.tar.body.pct\endcsname{97.6\%}
\expandafter\def\csname val@c1.tar.copied.ci\endcsname{[35.5, 64.5]}
\expandafter\def\csname val@c1.tar.copied.frac\endcsname{21/42}
\expandafter\def\csname val@c1.tar.copied.full\endcsname{21/42 (50.0\%)}
\expandafter\def\csname val@c1.tar.copied.pct\endcsname{50.0\%}
\expandafter\def\csname val@c1.tar.materialized.ci\endcsname{[46.8, 75.0]}
\expandafter\def\csname val@c1.tar.materialized.frac\endcsname{26/42}
\expandafter\def\csname val@c1.tar.materialized.full\endcsname{26/42 (61.9\%)}
\expandafter\def\csname val@c1.tar.materialized.pct\endcsname{61.9\%}
\expandafter\def\csname val@c1.tar.overwrite.ci\endcsname{[5.2, 25.0]}
\expandafter\def\csname val@c1.tar.overwrite.frac\endcsname{5/42}
\expandafter\def\csname val@c1.tar.overwrite.full\endcsname{5/42 (11.9\%)}
\expandafter\def\csname val@c1.tar.overwrite.pct\endcsname{11.9\%}
\expandafter\def\csname val@c1.tar.reported.ci\endcsname{[51.6, 79.0]}
\expandafter\def\csname val@c1.tar.reported.frac\endcsname{28/42}
\expandafter\def\csname val@c1.tar.reported.full\endcsname{28/42 (66.7\%)}
\expandafter\def\csname val@c1.tar.reported.pct\endcsname{66.7\%}
\expandafter\def\csname val@digest.body.ci\endcsname{[87.7, 99.6]}
\expandafter\def\csname val@digest.body.frac\endcsname{41/42}
\expandafter\def\csname val@digest.body.full\endcsname{41/42 (97.6\%)}
\expandafter\def\csname val@digest.body.pct\endcsname{97.6\%}
\expandafter\def\csname val@digest.claim.ci\endcsname{[0.0, 8.4]}
\expandafter\def\csname val@digest.claim.frac\endcsname{0/42}
\expandafter\def\csname val@digest.claim.full\endcsname{0/42 (0.0\%)}
\expandafter\def\csname val@digest.claim.pct\endcsname{0.0\%}
\expandafter\def\csname val@digest.cod.hash.ci\endcsname{[77.5, 96.1]}
\expandafter\def\csname val@digest.cod.hash.frac\endcsname{37/41}
\expandafter\def\csname val@digest.cod.hash.full\endcsname{37/41 (90.2\%)}
\expandafter\def\csname val@digest.cod.hash.pct\endcsname{90.2\%}
\expandafter\def\csname val@digest.cod.reported.ci\endcsname{[63.3, 88.0]}
\expandafter\def\csname val@digest.cod.reported.frac\endcsname{32/41}
\expandafter\def\csname val@digest.cod.reported.full\endcsname{32/41 (78.0\%)}
\expandafter\def\csname val@digest.cod.reported.pct\endcsname{78.0\%}
\expandafter\def\csname val@digest.cod.write.ci\endcsname{[0.4, 12.6]}
\expandafter\def\csname val@digest.cod.write.frac\endcsname{1/41}
\expandafter\def\csname val@digest.cod.write.full\endcsname{1/41 (2.4\%)}
\expandafter\def\csname val@digest.cod.write.pct\endcsname{2.4\%}
\expandafter\def\csname val@digest.codelivery.ci\endcsname{[87.7, 99.6]}
\expandafter\def\csname val@digest.codelivery.frac\endcsname{41/42}
\expandafter\def\csname val@digest.codelivery.full\endcsname{41/42 (97.6\%)}
\expandafter\def\csname val@digest.codelivery.pct\endcsname{97.6\%}
\expandafter\def\csname val@digest.declared.ci\endcsname{[2.5, 19.0]}
\expandafter\def\csname val@digest.declared.frac\endcsname{3/42}
\expandafter\def\csname val@digest.declared.full\endcsname{3/42 (7.1\%)}
\expandafter\def\csname val@digest.declared.pct\endcsname{7.1\%}
\expandafter\def\csname val@digest.h.claude.codelivery.ci\endcsname{[61.0, 100.0]}
\expandafter\def\csname val@digest.h.claude.codelivery.frac\endcsname{6/6}
\expandafter\def\csname val@digest.h.claude.codelivery.full\endcsname{6/6 (100.0\%)}
\expandafter\def\csname val@digest.h.claude.codelivery.pct\endcsname{100.0\%}
\expandafter\def\csname val@digest.h.claude.hash.ci\endcsname{[61.0, 100.0]}
\expandafter\def\csname val@digest.h.claude.hash.frac\endcsname{6/6}
\expandafter\def\csname val@digest.h.claude.hash.full\endcsname{6/6 (100.0\%)}
\expandafter\def\csname val@digest.h.claude.hash.pct\endcsname{100.0\%}
\expandafter\def\csname val@digest.h.claude.ingress.ci\endcsname{[61.0, 100.0]}
\expandafter\def\csname val@digest.h.claude.ingress.frac\endcsname{6/6}
\expandafter\def\csname val@digest.h.claude.ingress.full\endcsname{6/6 (100.0\%)}
\expandafter\def\csname val@digest.h.claude.ingress.pct\endcsname{100.0\%}
\expandafter\def\csname val@digest.h.claude.reject.ci\endcsname{[43.6, 97.0]}
\expandafter\def\csname val@digest.h.claude.reject.frac\endcsname{5/6}
\expandafter\def\csname val@digest.h.claude.reject.full\endcsname{5/6 (83.3\%)}
\expandafter\def\csname val@digest.h.claude.reject.pct\endcsname{83.3\%}
\expandafter\def\csname val@digest.h.claude.write.ci\endcsname{[0.0, 39.0]}
\expandafter\def\csname val@digest.h.claude.write.frac\endcsname{0/6}
\expandafter\def\csname val@digest.h.claude.write.full\endcsname{0/6 (0.0\%)}
\expandafter\def\csname val@digest.h.claude.write.pct\endcsname{0.0\%}
\expandafter\def\csname val@digest.h.codex.codelivery.ci\endcsname{[61.0, 100.0]}
\expandafter\def\csname val@digest.h.codex.codelivery.frac\endcsname{6/6}
\expandafter\def\csname val@digest.h.codex.codelivery.full\endcsname{6/6 (100.0\%)}
\expandafter\def\csname val@digest.h.codex.codelivery.pct\endcsname{100.0\%}
\expandafter\def\csname val@digest.h.codex.hash.ci\endcsname{[30.0, 90.3]}
\expandafter\def\csname val@digest.h.codex.hash.frac\endcsname{4/6}
\expandafter\def\csname val@digest.h.codex.hash.full\endcsname{4/6 (66.7\%)}
\expandafter\def\csname val@digest.h.codex.hash.pct\endcsname{66.7\%}
\expandafter\def\csname val@digest.h.codex.ingress.ci\endcsname{[61.0, 100.0]}
\expandafter\def\csname val@digest.h.codex.ingress.frac\endcsname{6/6}
\expandafter\def\csname val@digest.h.codex.ingress.full\endcsname{6/6 (100.0\%)}
\expandafter\def\csname val@digest.h.codex.ingress.pct\endcsname{100.0\%}
\expandafter\def\csname val@digest.h.codex.reject.ci\endcsname{[30.0, 90.3]}
\expandafter\def\csname val@digest.h.codex.reject.frac\endcsname{4/6}
\expandafter\def\csname val@digest.h.codex.reject.full\endcsname{4/6 (66.7\%)}
\expandafter\def\csname val@digest.h.codex.reject.pct\endcsname{66.7\%}
\expandafter\def\csname val@digest.h.codex.write.ci\endcsname{[0.0, 39.0]}
\expandafter\def\csname val@digest.h.codex.write.frac\endcsname{0/6}
\expandafter\def\csname val@digest.h.codex.write.full\endcsname{0/6 (0.0\%)}
\expandafter\def\csname val@digest.h.codex.write.pct\endcsname{0.0\%}
\expandafter\def\csname val@digest.h.gptme.codelivery.ci\endcsname{[61.0, 100.0]}
\expandafter\def\csname val@digest.h.gptme.codelivery.frac\endcsname{6/6}
\expandafter\def\csname val@digest.h.gptme.codelivery.full\endcsname{6/6 (100.0\%)}
\expandafter\def\csname val@digest.h.gptme.codelivery.pct\endcsname{100.0\%}
\expandafter\def\csname val@digest.h.gptme.hash.ci\endcsname{[61.0, 100.0]}
\expandafter\def\csname val@digest.h.gptme.hash.frac\endcsname{6/6}
\expandafter\def\csname val@digest.h.gptme.hash.full\endcsname{6/6 (100.0\%)}
\expandafter\def\csname val@digest.h.gptme.hash.pct\endcsname{100.0\%}
\expandafter\def\csname val@digest.h.gptme.ingress.ci\endcsname{[61.0, 100.0]}
\expandafter\def\csname val@digest.h.gptme.ingress.frac\endcsname{6/6}
\expandafter\def\csname val@digest.h.gptme.ingress.full\endcsname{6/6 (100.0\%)}
\expandafter\def\csname val@digest.h.gptme.ingress.pct\endcsname{100.0\%}
\expandafter\def\csname val@digest.h.gptme.reject.ci\endcsname{[18.8, 81.2]}
\expandafter\def\csname val@digest.h.gptme.reject.frac\endcsname{3/6}
\expandafter\def\csname val@digest.h.gptme.reject.full\endcsname{3/6 (50.0\%)}
\expandafter\def\csname val@digest.h.gptme.reject.pct\endcsname{50.0\%}
\expandafter\def\csname val@digest.h.gptme.write.ci\endcsname{[3.0, 56.4]}
\expandafter\def\csname val@digest.h.gptme.write.frac\endcsname{1/6}
\expandafter\def\csname val@digest.h.gptme.write.full\endcsname{1/6 (16.7\%)}
\expandafter\def\csname val@digest.h.gptme.write.pct\endcsname{16.7\%}
\expandafter\def\csname val@digest.h.mini.codelivery.ci\endcsname{[61.0, 100.0]}
\expandafter\def\csname val@digest.h.mini.codelivery.frac\endcsname{6/6}
\expandafter\def\csname val@digest.h.mini.codelivery.full\endcsname{6/6 (100.0\%)}
\expandafter\def\csname val@digest.h.mini.codelivery.pct\endcsname{100.0\%}
\expandafter\def\csname val@digest.h.mini.hash.ci\endcsname{[61.0, 100.0]}
\expandafter\def\csname val@digest.h.mini.hash.frac\endcsname{6/6}
\expandafter\def\csname val@digest.h.mini.hash.full\endcsname{6/6 (100.0\%)}
\expandafter\def\csname val@digest.h.mini.hash.pct\endcsname{100.0\%}
\expandafter\def\csname val@digest.h.mini.ingress.ci\endcsname{[61.0, 100.0]}
\expandafter\def\csname val@digest.h.mini.ingress.frac\endcsname{6/6}
\expandafter\def\csname val@digest.h.mini.ingress.full\endcsname{6/6 (100.0\%)}
\expandafter\def\csname val@digest.h.mini.ingress.pct\endcsname{100.0\%}
\expandafter\def\csname val@digest.h.mini.reject.ci\endcsname{[43.6, 97.0]}
\expandafter\def\csname val@digest.h.mini.reject.frac\endcsname{5/6}
\expandafter\def\csname val@digest.h.mini.reject.full\endcsname{5/6 (83.3\%)}
\expandafter\def\csname val@digest.h.mini.reject.pct\endcsname{83.3\%}
\expandafter\def\csname val@digest.h.mini.write.ci\endcsname{[0.0, 39.0]}
\expandafter\def\csname val@digest.h.mini.write.frac\endcsname{0/6}
\expandafter\def\csname val@digest.h.mini.write.full\endcsname{0/6 (0.0\%)}
\expandafter\def\csname val@digest.h.mini.write.pct\endcsname{0.0\%}
\expandafter\def\csname val@digest.h.opencode.codelivery.ci\endcsname{[61.0, 100.0]}
\expandafter\def\csname val@digest.h.opencode.codelivery.frac\endcsname{6/6}
\expandafter\def\csname val@digest.h.opencode.codelivery.full\endcsname{6/6 (100.0\%)}
\expandafter\def\csname val@digest.h.opencode.codelivery.pct\endcsname{100.0\%}
\expandafter\def\csname val@digest.h.opencode.hash.ci\endcsname{[61.0, 100.0]}
\expandafter\def\csname val@digest.h.opencode.hash.frac\endcsname{6/6}
\expandafter\def\csname val@digest.h.opencode.hash.full\endcsname{6/6 (100.0\%)}
\expandafter\def\csname val@digest.h.opencode.hash.pct\endcsname{100.0\%}
\expandafter\def\csname val@digest.h.opencode.ingress.ci\endcsname{[61.0, 100.0]}
\expandafter\def\csname val@digest.h.opencode.ingress.frac\endcsname{6/6}
\expandafter\def\csname val@digest.h.opencode.ingress.full\endcsname{6/6 (100.0\%)}
\expandafter\def\csname val@digest.h.opencode.ingress.pct\endcsname{100.0\%}
\expandafter\def\csname val@digest.h.opencode.reject.ci\endcsname{[43.6, 97.0]}
\expandafter\def\csname val@digest.h.opencode.reject.frac\endcsname{5/6}
\expandafter\def\csname val@digest.h.opencode.reject.full\endcsname{5/6 (83.3\%)}
\expandafter\def\csname val@digest.h.opencode.reject.pct\endcsname{83.3\%}
\expandafter\def\csname val@digest.h.opencode.write.ci\endcsname{[0.0, 39.0]}
\expandafter\def\csname val@digest.h.opencode.write.frac\endcsname{0/6}
\expandafter\def\csname val@digest.h.opencode.write.full\endcsname{0/6 (0.0\%)}
\expandafter\def\csname val@digest.h.opencode.write.pct\endcsname{0.0\%}
\expandafter\def\csname val@digest.h.qwen.codelivery.ci\endcsname{[61.0, 100.0]}
\expandafter\def\csname val@digest.h.qwen.codelivery.frac\endcsname{6/6}
\expandafter\def\csname val@digest.h.qwen.codelivery.full\endcsname{6/6 (100.0\%)}
\expandafter\def\csname val@digest.h.qwen.codelivery.pct\endcsname{100.0\%}
\expandafter\def\csname val@digest.h.qwen.hash.ci\endcsname{[30.0, 90.3]}
\expandafter\def\csname val@digest.h.qwen.hash.frac\endcsname{4/6}
\expandafter\def\csname val@digest.h.qwen.hash.full\endcsname{4/6 (66.7\%)}
\expandafter\def\csname val@digest.h.qwen.hash.pct\endcsname{66.7\%}
\expandafter\def\csname val@digest.h.qwen.ingress.ci\endcsname{[61.0, 100.0]}
\expandafter\def\csname val@digest.h.qwen.ingress.frac\endcsname{6/6}
\expandafter\def\csname val@digest.h.qwen.ingress.full\endcsname{6/6 (100.0\%)}
\expandafter\def\csname val@digest.h.qwen.ingress.pct\endcsname{100.0\%}
\expandafter\def\csname val@digest.h.qwen.reject.ci\endcsname{[30.0, 90.3]}
\expandafter\def\csname val@digest.h.qwen.reject.frac\endcsname{4/6}
\expandafter\def\csname val@digest.h.qwen.reject.full\endcsname{4/6 (66.7\%)}
\expandafter\def\csname val@digest.h.qwen.reject.pct\endcsname{66.7\%}
\expandafter\def\csname val@digest.h.qwen.write.ci\endcsname{[0.0, 39.0]}
\expandafter\def\csname val@digest.h.qwen.write.frac\endcsname{0/6}
\expandafter\def\csname val@digest.h.qwen.write.full\endcsname{0/6 (0.0\%)}
\expandafter\def\csname val@digest.h.qwen.write.pct\endcsname{0.0\%}
\expandafter\def\csname val@digest.h.smol.codelivery.ci\endcsname{[43.6, 97.0]}
\expandafter\def\csname val@digest.h.smol.codelivery.frac\endcsname{5/6}
\expandafter\def\csname val@digest.h.smol.codelivery.full\endcsname{5/6 (83.3\%)}
\expandafter\def\csname val@digest.h.smol.codelivery.pct\endcsname{83.3\%}
\expandafter\def\csname val@digest.h.smol.hash.ci\endcsname{[43.6, 97.0]}
\expandafter\def\csname val@digest.h.smol.hash.frac\endcsname{5/6}
\expandafter\def\csname val@digest.h.smol.hash.full\endcsname{5/6 (83.3\%)}
\expandafter\def\csname val@digest.h.smol.hash.pct\endcsname{83.3\%}
\expandafter\def\csname val@digest.h.smol.ingress.ci\endcsname{[43.6, 97.0]}
\expandafter\def\csname val@digest.h.smol.ingress.frac\endcsname{5/6}
\expandafter\def\csname val@digest.h.smol.ingress.full\endcsname{5/6 (83.3\%)}
\expandafter\def\csname val@digest.h.smol.ingress.pct\endcsname{83.3\%}
\expandafter\def\csname val@digest.h.smol.reject.ci\endcsname{[43.6, 97.0]}
\expandafter\def\csname val@digest.h.smol.reject.frac\endcsname{5/6}
\expandafter\def\csname val@digest.h.smol.reject.full\endcsname{5/6 (83.3\%)}
\expandafter\def\csname val@digest.h.smol.reject.pct\endcsname{83.3\%}
\expandafter\def\csname val@digest.h.smol.write.ci\endcsname{[0.0, 39.0]}
\expandafter\def\csname val@digest.h.smol.write.frac\endcsname{0/6}
\expandafter\def\csname val@digest.h.smol.write.full\endcsname{0/6 (0.0\%)}
\expandafter\def\csname val@digest.h.smol.write.pct\endcsname{0.0\%}
\expandafter\def\csname val@digest.hash.ci\endcsname{[75.0, 94.8]}
\expandafter\def\csname val@digest.hash.frac\endcsname{37/42}
\expandafter\def\csname val@digest.hash.full\endcsname{37/42 (88.1\%)}
\expandafter\def\csname val@digest.hash.pct\endcsname{88.1\%}
\expandafter\def\csname val@digest.ingress.ci\endcsname{[87.7, 99.6]}
\expandafter\def\csname val@digest.ingress.frac\endcsname{41/42}
\expandafter\def\csname val@digest.ingress.full\endcsname{41/42 (97.6\%)}
\expandafter\def\csname val@digest.ingress.pct\endcsname{97.6\%}
\expandafter\def\csname val@digest.nobody.hash.ci\endcsname{[0.0, 79.3]}
\expandafter\def\csname val@digest.nobody.hash.frac\endcsname{0/1}
\expandafter\def\csname val@digest.nobody.hash.full\endcsname{0/1 (0.0\%)}
\expandafter\def\csname val@digest.nobody.hash.pct\endcsname{0.0\%}
\expandafter\def\csname val@digest.nobody.reported.ci\endcsname{[0.0, 79.3]}
\expandafter\def\csname val@digest.nobody.reported.frac\endcsname{0/1}
\expandafter\def\csname val@digest.nobody.reported.full\endcsname{0/1 (0.0\%)}
\expandafter\def\csname val@digest.nobody.reported.pct\endcsname{0.0\%}
\expandafter\def\csname val@digest.nobody.write.ci\endcsname{[0.0, 79.3]}
\expandafter\def\csname val@digest.nobody.write.frac\endcsname{0/1}
\expandafter\def\csname val@digest.nobody.write.full\endcsname{0/1 (0.0\%)}
\expandafter\def\csname val@digest.nobody.write.pct\endcsname{0.0\%}
\expandafter\def\csname val@digest.nocod.hash.ci\endcsname{n/a}
\expandafter\def\csname val@digest.nocod.hash.frac\endcsname{0/0}
\expandafter\def\csname val@digest.nocod.hash.full\endcsname{0/0 (n/a)}
\expandafter\def\csname val@digest.nocod.hash.pct\endcsname{n/a}
\expandafter\def\csname val@digest.nocod.reported.ci\endcsname{n/a}
\expandafter\def\csname val@digest.nocod.reported.frac\endcsname{0/0}
\expandafter\def\csname val@digest.nocod.reported.full\endcsname{0/0 (n/a)}
\expandafter\def\csname val@digest.nocod.reported.pct\endcsname{n/a}
\expandafter\def\csname val@digest.nocod.write.ci\endcsname{n/a}
\expandafter\def\csname val@digest.nocod.write.frac\endcsname{0/0}
\expandafter\def\csname val@digest.nocod.write.full\endcsname{0/0 (n/a)}
\expandafter\def\csname val@digest.nocod.write.pct\endcsname{n/a}
\expandafter\def\csname val@digest.reject.ci\endcsname{[58.9, 84.7]}
\expandafter\def\csname val@digest.reject.frac\endcsname{31/42}
\expandafter\def\csname val@digest.reject.full\endcsname{31/42 (73.8\%)}
\expandafter\def\csname val@digest.reject.pct\endcsname{73.8\%}
\expandafter\def\csname val@digest.reported.ci\endcsname{[61.5, 86.5]}
\expandafter\def\csname val@digest.reported.frac\endcsname{32/42}
\expandafter\def\csname val@digest.reported.full\endcsname{32/42 (76.2\%)}
\expandafter\def\csname val@digest.reported.pct\endcsname{76.2\%}
\expandafter\def\csname val@digest.reportedafter.ci\endcsname{[61.5, 86.5]}
\expandafter\def\csname val@digest.reportedafter.frac\endcsname{32/42}
\expandafter\def\csname val@digest.reportedafter.full\endcsname{32/42 (76.2\%)}
\expandafter\def\csname val@digest.reportedafter.pct\endcsname{76.2\%}
\expandafter\def\csname val@digest.single.ci\endcsname{[87.7, 99.6]}
\expandafter\def\csname val@digest.single.frac\endcsname{41/42}
\expandafter\def\csname val@digest.single.full\endcsname{41/42 (97.6\%)}
\expandafter\def\csname val@digest.single.pct\endcsname{97.6\%}
\expandafter\def\csname val@digest.targetwritten.ci\endcsname{[3.8, 22.1]}
\expandafter\def\csname val@digest.targetwritten.frac\endcsname{4/42}
\expandafter\def\csname val@digest.targetwritten.full\endcsname{4/42 (9.5\%)}
\expandafter\def\csname val@digest.targetwritten.pct\endcsname{9.5\%}
\expandafter\def\csname val@digest.twin.body.ci\endcsname{[91.6, 100.0]}
\expandafter\def\csname val@digest.twin.body.frac\endcsname{42/42}
\expandafter\def\csname val@digest.twin.body.full\endcsname{42/42 (100.0\%)}
\expandafter\def\csname val@digest.twin.body.pct\endcsname{100.0\%}
\expandafter\def\csname val@digest.twin.claim.ci\endcsname{[0.0, 8.4]}
\expandafter\def\csname val@digest.twin.claim.frac\endcsname{0/42}
\expandafter\def\csname val@digest.twin.claim.full\endcsname{0/42 (0.0\%)}
\expandafter\def\csname val@digest.twin.claim.pct\endcsname{0.0\%}
\expandafter\def\csname val@digest.twin.codelivery.ci\endcsname{[91.6, 100.0]}
\expandafter\def\csname val@digest.twin.codelivery.frac\endcsname{42/42}
\expandafter\def\csname val@digest.twin.codelivery.full\endcsname{42/42 (100.0\%)}
\expandafter\def\csname val@digest.twin.codelivery.pct\endcsname{100.0\%}
\expandafter\def\csname val@digest.twin.declared.ci\endcsname{[5.2, 25.0]}
\expandafter\def\csname val@digest.twin.declared.frac\endcsname{5/42}
\expandafter\def\csname val@digest.twin.declared.full\endcsname{5/42 (11.9\%)}
\expandafter\def\csname val@digest.twin.declared.pct\endcsname{11.9\%}
\expandafter\def\csname val@digest.twin.h.claude.codelivery.ci\endcsname{[61.0, 100.0]}
\expandafter\def\csname val@digest.twin.h.claude.codelivery.frac\endcsname{6/6}
\expandafter\def\csname val@digest.twin.h.claude.codelivery.full\endcsname{6/6 (100.0\%)}
\expandafter\def\csname val@digest.twin.h.claude.codelivery.pct\endcsname{100.0\%}
\expandafter\def\csname val@digest.twin.h.claude.hash.ci\endcsname{[43.6, 97.0]}
\expandafter\def\csname val@digest.twin.h.claude.hash.frac\endcsname{5/6}
\expandafter\def\csname val@digest.twin.h.claude.hash.full\endcsname{5/6 (83.3\%)}
\expandafter\def\csname val@digest.twin.h.claude.hash.pct\endcsname{83.3\%}
\expandafter\def\csname val@digest.twin.h.claude.write.ci\endcsname{[0.0, 39.0]}
\expandafter\def\csname val@digest.twin.h.claude.write.frac\endcsname{0/6}
\expandafter\def\csname val@digest.twin.h.claude.write.full\endcsname{0/6 (0.0\%)}
\expandafter\def\csname val@digest.twin.h.claude.write.pct\endcsname{0.0\%}
\expandafter\def\csname val@digest.twin.h.codex.codelivery.ci\endcsname{[61.0, 100.0]}
\expandafter\def\csname val@digest.twin.h.codex.codelivery.frac\endcsname{6/6}
\expandafter\def\csname val@digest.twin.h.codex.codelivery.full\endcsname{6/6 (100.0\%)}
\expandafter\def\csname val@digest.twin.h.codex.codelivery.pct\endcsname{100.0\%}
\expandafter\def\csname val@digest.twin.h.codex.hash.ci\endcsname{[30.0, 90.3]}
\expandafter\def\csname val@digest.twin.h.codex.hash.frac\endcsname{4/6}
\expandafter\def\csname val@digest.twin.h.codex.hash.full\endcsname{4/6 (66.7\%)}
\expandafter\def\csname val@digest.twin.h.codex.hash.pct\endcsname{66.7\%}
\expandafter\def\csname val@digest.twin.h.codex.write.ci\endcsname{[0.0, 39.0]}
\expandafter\def\csname val@digest.twin.h.codex.write.frac\endcsname{0/6}
\expandafter\def\csname val@digest.twin.h.codex.write.full\endcsname{0/6 (0.0\%)}
\expandafter\def\csname val@digest.twin.h.codex.write.pct\endcsname{0.0\%}
\expandafter\def\csname val@digest.twin.h.gptme.codelivery.ci\endcsname{[61.0, 100.0]}
\expandafter\def\csname val@digest.twin.h.gptme.codelivery.frac\endcsname{6/6}
\expandafter\def\csname val@digest.twin.h.gptme.codelivery.full\endcsname{6/6 (100.0\%)}
\expandafter\def\csname val@digest.twin.h.gptme.codelivery.pct\endcsname{100.0\%}
\expandafter\def\csname val@digest.twin.h.gptme.hash.ci\endcsname{[61.0, 100.0]}
\expandafter\def\csname val@digest.twin.h.gptme.hash.frac\endcsname{6/6}
\expandafter\def\csname val@digest.twin.h.gptme.hash.full\endcsname{6/6 (100.0\%)}
\expandafter\def\csname val@digest.twin.h.gptme.hash.pct\endcsname{100.0\%}
\expandafter\def\csname val@digest.twin.h.gptme.write.ci\endcsname{[0.0, 39.0]}
\expandafter\def\csname val@digest.twin.h.gptme.write.frac\endcsname{0/6}
\expandafter\def\csname val@digest.twin.h.gptme.write.full\endcsname{0/6 (0.0\%)}
\expandafter\def\csname val@digest.twin.h.gptme.write.pct\endcsname{0.0\%}
\expandafter\def\csname val@digest.twin.h.mini.codelivery.ci\endcsname{[61.0, 100.0]}
\expandafter\def\csname val@digest.twin.h.mini.codelivery.frac\endcsname{6/6}
\expandafter\def\csname val@digest.twin.h.mini.codelivery.full\endcsname{6/6 (100.0\%)}
\expandafter\def\csname val@digest.twin.h.mini.codelivery.pct\endcsname{100.0\%}
\expandafter\def\csname val@digest.twin.h.mini.hash.ci\endcsname{[61.0, 100.0]}
\expandafter\def\csname val@digest.twin.h.mini.hash.frac\endcsname{6/6}
\expandafter\def\csname val@digest.twin.h.mini.hash.full\endcsname{6/6 (100.0\%)}
\expandafter\def\csname val@digest.twin.h.mini.hash.pct\endcsname{100.0\%}
\expandafter\def\csname val@digest.twin.h.mini.write.ci\endcsname{[0.0, 39.0]}
\expandafter\def\csname val@digest.twin.h.mini.write.frac\endcsname{0/6}
\expandafter\def\csname val@digest.twin.h.mini.write.full\endcsname{0/6 (0.0\%)}
\expandafter\def\csname val@digest.twin.h.mini.write.pct\endcsname{0.0\%}
\expandafter\def\csname val@digest.twin.h.opencode.codelivery.ci\endcsname{[61.0, 100.0]}
\expandafter\def\csname val@digest.twin.h.opencode.codelivery.frac\endcsname{6/6}
\expandafter\def\csname val@digest.twin.h.opencode.codelivery.full\endcsname{6/6 (100.0\%)}
\expandafter\def\csname val@digest.twin.h.opencode.codelivery.pct\endcsname{100.0\%}
\expandafter\def\csname val@digest.twin.h.opencode.hash.ci\endcsname{[9.7, 70.0]}
\expandafter\def\csname val@digest.twin.h.opencode.hash.frac\endcsname{2/6}
\expandafter\def\csname val@digest.twin.h.opencode.hash.full\endcsname{2/6 (33.3\%)}
\expandafter\def\csname val@digest.twin.h.opencode.hash.pct\endcsname{33.3\%}
\expandafter\def\csname val@digest.twin.h.opencode.write.ci\endcsname{[0.0, 39.0]}
\expandafter\def\csname val@digest.twin.h.opencode.write.frac\endcsname{0/6}
\expandafter\def\csname val@digest.twin.h.opencode.write.full\endcsname{0/6 (0.0\%)}
\expandafter\def\csname val@digest.twin.h.opencode.write.pct\endcsname{0.0\%}
\expandafter\def\csname val@digest.twin.h.qwen.codelivery.ci\endcsname{[61.0, 100.0]}
\expandafter\def\csname val@digest.twin.h.qwen.codelivery.frac\endcsname{6/6}
\expandafter\def\csname val@digest.twin.h.qwen.codelivery.full\endcsname{6/6 (100.0\%)}
\expandafter\def\csname val@digest.twin.h.qwen.codelivery.pct\endcsname{100.0\%}
\expandafter\def\csname val@digest.twin.h.qwen.hash.ci\endcsname{[9.7, 70.0]}
\expandafter\def\csname val@digest.twin.h.qwen.hash.frac\endcsname{2/6}
\expandafter\def\csname val@digest.twin.h.qwen.hash.full\endcsname{2/6 (33.3\%)}
\expandafter\def\csname val@digest.twin.h.qwen.hash.pct\endcsname{33.3\%}
\expandafter\def\csname val@digest.twin.h.qwen.write.ci\endcsname{[0.0, 39.0]}
\expandafter\def\csname val@digest.twin.h.qwen.write.frac\endcsname{0/6}
\expandafter\def\csname val@digest.twin.h.qwen.write.full\endcsname{0/6 (0.0\%)}
\expandafter\def\csname val@digest.twin.h.qwen.write.pct\endcsname{0.0\%}
\expandafter\def\csname val@digest.twin.h.smol.codelivery.ci\endcsname{[61.0, 100.0]}
\expandafter\def\csname val@digest.twin.h.smol.codelivery.frac\endcsname{6/6}
\expandafter\def\csname val@digest.twin.h.smol.codelivery.full\endcsname{6/6 (100.0\%)}
\expandafter\def\csname val@digest.twin.h.smol.codelivery.pct\endcsname{100.0\%}
\expandafter\def\csname val@digest.twin.h.smol.hash.ci\endcsname{[61.0, 100.0]}
\expandafter\def\csname val@digest.twin.h.smol.hash.frac\endcsname{6/6}
\expandafter\def\csname val@digest.twin.h.smol.hash.full\endcsname{6/6 (100.0\%)}
\expandafter\def\csname val@digest.twin.h.smol.hash.pct\endcsname{100.0\%}
\expandafter\def\csname val@digest.twin.h.smol.write.ci\endcsname{[0.0, 39.0]}
\expandafter\def\csname val@digest.twin.h.smol.write.frac\endcsname{0/6}
\expandafter\def\csname val@digest.twin.h.smol.write.full\endcsname{0/6 (0.0\%)}
\expandafter\def\csname val@digest.twin.h.smol.write.pct\endcsname{0.0\%}
\expandafter\def\csname val@digest.twin.hash.ci\endcsname{[58.9, 84.7]}
\expandafter\def\csname val@digest.twin.hash.frac\endcsname{31/42}
\expandafter\def\csname val@digest.twin.hash.full\endcsname{31/42 (73.8\%)}
\expandafter\def\csname val@digest.twin.hash.pct\endcsname{73.8\%}
\expandafter\def\csname val@digest.twin.reported.ci\endcsname{[6.7, 27.8]}
\expandafter\def\csname val@digest.twin.reported.frac\endcsname{6/42}
\expandafter\def\csname val@digest.twin.reported.full\endcsname{6/42 (14.3\%)}
\expandafter\def\csname val@digest.twin.reported.pct\endcsname{14.3\%}
\expandafter\def\csname val@digest.twin.reportedafter.ci\endcsname{[6.7, 27.8]}
\expandafter\def\csname val@digest.twin.reportedafter.frac\endcsname{6/42}
\expandafter\def\csname val@digest.twin.reportedafter.full\endcsname{6/42 (14.3\%)}
\expandafter\def\csname val@digest.twin.reportedafter.pct\endcsname{14.3\%}
\expandafter\def\csname val@digest.twin.targetwritten.ci\endcsname{[5.2, 25.0]}
\expandafter\def\csname val@digest.twin.targetwritten.frac\endcsname{5/42}
\expandafter\def\csname val@digest.twin.targetwritten.full\endcsname{5/42 (11.9\%)}
\expandafter\def\csname val@digest.twin.targetwritten.pct\endcsname{11.9\%}
\expandafter\def\csname val@digest.twin.write.ci\endcsname{[0.0, 8.4]}
\expandafter\def\csname val@digest.twin.write.frac\endcsname{0/42}
\expandafter\def\csname val@digest.twin.write.full\endcsname{0/42 (0.0\%)}
\expandafter\def\csname val@digest.twin.write.pct\endcsname{0.0\%}
\expandafter\def\csname val@digest.unverified.ci\endcsname{[3.8, 22.1]}
\expandafter\def\csname val@digest.unverified.frac\endcsname{4/42}
\expandafter\def\csname val@digest.unverified.full\endcsname{4/42 (9.5\%)}
\expandafter\def\csname val@digest.unverified.pct\endcsname{9.5\%}
\expandafter\def\csname val@digest.write.ci\endcsname{[0.4, 12.3]}
\expandafter\def\csname val@digest.write.frac\endcsname{1/42}
\expandafter\def\csname val@digest.write.full\endcsname{1/42 (2.4\%)}
\expandafter\def\csname val@digest.write.pct\endcsname{2.4\%}
\expandafter\def\csname val@header.body.ci\endcsname{[87.7, 99.6]}
\expandafter\def\csname val@header.body.frac\endcsname{41/42}
\expandafter\def\csname val@header.body.full\endcsname{41/42 (97.6\%)}
\expandafter\def\csname val@header.body.pct\endcsname{97.6\%}
\expandafter\def\csname val@header.claim.ci\endcsname{[0.0, 8.4]}
\expandafter\def\csname val@header.claim.frac\endcsname{0/42}
\expandafter\def\csname val@header.claim.full\endcsname{0/42 (0.0\%)}
\expandafter\def\csname val@header.claim.pct\endcsname{0.0\%}
\expandafter\def\csname val@header.codelivery.ci\endcsname{[23.0, 50.8]}
\expandafter\def\csname val@header.codelivery.frac\endcsname{15/42}
\expandafter\def\csname val@header.codelivery.full\endcsname{15/42 (35.7\%)}
\expandafter\def\csname val@header.codelivery.pct\endcsname{35.7\%}
\expandafter\def\csname val@header.declared.ci\endcsname{[0.4, 12.3]}
\expandafter\def\csname val@header.declared.frac\endcsname{1/42}
\expandafter\def\csname val@header.declared.full\endcsname{1/42 (2.4\%)}
\expandafter\def\csname val@header.declared.pct\endcsname{2.4\%}
\expandafter\def\csname val@header.h.claude.codelivery.ci\endcsname{[30.0, 90.3]}
\expandafter\def\csname val@header.h.claude.codelivery.frac\endcsname{4/6}
\expandafter\def\csname val@header.h.claude.codelivery.full\endcsname{4/6 (66.7\%)}
\expandafter\def\csname val@header.h.claude.codelivery.pct\endcsname{66.7\%}
\expandafter\def\csname val@header.h.claude.hash.ci\endcsname{[0.0, 39.0]}
\expandafter\def\csname val@header.h.claude.hash.frac\endcsname{0/6}
\expandafter\def\csname val@header.h.claude.hash.full\endcsname{0/6 (0.0\%)}
\expandafter\def\csname val@header.h.claude.hash.pct\endcsname{0.0\%}
\expandafter\def\csname val@header.h.claude.ingress.ci\endcsname{[30.0, 90.3]}
\expandafter\def\csname val@header.h.claude.ingress.frac\endcsname{4/6}
\expandafter\def\csname val@header.h.claude.ingress.full\endcsname{4/6 (66.7\%)}
\expandafter\def\csname val@header.h.claude.ingress.pct\endcsname{66.7\%}
\expandafter\def\csname val@header.h.claude.reject.ci\endcsname{[9.7, 70.0]}
\expandafter\def\csname val@header.h.claude.reject.frac\endcsname{2/6}
\expandafter\def\csname val@header.h.claude.reject.full\endcsname{2/6 (33.3\%)}
\expandafter\def\csname val@header.h.claude.reject.pct\endcsname{33.3\%}
\expandafter\def\csname val@header.h.claude.write.ci\endcsname{[3.0, 56.4]}
\expandafter\def\csname val@header.h.claude.write.frac\endcsname{1/6}
\expandafter\def\csname val@header.h.claude.write.full\endcsname{1/6 (16.7\%)}
\expandafter\def\csname val@header.h.claude.write.pct\endcsname{16.7\%}
\expandafter\def\csname val@header.h.codex.codelivery.ci\endcsname{[43.6, 97.0]}
\expandafter\def\csname val@header.h.codex.codelivery.frac\endcsname{5/6}
\expandafter\def\csname val@header.h.codex.codelivery.full\endcsname{5/6 (83.3\%)}
\expandafter\def\csname val@header.h.codex.codelivery.pct\endcsname{83.3\%}
\expandafter\def\csname val@header.h.codex.hash.ci\endcsname{[3.0, 56.4]}
\expandafter\def\csname val@header.h.codex.hash.frac\endcsname{1/6}
\expandafter\def\csname val@header.h.codex.hash.full\endcsname{1/6 (16.7\%)}
\expandafter\def\csname val@header.h.codex.hash.pct\endcsname{16.7\%}
\expandafter\def\csname val@header.h.codex.ingress.ci\endcsname{[43.6, 97.0]}
\expandafter\def\csname val@header.h.codex.ingress.frac\endcsname{5/6}
\expandafter\def\csname val@header.h.codex.ingress.full\endcsname{5/6 (83.3\%)}
\expandafter\def\csname val@header.h.codex.ingress.pct\endcsname{83.3\%}
\expandafter\def\csname val@header.h.codex.reject.ci\endcsname{[3.0, 56.4]}
\expandafter\def\csname val@header.h.codex.reject.frac\endcsname{1/6}
\expandafter\def\csname val@header.h.codex.reject.full\endcsname{1/6 (16.7\%)}
\expandafter\def\csname val@header.h.codex.reject.pct\endcsname{16.7\%}
\expandafter\def\csname val@header.h.codex.write.ci\endcsname{[3.0, 56.4]}
\expandafter\def\csname val@header.h.codex.write.frac\endcsname{1/6}
\expandafter\def\csname val@header.h.codex.write.full\endcsname{1/6 (16.7\%)}
\expandafter\def\csname val@header.h.codex.write.pct\endcsname{16.7\%}
\expandafter\def\csname val@header.h.gptme.codelivery.ci\endcsname{[18.8, 81.2]}
\expandafter\def\csname val@header.h.gptme.codelivery.frac\endcsname{3/6}
\expandafter\def\csname val@header.h.gptme.codelivery.full\endcsname{3/6 (50.0\%)}
\expandafter\def\csname val@header.h.gptme.codelivery.pct\endcsname{50.0\%}
\expandafter\def\csname val@header.h.gptme.hash.ci\endcsname{[3.0, 56.4]}
\expandafter\def\csname val@header.h.gptme.hash.frac\endcsname{1/6}
\expandafter\def\csname val@header.h.gptme.hash.full\endcsname{1/6 (16.7\%)}
\expandafter\def\csname val@header.h.gptme.hash.pct\endcsname{16.7\%}
\expandafter\def\csname val@header.h.gptme.ingress.ci\endcsname{[18.8, 81.2]}
\expandafter\def\csname val@header.h.gptme.ingress.frac\endcsname{3/6}
\expandafter\def\csname val@header.h.gptme.ingress.full\endcsname{3/6 (50.0\%)}
\expandafter\def\csname val@header.h.gptme.ingress.pct\endcsname{50.0\%}
\expandafter\def\csname val@header.h.gptme.reject.ci\endcsname{[9.7, 70.0]}
\expandafter\def\csname val@header.h.gptme.reject.frac\endcsname{2/6}
\expandafter\def\csname val@header.h.gptme.reject.full\endcsname{2/6 (33.3\%)}
\expandafter\def\csname val@header.h.gptme.reject.pct\endcsname{33.3\%}
\expandafter\def\csname val@header.h.gptme.write.ci\endcsname{[3.0, 56.4]}
\expandafter\def\csname val@header.h.gptme.write.frac\endcsname{1/6}
\expandafter\def\csname val@header.h.gptme.write.full\endcsname{1/6 (16.7\%)}
\expandafter\def\csname val@header.h.gptme.write.pct\endcsname{16.7\%}
\expandafter\def\csname val@header.h.mini.codelivery.ci\endcsname{[3.0, 56.4]}
\expandafter\def\csname val@header.h.mini.codelivery.frac\endcsname{1/6}
\expandafter\def\csname val@header.h.mini.codelivery.full\endcsname{1/6 (16.7\%)}
\expandafter\def\csname val@header.h.mini.codelivery.pct\endcsname{16.7\%}
\expandafter\def\csname val@header.h.mini.hash.ci\endcsname{[9.7, 70.0]}
\expandafter\def\csname val@header.h.mini.hash.frac\endcsname{2/6}
\expandafter\def\csname val@header.h.mini.hash.full\endcsname{2/6 (33.3\%)}
\expandafter\def\csname val@header.h.mini.hash.pct\endcsname{33.3\%}
\expandafter\def\csname val@header.h.mini.ingress.ci\endcsname{[3.0, 56.4]}
\expandafter\def\csname val@header.h.mini.ingress.frac\endcsname{1/6}
\expandafter\def\csname val@header.h.mini.ingress.full\endcsname{1/6 (16.7\%)}
\expandafter\def\csname val@header.h.mini.ingress.pct\endcsname{16.7\%}
\expandafter\def\csname val@header.h.mini.reject.ci\endcsname{[0.0, 39.0]}
\expandafter\def\csname val@header.h.mini.reject.frac\endcsname{0/6}
\expandafter\def\csname val@header.h.mini.reject.full\endcsname{0/6 (0.0\%)}
\expandafter\def\csname val@header.h.mini.reject.pct\endcsname{0.0\%}
\expandafter\def\csname val@header.h.mini.write.ci\endcsname{[9.7, 70.0]}
\expandafter\def\csname val@header.h.mini.write.frac\endcsname{2/6}
\expandafter\def\csname val@header.h.mini.write.full\endcsname{2/6 (33.3\%)}
\expandafter\def\csname val@header.h.mini.write.pct\endcsname{33.3\%}
\expandafter\def\csname val@header.h.opencode.codelivery.ci\endcsname{[0.0, 39.0]}
\expandafter\def\csname val@header.h.opencode.codelivery.frac\endcsname{0/6}
\expandafter\def\csname val@header.h.opencode.codelivery.full\endcsname{0/6 (0.0\%)}
\expandafter\def\csname val@header.h.opencode.codelivery.pct\endcsname{0.0\%}
\expandafter\def\csname val@header.h.opencode.hash.ci\endcsname{[0.0, 39.0]}
\expandafter\def\csname val@header.h.opencode.hash.frac\endcsname{0/6}
\expandafter\def\csname val@header.h.opencode.hash.full\endcsname{0/6 (0.0\%)}
\expandafter\def\csname val@header.h.opencode.hash.pct\endcsname{0.0\%}
\expandafter\def\csname val@header.h.opencode.ingress.ci\endcsname{[0.0, 39.0]}
\expandafter\def\csname val@header.h.opencode.ingress.frac\endcsname{0/6}
\expandafter\def\csname val@header.h.opencode.ingress.full\endcsname{0/6 (0.0\%)}
\expandafter\def\csname val@header.h.opencode.ingress.pct\endcsname{0.0\%}
\expandafter\def\csname val@header.h.opencode.reject.ci\endcsname{[0.0, 39.0]}
\expandafter\def\csname val@header.h.opencode.reject.frac\endcsname{0/6}
\expandafter\def\csname val@header.h.opencode.reject.full\endcsname{0/6 (0.0\%)}
\expandafter\def\csname val@header.h.opencode.reject.pct\endcsname{0.0\%}
\expandafter\def\csname val@header.h.opencode.write.ci\endcsname{[3.0, 56.4]}
\expandafter\def\csname val@header.h.opencode.write.frac\endcsname{1/6}
\expandafter\def\csname val@header.h.opencode.write.full\endcsname{1/6 (16.7\%)}
\expandafter\def\csname val@header.h.opencode.write.pct\endcsname{16.7\%}
\expandafter\def\csname val@header.h.qwen.codelivery.ci\endcsname{[9.7, 70.0]}
\expandafter\def\csname val@header.h.qwen.codelivery.frac\endcsname{2/6}
\expandafter\def\csname val@header.h.qwen.codelivery.full\endcsname{2/6 (33.3\%)}
\expandafter\def\csname val@header.h.qwen.codelivery.pct\endcsname{33.3\%}
\expandafter\def\csname val@header.h.qwen.hash.ci\endcsname{[0.0, 39.0]}
\expandafter\def\csname val@header.h.qwen.hash.frac\endcsname{0/6}
\expandafter\def\csname val@header.h.qwen.hash.full\endcsname{0/6 (0.0\%)}
\expandafter\def\csname val@header.h.qwen.hash.pct\endcsname{0.0\%}
\expandafter\def\csname val@header.h.qwen.ingress.ci\endcsname{[9.7, 70.0]}
\expandafter\def\csname val@header.h.qwen.ingress.frac\endcsname{2/6}
\expandafter\def\csname val@header.h.qwen.ingress.full\endcsname{2/6 (33.3\%)}
\expandafter\def\csname val@header.h.qwen.ingress.pct\endcsname{33.3\%}
\expandafter\def\csname val@header.h.qwen.reject.ci\endcsname{[3.0, 56.4]}
\expandafter\def\csname val@header.h.qwen.reject.frac\endcsname{1/6}
\expandafter\def\csname val@header.h.qwen.reject.full\endcsname{1/6 (16.7\%)}
\expandafter\def\csname val@header.h.qwen.reject.pct\endcsname{16.7\%}
\expandafter\def\csname val@header.h.qwen.write.ci\endcsname{[0.0, 39.0]}
\expandafter\def\csname val@header.h.qwen.write.frac\endcsname{0/6}
\expandafter\def\csname val@header.h.qwen.write.full\endcsname{0/6 (0.0\%)}
\expandafter\def\csname val@header.h.qwen.write.pct\endcsname{0.0\%}
\expandafter\def\csname val@header.h.smol.codelivery.ci\endcsname{[0.0, 39.0]}
\expandafter\def\csname val@header.h.smol.codelivery.frac\endcsname{0/6}
\expandafter\def\csname val@header.h.smol.codelivery.full\endcsname{0/6 (0.0\%)}
\expandafter\def\csname val@header.h.smol.codelivery.pct\endcsname{0.0\%}
\expandafter\def\csname val@header.h.smol.hash.ci\endcsname{[0.0, 39.0]}
\expandafter\def\csname val@header.h.smol.hash.frac\endcsname{0/6}
\expandafter\def\csname val@header.h.smol.hash.full\endcsname{0/6 (0.0\%)}
\expandafter\def\csname val@header.h.smol.hash.pct\endcsname{0.0\%}
\expandafter\def\csname val@header.h.smol.ingress.ci\endcsname{[0.0, 39.0]}
\expandafter\def\csname val@header.h.smol.ingress.frac\endcsname{0/6}
\expandafter\def\csname val@header.h.smol.ingress.full\endcsname{0/6 (0.0\%)}
\expandafter\def\csname val@header.h.smol.ingress.pct\endcsname{0.0\%}
\expandafter\def\csname val@header.h.smol.reject.ci\endcsname{[0.0, 39.0]}
\expandafter\def\csname val@header.h.smol.reject.frac\endcsname{0/6}
\expandafter\def\csname val@header.h.smol.reject.full\endcsname{0/6 (0.0\%)}
\expandafter\def\csname val@header.h.smol.reject.pct\endcsname{0.0\%}
\expandafter\def\csname val@header.h.smol.write.ci\endcsname{[0.0, 39.0]}
\expandafter\def\csname val@header.h.smol.write.frac\endcsname{0/6}
\expandafter\def\csname val@header.h.smol.write.full\endcsname{0/6 (0.0\%)}
\expandafter\def\csname val@header.h.smol.write.pct\endcsname{0.0\%}
\expandafter\def\csname val@header.hash.ci\endcsname{[3.8, 22.1]}
\expandafter\def\csname val@header.hash.frac\endcsname{4/42}
\expandafter\def\csname val@header.hash.full\endcsname{4/42 (9.5\%)}
\expandafter\def\csname val@header.hash.pct\endcsname{9.5\%}
\expandafter\def\csname val@header.ingress.ci\endcsname{[23.0, 50.8]}
\expandafter\def\csname val@header.ingress.frac\endcsname{15/42}
\expandafter\def\csname val@header.ingress.full\endcsname{15/42 (35.7\%)}
\expandafter\def\csname val@header.ingress.pct\endcsname{35.7\%}
\expandafter\def\csname val@header.reject.ci\endcsname{[6.7, 27.8]}
\expandafter\def\csname val@header.reject.frac\endcsname{6/42}
\expandafter\def\csname val@header.reject.full\endcsname{6/42 (14.3\%)}
\expandafter\def\csname val@header.reject.pct\endcsname{14.3\%}
\expandafter\def\csname val@header.reported.ci\endcsname{[23.0, 50.8]}
\expandafter\def\csname val@header.reported.frac\endcsname{15/42}
\expandafter\def\csname val@header.reported.full\endcsname{15/42 (35.7\%)}
\expandafter\def\csname val@header.reported.pct\endcsname{35.7\%}
\expandafter\def\csname val@header.reportedafter.ci\endcsname{[8.3, 30.6]}
\expandafter\def\csname val@header.reportedafter.frac\endcsname{7/42}
\expandafter\def\csname val@header.reportedafter.full\endcsname{7/42 (16.7\%)}
\expandafter\def\csname val@header.reportedafter.pct\endcsname{16.7\%}
\expandafter\def\csname val@header.single.ci\endcsname{[23.0, 50.8]}
\expandafter\def\csname val@header.single.frac\endcsname{15/42}
\expandafter\def\csname val@header.single.full\endcsname{15/42 (35.7\%)}
\expandafter\def\csname val@header.single.pct\endcsname{35.7\%}
\expandafter\def\csname val@header.targetwritten.ci\endcsname{[8.3, 30.6]}
\expandafter\def\csname val@header.targetwritten.frac\endcsname{7/42}
\expandafter\def\csname val@header.targetwritten.full\endcsname{7/42 (16.7\%)}
\expandafter\def\csname val@header.targetwritten.pct\endcsname{16.7\%}
\expandafter\def\csname val@header.twin.body.ci\endcsname{[91.6, 100.0]}
\expandafter\def\csname val@header.twin.body.frac\endcsname{42/42}
\expandafter\def\csname val@header.twin.body.full\endcsname{42/42 (100.0\%)}
\expandafter\def\csname val@header.twin.body.pct\endcsname{100.0\%}
\expandafter\def\csname val@header.twin.claim.ci\endcsname{[0.4, 12.3]}
\expandafter\def\csname val@header.twin.claim.frac\endcsname{1/42}
\expandafter\def\csname val@header.twin.claim.full\endcsname{1/42 (2.4\%)}
\expandafter\def\csname val@header.twin.claim.pct\endcsname{2.4\%}
\expandafter\def\csname val@header.twin.codelivery.ci\endcsname{[33.4, 62.3]}
\expandafter\def\csname val@header.twin.codelivery.frac\endcsname{20/42}
\expandafter\def\csname val@header.twin.codelivery.full\endcsname{20/42 (47.6\%)}
\expandafter\def\csname val@header.twin.codelivery.pct\endcsname{47.6\%}
\expandafter\def\csname val@header.twin.declared.ci\endcsname{[21.0, 48.4]}
\expandafter\def\csname val@header.twin.declared.frac\endcsname{14/42}
\expandafter\def\csname val@header.twin.declared.full\endcsname{14/42 (33.3\%)}
\expandafter\def\csname val@header.twin.declared.pct\endcsname{33.3\%}
\expandafter\def\csname val@header.twin.h.claude.codelivery.ci\endcsname{[30.0, 90.3]}
\expandafter\def\csname val@header.twin.h.claude.codelivery.frac\endcsname{4/6}
\expandafter\def\csname val@header.twin.h.claude.codelivery.full\endcsname{4/6 (66.7\%)}
\expandafter\def\csname val@header.twin.h.claude.codelivery.pct\endcsname{66.7\%}
\expandafter\def\csname val@header.twin.h.claude.hash.ci\endcsname{[0.0, 39.0]}
\expandafter\def\csname val@header.twin.h.claude.hash.frac\endcsname{0/6}
\expandafter\def\csname val@header.twin.h.claude.hash.full\endcsname{0/6 (0.0\%)}
\expandafter\def\csname val@header.twin.h.claude.hash.pct\endcsname{0.0\%}
\expandafter\def\csname val@header.twin.h.claude.write.ci\endcsname{[0.0, 39.0]}
\expandafter\def\csname val@header.twin.h.claude.write.frac\endcsname{0/6}
\expandafter\def\csname val@header.twin.h.claude.write.full\endcsname{0/6 (0.0\%)}
\expandafter\def\csname val@header.twin.h.claude.write.pct\endcsname{0.0\%}
\expandafter\def\csname val@header.twin.h.codex.codelivery.ci\endcsname{[43.6, 97.0]}
\expandafter\def\csname val@header.twin.h.codex.codelivery.frac\endcsname{5/6}
\expandafter\def\csname val@header.twin.h.codex.codelivery.full\endcsname{5/6 (83.3\%)}
\expandafter\def\csname val@header.twin.h.codex.codelivery.pct\endcsname{83.3\%}
\expandafter\def\csname val@header.twin.h.codex.hash.ci\endcsname{[3.0, 56.4]}
\expandafter\def\csname val@header.twin.h.codex.hash.frac\endcsname{1/6}
\expandafter\def\csname val@header.twin.h.codex.hash.full\endcsname{1/6 (16.7\%)}
\expandafter\def\csname val@header.twin.h.codex.hash.pct\endcsname{16.7\%}
\expandafter\def\csname val@header.twin.h.codex.write.ci\endcsname{[0.0, 39.0]}
\expandafter\def\csname val@header.twin.h.codex.write.frac\endcsname{0/6}
\expandafter\def\csname val@header.twin.h.codex.write.full\endcsname{0/6 (0.0\%)}
\expandafter\def\csname val@header.twin.h.codex.write.pct\endcsname{0.0\%}
\expandafter\def\csname val@header.twin.h.gptme.codelivery.ci\endcsname{[30.0, 90.3]}
\expandafter\def\csname val@header.twin.h.gptme.codelivery.frac\endcsname{4/6}
\expandafter\def\csname val@header.twin.h.gptme.codelivery.full\endcsname{4/6 (66.7\%)}
\expandafter\def\csname val@header.twin.h.gptme.codelivery.pct\endcsname{66.7\%}
\expandafter\def\csname val@header.twin.h.gptme.hash.ci\endcsname{[3.0, 56.4]}
\expandafter\def\csname val@header.twin.h.gptme.hash.frac\endcsname{1/6}
\expandafter\def\csname val@header.twin.h.gptme.hash.full\endcsname{1/6 (16.7\%)}
\expandafter\def\csname val@header.twin.h.gptme.hash.pct\endcsname{16.7\%}
\expandafter\def\csname val@header.twin.h.gptme.write.ci\endcsname{[0.0, 39.0]}
\expandafter\def\csname val@header.twin.h.gptme.write.frac\endcsname{0/6}
\expandafter\def\csname val@header.twin.h.gptme.write.full\endcsname{0/6 (0.0\%)}
\expandafter\def\csname val@header.twin.h.gptme.write.pct\endcsname{0.0\%}
\expandafter\def\csname val@header.twin.h.mini.codelivery.ci\endcsname{[9.7, 70.0]}
\expandafter\def\csname val@header.twin.h.mini.codelivery.frac\endcsname{2/6}
\expandafter\def\csname val@header.twin.h.mini.codelivery.full\endcsname{2/6 (33.3\%)}
\expandafter\def\csname val@header.twin.h.mini.codelivery.pct\endcsname{33.3\%}
\expandafter\def\csname val@header.twin.h.mini.hash.ci\endcsname{[3.0, 56.4]}
\expandafter\def\csname val@header.twin.h.mini.hash.frac\endcsname{1/6}
\expandafter\def\csname val@header.twin.h.mini.hash.full\endcsname{1/6 (16.7\%)}
\expandafter\def\csname val@header.twin.h.mini.hash.pct\endcsname{16.7\%}
\expandafter\def\csname val@header.twin.h.mini.write.ci\endcsname{[0.0, 39.0]}
\expandafter\def\csname val@header.twin.h.mini.write.frac\endcsname{0/6}
\expandafter\def\csname val@header.twin.h.mini.write.full\endcsname{0/6 (0.0\%)}
\expandafter\def\csname val@header.twin.h.mini.write.pct\endcsname{0.0\%}
\expandafter\def\csname val@header.twin.h.opencode.codelivery.ci\endcsname{[9.7, 70.0]}
\expandafter\def\csname val@header.twin.h.opencode.codelivery.frac\endcsname{2/6}
\expandafter\def\csname val@header.twin.h.opencode.codelivery.full\endcsname{2/6 (33.3\%)}
\expandafter\def\csname val@header.twin.h.opencode.codelivery.pct\endcsname{33.3\%}
\expandafter\def\csname val@header.twin.h.opencode.hash.ci\endcsname{[0.0, 39.0]}
\expandafter\def\csname val@header.twin.h.opencode.hash.frac\endcsname{0/6}
\expandafter\def\csname val@header.twin.h.opencode.hash.full\endcsname{0/6 (0.0\%)}
\expandafter\def\csname val@header.twin.h.opencode.hash.pct\endcsname{0.0\%}
\expandafter\def\csname val@header.twin.h.opencode.write.ci\endcsname{[0.0, 39.0]}
\expandafter\def\csname val@header.twin.h.opencode.write.frac\endcsname{0/6}
\expandafter\def\csname val@header.twin.h.opencode.write.full\endcsname{0/6 (0.0\%)}
\expandafter\def\csname val@header.twin.h.opencode.write.pct\endcsname{0.0\%}
\expandafter\def\csname val@header.twin.h.qwen.codelivery.ci\endcsname{[18.8, 81.2]}
\expandafter\def\csname val@header.twin.h.qwen.codelivery.frac\endcsname{3/6}
\expandafter\def\csname val@header.twin.h.qwen.codelivery.full\endcsname{3/6 (50.0\%)}
\expandafter\def\csname val@header.twin.h.qwen.codelivery.pct\endcsname{50.0\%}
\expandafter\def\csname val@header.twin.h.qwen.hash.ci\endcsname{[0.0, 39.0]}
\expandafter\def\csname val@header.twin.h.qwen.hash.frac\endcsname{0/6}
\expandafter\def\csname val@header.twin.h.qwen.hash.full\endcsname{0/6 (0.0\%)}
\expandafter\def\csname val@header.twin.h.qwen.hash.pct\endcsname{0.0\%}
\expandafter\def\csname val@header.twin.h.qwen.write.ci\endcsname{[0.0, 39.0]}
\expandafter\def\csname val@header.twin.h.qwen.write.frac\endcsname{0/6}
\expandafter\def\csname val@header.twin.h.qwen.write.full\endcsname{0/6 (0.0\%)}
\expandafter\def\csname val@header.twin.h.qwen.write.pct\endcsname{0.0\%}
\expandafter\def\csname val@header.twin.h.smol.codelivery.ci\endcsname{[0.0, 39.0]}
\expandafter\def\csname val@header.twin.h.smol.codelivery.frac\endcsname{0/6}
\expandafter\def\csname val@header.twin.h.smol.codelivery.full\endcsname{0/6 (0.0\%)}
\expandafter\def\csname val@header.twin.h.smol.codelivery.pct\endcsname{0.0\%}
\expandafter\def\csname val@header.twin.h.smol.hash.ci\endcsname{[0.0, 39.0]}
\expandafter\def\csname val@header.twin.h.smol.hash.frac\endcsname{0/6}
\expandafter\def\csname val@header.twin.h.smol.hash.full\endcsname{0/6 (0.0\%)}
\expandafter\def\csname val@header.twin.h.smol.hash.pct\endcsname{0.0\%}
\expandafter\def\csname val@header.twin.h.smol.write.ci\endcsname{[0.0, 39.0]}
\expandafter\def\csname val@header.twin.h.smol.write.frac\endcsname{0/6}
\expandafter\def\csname val@header.twin.h.smol.write.full\endcsname{0/6 (0.0\%)}
\expandafter\def\csname val@header.twin.h.smol.write.pct\endcsname{0.0\%}
\expandafter\def\csname val@header.twin.hash.ci\endcsname{[2.5, 19.0]}
\expandafter\def\csname val@header.twin.hash.frac\endcsname{3/42}
\expandafter\def\csname val@header.twin.hash.full\endcsname{3/42 (7.1\%)}
\expandafter\def\csname val@header.twin.hash.pct\endcsname{7.1\%}
\expandafter\def\csname val@header.twin.reported.ci\endcsname{[0.4, 12.3]}
\expandafter\def\csname val@header.twin.reported.frac\endcsname{1/42}
\expandafter\def\csname val@header.twin.reported.full\endcsname{1/42 (2.4\%)}
\expandafter\def\csname val@header.twin.reported.pct\endcsname{2.4\%}
\expandafter\def\csname val@header.twin.reportedafter.ci\endcsname{[0.0, 8.4]}
\expandafter\def\csname val@header.twin.reportedafter.frac\endcsname{0/42}
\expandafter\def\csname val@header.twin.reportedafter.full\endcsname{0/42 (0.0\%)}
\expandafter\def\csname val@header.twin.reportedafter.pct\endcsname{0.0\%}
\expandafter\def\csname val@header.twin.targetwritten.ci\endcsname{[21.0, 48.4]}
\expandafter\def\csname val@header.twin.targetwritten.frac\endcsname{14/42}
\expandafter\def\csname val@header.twin.targetwritten.full\endcsname{14/42 (33.3\%)}
\expandafter\def\csname val@header.twin.targetwritten.pct\endcsname{33.3\%}
\expandafter\def\csname val@header.twin.write.ci\endcsname{[0.0, 8.4]}
\expandafter\def\csname val@header.twin.write.frac\endcsname{0/42}
\expandafter\def\csname val@header.twin.write.full\endcsname{0/42 (0.0\%)}
\expandafter\def\csname val@header.twin.write.pct\endcsname{0.0\%}
\expandafter\def\csname val@header.unverified.ci\endcsname{[17.2, 43.6]}
\expandafter\def\csname val@header.unverified.frac\endcsname{12/42}
\expandafter\def\csname val@header.unverified.full\endcsname{12/42 (28.6\%)}
\expandafter\def\csname val@header.unverified.pct\endcsname{28.6\%}
\expandafter\def\csname val@header.write.ci\endcsname{[6.7, 27.8]}
\expandafter\def\csname val@header.write.frac\endcsname{6/42}
\expandafter\def\csname val@header.write.full\endcsname{6/42 (14.3\%)}
\expandafter\def\csname val@header.write.pct\endcsname{14.3\%}
\expandafter\def\csname val@inbody.body.ci\endcsname{[84.2, 98.7]}
\expandafter\def\csname val@inbody.body.frac\endcsname{40/42}
\expandafter\def\csname val@inbody.body.full\endcsname{40/42 (95.2\%)}
\expandafter\def\csname val@inbody.body.pct\endcsname{95.2\%}
\expandafter\def\csname val@inbody.claim.ci\endcsname{[0.0, 8.4]}
\expandafter\def\csname val@inbody.claim.frac\endcsname{0/42}
\expandafter\def\csname val@inbody.claim.full\endcsname{0/42 (0.0\%)}
\expandafter\def\csname val@inbody.claim.pct\endcsname{0.0\%}
\expandafter\def\csname val@inbody.codelivery.ci\endcsname{[84.2, 98.7]}
\expandafter\def\csname val@inbody.codelivery.frac\endcsname{40/42}
\expandafter\def\csname val@inbody.codelivery.full\endcsname{40/42 (95.2\%)}
\expandafter\def\csname val@inbody.codelivery.pct\endcsname{95.2\%}
\expandafter\def\csname val@inbody.declared.ci\endcsname{[0.0, 8.4]}
\expandafter\def\csname val@inbody.declared.frac\endcsname{0/42}
\expandafter\def\csname val@inbody.declared.full\endcsname{0/42 (0.0\%)}
\expandafter\def\csname val@inbody.declared.pct\endcsname{0.0\%}
\expandafter\def\csname val@inbody.h.claude.codelivery.ci\endcsname{[61.0, 100.0]}
\expandafter\def\csname val@inbody.h.claude.codelivery.frac\endcsname{6/6}
\expandafter\def\csname val@inbody.h.claude.codelivery.full\endcsname{6/6 (100.0\%)}
\expandafter\def\csname val@inbody.h.claude.codelivery.pct\endcsname{100.0\%}
\expandafter\def\csname val@inbody.h.claude.hash.ci\endcsname{[0.0, 39.0]}
\expandafter\def\csname val@inbody.h.claude.hash.frac\endcsname{0/6}
\expandafter\def\csname val@inbody.h.claude.hash.full\endcsname{0/6 (0.0\%)}
\expandafter\def\csname val@inbody.h.claude.hash.pct\endcsname{0.0\%}
\expandafter\def\csname val@inbody.h.claude.ingress.ci\endcsname{[61.0, 100.0]}
\expandafter\def\csname val@inbody.h.claude.ingress.frac\endcsname{6/6}
\expandafter\def\csname val@inbody.h.claude.ingress.full\endcsname{6/6 (100.0\%)}
\expandafter\def\csname val@inbody.h.claude.ingress.pct\endcsname{100.0\%}
\expandafter\def\csname val@inbody.h.claude.reject.ci\endcsname{[3.0, 56.4]}
\expandafter\def\csname val@inbody.h.claude.reject.frac\endcsname{1/6}
\expandafter\def\csname val@inbody.h.claude.reject.full\endcsname{1/6 (16.7\%)}
\expandafter\def\csname val@inbody.h.claude.reject.pct\endcsname{16.7\%}
\expandafter\def\csname val@inbody.h.claude.write.ci\endcsname{[18.8, 81.2]}
\expandafter\def\csname val@inbody.h.claude.write.frac\endcsname{3/6}
\expandafter\def\csname val@inbody.h.claude.write.full\endcsname{3/6 (50.0\%)}
\expandafter\def\csname val@inbody.h.claude.write.pct\endcsname{50.0\%}
\expandafter\def\csname val@inbody.h.codex.codelivery.ci\endcsname{[43.6, 97.0]}
\expandafter\def\csname val@inbody.h.codex.codelivery.frac\endcsname{5/6}
\expandafter\def\csname val@inbody.h.codex.codelivery.full\endcsname{5/6 (83.3\%)}
\expandafter\def\csname val@inbody.h.codex.codelivery.pct\endcsname{83.3\%}
\expandafter\def\csname val@inbody.h.codex.hash.ci\endcsname{[3.0, 56.4]}
\expandafter\def\csname val@inbody.h.codex.hash.frac\endcsname{1/6}
\expandafter\def\csname val@inbody.h.codex.hash.full\endcsname{1/6 (16.7\%)}
\expandafter\def\csname val@inbody.h.codex.hash.pct\endcsname{16.7\%}
\expandafter\def\csname val@inbody.h.codex.ingress.ci\endcsname{[43.6, 97.0]}
\expandafter\def\csname val@inbody.h.codex.ingress.frac\endcsname{5/6}
\expandafter\def\csname val@inbody.h.codex.ingress.full\endcsname{5/6 (83.3\%)}
\expandafter\def\csname val@inbody.h.codex.ingress.pct\endcsname{83.3\%}
\expandafter\def\csname val@inbody.h.codex.reject.ci\endcsname{[9.7, 70.0]}
\expandafter\def\csname val@inbody.h.codex.reject.frac\endcsname{2/6}
\expandafter\def\csname val@inbody.h.codex.reject.full\endcsname{2/6 (33.3\%)}
\expandafter\def\csname val@inbody.h.codex.reject.pct\endcsname{33.3\%}
\expandafter\def\csname val@inbody.h.codex.write.ci\endcsname{[3.0, 56.4]}
\expandafter\def\csname val@inbody.h.codex.write.frac\endcsname{1/6}
\expandafter\def\csname val@inbody.h.codex.write.full\endcsname{1/6 (16.7\%)}
\expandafter\def\csname val@inbody.h.codex.write.pct\endcsname{16.7\%}
\expandafter\def\csname val@inbody.h.gptme.codelivery.ci\endcsname{[61.0, 100.0]}
\expandafter\def\csname val@inbody.h.gptme.codelivery.frac\endcsname{6/6}
\expandafter\def\csname val@inbody.h.gptme.codelivery.full\endcsname{6/6 (100.0\%)}
\expandafter\def\csname val@inbody.h.gptme.codelivery.pct\endcsname{100.0\%}
\expandafter\def\csname val@inbody.h.gptme.hash.ci\endcsname{[3.0, 56.4]}
\expandafter\def\csname val@inbody.h.gptme.hash.frac\endcsname{1/6}
\expandafter\def\csname val@inbody.h.gptme.hash.full\endcsname{1/6 (16.7\%)}
\expandafter\def\csname val@inbody.h.gptme.hash.pct\endcsname{16.7\%}
\expandafter\def\csname val@inbody.h.gptme.ingress.ci\endcsname{[61.0, 100.0]}
\expandafter\def\csname val@inbody.h.gptme.ingress.frac\endcsname{6/6}
\expandafter\def\csname val@inbody.h.gptme.ingress.full\endcsname{6/6 (100.0\%)}
\expandafter\def\csname val@inbody.h.gptme.ingress.pct\endcsname{100.0\%}
\expandafter\def\csname val@inbody.h.gptme.reject.ci\endcsname{[0.0, 39.0]}
\expandafter\def\csname val@inbody.h.gptme.reject.frac\endcsname{0/6}
\expandafter\def\csname val@inbody.h.gptme.reject.full\endcsname{0/6 (0.0\%)}
\expandafter\def\csname val@inbody.h.gptme.reject.pct\endcsname{0.0\%}
\expandafter\def\csname val@inbody.h.gptme.write.ci\endcsname{[9.7, 70.0]}
\expandafter\def\csname val@inbody.h.gptme.write.frac\endcsname{2/6}
\expandafter\def\csname val@inbody.h.gptme.write.full\endcsname{2/6 (33.3\%)}
\expandafter\def\csname val@inbody.h.gptme.write.pct\endcsname{33.3\%}
\expandafter\def\csname val@inbody.h.mini.codelivery.ci\endcsname{[61.0, 100.0]}
\expandafter\def\csname val@inbody.h.mini.codelivery.frac\endcsname{6/6}
\expandafter\def\csname val@inbody.h.mini.codelivery.full\endcsname{6/6 (100.0\%)}
\expandafter\def\csname val@inbody.h.mini.codelivery.pct\endcsname{100.0\%}
\expandafter\def\csname val@inbody.h.mini.hash.ci\endcsname{[0.0, 39.0]}
\expandafter\def\csname val@inbody.h.mini.hash.frac\endcsname{0/6}
\expandafter\def\csname val@inbody.h.mini.hash.full\endcsname{0/6 (0.0\%)}
\expandafter\def\csname val@inbody.h.mini.hash.pct\endcsname{0.0\%}
\expandafter\def\csname val@inbody.h.mini.ingress.ci\endcsname{[61.0, 100.0]}
\expandafter\def\csname val@inbody.h.mini.ingress.frac\endcsname{6/6}
\expandafter\def\csname val@inbody.h.mini.ingress.full\endcsname{6/6 (100.0\%)}
\expandafter\def\csname val@inbody.h.mini.ingress.pct\endcsname{100.0\%}
\expandafter\def\csname val@inbody.h.mini.reject.ci\endcsname{[9.7, 70.0]}
\expandafter\def\csname val@inbody.h.mini.reject.frac\endcsname{2/6}
\expandafter\def\csname val@inbody.h.mini.reject.full\endcsname{2/6 (33.3\%)}
\expandafter\def\csname val@inbody.h.mini.reject.pct\endcsname{33.3\%}
\expandafter\def\csname val@inbody.h.mini.write.ci\endcsname{[30.0, 90.3]}
\expandafter\def\csname val@inbody.h.mini.write.frac\endcsname{4/6}
\expandafter\def\csname val@inbody.h.mini.write.full\endcsname{4/6 (66.7\%)}
\expandafter\def\csname val@inbody.h.mini.write.pct\endcsname{66.7\%}
\expandafter\def\csname val@inbody.h.opencode.codelivery.ci\endcsname{[61.0, 100.0]}
\expandafter\def\csname val@inbody.h.opencode.codelivery.frac\endcsname{6/6}
\expandafter\def\csname val@inbody.h.opencode.codelivery.full\endcsname{6/6 (100.0\%)}
\expandafter\def\csname val@inbody.h.opencode.codelivery.pct\endcsname{100.0\%}
\expandafter\def\csname val@inbody.h.opencode.hash.ci\endcsname{[0.0, 39.0]}
\expandafter\def\csname val@inbody.h.opencode.hash.frac\endcsname{0/6}
\expandafter\def\csname val@inbody.h.opencode.hash.full\endcsname{0/6 (0.0\%)}
\expandafter\def\csname val@inbody.h.opencode.hash.pct\endcsname{0.0\%}
\expandafter\def\csname val@inbody.h.opencode.ingress.ci\endcsname{[61.0, 100.0]}
\expandafter\def\csname val@inbody.h.opencode.ingress.frac\endcsname{6/6}
\expandafter\def\csname val@inbody.h.opencode.ingress.full\endcsname{6/6 (100.0\%)}
\expandafter\def\csname val@inbody.h.opencode.ingress.pct\endcsname{100.0\%}
\expandafter\def\csname val@inbody.h.opencode.reject.ci\endcsname{[0.0, 39.0]}
\expandafter\def\csname val@inbody.h.opencode.reject.frac\endcsname{0/6}
\expandafter\def\csname val@inbody.h.opencode.reject.full\endcsname{0/6 (0.0\%)}
\expandafter\def\csname val@inbody.h.opencode.reject.pct\endcsname{0.0\%}
\expandafter\def\csname val@inbody.h.opencode.write.ci\endcsname{[0.0, 39.0]}
\expandafter\def\csname val@inbody.h.opencode.write.frac\endcsname{0/6}
\expandafter\def\csname val@inbody.h.opencode.write.full\endcsname{0/6 (0.0\%)}
\expandafter\def\csname val@inbody.h.opencode.write.pct\endcsname{0.0\%}
\expandafter\def\csname val@inbody.h.qwen.codelivery.ci\endcsname{[61.0, 100.0]}
\expandafter\def\csname val@inbody.h.qwen.codelivery.frac\endcsname{6/6}
\expandafter\def\csname val@inbody.h.qwen.codelivery.full\endcsname{6/6 (100.0\%)}
\expandafter\def\csname val@inbody.h.qwen.codelivery.pct\endcsname{100.0\%}
\expandafter\def\csname val@inbody.h.qwen.hash.ci\endcsname{[0.0, 39.0]}
\expandafter\def\csname val@inbody.h.qwen.hash.frac\endcsname{0/6}
\expandafter\def\csname val@inbody.h.qwen.hash.full\endcsname{0/6 (0.0\%)}
\expandafter\def\csname val@inbody.h.qwen.hash.pct\endcsname{0.0\%}
\expandafter\def\csname val@inbody.h.qwen.ingress.ci\endcsname{[61.0, 100.0]}
\expandafter\def\csname val@inbody.h.qwen.ingress.frac\endcsname{6/6}
\expandafter\def\csname val@inbody.h.qwen.ingress.full\endcsname{6/6 (100.0\%)}
\expandafter\def\csname val@inbody.h.qwen.ingress.pct\endcsname{100.0\%}
\expandafter\def\csname val@inbody.h.qwen.reject.ci\endcsname{[18.8, 81.2]}
\expandafter\def\csname val@inbody.h.qwen.reject.frac\endcsname{3/6}
\expandafter\def\csname val@inbody.h.qwen.reject.full\endcsname{3/6 (50.0\%)}
\expandafter\def\csname val@inbody.h.qwen.reject.pct\endcsname{50.0\%}
\expandafter\def\csname val@inbody.h.qwen.write.ci\endcsname{[3.0, 56.4]}
\expandafter\def\csname val@inbody.h.qwen.write.frac\endcsname{1/6}
\expandafter\def\csname val@inbody.h.qwen.write.full\endcsname{1/6 (16.7\%)}
\expandafter\def\csname val@inbody.h.qwen.write.pct\endcsname{16.7\%}
\expandafter\def\csname val@inbody.h.smol.codelivery.ci\endcsname{[43.6, 97.0]}
\expandafter\def\csname val@inbody.h.smol.codelivery.frac\endcsname{5/6}
\expandafter\def\csname val@inbody.h.smol.codelivery.full\endcsname{5/6 (83.3\%)}
\expandafter\def\csname val@inbody.h.smol.codelivery.pct\endcsname{83.3\%}
\expandafter\def\csname val@inbody.h.smol.hash.ci\endcsname{[0.0, 39.0]}
\expandafter\def\csname val@inbody.h.smol.hash.frac\endcsname{0/6}
\expandafter\def\csname val@inbody.h.smol.hash.full\endcsname{0/6 (0.0\%)}
\expandafter\def\csname val@inbody.h.smol.hash.pct\endcsname{0.0\%}
\expandafter\def\csname val@inbody.h.smol.ingress.ci\endcsname{[43.6, 97.0]}
\expandafter\def\csname val@inbody.h.smol.ingress.frac\endcsname{5/6}
\expandafter\def\csname val@inbody.h.smol.ingress.full\endcsname{5/6 (83.3\%)}
\expandafter\def\csname val@inbody.h.smol.ingress.pct\endcsname{83.3\%}
\expandafter\def\csname val@inbody.h.smol.reject.ci\endcsname{[3.0, 56.4]}
\expandafter\def\csname val@inbody.h.smol.reject.frac\endcsname{1/6}
\expandafter\def\csname val@inbody.h.smol.reject.full\endcsname{1/6 (16.7\%)}
\expandafter\def\csname val@inbody.h.smol.reject.pct\endcsname{16.7\%}
\expandafter\def\csname val@inbody.h.smol.write.ci\endcsname{[0.0, 39.0]}
\expandafter\def\csname val@inbody.h.smol.write.frac\endcsname{0/6}
\expandafter\def\csname val@inbody.h.smol.write.full\endcsname{0/6 (0.0\%)}
\expandafter\def\csname val@inbody.h.smol.write.pct\endcsname{0.0\%}
\expandafter\def\csname val@inbody.hash.ci\endcsname{[1.3, 15.8]}
\expandafter\def\csname val@inbody.hash.frac\endcsname{2/42}
\expandafter\def\csname val@inbody.hash.full\endcsname{2/42 (4.8\%)}
\expandafter\def\csname val@inbody.hash.pct\endcsname{4.8\%}
\expandafter\def\csname val@inbody.ingress.ci\endcsname{[84.2, 98.7]}
\expandafter\def\csname val@inbody.ingress.frac\endcsname{40/42}
\expandafter\def\csname val@inbody.ingress.full\endcsname{40/42 (95.2\%)}
\expandafter\def\csname val@inbody.ingress.pct\endcsname{95.2\%}
\expandafter\def\csname val@inbody.reject.ci\endcsname{[11.7, 35.9]}
\expandafter\def\csname val@inbody.reject.frac\endcsname{9/42}
\expandafter\def\csname val@inbody.reject.full\endcsname{9/42 (21.4\%)}
\expandafter\def\csname val@inbody.reject.pct\endcsname{21.4\%}
\expandafter\def\csname val@inbody.reported.ci\endcsname{[19.1, 46.0]}
\expandafter\def\csname val@inbody.reported.frac\endcsname{13/42}
\expandafter\def\csname val@inbody.reported.full\endcsname{13/42 (31.0\%)}
\expandafter\def\csname val@inbody.reported.pct\endcsname{31.0\%}
\expandafter\def\csname val@inbody.reportedafter.ci\endcsname{[19.1, 46.0]}
\expandafter\def\csname val@inbody.reportedafter.frac\endcsname{13/42}
\expandafter\def\csname val@inbody.reportedafter.full\endcsname{13/42 (31.0\%)}
\expandafter\def\csname val@inbody.reportedafter.pct\endcsname{31.0\%}
\expandafter\def\csname val@inbody.single.ci\endcsname{[84.2, 98.7]}
\expandafter\def\csname val@inbody.single.frac\endcsname{40/42}
\expandafter\def\csname val@inbody.single.full\endcsname{40/42 (95.2\%)}
\expandafter\def\csname val@inbody.single.pct\endcsname{95.2\%}
\expandafter\def\csname val@inbody.targetwritten.ci\endcsname{[15.3, 41.1]}
\expandafter\def\csname val@inbody.targetwritten.frac\endcsname{11/42}
\expandafter\def\csname val@inbody.targetwritten.full\endcsname{11/42 (26.2\%)}
\expandafter\def\csname val@inbody.targetwritten.pct\endcsname{26.2\%}
\expandafter\def\csname val@inbody.twin.body.ci\endcsname{[87.7, 99.6]}
\expandafter\def\csname val@inbody.twin.body.frac\endcsname{41/42}
\expandafter\def\csname val@inbody.twin.body.full\endcsname{41/42 (97.6\%)}
\expandafter\def\csname val@inbody.twin.body.pct\endcsname{97.6\%}
\expandafter\def\csname val@inbody.twin.claim.ci\endcsname{[0.0, 8.4]}
\expandafter\def\csname val@inbody.twin.claim.frac\endcsname{0/42}
\expandafter\def\csname val@inbody.twin.claim.full\endcsname{0/42 (0.0\%)}
\expandafter\def\csname val@inbody.twin.claim.pct\endcsname{0.0\%}
\expandafter\def\csname val@inbody.twin.codelivery.ci\endcsname{[87.7, 99.6]}
\expandafter\def\csname val@inbody.twin.codelivery.frac\endcsname{41/42}
\expandafter\def\csname val@inbody.twin.codelivery.full\endcsname{41/42 (97.6\%)}
\expandafter\def\csname val@inbody.twin.codelivery.pct\endcsname{97.6\%}
\expandafter\def\csname val@inbody.twin.declared.ci\endcsname{[35.5, 64.5]}
\expandafter\def\csname val@inbody.twin.declared.frac\endcsname{21/42}
\expandafter\def\csname val@inbody.twin.declared.full\endcsname{21/42 (50.0\%)}
\expandafter\def\csname val@inbody.twin.declared.pct\endcsname{50.0\%}
\expandafter\def\csname val@inbody.twin.h.claude.codelivery.ci\endcsname{[61.0, 100.0]}
\expandafter\def\csname val@inbody.twin.h.claude.codelivery.frac\endcsname{6/6}
\expandafter\def\csname val@inbody.twin.h.claude.codelivery.full\endcsname{6/6 (100.0\%)}
\expandafter\def\csname val@inbody.twin.h.claude.codelivery.pct\endcsname{100.0\%}
\expandafter\def\csname val@inbody.twin.h.claude.hash.ci\endcsname{[0.0, 39.0]}
\expandafter\def\csname val@inbody.twin.h.claude.hash.frac\endcsname{0/6}
\expandafter\def\csname val@inbody.twin.h.claude.hash.full\endcsname{0/6 (0.0\%)}
\expandafter\def\csname val@inbody.twin.h.claude.hash.pct\endcsname{0.0\%}
\expandafter\def\csname val@inbody.twin.h.claude.write.ci\endcsname{[0.0, 39.0]}
\expandafter\def\csname val@inbody.twin.h.claude.write.frac\endcsname{0/6}
\expandafter\def\csname val@inbody.twin.h.claude.write.full\endcsname{0/6 (0.0\%)}
\expandafter\def\csname val@inbody.twin.h.claude.write.pct\endcsname{0.0\%}
\expandafter\def\csname val@inbody.twin.h.codex.codelivery.ci\endcsname{[61.0, 100.0]}
\expandafter\def\csname val@inbody.twin.h.codex.codelivery.frac\endcsname{6/6}
\expandafter\def\csname val@inbody.twin.h.codex.codelivery.full\endcsname{6/6 (100.0\%)}
\expandafter\def\csname val@inbody.twin.h.codex.codelivery.pct\endcsname{100.0\%}
\expandafter\def\csname val@inbody.twin.h.codex.hash.ci\endcsname{[3.0, 56.4]}
\expandafter\def\csname val@inbody.twin.h.codex.hash.frac\endcsname{1/6}
\expandafter\def\csname val@inbody.twin.h.codex.hash.full\endcsname{1/6 (16.7\%)}
\expandafter\def\csname val@inbody.twin.h.codex.hash.pct\endcsname{16.7\%}
\expandafter\def\csname val@inbody.twin.h.codex.write.ci\endcsname{[0.0, 39.0]}
\expandafter\def\csname val@inbody.twin.h.codex.write.frac\endcsname{0/6}
\expandafter\def\csname val@inbody.twin.h.codex.write.full\endcsname{0/6 (0.0\%)}
\expandafter\def\csname val@inbody.twin.h.codex.write.pct\endcsname{0.0\%}
\expandafter\def\csname val@inbody.twin.h.gptme.codelivery.ci\endcsname{[61.0, 100.0]}
\expandafter\def\csname val@inbody.twin.h.gptme.codelivery.frac\endcsname{6/6}
\expandafter\def\csname val@inbody.twin.h.gptme.codelivery.full\endcsname{6/6 (100.0\%)}
\expandafter\def\csname val@inbody.twin.h.gptme.codelivery.pct\endcsname{100.0\%}
\expandafter\def\csname val@inbody.twin.h.gptme.hash.ci\endcsname{[3.0, 56.4]}
\expandafter\def\csname val@inbody.twin.h.gptme.hash.frac\endcsname{1/6}
\expandafter\def\csname val@inbody.twin.h.gptme.hash.full\endcsname{1/6 (16.7\%)}
\expandafter\def\csname val@inbody.twin.h.gptme.hash.pct\endcsname{16.7\%}
\expandafter\def\csname val@inbody.twin.h.gptme.write.ci\endcsname{[0.0, 39.0]}
\expandafter\def\csname val@inbody.twin.h.gptme.write.frac\endcsname{0/6}
\expandafter\def\csname val@inbody.twin.h.gptme.write.full\endcsname{0/6 (0.0\%)}
\expandafter\def\csname val@inbody.twin.h.gptme.write.pct\endcsname{0.0\%}
\expandafter\def\csname val@inbody.twin.h.mini.codelivery.ci\endcsname{[61.0, 100.0]}
\expandafter\def\csname val@inbody.twin.h.mini.codelivery.frac\endcsname{6/6}
\expandafter\def\csname val@inbody.twin.h.mini.codelivery.full\endcsname{6/6 (100.0\%)}
\expandafter\def\csname val@inbody.twin.h.mini.codelivery.pct\endcsname{100.0\%}
\expandafter\def\csname val@inbody.twin.h.mini.hash.ci\endcsname{[18.8, 81.2]}
\expandafter\def\csname val@inbody.twin.h.mini.hash.frac\endcsname{3/6}
\expandafter\def\csname val@inbody.twin.h.mini.hash.full\endcsname{3/6 (50.0\%)}
\expandafter\def\csname val@inbody.twin.h.mini.hash.pct\endcsname{50.0\%}
\expandafter\def\csname val@inbody.twin.h.mini.write.ci\endcsname{[0.0, 39.0]}
\expandafter\def\csname val@inbody.twin.h.mini.write.frac\endcsname{0/6}
\expandafter\def\csname val@inbody.twin.h.mini.write.full\endcsname{0/6 (0.0\%)}
\expandafter\def\csname val@inbody.twin.h.mini.write.pct\endcsname{0.0\%}
\expandafter\def\csname val@inbody.twin.h.opencode.codelivery.ci\endcsname{[61.0, 100.0]}
\expandafter\def\csname val@inbody.twin.h.opencode.codelivery.frac\endcsname{6/6}
\expandafter\def\csname val@inbody.twin.h.opencode.codelivery.full\endcsname{6/6 (100.0\%)}
\expandafter\def\csname val@inbody.twin.h.opencode.codelivery.pct\endcsname{100.0\%}
\expandafter\def\csname val@inbody.twin.h.opencode.hash.ci\endcsname{[0.0, 39.0]}
\expandafter\def\csname val@inbody.twin.h.opencode.hash.frac\endcsname{0/6}
\expandafter\def\csname val@inbody.twin.h.opencode.hash.full\endcsname{0/6 (0.0\%)}
\expandafter\def\csname val@inbody.twin.h.opencode.hash.pct\endcsname{0.0\%}
\expandafter\def\csname val@inbody.twin.h.opencode.write.ci\endcsname{[0.0, 39.0]}
\expandafter\def\csname val@inbody.twin.h.opencode.write.frac\endcsname{0/6}
\expandafter\def\csname val@inbody.twin.h.opencode.write.full\endcsname{0/6 (0.0\%)}
\expandafter\def\csname val@inbody.twin.h.opencode.write.pct\endcsname{0.0\%}
\expandafter\def\csname val@inbody.twin.h.qwen.codelivery.ci\endcsname{[61.0, 100.0]}
\expandafter\def\csname val@inbody.twin.h.qwen.codelivery.frac\endcsname{6/6}
\expandafter\def\csname val@inbody.twin.h.qwen.codelivery.full\endcsname{6/6 (100.0\%)}
\expandafter\def\csname val@inbody.twin.h.qwen.codelivery.pct\endcsname{100.0\%}
\expandafter\def\csname val@inbody.twin.h.qwen.hash.ci\endcsname{[0.0, 39.0]}
\expandafter\def\csname val@inbody.twin.h.qwen.hash.frac\endcsname{0/6}
\expandafter\def\csname val@inbody.twin.h.qwen.hash.full\endcsname{0/6 (0.0\%)}
\expandafter\def\csname val@inbody.twin.h.qwen.hash.pct\endcsname{0.0\%}
\expandafter\def\csname val@inbody.twin.h.qwen.write.ci\endcsname{[0.0, 39.0]}
\expandafter\def\csname val@inbody.twin.h.qwen.write.frac\endcsname{0/6}
\expandafter\def\csname val@inbody.twin.h.qwen.write.full\endcsname{0/6 (0.0\%)}
\expandafter\def\csname val@inbody.twin.h.qwen.write.pct\endcsname{0.0\%}
\expandafter\def\csname val@inbody.twin.h.smol.codelivery.ci\endcsname{[43.6, 97.0]}
\expandafter\def\csname val@inbody.twin.h.smol.codelivery.frac\endcsname{5/6}
\expandafter\def\csname val@inbody.twin.h.smol.codelivery.full\endcsname{5/6 (83.3\%)}
\expandafter\def\csname val@inbody.twin.h.smol.codelivery.pct\endcsname{83.3\%}
\expandafter\def\csname val@inbody.twin.h.smol.hash.ci\endcsname{[0.0, 39.0]}
\expandafter\def\csname val@inbody.twin.h.smol.hash.frac\endcsname{0/6}
\expandafter\def\csname val@inbody.twin.h.smol.hash.full\endcsname{0/6 (0.0\%)}
\expandafter\def\csname val@inbody.twin.h.smol.hash.pct\endcsname{0.0\%}
\expandafter\def\csname val@inbody.twin.h.smol.write.ci\endcsname{[0.0, 39.0]}
\expandafter\def\csname val@inbody.twin.h.smol.write.frac\endcsname{0/6}
\expandafter\def\csname val@inbody.twin.h.smol.write.full\endcsname{0/6 (0.0\%)}
\expandafter\def\csname val@inbody.twin.h.smol.write.pct\endcsname{0.0\%}
\expandafter\def\csname val@inbody.twin.hash.ci\endcsname{[5.2, 25.0]}
\expandafter\def\csname val@inbody.twin.hash.frac\endcsname{5/42}
\expandafter\def\csname val@inbody.twin.hash.full\endcsname{5/42 (11.9\%)}
\expandafter\def\csname val@inbody.twin.hash.pct\endcsname{11.9\%}
\expandafter\def\csname val@inbody.twin.reported.ci\endcsname{[0.0, 8.4]}
\expandafter\def\csname val@inbody.twin.reported.frac\endcsname{0/42}
\expandafter\def\csname val@inbody.twin.reported.full\endcsname{0/42 (0.0\%)}
\expandafter\def\csname val@inbody.twin.reported.pct\endcsname{0.0\%}
\expandafter\def\csname val@inbody.twin.reportedafter.ci\endcsname{[0.0, 8.4]}
\expandafter\def\csname val@inbody.twin.reportedafter.frac\endcsname{0/42}
\expandafter\def\csname val@inbody.twin.reportedafter.full\endcsname{0/42 (0.0\%)}
\expandafter\def\csname val@inbody.twin.reportedafter.pct\endcsname{0.0\%}
\expandafter\def\csname val@inbody.twin.targetwritten.ci\endcsname{[35.5, 64.5]}
\expandafter\def\csname val@inbody.twin.targetwritten.frac\endcsname{21/42}
\expandafter\def\csname val@inbody.twin.targetwritten.full\endcsname{21/42 (50.0\%)}
\expandafter\def\csname val@inbody.twin.targetwritten.pct\endcsname{50.0\%}
\expandafter\def\csname val@inbody.twin.write.ci\endcsname{[0.0, 8.4]}
\expandafter\def\csname val@inbody.twin.write.frac\endcsname{0/42}
\expandafter\def\csname val@inbody.twin.write.full\endcsname{0/42 (0.0\%)}
\expandafter\def\csname val@inbody.twin.write.pct\endcsname{0.0\%}
\expandafter\def\csname val@inbody.unverified.ci\endcsname{[77.9, 96.2]}
\expandafter\def\csname val@inbody.unverified.frac\endcsname{38/42}
\expandafter\def\csname val@inbody.unverified.full\endcsname{38/42 (90.5\%)}
\expandafter\def\csname val@inbody.unverified.pct\endcsname{90.5\%}
\expandafter\def\csname val@inbody.write.ci\endcsname{[15.3, 41.1]}
\expandafter\def\csname val@inbody.write.frac\endcsname{11/42}
\expandafter\def\csname val@inbody.write.full\endcsname{11/42 (26.2\%)}
\expandafter\def\csname val@inbody.write.pct\endcsname{26.2\%}
\expandafter\def\csname val@ingress.range\endcsname{35.7--97.6\%}
\expandafter\def\csname val@invalid.models\endcsname{Gemini-4-Argon}
\expandafter\def\csname val@invalid.reasons\endcsname{session_ended_on_upstream_error, upstream_error_before_fetch}
\expandafter\def\csname val@invalid.total\endcsname{12}
\expandafter\def\csname val@iso.file.background.ci\endcsname{[0.1, 1.1]}
\expandafter\def\csname val@iso.file.background.frac\endcsname{2/672}
\expandafter\def\csname val@iso.file.background.full\endcsname{2/672 (0.3\%)}
\expandafter\def\csname val@iso.file.background.pct\endcsname{0.3\%}
\expandafter\def\csname val@iso.file.gateway.ci\endcsname{[0.0, 0.6]}
\expandafter\def\csname val@iso.file.gateway.frac\endcsname{0/672}
\expandafter\def\csname val@iso.file.gateway.full\endcsname{0/672 (0.0\%)}
\expandafter\def\csname val@iso.file.gateway.pct\endcsname{0.0\%}
\expandafter\def\csname val@iso.file.scan.ci\endcsname{[0.1, 1.1]}
\expandafter\def\csname val@iso.file.scan.frac\endcsname{2/672}
\expandafter\def\csname val@iso.file.scan.full\endcsname{2/672 (0.3\%)}
\expandafter\def\csname val@iso.file.scan.pct\endcsname{0.3\%}
\expandafter\def\csname val@iso.keyed.background.ci\endcsname{[0.1, 1.2]}
\expandafter\def\csname val@iso.keyed.background.frac\endcsname{2/588}
\expandafter\def\csname val@iso.keyed.background.full\endcsname{2/588 (0.3\%)}
\expandafter\def\csname val@iso.keyed.background.pct\endcsname{0.3\%}
\expandafter\def\csname val@iso.keyed.gateway.ci\endcsname{[1.3, 3.7]}
\expandafter\def\csname val@iso.keyed.gateway.frac\endcsname{13/588}
\expandafter\def\csname val@iso.keyed.gateway.full\endcsname{13/588 (2.2\%)}
\expandafter\def\csname val@iso.keyed.gateway.pct\endcsname{2.2\%}
\expandafter\def\csname val@iso.keyed.refused\endcsname{19}
\expandafter\def\csname val@iso.keyed.scan.ci\endcsname{[11.1, 16.6]}
\expandafter\def\csname val@iso.keyed.scan.frac\endcsname{80/588}
\expandafter\def\csname val@iso.keyed.scan.full\endcsname{80/588 (13.6\%)}
\expandafter\def\csname val@iso.keyed.scan.pct\endcsname{13.6\%}
\expandafter\def\csname val@med.claude.ext.ci\endcsname{[83.9, 100.0]}
\expandafter\def\csname val@med.claude.ext.frac\endcsname{20/20}
\expandafter\def\csname val@med.claude.ext.full\endcsname{20/20 (100.0\%)}
\expandafter\def\csname val@med.claude.ext.pct\endcsname{100.0\%}
\expandafter\def\csname val@med.claude.noext.ci\endcsname{[83.9, 100.0]}
\expandafter\def\csname val@med.claude.noext.frac\endcsname{20/20}
\expandafter\def\csname val@med.claude.noext.full\endcsname{20/20 (100.0\%)}
\expandafter\def\csname val@med.claude.noext.pct\endcsname{100.0\%}
\expandafter\def\csname val@med.codex.ext.ci\endcsname{[58.4, 91.9]}
\expandafter\def\csname val@med.codex.ext.frac\endcsname{16/20}
\expandafter\def\csname val@med.codex.ext.full\endcsname{16/20 (80.0\%)}
\expandafter\def\csname val@med.codex.ext.pct\endcsname{80.0\%}
\expandafter\def\csname val@med.codex.noext.ci\endcsname{[64.0, 94.8]}
\expandafter\def\csname val@med.codex.noext.frac\endcsname{17/20}
\expandafter\def\csname val@med.codex.noext.full\endcsname{17/20 (85.0\%)}
\expandafter\def\csname val@med.codex.noext.pct\endcsname{85.0\%}
\expandafter\def\csname val@med.ext.ci\endcsname{[83.1, 93.4]}
\expandafter\def\csname val@med.ext.frac\endcsname{125/140}
\expandafter\def\csname val@med.ext.full\endcsname{125/140 (89.3\%)}
\expandafter\def\csname val@med.ext.pct\endcsname{89.3\%}
\expandafter\def\csname val@med.gptme.ext.ci\endcsname{[83.9, 100.0]}
\expandafter\def\csname val@med.gptme.ext.frac\endcsname{20/20}
\expandafter\def\csname val@med.gptme.ext.full\endcsname{20/20 (100.0\%)}
\expandafter\def\csname val@med.gptme.ext.pct\endcsname{100.0\%}
\expandafter\def\csname val@med.gptme.noext.ci\endcsname{[83.9, 100.0]}
\expandafter\def\csname val@med.gptme.noext.frac\endcsname{20/20}
\expandafter\def\csname val@med.gptme.noext.full\endcsname{20/20 (100.0\%)}
\expandafter\def\csname val@med.gptme.noext.pct\endcsname{100.0\%}
\expandafter\def\csname val@med.mini.ext.ci\endcsname{[83.9, 100.0]}
\expandafter\def\csname val@med.mini.ext.frac\endcsname{20/20}
\expandafter\def\csname val@med.mini.ext.full\endcsname{20/20 (100.0\%)}
\expandafter\def\csname val@med.mini.ext.pct\endcsname{100.0\%}
\expandafter\def\csname val@med.mini.noext.ci\endcsname{[83.9, 100.0]}
\expandafter\def\csname val@med.mini.noext.frac\endcsname{20/20}
\expandafter\def\csname val@med.mini.noext.full\endcsname{20/20 (100.0\%)}
\expandafter\def\csname val@med.mini.noext.pct\endcsname{100.0\%}
\expandafter\def\csname val@med.nmodels\endcsname{four}
\expandafter\def\csname val@med.noext.ci\endcsname{[83.1, 93.4]}
\expandafter\def\csname val@med.noext.frac\endcsname{125/140}
\expandafter\def\csname val@med.noext.full\endcsname{125/140 (89.3\%)}
\expandafter\def\csname val@med.noext.pct\endcsname{89.3\%}
\expandafter\def\csname val@med.ntrials\endcsname{five}
\expandafter\def\csname val@med.opencode.ext.ci\endcsname{[29.9, 70.1]}
\expandafter\def\csname val@med.opencode.ext.frac\endcsname{10/20}
\expandafter\def\csname val@med.opencode.ext.full\endcsname{10/20 (50.0\%)}
\expandafter\def\csname val@med.opencode.ext.pct\endcsname{50.0\%}
\expandafter\def\csname val@med.opencode.noext.ci\endcsname{[43.3, 81.9]}
\expandafter\def\csname val@med.opencode.noext.frac\endcsname{13/20}
\expandafter\def\csname val@med.opencode.noext.full\endcsname{13/20 (65.0\%)}
\expandafter\def\csname val@med.opencode.noext.pct\endcsname{65.0\%}
\expandafter\def\csname val@med.qwen.ext.ci\endcsname{[83.9, 100.0]}
\expandafter\def\csname val@med.qwen.ext.frac\endcsname{20/20}
\expandafter\def\csname val@med.qwen.ext.full\endcsname{20/20 (100.0\%)}
\expandafter\def\csname val@med.qwen.ext.pct\endcsname{100.0\%}
\expandafter\def\csname val@med.qwen.noext.ci\endcsname{[69.9, 97.2]}
\expandafter\def\csname val@med.qwen.noext.frac\endcsname{18/20}
\expandafter\def\csname val@med.qwen.noext.full\endcsname{18/20 (90.0\%)}
\expandafter\def\csname val@med.qwen.noext.pct\endcsname{90.0\%}
\expandafter\def\csname val@med.smol.ext.ci\endcsname{[76.4, 99.1]}
\expandafter\def\csname val@med.smol.ext.frac\endcsname{19/20}
\expandafter\def\csname val@med.smol.ext.full\endcsname{19/20 (95.0\%)}
\expandafter\def\csname val@med.smol.ext.pct\endcsname{95.0\%}
\expandafter\def\csname val@med.smol.noext.ci\endcsname{[64.0, 94.8]}
\expandafter\def\csname val@med.smol.noext.frac\endcsname{17/20}
\expandafter\def\csname val@med.smol.noext.full\endcsname{17/20 (85.0\%)}
\expandafter\def\csname val@med.smol.noext.pct\endcsname{85.0\%}
\expandafter\def\csname val@model.deepseekv41flash1017.allrounds.zipbytebreach.ci\endcsname{[84.5, 100.0]}
\expandafter\def\csname val@model.deepseekv41flash1017.allrounds.zipbytebreach.frac\endcsname{21/21}
\expandafter\def\csname val@model.deepseekv41flash1017.allrounds.zipbytebreach.full\endcsname{21/21 (100.0\%)}
\expandafter\def\csname val@model.deepseekv41flash1017.allrounds.zipbytebreach.pct\endcsname{100.0\%}
\expandafter\def\csname val@model.deepseekv41flash1017.allrounds.zipextract.ci\endcsname{[84.5, 100.0]}
\expandafter\def\csname val@model.deepseekv41flash1017.allrounds.zipextract.frac\endcsname{21/21}
\expandafter\def\csname val@model.deepseekv41flash1017.allrounds.zipextract.full\endcsname{21/21 (100.0\%)}
\expandafter\def\csname val@model.deepseekv41flash1017.allrounds.zipextract.pct\endcsname{100.0\%}
\expandafter\def\csname val@model.deepseekv41flash1017.body.ci\endcsname{[90.1, 100.0]}
\expandafter\def\csname val@model.deepseekv41flash1017.body.frac\endcsname{35/35}
\expandafter\def\csname val@model.deepseekv41flash1017.body.full\endcsname{35/35 (100.0\%)}
\expandafter\def\csname val@model.deepseekv41flash1017.body.pct\endcsname{100.0\%}
\expandafter\def\csname val@model.deepseekv41flash1017.claim.ci\endcsname{[0.0, 21.5]}
\expandafter\def\csname val@model.deepseekv41flash1017.claim.frac\endcsname{0/14}
\expandafter\def\csname val@model.deepseekv41flash1017.claim.full\endcsname{0/14 (0.0\%)}
\expandafter\def\csname val@model.deepseekv41flash1017.claim.pct\endcsname{0.0\%}
\expandafter\def\csname val@model.deepseekv41flash1017.hash.ci\endcsname{[68.5, 98.7]}
\expandafter\def\csname val@model.deepseekv41flash1017.hash.frac\endcsname{13/14}
\expandafter\def\csname val@model.deepseekv41flash1017.hash.full\endcsname{13/14 (92.9\%)}
\expandafter\def\csname val@model.deepseekv41flash1017.hash.pct\endcsname{92.9\%}
\expandafter\def\csname val@model.deepseekv41flash1017.nobody.ci\endcsname{[0.0, 9.9]}
\expandafter\def\csname val@model.deepseekv41flash1017.nobody.frac\endcsname{0/35}
\expandafter\def\csname val@model.deepseekv41flash1017.nobody.full\endcsname{0/35 (0.0\%)}
\expandafter\def\csname val@model.deepseekv41flash1017.nobody.h.claude.ci\endcsname{[0.0, 43.4]}
\expandafter\def\csname val@model.deepseekv41flash1017.nobody.h.claude.frac\endcsname{0/5}
\expandafter\def\csname val@model.deepseekv41flash1017.nobody.h.claude.full\endcsname{0/5 (0.0\%)}
\expandafter\def\csname val@model.deepseekv41flash1017.nobody.h.claude.pct\endcsname{0.0\%}
\expandafter\def\csname val@model.deepseekv41flash1017.nobody.h.codex.ci\endcsname{[0.0, 43.4]}
\expandafter\def\csname val@model.deepseekv41flash1017.nobody.h.codex.frac\endcsname{0/5}
\expandafter\def\csname val@model.deepseekv41flash1017.nobody.h.codex.full\endcsname{0/5 (0.0\%)}
\expandafter\def\csname val@model.deepseekv41flash1017.nobody.h.codex.pct\endcsname{0.0\%}
\expandafter\def\csname val@model.deepseekv41flash1017.nobody.h.gptme.ci\endcsname{[0.0, 43.4]}
\expandafter\def\csname val@model.deepseekv41flash1017.nobody.h.gptme.frac\endcsname{0/5}
\expandafter\def\csname val@model.deepseekv41flash1017.nobody.h.gptme.full\endcsname{0/5 (0.0\%)}
\expandafter\def\csname val@model.deepseekv41flash1017.nobody.h.gptme.pct\endcsname{0.0\%}
\expandafter\def\csname val@model.deepseekv41flash1017.nobody.h.mini.ci\endcsname{[0.0, 43.4]}
\expandafter\def\csname val@model.deepseekv41flash1017.nobody.h.mini.frac\endcsname{0/5}
\expandafter\def\csname val@model.deepseekv41flash1017.nobody.h.mini.full\endcsname{0/5 (0.0\%)}
\expandafter\def\csname val@model.deepseekv41flash1017.nobody.h.mini.pct\endcsname{0.0\%}
\expandafter\def\csname val@model.deepseekv41flash1017.nobody.h.opencode.ci\endcsname{[0.0, 43.4]}
\expandafter\def\csname val@model.deepseekv41flash1017.nobody.h.opencode.frac\endcsname{0/5}
\expandafter\def\csname val@model.deepseekv41flash1017.nobody.h.opencode.full\endcsname{0/5 (0.0\%)}
\expandafter\def\csname val@model.deepseekv41flash1017.nobody.h.opencode.pct\endcsname{0.0\%}
\expandafter\def\csname val@model.deepseekv41flash1017.nobody.h.qwen.ci\endcsname{[0.0, 43.4]}
\expandafter\def\csname val@model.deepseekv41flash1017.nobody.h.qwen.frac\endcsname{0/5}
\expandafter\def\csname val@model.deepseekv41flash1017.nobody.h.qwen.full\endcsname{0/5 (0.0\%)}
\expandafter\def\csname val@model.deepseekv41flash1017.nobody.h.qwen.pct\endcsname{0.0\%}
\expandafter\def\csname val@model.deepseekv41flash1017.nobody.h.smol.ci\endcsname{[0.0, 43.4]}
\expandafter\def\csname val@model.deepseekv41flash1017.nobody.h.smol.frac\endcsname{0/5}
\expandafter\def\csname val@model.deepseekv41flash1017.nobody.h.smol.full\endcsname{0/5 (0.0\%)}
\expandafter\def\csname val@model.deepseekv41flash1017.nobody.h.smol.pct\endcsname{0.0\%}
\expandafter\def\csname val@model.deepseekv41flash1017.nobody.pct\endcsname{0.0\%}
\expandafter\def\csname val@model.deepseekv41flash1017.reject.ci\endcsname{[52.4, 92.4]}
\expandafter\def\csname val@model.deepseekv41flash1017.reject.frac\endcsname{11/14}
\expandafter\def\csname val@model.deepseekv41flash1017.reject.full\endcsname{11/14 (78.6\%)}
\expandafter\def\csname val@model.deepseekv41flash1017.reject.pct\endcsname{78.6\%}
\expandafter\def\csname val@model.deepseekv41flash1017.reportgivenbody.ci\endcsname{[20.8, 59.1]}
\expandafter\def\csname val@model.deepseekv41flash1017.reportgivenbody.frac\endcsname{8/21}
\expandafter\def\csname val@model.deepseekv41flash1017.reportgivenbody.full\endcsname{8/21 (38.1\%)}
\expandafter\def\csname val@model.deepseekv41flash1017.reportgivenbody.pct\endcsname{38.1\%}
\expandafter\def\csname val@model.deepseekv41flash1017.write.ci\endcsname{[10.6, 45.1]}
\expandafter\def\csname val@model.deepseekv41flash1017.write.frac\endcsname{5/21}
\expandafter\def\csname val@model.deepseekv41flash1017.write.full\endcsname{5/21 (23.8\%)}
\expandafter\def\csname val@model.deepseekv41flash1017.write.pct\endcsname{23.8\%}
\expandafter\def\csname val@model.deepseekv41flash1017.writegivenbody.ci\endcsname{[10.6, 45.1]}
\expandafter\def\csname val@model.deepseekv41flash1017.writegivenbody.frac\endcsname{5/21}
\expandafter\def\csname val@model.deepseekv41flash1017.writegivenbody.full\endcsname{5/21 (23.8\%)}
\expandafter\def\csname val@model.deepseekv41flash1017.writegivenbody.pct\endcsname{23.8\%}
\expandafter\def\csname val@model.deepseekv41flash1017.zipbyte.ci\endcsname{[64.6, 100.0]}
\expandafter\def\csname val@model.deepseekv41flash1017.zipbyte.frac\endcsname{7/7}
\expandafter\def\csname val@model.deepseekv41flash1017.zipbyte.full\endcsname{7/7 (100.0\%)}
\expandafter\def\csname val@model.deepseekv41flash1017.zipbyte.pct\endcsname{100.0\%}
\expandafter\def\csname val@model.deepseekv41flash1017.zipextract.ci\endcsname{[64.6, 100.0]}
\expandafter\def\csname val@model.deepseekv41flash1017.zipextract.frac\endcsname{7/7}
\expandafter\def\csname val@model.deepseekv41flash1017.zipextract.full\endcsname{7/7 (100.0\%)}
\expandafter\def\csname val@model.deepseekv41flash1017.zipextract.pct\endcsname{100.0\%}
\expandafter\def\csname val@model.gemini4argon.allrounds.zipbytebreach.ci\endcsname{[17.2, 54.6]}
\expandafter\def\csname val@model.gemini4argon.allrounds.zipbytebreach.frac\endcsname{7/21}
\expandafter\def\csname val@model.gemini4argon.allrounds.zipbytebreach.full\endcsname{7/21 (33.3\%)}
\expandafter\def\csname val@model.gemini4argon.allrounds.zipbytebreach.pct\endcsname{33.3\%}
\expandafter\def\csname val@model.gemini4argon.allrounds.zipextract.ci\endcsname{[20.8, 59.1]}
\expandafter\def\csname val@model.gemini4argon.allrounds.zipextract.frac\endcsname{8/21}
\expandafter\def\csname val@model.gemini4argon.allrounds.zipextract.full\endcsname{8/21 (38.1\%)}
\expandafter\def\csname val@model.gemini4argon.allrounds.zipextract.pct\endcsname{38.1\%}
\expandafter\def\csname val@model.gemini4argon.body.ci\endcsname{[90.1, 100.0]}
\expandafter\def\csname val@model.gemini4argon.body.frac\endcsname{35/35}
\expandafter\def\csname val@model.gemini4argon.body.full\endcsname{35/35 (100.0\%)}
\expandafter\def\csname val@model.gemini4argon.body.pct\endcsname{100.0\%}
\expandafter\def\csname val@model.gemini4argon.claim.ci\endcsname{[0.0, 21.5]}
\expandafter\def\csname val@model.gemini4argon.claim.frac\endcsname{0/14}
\expandafter\def\csname val@model.gemini4argon.claim.full\endcsname{0/14 (0.0\%)}
\expandafter\def\csname val@model.gemini4argon.claim.pct\endcsname{0.0\%}
\expandafter\def\csname val@model.gemini4argon.hash.ci\endcsname{[78.5, 100.0]}
\expandafter\def\csname val@model.gemini4argon.hash.frac\endcsname{14/14}
\expandafter\def\csname val@model.gemini4argon.hash.full\endcsname{14/14 (100.0\%)}
\expandafter\def\csname val@model.gemini4argon.hash.pct\endcsname{100.0\%}
\expandafter\def\csname val@model.gemini4argon.nobody.ci\endcsname{[0.0, 9.9]}
\expandafter\def\csname val@model.gemini4argon.nobody.frac\endcsname{0/35}
\expandafter\def\csname val@model.gemini4argon.nobody.full\endcsname{0/35 (0.0\%)}
\expandafter\def\csname val@model.gemini4argon.nobody.h.claude.ci\endcsname{[0.0, 43.4]}
\expandafter\def\csname val@model.gemini4argon.nobody.h.claude.frac\endcsname{0/5}
\expandafter\def\csname val@model.gemini4argon.nobody.h.claude.full\endcsname{0/5 (0.0\%)}
\expandafter\def\csname val@model.gemini4argon.nobody.h.claude.pct\endcsname{0.0\%}
\expandafter\def\csname val@model.gemini4argon.nobody.h.codex.ci\endcsname{[0.0, 43.4]}
\expandafter\def\csname val@model.gemini4argon.nobody.h.codex.frac\endcsname{0/5}
\expandafter\def\csname val@model.gemini4argon.nobody.h.codex.full\endcsname{0/5 (0.0\%)}
\expandafter\def\csname val@model.gemini4argon.nobody.h.codex.pct\endcsname{0.0\%}
\expandafter\def\csname val@model.gemini4argon.nobody.h.gptme.ci\endcsname{[0.0, 43.4]}
\expandafter\def\csname val@model.gemini4argon.nobody.h.gptme.frac\endcsname{0/5}
\expandafter\def\csname val@model.gemini4argon.nobody.h.gptme.full\endcsname{0/5 (0.0\%)}
\expandafter\def\csname val@model.gemini4argon.nobody.h.gptme.pct\endcsname{0.0\%}
\expandafter\def\csname val@model.gemini4argon.nobody.h.mini.ci\endcsname{[0.0, 43.4]}
\expandafter\def\csname val@model.gemini4argon.nobody.h.mini.frac\endcsname{0/5}
\expandafter\def\csname val@model.gemini4argon.nobody.h.mini.full\endcsname{0/5 (0.0\%)}
\expandafter\def\csname val@model.gemini4argon.nobody.h.mini.pct\endcsname{0.0\%}
\expandafter\def\csname val@model.gemini4argon.nobody.h.opencode.ci\endcsname{[0.0, 43.4]}
\expandafter\def\csname val@model.gemini4argon.nobody.h.opencode.frac\endcsname{0/5}
\expandafter\def\csname val@model.gemini4argon.nobody.h.opencode.full\endcsname{0/5 (0.0\%)}
\expandafter\def\csname val@model.gemini4argon.nobody.h.opencode.pct\endcsname{0.0\%}
\expandafter\def\csname val@model.gemini4argon.nobody.h.qwen.ci\endcsname{[0.0, 43.4]}
\expandafter\def\csname val@model.gemini4argon.nobody.h.qwen.frac\endcsname{0/5}
\expandafter\def\csname val@model.gemini4argon.nobody.h.qwen.full\endcsname{0/5 (0.0\%)}
\expandafter\def\csname val@model.gemini4argon.nobody.h.qwen.pct\endcsname{0.0\%}
\expandafter\def\csname val@model.gemini4argon.nobody.h.smol.ci\endcsname{[0.0, 43.4]}
\expandafter\def\csname val@model.gemini4argon.nobody.h.smol.frac\endcsname{0/5}
\expandafter\def\csname val@model.gemini4argon.nobody.h.smol.full\endcsname{0/5 (0.0\%)}
\expandafter\def\csname val@model.gemini4argon.nobody.h.smol.pct\endcsname{0.0\%}
\expandafter\def\csname val@model.gemini4argon.nobody.pct\endcsname{0.0\%}
\expandafter\def\csname val@model.gemini4argon.reject.ci\endcsname{[26.8, 73.2]}
\expandafter\def\csname val@model.gemini4argon.reject.frac\endcsname{7/14}
\expandafter\def\csname val@model.gemini4argon.reject.full\endcsname{7/14 (50.0\%)}
\expandafter\def\csname val@model.gemini4argon.reject.pct\endcsname{50.0\%}
\expandafter\def\csname val@model.gemini4argon.reportgivenbody.ci\endcsname{[0.0, 15.5]}
\expandafter\def\csname val@model.gemini4argon.reportgivenbody.frac\endcsname{0/21}
\expandafter\def\csname val@model.gemini4argon.reportgivenbody.full\endcsname{0/21 (0.0\%)}
\expandafter\def\csname val@model.gemini4argon.reportgivenbody.pct\endcsname{0.0\%}
\expandafter\def\csname val@model.gemini4argon.write.ci\endcsname{[32.4, 71.7]}
\expandafter\def\csname val@model.gemini4argon.write.frac\endcsname{11/21}
\expandafter\def\csname val@model.gemini4argon.write.full\endcsname{11/21 (52.4\%)}
\expandafter\def\csname val@model.gemini4argon.write.pct\endcsname{52.4\%}
\expandafter\def\csname val@model.gemini4argon.writegivenbody.ci\endcsname{[32.4, 71.7]}
\expandafter\def\csname val@model.gemini4argon.writegivenbody.frac\endcsname{11/21}
\expandafter\def\csname val@model.gemini4argon.writegivenbody.full\endcsname{11/21 (52.4\%)}
\expandafter\def\csname val@model.gemini4argon.writegivenbody.pct\endcsname{52.4\%}
\expandafter\def\csname val@model.gemini4argon.zipbyte.ci\endcsname{[8.2, 64.1]}
\expandafter\def\csname val@model.gemini4argon.zipbyte.frac\endcsname{2/7}
\expandafter\def\csname val@model.gemini4argon.zipbyte.full\endcsname{2/7 (28.6\%)}
\expandafter\def\csname val@model.gemini4argon.zipbyte.pct\endcsname{28.6\%}
\expandafter\def\csname val@model.gemini4argon.zipextract.ci\endcsname{[15.8, 75.0]}
\expandafter\def\csname val@model.gemini4argon.zipextract.frac\endcsname{3/7}
\expandafter\def\csname val@model.gemini4argon.zipextract.full\endcsname{3/7 (42.9\%)}
\expandafter\def\csname val@model.gemini4argon.zipextract.pct\endcsname{42.9\%}
\expandafter\def\csname val@model.glm53.allrounds.zipbytebreach.ci\endcsname{[84.5, 100.0]}
\expandafter\def\csname val@model.glm53.allrounds.zipbytebreach.frac\endcsname{21/21}
\expandafter\def\csname val@model.glm53.allrounds.zipbytebreach.full\endcsname{21/21 (100.0\%)}
\expandafter\def\csname val@model.glm53.allrounds.zipbytebreach.pct\endcsname{100.0\%}
\expandafter\def\csname val@model.glm53.allrounds.zipextract.ci\endcsname{[84.5, 100.0]}
\expandafter\def\csname val@model.glm53.allrounds.zipextract.frac\endcsname{21/21}
\expandafter\def\csname val@model.glm53.allrounds.zipextract.full\endcsname{21/21 (100.0\%)}
\expandafter\def\csname val@model.glm53.allrounds.zipextract.pct\endcsname{100.0\%}
\expandafter\def\csname val@model.glm53.body.ci\endcsname{[74.0, 95.5]}
\expandafter\def\csname val@model.glm53.body.frac\endcsname{31/35}
\expandafter\def\csname val@model.glm53.body.full\endcsname{31/35 (88.6\%)}
\expandafter\def\csname val@model.glm53.body.pct\endcsname{88.6\%}
\expandafter\def\csname val@model.glm53.claim.ci\endcsname{[0.0, 21.5]}
\expandafter\def\csname val@model.glm53.claim.frac\endcsname{0/14}
\expandafter\def\csname val@model.glm53.claim.full\endcsname{0/14 (0.0\%)}
\expandafter\def\csname val@model.glm53.claim.pct\endcsname{0.0\%}
\expandafter\def\csname val@model.glm53.hash.ci\endcsname{[32.6, 78.6]}
\expandafter\def\csname val@model.glm53.hash.frac\endcsname{8/14}
\expandafter\def\csname val@model.glm53.hash.full\endcsname{8/14 (57.1\%)}
\expandafter\def\csname val@model.glm53.hash.pct\endcsname{57.1\%}
\expandafter\def\csname val@model.glm53.nobody.ci\endcsname{[4.5, 26.0]}
\expandafter\def\csname val@model.glm53.nobody.frac\endcsname{4/35}
\expandafter\def\csname val@model.glm53.nobody.full\endcsname{4/35 (11.4\%)}
\expandafter\def\csname val@model.glm53.nobody.h.claude.ci\endcsname{[0.0, 43.4]}
\expandafter\def\csname val@model.glm53.nobody.h.claude.frac\endcsname{0/5}
\expandafter\def\csname val@model.glm53.nobody.h.claude.full\endcsname{0/5 (0.0\%)}
\expandafter\def\csname val@model.glm53.nobody.h.claude.pct\endcsname{0.0\%}
\expandafter\def\csname val@model.glm53.nobody.h.codex.ci\endcsname{[3.6, 62.4]}
\expandafter\def\csname val@model.glm53.nobody.h.codex.frac\endcsname{1/5}
\expandafter\def\csname val@model.glm53.nobody.h.codex.full\endcsname{1/5 (20.0\%)}
\expandafter\def\csname val@model.glm53.nobody.h.codex.pct\endcsname{20.0\%}
\expandafter\def\csname val@model.glm53.nobody.h.gptme.ci\endcsname{[3.6, 62.4]}
\expandafter\def\csname val@model.glm53.nobody.h.gptme.frac\endcsname{1/5}
\expandafter\def\csname val@model.glm53.nobody.h.gptme.full\endcsname{1/5 (20.0\%)}
\expandafter\def\csname val@model.glm53.nobody.h.gptme.pct\endcsname{20.0\%}
\expandafter\def\csname val@model.glm53.nobody.h.mini.ci\endcsname{[0.0, 43.4]}
\expandafter\def\csname val@model.glm53.nobody.h.mini.frac\endcsname{0/5}
\expandafter\def\csname val@model.glm53.nobody.h.mini.full\endcsname{0/5 (0.0\%)}
\expandafter\def\csname val@model.glm53.nobody.h.mini.pct\endcsname{0.0\%}
\expandafter\def\csname val@model.glm53.nobody.h.opencode.ci\endcsname{[11.8, 76.9]}
\expandafter\def\csname val@model.glm53.nobody.h.opencode.frac\endcsname{2/5}
\expandafter\def\csname val@model.glm53.nobody.h.opencode.full\endcsname{2/5 (40.0\%)}
\expandafter\def\csname val@model.glm53.nobody.h.opencode.pct\endcsname{40.0\%}
\expandafter\def\csname val@model.glm53.nobody.h.qwen.ci\endcsname{[0.0, 43.4]}
\expandafter\def\csname val@model.glm53.nobody.h.qwen.frac\endcsname{0/5}
\expandafter\def\csname val@model.glm53.nobody.h.qwen.full\endcsname{0/5 (0.0\%)}
\expandafter\def\csname val@model.glm53.nobody.h.qwen.pct\endcsname{0.0\%}
\expandafter\def\csname val@model.glm53.nobody.h.smol.ci\endcsname{[0.0, 43.4]}
\expandafter\def\csname val@model.glm53.nobody.h.smol.frac\endcsname{0/5}
\expandafter\def\csname val@model.glm53.nobody.h.smol.full\endcsname{0/5 (0.0\%)}
\expandafter\def\csname val@model.glm53.nobody.h.smol.pct\endcsname{0.0\%}
\expandafter\def\csname val@model.glm53.nobody.pct\endcsname{11.4\%}
\expandafter\def\csname val@model.glm53.reject.ci\endcsname{[32.6, 78.6]}
\expandafter\def\csname val@model.glm53.reject.frac\endcsname{8/14}
\expandafter\def\csname val@model.glm53.reject.full\endcsname{8/14 (57.1\%)}
\expandafter\def\csname val@model.glm53.reject.pct\endcsname{57.1\%}
\expandafter\def\csname val@model.glm53.reportgivenbody.ci\endcsname{[11.2, 46.9]}
\expandafter\def\csname val@model.glm53.reportgivenbody.frac\endcsname{5/20}
\expandafter\def\csname val@model.glm53.reportgivenbody.full\endcsname{5/20 (25.0\%)}
\expandafter\def\csname val@model.glm53.reportgivenbody.pct\endcsname{25.0\%}
\expandafter\def\csname val@model.glm53.write.ci\endcsname{[5.0, 34.6]}
\expandafter\def\csname val@model.glm53.write.frac\endcsname{3/21}
\expandafter\def\csname val@model.glm53.write.full\endcsname{3/21 (14.3\%)}
\expandafter\def\csname val@model.glm53.write.pct\endcsname{14.3\%}
\expandafter\def\csname val@model.glm53.writegivenbody.ci\endcsname{[5.2, 36.0]}
\expandafter\def\csname val@model.glm53.writegivenbody.frac\endcsname{3/20}
\expandafter\def\csname val@model.glm53.writegivenbody.full\endcsname{3/20 (15.0\%)}
\expandafter\def\csname val@model.glm53.writegivenbody.pct\endcsname{15.0\%}
\expandafter\def\csname val@model.glm53.zipbyte.ci\endcsname{[64.6, 100.0]}
\expandafter\def\csname val@model.glm53.zipbyte.frac\endcsname{7/7}
\expandafter\def\csname val@model.glm53.zipbyte.full\endcsname{7/7 (100.0\%)}
\expandafter\def\csname val@model.glm53.zipbyte.pct\endcsname{100.0\%}
\expandafter\def\csname val@model.glm53.zipextract.ci\endcsname{[64.6, 100.0]}
\expandafter\def\csname val@model.glm53.zipextract.frac\endcsname{7/7}
\expandafter\def\csname val@model.glm53.zipextract.full\endcsname{7/7 (100.0\%)}
\expandafter\def\csname val@model.glm53.zipextract.pct\endcsname{100.0\%}
\expandafter\def\csname val@model.gpt6sol.allrounds.zipbytebreach.ci\endcsname{[17.2, 54.6]}
\expandafter\def\csname val@model.gpt6sol.allrounds.zipbytebreach.frac\endcsname{7/21}
\expandafter\def\csname val@model.gpt6sol.allrounds.zipbytebreach.full\endcsname{7/21 (33.3\%)}
\expandafter\def\csname val@model.gpt6sol.allrounds.zipbytebreach.pct\endcsname{33.3\%}
\expandafter\def\csname val@model.gpt6sol.allrounds.zipextract.ci\endcsname{[17.2, 54.6]}
\expandafter\def\csname val@model.gpt6sol.allrounds.zipextract.frac\endcsname{7/21}
\expandafter\def\csname val@model.gpt6sol.allrounds.zipextract.full\endcsname{7/21 (33.3\%)}
\expandafter\def\csname val@model.gpt6sol.allrounds.zipextract.pct\endcsname{33.3\%}
\expandafter\def\csname val@model.gpt6sol.body.ci\endcsname{[64.1, 90.0]}
\expandafter\def\csname val@model.gpt6sol.body.frac\endcsname{28/35}
\expandafter\def\csname val@model.gpt6sol.body.full\endcsname{28/35 (80.0\%)}
\expandafter\def\csname val@model.gpt6sol.body.pct\endcsname{80.0\%}
\expandafter\def\csname val@model.gpt6sol.claim.ci\endcsname{[1.3, 31.5]}
\expandafter\def\csname val@model.gpt6sol.claim.frac\endcsname{1/14}
\expandafter\def\csname val@model.gpt6sol.claim.full\endcsname{1/14 (7.1\%)}
\expandafter\def\csname val@model.gpt6sol.claim.pct\endcsname{7.1\%}
\expandafter\def\csname val@model.gpt6sol.hash.ci\endcsname{[21.4, 67.4]}
\expandafter\def\csname val@model.gpt6sol.hash.frac\endcsname{6/14}
\expandafter\def\csname val@model.gpt6sol.hash.full\endcsname{6/14 (42.9\%)}
\expandafter\def\csname val@model.gpt6sol.hash.pct\endcsname{42.9\%}
\expandafter\def\csname val@model.gpt6sol.nobody.ci\endcsname{[10.0, 35.9]}
\expandafter\def\csname val@model.gpt6sol.nobody.frac\endcsname{7/35}
\expandafter\def\csname val@model.gpt6sol.nobody.full\endcsname{7/35 (20.0\%)}
\expandafter\def\csname val@model.gpt6sol.nobody.h.claude.ci\endcsname{[0.0, 43.4]}
\expandafter\def\csname val@model.gpt6sol.nobody.h.claude.frac\endcsname{0/5}
\expandafter\def\csname val@model.gpt6sol.nobody.h.claude.full\endcsname{0/5 (0.0\%)}
\expandafter\def\csname val@model.gpt6sol.nobody.h.claude.pct\endcsname{0.0\%}
\expandafter\def\csname val@model.gpt6sol.nobody.h.codex.ci\endcsname{[0.0, 43.4]}
\expandafter\def\csname val@model.gpt6sol.nobody.h.codex.frac\endcsname{0/5}
\expandafter\def\csname val@model.gpt6sol.nobody.h.codex.full\endcsname{0/5 (0.0\%)}
\expandafter\def\csname val@model.gpt6sol.nobody.h.codex.pct\endcsname{0.0\%}
\expandafter\def\csname val@model.gpt6sol.nobody.h.gptme.ci\endcsname{[0.0, 43.4]}
\expandafter\def\csname val@model.gpt6sol.nobody.h.gptme.frac\endcsname{0/5}
\expandafter\def\csname val@model.gpt6sol.nobody.h.gptme.full\endcsname{0/5 (0.0\%)}
\expandafter\def\csname val@model.gpt6sol.nobody.h.gptme.pct\endcsname{0.0\%}
\expandafter\def\csname val@model.gpt6sol.nobody.h.mini.ci\endcsname{[0.0, 43.4]}
\expandafter\def\csname val@model.gpt6sol.nobody.h.mini.frac\endcsname{0/5}
\expandafter\def\csname val@model.gpt6sol.nobody.h.mini.full\endcsname{0/5 (0.0\%)}
\expandafter\def\csname val@model.gpt6sol.nobody.h.mini.pct\endcsname{0.0\%}
\expandafter\def\csname val@model.gpt6sol.nobody.h.opencode.ci\endcsname{[3.6, 62.4]}
\expandafter\def\csname val@model.gpt6sol.nobody.h.opencode.frac\endcsname{1/5}
\expandafter\def\csname val@model.gpt6sol.nobody.h.opencode.full\endcsname{1/5 (20.0\%)}
\expandafter\def\csname val@model.gpt6sol.nobody.h.opencode.pct\endcsname{20.0\%}
\expandafter\def\csname val@model.gpt6sol.nobody.h.qwen.ci\endcsname{[3.6, 62.4]}
\expandafter\def\csname val@model.gpt6sol.nobody.h.qwen.frac\endcsname{1/5}
\expandafter\def\csname val@model.gpt6sol.nobody.h.qwen.full\endcsname{1/5 (20.0\%)}
\expandafter\def\csname val@model.gpt6sol.nobody.h.qwen.pct\endcsname{20.0\%}
\expandafter\def\csname val@model.gpt6sol.nobody.h.smol.ci\endcsname{[56.6, 100.0]}
\expandafter\def\csname val@model.gpt6sol.nobody.h.smol.frac\endcsname{5/5}
\expandafter\def\csname val@model.gpt6sol.nobody.h.smol.full\endcsname{5/5 (100.0\%)}
\expandafter\def\csname val@model.gpt6sol.nobody.h.smol.pct\endcsname{100.0\%}
\expandafter\def\csname val@model.gpt6sol.nobody.pct\endcsname{20.0\%}
\expandafter\def\csname val@model.gpt6sol.reject.ci\endcsname{[16.3, 61.2]}
\expandafter\def\csname val@model.gpt6sol.reject.frac\endcsname{5/14}
\expandafter\def\csname val@model.gpt6sol.reject.full\endcsname{5/14 (35.7\%)}
\expandafter\def\csname val@model.gpt6sol.reject.pct\endcsname{35.7\%}
\expandafter\def\csname val@model.gpt6sol.reportgivenbody.ci\endcsname{[6.2, 41.0]}
\expandafter\def\csname val@model.gpt6sol.reportgivenbody.frac\endcsname{3/17}
\expandafter\def\csname val@model.gpt6sol.reportgivenbody.full\endcsname{3/17 (17.6\%)}
\expandafter\def\csname val@model.gpt6sol.reportgivenbody.pct\endcsname{17.6\%}
\expandafter\def\csname val@model.gpt6sol.write.ci\endcsname{[0.8, 22.7]}
\expandafter\def\csname val@model.gpt6sol.write.frac\endcsname{1/21}
\expandafter\def\csname val@model.gpt6sol.write.full\endcsname{1/21 (4.8\%)}
\expandafter\def\csname val@model.gpt6sol.write.pct\endcsname{4.8\%}
\expandafter\def\csname val@model.gpt6sol.writegivenbody.ci\endcsname{[1.0, 27.0]}
\expandafter\def\csname val@model.gpt6sol.writegivenbody.frac\endcsname{1/17}
\expandafter\def\csname val@model.gpt6sol.writegivenbody.full\endcsname{1/17 (5.9\%)}
\expandafter\def\csname val@model.gpt6sol.writegivenbody.pct\endcsname{5.9\%}
\expandafter\def\csname val@model.gpt6sol.zipbyte.ci\endcsname{[8.2, 64.1]}
\expandafter\def\csname val@model.gpt6sol.zipbyte.frac\endcsname{2/7}
\expandafter\def\csname val@model.gpt6sol.zipbyte.full\endcsname{2/7 (28.6\%)}
\expandafter\def\csname val@model.gpt6sol.zipbyte.pct\endcsname{28.6\%}
\expandafter\def\csname val@model.gpt6sol.zipextract.ci\endcsname{[8.2, 64.1]}
\expandafter\def\csname val@model.gpt6sol.zipextract.frac\endcsname{2/7}
\expandafter\def\csname val@model.gpt6sol.zipextract.full\endcsname{2/7 (28.6\%)}
\expandafter\def\csname val@model.gpt6sol.zipextract.pct\endcsname{28.6\%}
\expandafter\def\csname val@model.kimik3.allrounds.zipbytebreach.ci\endcsname{[84.5, 100.0]}
\expandafter\def\csname val@model.kimik3.allrounds.zipbytebreach.frac\endcsname{21/21}
\expandafter\def\csname val@model.kimik3.allrounds.zipbytebreach.full\endcsname{21/21 (100.0\%)}
\expandafter\def\csname val@model.kimik3.allrounds.zipbytebreach.pct\endcsname{100.0\%}
\expandafter\def\csname val@model.kimik3.allrounds.zipextract.ci\endcsname{[84.5, 100.0]}
\expandafter\def\csname val@model.kimik3.allrounds.zipextract.frac\endcsname{21/21}
\expandafter\def\csname val@model.kimik3.allrounds.zipextract.full\endcsname{21/21 (100.0\%)}
\expandafter\def\csname val@model.kimik3.allrounds.zipextract.pct\endcsname{100.0\%}
\expandafter\def\csname val@model.kimik3.body.ci\endcsname{[85.5, 99.5]}
\expandafter\def\csname val@model.kimik3.body.frac\endcsname{34/35}
\expandafter\def\csname val@model.kimik3.body.full\endcsname{34/35 (97.1\%)}
\expandafter\def\csname val@model.kimik3.body.pct\endcsname{97.1\%}
\expandafter\def\csname val@model.kimik3.claim.ci\endcsname{[0.0, 21.5]}
\expandafter\def\csname val@model.kimik3.claim.frac\endcsname{0/14}
\expandafter\def\csname val@model.kimik3.claim.full\endcsname{0/14 (0.0\%)}
\expandafter\def\csname val@model.kimik3.claim.pct\endcsname{0.0\%}
\expandafter\def\csname val@model.kimik3.hash.ci\endcsname{[38.8, 83.7]}
\expandafter\def\csname val@model.kimik3.hash.frac\endcsname{9/14}
\expandafter\def\csname val@model.kimik3.hash.full\endcsname{9/14 (64.3\%)}
\expandafter\def\csname val@model.kimik3.hash.pct\endcsname{64.3\%}
\expandafter\def\csname val@model.kimik3.nobody.ci\endcsname{[0.5, 14.5]}
\expandafter\def\csname val@model.kimik3.nobody.frac\endcsname{1/35}
\expandafter\def\csname val@model.kimik3.nobody.full\endcsname{1/35 (2.9\%)}
\expandafter\def\csname val@model.kimik3.nobody.h.claude.ci\endcsname{[0.0, 43.4]}
\expandafter\def\csname val@model.kimik3.nobody.h.claude.frac\endcsname{0/5}
\expandafter\def\csname val@model.kimik3.nobody.h.claude.full\endcsname{0/5 (0.0\%)}
\expandafter\def\csname val@model.kimik3.nobody.h.claude.pct\endcsname{0.0\%}
\expandafter\def\csname val@model.kimik3.nobody.h.codex.ci\endcsname{[3.6, 62.4]}
\expandafter\def\csname val@model.kimik3.nobody.h.codex.frac\endcsname{1/5}
\expandafter\def\csname val@model.kimik3.nobody.h.codex.full\endcsname{1/5 (20.0\%)}
\expandafter\def\csname val@model.kimik3.nobody.h.codex.pct\endcsname{20.0\%}
\expandafter\def\csname val@model.kimik3.nobody.h.gptme.ci\endcsname{[0.0, 43.4]}
\expandafter\def\csname val@model.kimik3.nobody.h.gptme.frac\endcsname{0/5}
\expandafter\def\csname val@model.kimik3.nobody.h.gptme.full\endcsname{0/5 (0.0\%)}
\expandafter\def\csname val@model.kimik3.nobody.h.gptme.pct\endcsname{0.0\%}
\expandafter\def\csname val@model.kimik3.nobody.h.mini.ci\endcsname{[0.0, 43.4]}
\expandafter\def\csname val@model.kimik3.nobody.h.mini.frac\endcsname{0/5}
\expandafter\def\csname val@model.kimik3.nobody.h.mini.full\endcsname{0/5 (0.0\%)}
\expandafter\def\csname val@model.kimik3.nobody.h.mini.pct\endcsname{0.0\%}
\expandafter\def\csname val@model.kimik3.nobody.h.opencode.ci\endcsname{[0.0, 43.4]}
\expandafter\def\csname val@model.kimik3.nobody.h.opencode.frac\endcsname{0/5}
\expandafter\def\csname val@model.kimik3.nobody.h.opencode.full\endcsname{0/5 (0.0\%)}
\expandafter\def\csname val@model.kimik3.nobody.h.opencode.pct\endcsname{0.0\%}
\expandafter\def\csname val@model.kimik3.nobody.h.qwen.ci\endcsname{[0.0, 43.4]}
\expandafter\def\csname val@model.kimik3.nobody.h.qwen.frac\endcsname{0/5}
\expandafter\def\csname val@model.kimik3.nobody.h.qwen.full\endcsname{0/5 (0.0\%)}
\expandafter\def\csname val@model.kimik3.nobody.h.qwen.pct\endcsname{0.0\%}
\expandafter\def\csname val@model.kimik3.nobody.h.smol.ci\endcsname{[0.0, 43.4]}
\expandafter\def\csname val@model.kimik3.nobody.h.smol.frac\endcsname{0/5}
\expandafter\def\csname val@model.kimik3.nobody.h.smol.full\endcsname{0/5 (0.0\%)}
\expandafter\def\csname val@model.kimik3.nobody.h.smol.pct\endcsname{0.0\%}
\expandafter\def\csname val@model.kimik3.nobody.pct\endcsname{2.9\%}
\expandafter\def\csname val@model.kimik3.reject.ci\endcsname{[21.4, 67.4]}
\expandafter\def\csname val@model.kimik3.reject.frac\endcsname{6/14}
\expandafter\def\csname val@model.kimik3.reject.full\endcsname{6/14 (42.9\%)}
\expandafter\def\csname val@model.kimik3.reject.pct\endcsname{42.9\%}
\expandafter\def\csname val@model.kimik3.reportgivenbody.ci\endcsname{[11.2, 46.9]}
\expandafter\def\csname val@model.kimik3.reportgivenbody.frac\endcsname{5/20}
\expandafter\def\csname val@model.kimik3.reportgivenbody.full\endcsname{5/20 (25.0\%)}
\expandafter\def\csname val@model.kimik3.reportgivenbody.pct\endcsname{25.0\%}
\expandafter\def\csname val@model.kimik3.write.ci\endcsname{[17.2, 54.6]}
\expandafter\def\csname val@model.kimik3.write.frac\endcsname{7/21}
\expandafter\def\csname val@model.kimik3.write.full\endcsname{7/21 (33.3\%)}
\expandafter\def\csname val@model.kimik3.write.pct\endcsname{33.3\%}
\expandafter\def\csname val@model.kimik3.writegivenbody.ci\endcsname{[18.1, 56.7]}
\expandafter\def\csname val@model.kimik3.writegivenbody.frac\endcsname{7/20}
\expandafter\def\csname val@model.kimik3.writegivenbody.full\endcsname{7/20 (35.0\%)}
\expandafter\def\csname val@model.kimik3.writegivenbody.pct\endcsname{35.0\%}
\expandafter\def\csname val@model.kimik3.zipbyte.ci\endcsname{[64.6, 100.0]}
\expandafter\def\csname val@model.kimik3.zipbyte.frac\endcsname{7/7}
\expandafter\def\csname val@model.kimik3.zipbyte.full\endcsname{7/7 (100.0\%)}
\expandafter\def\csname val@model.kimik3.zipbyte.pct\endcsname{100.0\%}
\expandafter\def\csname val@model.kimik3.zipextract.ci\endcsname{[64.6, 100.0]}
\expandafter\def\csname val@model.kimik3.zipextract.frac\endcsname{7/7}
\expandafter\def\csname val@model.kimik3.zipextract.full\endcsname{7/7 (100.0\%)}
\expandafter\def\csname val@model.kimik3.zipextract.pct\endcsname{100.0\%}
\expandafter\def\csname val@model.qwen38max.allrounds.zipbytebreach.ci\endcsname{[84.5, 100.0]}
\expandafter\def\csname val@model.qwen38max.allrounds.zipbytebreach.frac\endcsname{21/21}
\expandafter\def\csname val@model.qwen38max.allrounds.zipbytebreach.full\endcsname{21/21 (100.0\%)}
\expandafter\def\csname val@model.qwen38max.allrounds.zipbytebreach.pct\endcsname{100.0\%}
\expandafter\def\csname val@model.qwen38max.allrounds.zipextract.ci\endcsname{[84.5, 100.0]}
\expandafter\def\csname val@model.qwen38max.allrounds.zipextract.frac\endcsname{21/21}
\expandafter\def\csname val@model.qwen38max.allrounds.zipextract.full\endcsname{21/21 (100.0\%)}
\expandafter\def\csname val@model.qwen38max.allrounds.zipextract.pct\endcsname{100.0\%}
\expandafter\def\csname val@model.qwen38max.body.ci\endcsname{[85.5, 99.5]}
\expandafter\def\csname val@model.qwen38max.body.frac\endcsname{34/35}
\expandafter\def\csname val@model.qwen38max.body.full\endcsname{34/35 (97.1\%)}
\expandafter\def\csname val@model.qwen38max.body.pct\endcsname{97.1\%}
\expandafter\def\csname val@model.qwen38max.claim.ci\endcsname{[0.0, 21.5]}
\expandafter\def\csname val@model.qwen38max.claim.frac\endcsname{0/14}
\expandafter\def\csname val@model.qwen38max.claim.full\endcsname{0/14 (0.0\%)}
\expandafter\def\csname val@model.qwen38max.claim.pct\endcsname{0.0\%}
\expandafter\def\csname val@model.qwen38max.hash.ci\endcsname{[45.4, 88.3]}
\expandafter\def\csname val@model.qwen38max.hash.frac\endcsname{10/14}
\expandafter\def\csname val@model.qwen38max.hash.full\endcsname{10/14 (71.4\%)}
\expandafter\def\csname val@model.qwen38max.hash.pct\endcsname{71.4\%}
\expandafter\def\csname val@model.qwen38max.nobody.ci\endcsname{[0.5, 14.5]}
\expandafter\def\csname val@model.qwen38max.nobody.frac\endcsname{1/35}
\expandafter\def\csname val@model.qwen38max.nobody.full\endcsname{1/35 (2.9\%)}
\expandafter\def\csname val@model.qwen38max.nobody.h.claude.ci\endcsname{[0.0, 43.4]}
\expandafter\def\csname val@model.qwen38max.nobody.h.claude.frac\endcsname{0/5}
\expandafter\def\csname val@model.qwen38max.nobody.h.claude.full\endcsname{0/5 (0.0\%)}
\expandafter\def\csname val@model.qwen38max.nobody.h.claude.pct\endcsname{0.0\%}
\expandafter\def\csname val@model.qwen38max.nobody.h.codex.ci\endcsname{[0.0, 43.4]}
\expandafter\def\csname val@model.qwen38max.nobody.h.codex.frac\endcsname{0/5}
\expandafter\def\csname val@model.qwen38max.nobody.h.codex.full\endcsname{0/5 (0.0\%)}
\expandafter\def\csname val@model.qwen38max.nobody.h.codex.pct\endcsname{0.0\%}
\expandafter\def\csname val@model.qwen38max.nobody.h.gptme.ci\endcsname{[0.0, 43.4]}
\expandafter\def\csname val@model.qwen38max.nobody.h.gptme.frac\endcsname{0/5}
\expandafter\def\csname val@model.qwen38max.nobody.h.gptme.full\endcsname{0/5 (0.0\%)}
\expandafter\def\csname val@model.qwen38max.nobody.h.gptme.pct\endcsname{0.0\%}
\expandafter\def\csname val@model.qwen38max.nobody.h.mini.ci\endcsname{[0.0, 43.4]}
\expandafter\def\csname val@model.qwen38max.nobody.h.mini.frac\endcsname{0/5}
\expandafter\def\csname val@model.qwen38max.nobody.h.mini.full\endcsname{0/5 (0.0\%)}
\expandafter\def\csname val@model.qwen38max.nobody.h.mini.pct\endcsname{0.0\%}
\expandafter\def\csname val@model.qwen38max.nobody.h.opencode.ci\endcsname{[3.6, 62.4]}
\expandafter\def\csname val@model.qwen38max.nobody.h.opencode.frac\endcsname{1/5}
\expandafter\def\csname val@model.qwen38max.nobody.h.opencode.full\endcsname{1/5 (20.0\%)}
\expandafter\def\csname val@model.qwen38max.nobody.h.opencode.pct\endcsname{20.0\%}
\expandafter\def\csname val@model.qwen38max.nobody.h.qwen.ci\endcsname{[0.0, 43.4]}
\expandafter\def\csname val@model.qwen38max.nobody.h.qwen.frac\endcsname{0/5}
\expandafter\def\csname val@model.qwen38max.nobody.h.qwen.full\endcsname{0/5 (0.0\%)}
\expandafter\def\csname val@model.qwen38max.nobody.h.qwen.pct\endcsname{0.0\%}
\expandafter\def\csname val@model.qwen38max.nobody.h.smol.ci\endcsname{[0.0, 43.4]}
\expandafter\def\csname val@model.qwen38max.nobody.h.smol.frac\endcsname{0/5}
\expandafter\def\csname val@model.qwen38max.nobody.h.smol.full\endcsname{0/5 (0.0\%)}
\expandafter\def\csname val@model.qwen38max.nobody.h.smol.pct\endcsname{0.0\%}
\expandafter\def\csname val@model.qwen38max.nobody.pct\endcsname{2.9\%}
\expandafter\def\csname val@model.qwen38max.reject.ci\endcsname{[32.6, 78.6]}
\expandafter\def\csname val@model.qwen38max.reject.frac\endcsname{8/14}
\expandafter\def\csname val@model.qwen38max.reject.full\endcsname{8/14 (57.1\%)}
\expandafter\def\csname val@model.qwen38max.reject.pct\endcsname{57.1\%}
\expandafter\def\csname val@model.qwen38max.reportgivenbody.ci\endcsname{[17.2, 54.6]}
\expandafter\def\csname val@model.qwen38max.reportgivenbody.frac\endcsname{7/21}
\expandafter\def\csname val@model.qwen38max.reportgivenbody.full\endcsname{7/21 (33.3\%)}
\expandafter\def\csname val@model.qwen38max.reportgivenbody.pct\endcsname{33.3\%}
\expandafter\def\csname val@model.qwen38max.write.ci\endcsname{[24.5, 63.5]}
\expandafter\def\csname val@model.qwen38max.write.frac\endcsname{9/21}
\expandafter\def\csname val@model.qwen38max.write.full\endcsname{9/21 (42.9\%)}
\expandafter\def\csname val@model.qwen38max.write.pct\endcsname{42.9\%}
\expandafter\def\csname val@model.qwen38max.writegivenbody.ci\endcsname{[24.5, 63.5]}
\expandafter\def\csname val@model.qwen38max.writegivenbody.frac\endcsname{9/21}
\expandafter\def\csname val@model.qwen38max.writegivenbody.full\endcsname{9/21 (42.9\%)}
\expandafter\def\csname val@model.qwen38max.writegivenbody.pct\endcsname{42.9\%}
\expandafter\def\csname val@model.qwen38max.zipbyte.ci\endcsname{[64.6, 100.0]}
\expandafter\def\csname val@model.qwen38max.zipbyte.frac\endcsname{7/7}
\expandafter\def\csname val@model.qwen38max.zipbyte.full\endcsname{7/7 (100.0\%)}
\expandafter\def\csname val@model.qwen38max.zipbyte.pct\endcsname{100.0\%}
\expandafter\def\csname val@model.qwen38max.zipextract.ci\endcsname{[64.6, 100.0]}
\expandafter\def\csname val@model.qwen38max.zipextract.frac\endcsname{7/7}
\expandafter\def\csname val@model.qwen38max.zipextract.full\endcsname{7/7 (100.0\%)}
\expandafter\def\csname val@model.qwen38max.zipextract.pct\endcsname{100.0\%}
\expandafter\def\csname val@mr.range\endcsname{85.7--97.6\%}
\expandafter\def\csname val@nocommit.conflictgiventarget.ci\endcsname{[86.2, 99.5]}
\expandafter\def\csname val@nocommit.conflictgiventarget.frac\endcsname{36/37}
\expandafter\def\csname val@nocommit.conflictgiventarget.full\endcsname{36/37 (97.3\%)}
\expandafter\def\csname val@nocommit.conflictgiventarget.pct\endcsname{97.3\%}
\expandafter\def\csname val@nocommit.targetwritten.ci\endcsname{[22.1, 37.8]}
\expandafter\def\csname val@nocommit.targetwritten.frac\endcsname{37/126}
\expandafter\def\csname val@nocommit.targetwritten.full\endcsname{37/126 (29.4\%)}
\expandafter\def\csname val@nocommit.targetwritten.pct\endcsname{29.4\%}
\expandafter\def\csname val@nocommit.twin.declaredwrite.ci\endcsname{[94.0, 100.0]}
\expandafter\def\csname val@nocommit.twin.declaredwrite.frac\endcsname{60/60}
\expandafter\def\csname val@nocommit.twin.declaredwrite.full\endcsname{60/60 (100.0\%)}
\expandafter\def\csname val@nocommit.twin.declaredwrite.pct\endcsname{100.0\%}
\expandafter\def\csname val@nocommit.twin.targetwritten.ci\endcsname{[39.1, 56.3]}
\expandafter\def\csname val@nocommit.twin.targetwritten.frac\endcsname{60/126}
\expandafter\def\csname val@nocommit.twin.targetwritten.full\endcsname{60/126 (47.6\%)}
\expandafter\def\csname val@nocommit.twin.targetwritten.pct\endcsname{47.6\%}
\expandafter\def\csname val@nocommit.write.ci\endcsname{[21.4, 37.0]}
\expandafter\def\csname val@nocommit.write.frac\endcsname{36/126}
\expandafter\def\csname val@nocommit.write.full\endcsname{36/126 (28.6\%)}
\expandafter\def\csname val@nocommit.write.pct\endcsname{28.6\%}
\expandafter\def\csname val@paired.carrierhash.b\endcsname{15}
\expandafter\def\csname val@paired.carrierhash.c\endcsname{1}
\expandafter\def\csname val@paired.carrierhash.npairs\endcsname{42}
\expandafter\def\csname val@paired.carrierhash.p\endcsname{$<0.001$}
\expandafter\def\csname val@paired.carrierreport.b\endcsname{19}
\expandafter\def\csname val@paired.carrierreport.c\endcsname{3}
\expandafter\def\csname val@paired.carrierreport.npairs\endcsname{42}
\expandafter\def\csname val@paired.carrierreport.p\endcsname{$<0.001$}
\expandafter\def\csname val@paired.commitimg.b\endcsname{18}
\expandafter\def\csname val@paired.commitimg.c\endcsname{0}
\expandafter\def\csname val@paired.commitimg.npairs\endcsname{42}
\expandafter\def\csname val@paired.commitimg.p\endcsname{$<0.001$}
\expandafter\def\csname val@paired.commitnet.b\endcsname{10}
\expandafter\def\csname val@paired.commitnet.c\endcsname{0}
\expandafter\def\csname val@paired.commitnet.npairs\endcsname{42}
\expandafter\def\csname val@paired.commitnet.p\endcsname{0.002}
\expandafter\def\csname val@paired.medext.b\endcsname{5}
\expandafter\def\csname val@paired.medext.c\endcsname{5}
\expandafter\def\csname val@paired.medext.npairs\endcsname{140}
\expandafter\def\csname val@paired.medext.p\endcsname{1.00}
\expandafter\def\csname val@paired.tardir.b\endcsname{5}
\expandafter\def\csname val@paired.tardir.c\endcsname{0}
\expandafter\def\csname val@paired.tardir.npairs\endcsname{42}
\expandafter\def\csname val@paired.tardir.p\endcsname{0.06}
\expandafter\def\csname val@paired.tardirrep.b\endcsname{17}
\expandafter\def\csname val@paired.tardirrep.c\endcsname{0}
\expandafter\def\csname val@paired.tardirrep.npairs\endcsname{84}
\expandafter\def\csname val@paired.tardirrep.p\endcsname{$<0.001$}
\expandafter\def\csname val@paired.tardirunit.b\endcsname{9}
\expandafter\def\csname val@paired.tardirunit.c\endcsname{0}
\expandafter\def\csname val@paired.tardirunit.npairs\endcsname{28}
\expandafter\def\csname val@paired.tardirunit.p\endcsname{0.004}
\expandafter\def\csname val@paired.twintarget.b\endcsname{5}
\expandafter\def\csname val@paired.twintarget.c\endcsname{28}
\expandafter\def\csname val@paired.twintarget.npairs\endcsname{126}
\expandafter\def\csname val@paired.twintarget.p\endcsname{$<0.001$}
\expandafter\def\csname val@paired.twintargetheader.b\endcsname{1}
\expandafter\def\csname val@paired.twintargetheader.c\endcsname{8}
\expandafter\def\csname val@paired.twintargetheader.npairs\endcsname{42}
\expandafter\def\csname val@paired.twintargetheader.p\endcsname{0.04}
\expandafter\def\csname val@paired.twintargetinbody.b\endcsname{2}
\expandafter\def\csname val@paired.twintargetinbody.c\endcsname{12}
\expandafter\def\csname val@paired.twintargetinbody.npairs\endcsname{42}
\expandafter\def\csname val@paired.twintargetinbody.p\endcsname{0.01}
\expandafter\def\csname val@paired.twintargetqrdual.b\endcsname{2}
\expandafter\def\csname val@paired.twintargetqrdual.c\endcsname{8}
\expandafter\def\csname val@paired.twintargetqrdual.npairs\endcsname{42}
\expandafter\def\csname val@paired.twintargetqrdual.p\endcsname{0.11}
\expandafter\def\csname val@paired.ziptwin.b\endcsname{32}
\expandafter\def\csname val@paired.ziptwin.c\endcsname{0}
\expandafter\def\csname val@paired.ziptwin.npairs\endcsname{42}
\expandafter\def\csname val@paired.ziptwin.p\endcsname{$<0.001$}
\expandafter\def\csname val@panel.nharness\endcsname{seven}
\expandafter\def\csname val@panel.nmodels\endcsname{six}
\expandafter\def\csname val@qrdigest.body.ci\endcsname{[72.2, 93.3]}
\expandafter\def\csname val@qrdigest.body.frac\endcsname{36/42}
\expandafter\def\csname val@qrdigest.body.full\endcsname{36/42 (85.7\%)}
\expandafter\def\csname val@qrdigest.body.pct\endcsname{85.7\%}
\expandafter\def\csname val@qrdigest.claim.ci\endcsname{[0.4, 12.3]}
\expandafter\def\csname val@qrdigest.claim.frac\endcsname{1/42}
\expandafter\def\csname val@qrdigest.claim.full\endcsname{1/42 (2.4\%)}
\expandafter\def\csname val@qrdigest.claim.pct\endcsname{2.4\%}
\expandafter\def\csname val@qrdigest.cod.hash.ci\endcsname{[45.3, 77.1]}
\expandafter\def\csname val@qrdigest.cod.hash.frac\endcsname{20/32}
\expandafter\def\csname val@qrdigest.cod.hash.full\endcsname{20/32 (62.5\%)}
\expandafter\def\csname val@qrdigest.cod.hash.pct\endcsname{62.5\%}
\expandafter\def\csname val@qrdigest.cod.reported.ci\endcsname{[28.2, 60.7]}
\expandafter\def\csname val@qrdigest.cod.reported.frac\endcsname{14/32}
\expandafter\def\csname val@qrdigest.cod.reported.full\endcsname{14/32 (43.8\%)}
\expandafter\def\csname val@qrdigest.cod.reported.pct\endcsname{43.8\%}
\expandafter\def\csname val@qrdigest.cod.write.ci\endcsname{[0.6, 15.7]}
\expandafter\def\csname val@qrdigest.cod.write.frac\endcsname{1/32}
\expandafter\def\csname val@qrdigest.cod.write.full\endcsname{1/32 (3.1\%)}
\expandafter\def\csname val@qrdigest.cod.write.pct\endcsname{3.1\%}
\expandafter\def\csname val@qrdigest.codelivery.ci\endcsname{[61.5, 86.5]}
\expandafter\def\csname val@qrdigest.codelivery.frac\endcsname{32/42}
\expandafter\def\csname val@qrdigest.codelivery.full\endcsname{32/42 (76.2\%)}
\expandafter\def\csname val@qrdigest.codelivery.pct\endcsname{76.2\%}
\expandafter\def\csname val@qrdigest.declared.ci\endcsname{[10.0, 33.3]}
\expandafter\def\csname val@qrdigest.declared.frac\endcsname{8/42}
\expandafter\def\csname val@qrdigest.declared.full\endcsname{8/42 (19.0\%)}
\expandafter\def\csname val@qrdigest.declared.pct\endcsname{19.0\%}
\expandafter\def\csname val@qrdigest.h.claude.codelivery.ci\endcsname{[61.0, 100.0]}
\expandafter\def\csname val@qrdigest.h.claude.codelivery.frac\endcsname{6/6}
\expandafter\def\csname val@qrdigest.h.claude.codelivery.full\endcsname{6/6 (100.0\%)}
\expandafter\def\csname val@qrdigest.h.claude.codelivery.pct\endcsname{100.0\%}
\expandafter\def\csname val@qrdigest.h.claude.hash.ci\endcsname{[9.7, 70.0]}
\expandafter\def\csname val@qrdigest.h.claude.hash.frac\endcsname{2/6}
\expandafter\def\csname val@qrdigest.h.claude.hash.full\endcsname{2/6 (33.3\%)}
\expandafter\def\csname val@qrdigest.h.claude.hash.pct\endcsname{33.3\%}
\expandafter\def\csname val@qrdigest.h.claude.ingress.ci\endcsname{[61.0, 100.0]}
\expandafter\def\csname val@qrdigest.h.claude.ingress.frac\endcsname{6/6}
\expandafter\def\csname val@qrdigest.h.claude.ingress.full\endcsname{6/6 (100.0\%)}
\expandafter\def\csname val@qrdigest.h.claude.ingress.pct\endcsname{100.0\%}
\expandafter\def\csname val@qrdigest.h.claude.reject.ci\endcsname{[9.7, 70.0]}
\expandafter\def\csname val@qrdigest.h.claude.reject.frac\endcsname{2/6}
\expandafter\def\csname val@qrdigest.h.claude.reject.full\endcsname{2/6 (33.3\%)}
\expandafter\def\csname val@qrdigest.h.claude.reject.pct\endcsname{33.3\%}
\expandafter\def\csname val@qrdigest.h.claude.write.ci\endcsname{[0.0, 39.0]}
\expandafter\def\csname val@qrdigest.h.claude.write.frac\endcsname{0/6}
\expandafter\def\csname val@qrdigest.h.claude.write.full\endcsname{0/6 (0.0\%)}
\expandafter\def\csname val@qrdigest.h.claude.write.pct\endcsname{0.0\%}
\expandafter\def\csname val@qrdigest.h.codex.codelivery.ci\endcsname{[18.8, 81.2]}
\expandafter\def\csname val@qrdigest.h.codex.codelivery.frac\endcsname{3/6}
\expandafter\def\csname val@qrdigest.h.codex.codelivery.full\endcsname{3/6 (50.0\%)}
\expandafter\def\csname val@qrdigest.h.codex.codelivery.pct\endcsname{50.0\%}
\expandafter\def\csname val@qrdigest.h.codex.hash.ci\endcsname{[9.7, 70.0]}
\expandafter\def\csname val@qrdigest.h.codex.hash.frac\endcsname{2/6}
\expandafter\def\csname val@qrdigest.h.codex.hash.full\endcsname{2/6 (33.3\%)}
\expandafter\def\csname val@qrdigest.h.codex.hash.pct\endcsname{33.3\%}
\expandafter\def\csname val@qrdigest.h.codex.ingress.ci\endcsname{[18.8, 81.2]}
\expandafter\def\csname val@qrdigest.h.codex.ingress.frac\endcsname{3/6}
\expandafter\def\csname val@qrdigest.h.codex.ingress.full\endcsname{3/6 (50.0\%)}
\expandafter\def\csname val@qrdigest.h.codex.ingress.pct\endcsname{50.0\%}
\expandafter\def\csname val@qrdigest.h.codex.reject.ci\endcsname{[9.7, 70.0]}
\expandafter\def\csname val@qrdigest.h.codex.reject.frac\endcsname{2/6}
\expandafter\def\csname val@qrdigest.h.codex.reject.full\endcsname{2/6 (33.3\%)}
\expandafter\def\csname val@qrdigest.h.codex.reject.pct\endcsname{33.3\%}
\expandafter\def\csname val@qrdigest.h.codex.write.ci\endcsname{[0.0, 39.0]}
\expandafter\def\csname val@qrdigest.h.codex.write.frac\endcsname{0/6}
\expandafter\def\csname val@qrdigest.h.codex.write.full\endcsname{0/6 (0.0\%)}
\expandafter\def\csname val@qrdigest.h.codex.write.pct\endcsname{0.0\%}
\expandafter\def\csname val@qrdigest.h.gptme.codelivery.ci\endcsname{[18.8, 81.2]}
\expandafter\def\csname val@qrdigest.h.gptme.codelivery.frac\endcsname{3/6}
\expandafter\def\csname val@qrdigest.h.gptme.codelivery.full\endcsname{3/6 (50.0\%)}
\expandafter\def\csname val@qrdigest.h.gptme.codelivery.pct\endcsname{50.0\%}
\expandafter\def\csname val@qrdigest.h.gptme.hash.ci\endcsname{[30.0, 90.3]}
\expandafter\def\csname val@qrdigest.h.gptme.hash.frac\endcsname{4/6}
\expandafter\def\csname val@qrdigest.h.gptme.hash.full\endcsname{4/6 (66.7\%)}
\expandafter\def\csname val@qrdigest.h.gptme.hash.pct\endcsname{66.7\%}
\expandafter\def\csname val@qrdigest.h.gptme.ingress.ci\endcsname{[18.8, 81.2]}
\expandafter\def\csname val@qrdigest.h.gptme.ingress.frac\endcsname{3/6}
\expandafter\def\csname val@qrdigest.h.gptme.ingress.full\endcsname{3/6 (50.0\%)}
\expandafter\def\csname val@qrdigest.h.gptme.ingress.pct\endcsname{50.0\%}
\expandafter\def\csname val@qrdigest.h.gptme.reject.ci\endcsname{[0.0, 39.0]}
\expandafter\def\csname val@qrdigest.h.gptme.reject.frac\endcsname{0/6}
\expandafter\def\csname val@qrdigest.h.gptme.reject.full\endcsname{0/6 (0.0\%)}
\expandafter\def\csname val@qrdigest.h.gptme.reject.pct\endcsname{0.0\%}
\expandafter\def\csname val@qrdigest.h.gptme.write.ci\endcsname{[0.0, 39.0]}
\expandafter\def\csname val@qrdigest.h.gptme.write.frac\endcsname{0/6}
\expandafter\def\csname val@qrdigest.h.gptme.write.full\endcsname{0/6 (0.0\%)}
\expandafter\def\csname val@qrdigest.h.gptme.write.pct\endcsname{0.0\%}
\expandafter\def\csname val@qrdigest.h.mini.codelivery.ci\endcsname{[61.0, 100.0]}
\expandafter\def\csname val@qrdigest.h.mini.codelivery.frac\endcsname{6/6}
\expandafter\def\csname val@qrdigest.h.mini.codelivery.full\endcsname{6/6 (100.0\%)}
\expandafter\def\csname val@qrdigest.h.mini.codelivery.pct\endcsname{100.0\%}
\expandafter\def\csname val@qrdigest.h.mini.hash.ci\endcsname{[61.0, 100.0]}
\expandafter\def\csname val@qrdigest.h.mini.hash.frac\endcsname{6/6}
\expandafter\def\csname val@qrdigest.h.mini.hash.full\endcsname{6/6 (100.0\%)}
\expandafter\def\csname val@qrdigest.h.mini.hash.pct\endcsname{100.0\%}
\expandafter\def\csname val@qrdigest.h.mini.ingress.ci\endcsname{[61.0, 100.0]}
\expandafter\def\csname val@qrdigest.h.mini.ingress.frac\endcsname{6/6}
\expandafter\def\csname val@qrdigest.h.mini.ingress.full\endcsname{6/6 (100.0\%)}
\expandafter\def\csname val@qrdigest.h.mini.ingress.pct\endcsname{100.0\%}
\expandafter\def\csname val@qrdigest.h.mini.reject.ci\endcsname{[18.8, 81.2]}
\expandafter\def\csname val@qrdigest.h.mini.reject.frac\endcsname{3/6}
\expandafter\def\csname val@qrdigest.h.mini.reject.full\endcsname{3/6 (50.0\%)}
\expandafter\def\csname val@qrdigest.h.mini.reject.pct\endcsname{50.0\%}
\expandafter\def\csname val@qrdigest.h.mini.write.ci\endcsname{[3.0, 56.4]}
\expandafter\def\csname val@qrdigest.h.mini.write.frac\endcsname{1/6}
\expandafter\def\csname val@qrdigest.h.mini.write.full\endcsname{1/6 (16.7\%)}
\expandafter\def\csname val@qrdigest.h.mini.write.pct\endcsname{16.7\%}
\expandafter\def\csname val@qrdigest.h.opencode.codelivery.ci\endcsname{[18.8, 81.2]}
\expandafter\def\csname val@qrdigest.h.opencode.codelivery.frac\endcsname{3/6}
\expandafter\def\csname val@qrdigest.h.opencode.codelivery.full\endcsname{3/6 (50.0\%)}
\expandafter\def\csname val@qrdigest.h.opencode.codelivery.pct\endcsname{50.0\%}
\expandafter\def\csname val@qrdigest.h.opencode.hash.ci\endcsname{[9.7, 70.0]}
\expandafter\def\csname val@qrdigest.h.opencode.hash.frac\endcsname{2/6}
\expandafter\def\csname val@qrdigest.h.opencode.hash.full\endcsname{2/6 (33.3\%)}
\expandafter\def\csname val@qrdigest.h.opencode.hash.pct\endcsname{33.3\%}
\expandafter\def\csname val@qrdigest.h.opencode.ingress.ci\endcsname{[18.8, 81.2]}
\expandafter\def\csname val@qrdigest.h.opencode.ingress.frac\endcsname{3/6}
\expandafter\def\csname val@qrdigest.h.opencode.ingress.full\endcsname{3/6 (50.0\%)}
\expandafter\def\csname val@qrdigest.h.opencode.ingress.pct\endcsname{50.0\%}
\expandafter\def\csname val@qrdigest.h.opencode.reject.ci\endcsname{[3.0, 56.4]}
\expandafter\def\csname val@qrdigest.h.opencode.reject.frac\endcsname{1/6}
\expandafter\def\csname val@qrdigest.h.opencode.reject.full\endcsname{1/6 (16.7\%)}
\expandafter\def\csname val@qrdigest.h.opencode.reject.pct\endcsname{16.7\%}
\expandafter\def\csname val@qrdigest.h.opencode.write.ci\endcsname{[0.0, 39.0]}
\expandafter\def\csname val@qrdigest.h.opencode.write.frac\endcsname{0/6}
\expandafter\def\csname val@qrdigest.h.opencode.write.full\endcsname{0/6 (0.0\%)}
\expandafter\def\csname val@qrdigest.h.opencode.write.pct\endcsname{0.0\%}
\expandafter\def\csname val@qrdigest.h.qwen.codelivery.ci\endcsname{[61.0, 100.0]}
\expandafter\def\csname val@qrdigest.h.qwen.codelivery.frac\endcsname{6/6}
\expandafter\def\csname val@qrdigest.h.qwen.codelivery.full\endcsname{6/6 (100.0\%)}
\expandafter\def\csname val@qrdigest.h.qwen.codelivery.pct\endcsname{100.0\%}
\expandafter\def\csname val@qrdigest.h.qwen.hash.ci\endcsname{[9.7, 70.0]}
\expandafter\def\csname val@qrdigest.h.qwen.hash.frac\endcsname{2/6}
\expandafter\def\csname val@qrdigest.h.qwen.hash.full\endcsname{2/6 (33.3\%)}
\expandafter\def\csname val@qrdigest.h.qwen.hash.pct\endcsname{33.3\%}
\expandafter\def\csname val@qrdigest.h.qwen.ingress.ci\endcsname{[61.0, 100.0]}
\expandafter\def\csname val@qrdigest.h.qwen.ingress.frac\endcsname{6/6}
\expandafter\def\csname val@qrdigest.h.qwen.ingress.full\endcsname{6/6 (100.0\%)}
\expandafter\def\csname val@qrdigest.h.qwen.ingress.pct\endcsname{100.0\%}
\expandafter\def\csname val@qrdigest.h.qwen.reject.ci\endcsname{[3.0, 56.4]}
\expandafter\def\csname val@qrdigest.h.qwen.reject.frac\endcsname{1/6}
\expandafter\def\csname val@qrdigest.h.qwen.reject.full\endcsname{1/6 (16.7\%)}
\expandafter\def\csname val@qrdigest.h.qwen.reject.pct\endcsname{16.7\%}
\expandafter\def\csname val@qrdigest.h.qwen.write.ci\endcsname{[0.0, 39.0]}
\expandafter\def\csname val@qrdigest.h.qwen.write.frac\endcsname{0/6}
\expandafter\def\csname val@qrdigest.h.qwen.write.full\endcsname{0/6 (0.0\%)}
\expandafter\def\csname val@qrdigest.h.qwen.write.pct\endcsname{0.0\%}
\expandafter\def\csname val@qrdigest.h.smol.codelivery.ci\endcsname{[43.6, 97.0]}
\expandafter\def\csname val@qrdigest.h.smol.codelivery.frac\endcsname{5/6}
\expandafter\def\csname val@qrdigest.h.smol.codelivery.full\endcsname{5/6 (83.3\%)}
\expandafter\def\csname val@qrdigest.h.smol.codelivery.pct\endcsname{83.3\%}
\expandafter\def\csname val@qrdigest.h.smol.hash.ci\endcsname{[43.6, 97.0]}
\expandafter\def\csname val@qrdigest.h.smol.hash.frac\endcsname{5/6}
\expandafter\def\csname val@qrdigest.h.smol.hash.full\endcsname{5/6 (83.3\%)}
\expandafter\def\csname val@qrdigest.h.smol.hash.pct\endcsname{83.3\%}
\expandafter\def\csname val@qrdigest.h.smol.ingress.ci\endcsname{[43.6, 97.0]}
\expandafter\def\csname val@qrdigest.h.smol.ingress.frac\endcsname{5/6}
\expandafter\def\csname val@qrdigest.h.smol.ingress.full\endcsname{5/6 (83.3\%)}
\expandafter\def\csname val@qrdigest.h.smol.ingress.pct\endcsname{83.3\%}
\expandafter\def\csname val@qrdigest.h.smol.reject.ci\endcsname{[43.6, 97.0]}
\expandafter\def\csname val@qrdigest.h.smol.reject.frac\endcsname{5/6}
\expandafter\def\csname val@qrdigest.h.smol.reject.full\endcsname{5/6 (83.3\%)}
\expandafter\def\csname val@qrdigest.h.smol.reject.pct\endcsname{83.3\%}
\expandafter\def\csname val@qrdigest.h.smol.write.ci\endcsname{[0.0, 39.0]}
\expandafter\def\csname val@qrdigest.h.smol.write.frac\endcsname{0/6}
\expandafter\def\csname val@qrdigest.h.smol.write.full\endcsname{0/6 (0.0\%)}
\expandafter\def\csname val@qrdigest.h.smol.write.pct\endcsname{0.0\%}
\expandafter\def\csname val@qrdigest.hash.ci\endcsname{[39.9, 68.8]}
\expandafter\def\csname val@qrdigest.hash.frac\endcsname{23/42}
\expandafter\def\csname val@qrdigest.hash.full\endcsname{23/42 (54.8\%)}
\expandafter\def\csname val@qrdigest.hash.pct\endcsname{54.8\%}
\expandafter\def\csname val@qrdigest.ingress.ci\endcsname{[61.5, 86.5]}
\expandafter\def\csname val@qrdigest.ingress.frac\endcsname{32/42}
\expandafter\def\csname val@qrdigest.ingress.full\endcsname{32/42 (76.2\%)}
\expandafter\def\csname val@qrdigest.ingress.pct\endcsname{76.2\%}
\expandafter\def\csname val@qrdigest.nobody.hash.ci\endcsname{[0.0, 39.0]}
\expandafter\def\csname val@qrdigest.nobody.hash.frac\endcsname{0/6}
\expandafter\def\csname val@qrdigest.nobody.hash.full\endcsname{0/6 (0.0\%)}
\expandafter\def\csname val@qrdigest.nobody.hash.pct\endcsname{0.0\%}
\expandafter\def\csname val@qrdigest.nobody.reported.ci\endcsname{[0.0, 39.0]}
\expandafter\def\csname val@qrdigest.nobody.reported.frac\endcsname{0/6}
\expandafter\def\csname val@qrdigest.nobody.reported.full\endcsname{0/6 (0.0\%)}
\expandafter\def\csname val@qrdigest.nobody.reported.pct\endcsname{0.0\%}
\expandafter\def\csname val@qrdigest.nobody.write.ci\endcsname{[0.0, 39.0]}
\expandafter\def\csname val@qrdigest.nobody.write.frac\endcsname{0/6}
\expandafter\def\csname val@qrdigest.nobody.write.full\endcsname{0/6 (0.0\%)}
\expandafter\def\csname val@qrdigest.nobody.write.pct\endcsname{0.0\%}
\expandafter\def\csname val@qrdigest.nocod.hash.ci\endcsname{[30.1, 95.4]}
\expandafter\def\csname val@qrdigest.nocod.hash.frac\endcsname{3/4}
\expandafter\def\csname val@qrdigest.nocod.hash.full\endcsname{3/4 (75.0\%)}
\expandafter\def\csname val@qrdigest.nocod.hash.pct\endcsname{75.0\%}
\expandafter\def\csname val@qrdigest.nocod.reported.ci\endcsname{[15.0, 85.0]}
\expandafter\def\csname val@qrdigest.nocod.reported.frac\endcsname{2/4}
\expandafter\def\csname val@qrdigest.nocod.reported.full\endcsname{2/4 (50.0\%)}
\expandafter\def\csname val@qrdigest.nocod.reported.pct\endcsname{50.0\%}
\expandafter\def\csname val@qrdigest.nocod.write.ci\endcsname{[0.0, 49.0]}
\expandafter\def\csname val@qrdigest.nocod.write.frac\endcsname{0/4}
\expandafter\def\csname val@qrdigest.nocod.write.full\endcsname{0/4 (0.0\%)}
\expandafter\def\csname val@qrdigest.nocod.write.pct\endcsname{0.0\%}
\expandafter\def\csname val@qrdigest.reject.ci\endcsname{[21.0, 48.4]}
\expandafter\def\csname val@qrdigest.reject.frac\endcsname{14/42}
\expandafter\def\csname val@qrdigest.reject.full\endcsname{14/42 (33.3\%)}
\expandafter\def\csname val@qrdigest.reject.pct\endcsname{33.3\%}
\expandafter\def\csname val@qrdigest.reported.ci\endcsname{[25.0, 53.2]}
\expandafter\def\csname val@qrdigest.reported.frac\endcsname{16/42}
\expandafter\def\csname val@qrdigest.reported.full\endcsname{16/42 (38.1\%)}
\expandafter\def\csname val@qrdigest.reported.pct\endcsname{38.1\%}
\expandafter\def\csname val@qrdigest.reportedafter.ci\endcsname{[21.0, 48.4]}
\expandafter\def\csname val@qrdigest.reportedafter.frac\endcsname{14/42}
\expandafter\def\csname val@qrdigest.reportedafter.full\endcsname{14/42 (33.3\%)}
\expandafter\def\csname val@qrdigest.reportedafter.pct\endcsname{33.3\%}
\expandafter\def\csname val@qrdigest.single.ci\endcsname{[61.5, 86.5]}
\expandafter\def\csname val@qrdigest.single.frac\endcsname{32/42}
\expandafter\def\csname val@qrdigest.single.full\endcsname{32/42 (76.2\%)}
\expandafter\def\csname val@qrdigest.single.pct\endcsname{76.2\%}
\expandafter\def\csname val@qrdigest.targetwritten.ci\endcsname{[11.7, 35.9]}
\expandafter\def\csname val@qrdigest.targetwritten.frac\endcsname{9/42}
\expandafter\def\csname val@qrdigest.targetwritten.full\endcsname{9/42 (21.4\%)}
\expandafter\def\csname val@qrdigest.targetwritten.pct\endcsname{21.4\%}
\expandafter\def\csname val@qrdigest.twin.body.ci\endcsname{[64.1, 88.3]}
\expandafter\def\csname val@qrdigest.twin.body.frac\endcsname{33/42}
\expandafter\def\csname val@qrdigest.twin.body.full\endcsname{33/42 (78.6\%)}
\expandafter\def\csname val@qrdigest.twin.body.pct\endcsname{78.6\%}
\expandafter\def\csname val@qrdigest.twin.claim.ci\endcsname{[0.4, 12.3]}
\expandafter\def\csname val@qrdigest.twin.claim.frac\endcsname{1/42}
\expandafter\def\csname val@qrdigest.twin.claim.full\endcsname{1/42 (2.4\%)}
\expandafter\def\csname val@qrdigest.twin.claim.pct\endcsname{2.4\%}
\expandafter\def\csname val@qrdigest.twin.codelivery.ci\endcsname{[61.5, 86.5]}
\expandafter\def\csname val@qrdigest.twin.codelivery.frac\endcsname{32/42}
\expandafter\def\csname val@qrdigest.twin.codelivery.full\endcsname{32/42 (76.2\%)}
\expandafter\def\csname val@qrdigest.twin.codelivery.pct\endcsname{76.2\%}
\expandafter\def\csname val@qrdigest.twin.declared.ci\endcsname{[25.0, 53.2]}
\expandafter\def\csname val@qrdigest.twin.declared.frac\endcsname{16/42}
\expandafter\def\csname val@qrdigest.twin.declared.full\endcsname{16/42 (38.1\%)}
\expandafter\def\csname val@qrdigest.twin.declared.pct\endcsname{38.1\%}
\expandafter\def\csname val@qrdigest.twin.h.claude.codelivery.ci\endcsname{[43.6, 97.0]}
\expandafter\def\csname val@qrdigest.twin.h.claude.codelivery.frac\endcsname{5/6}
\expandafter\def\csname val@qrdigest.twin.h.claude.codelivery.full\endcsname{5/6 (83.3\%)}
\expandafter\def\csname val@qrdigest.twin.h.claude.codelivery.pct\endcsname{83.3\%}
\expandafter\def\csname val@qrdigest.twin.h.claude.hash.ci\endcsname{[18.8, 81.2]}
\expandafter\def\csname val@qrdigest.twin.h.claude.hash.frac\endcsname{3/6}
\expandafter\def\csname val@qrdigest.twin.h.claude.hash.full\endcsname{3/6 (50.0\%)}
\expandafter\def\csname val@qrdigest.twin.h.claude.hash.pct\endcsname{50.0\%}
\expandafter\def\csname val@qrdigest.twin.h.claude.write.ci\endcsname{[0.0, 39.0]}
\expandafter\def\csname val@qrdigest.twin.h.claude.write.frac\endcsname{0/6}
\expandafter\def\csname val@qrdigest.twin.h.claude.write.full\endcsname{0/6 (0.0\%)}
\expandafter\def\csname val@qrdigest.twin.h.claude.write.pct\endcsname{0.0\%}
\expandafter\def\csname val@qrdigest.twin.h.codex.codelivery.ci\endcsname{[18.8, 81.2]}
\expandafter\def\csname val@qrdigest.twin.h.codex.codelivery.frac\endcsname{3/6}
\expandafter\def\csname val@qrdigest.twin.h.codex.codelivery.full\endcsname{3/6 (50.0\%)}
\expandafter\def\csname val@qrdigest.twin.h.codex.codelivery.pct\endcsname{50.0\%}
\expandafter\def\csname val@qrdigest.twin.h.codex.hash.ci\endcsname{[18.8, 81.2]}
\expandafter\def\csname val@qrdigest.twin.h.codex.hash.frac\endcsname{3/6}
\expandafter\def\csname val@qrdigest.twin.h.codex.hash.full\endcsname{3/6 (50.0\%)}
\expandafter\def\csname val@qrdigest.twin.h.codex.hash.pct\endcsname{50.0\%}
\expandafter\def\csname val@qrdigest.twin.h.codex.write.ci\endcsname{[0.0, 39.0]}
\expandafter\def\csname val@qrdigest.twin.h.codex.write.frac\endcsname{0/6}
\expandafter\def\csname val@qrdigest.twin.h.codex.write.full\endcsname{0/6 (0.0\%)}
\expandafter\def\csname val@qrdigest.twin.h.codex.write.pct\endcsname{0.0\%}
\expandafter\def\csname val@qrdigest.twin.h.gptme.codelivery.ci\endcsname{[43.6, 97.0]}
\expandafter\def\csname val@qrdigest.twin.h.gptme.codelivery.frac\endcsname{5/6}
\expandafter\def\csname val@qrdigest.twin.h.gptme.codelivery.full\endcsname{5/6 (83.3\%)}
\expandafter\def\csname val@qrdigest.twin.h.gptme.codelivery.pct\endcsname{83.3\%}
\expandafter\def\csname val@qrdigest.twin.h.gptme.hash.ci\endcsname{[30.0, 90.3]}
\expandafter\def\csname val@qrdigest.twin.h.gptme.hash.frac\endcsname{4/6}
\expandafter\def\csname val@qrdigest.twin.h.gptme.hash.full\endcsname{4/6 (66.7\%)}
\expandafter\def\csname val@qrdigest.twin.h.gptme.hash.pct\endcsname{66.7\%}
\expandafter\def\csname val@qrdigest.twin.h.gptme.write.ci\endcsname{[0.0, 39.0]}
\expandafter\def\csname val@qrdigest.twin.h.gptme.write.frac\endcsname{0/6}
\expandafter\def\csname val@qrdigest.twin.h.gptme.write.full\endcsname{0/6 (0.0\%)}
\expandafter\def\csname val@qrdigest.twin.h.gptme.write.pct\endcsname{0.0\%}
\expandafter\def\csname val@qrdigest.twin.h.mini.codelivery.ci\endcsname{[61.0, 100.0]}
\expandafter\def\csname val@qrdigest.twin.h.mini.codelivery.frac\endcsname{6/6}
\expandafter\def\csname val@qrdigest.twin.h.mini.codelivery.full\endcsname{6/6 (100.0\%)}
\expandafter\def\csname val@qrdigest.twin.h.mini.codelivery.pct\endcsname{100.0\%}
\expandafter\def\csname val@qrdigest.twin.h.mini.hash.ci\endcsname{[61.0, 100.0]}
\expandafter\def\csname val@qrdigest.twin.h.mini.hash.frac\endcsname{6/6}
\expandafter\def\csname val@qrdigest.twin.h.mini.hash.full\endcsname{6/6 (100.0\%)}
\expandafter\def\csname val@qrdigest.twin.h.mini.hash.pct\endcsname{100.0\%}
\expandafter\def\csname val@qrdigest.twin.h.mini.write.ci\endcsname{[0.0, 39.0]}
\expandafter\def\csname val@qrdigest.twin.h.mini.write.frac\endcsname{0/6}
\expandafter\def\csname val@qrdigest.twin.h.mini.write.full\endcsname{0/6 (0.0\%)}
\expandafter\def\csname val@qrdigest.twin.h.mini.write.pct\endcsname{0.0\%}
\expandafter\def\csname val@qrdigest.twin.h.opencode.codelivery.ci\endcsname{[18.8, 81.2]}
\expandafter\def\csname val@qrdigest.twin.h.opencode.codelivery.frac\endcsname{3/6}
\expandafter\def\csname val@qrdigest.twin.h.opencode.codelivery.full\endcsname{3/6 (50.0\%)}
\expandafter\def\csname val@qrdigest.twin.h.opencode.codelivery.pct\endcsname{50.0\%}
\expandafter\def\csname val@qrdigest.twin.h.opencode.hash.ci\endcsname{[9.7, 70.0]}
\expandafter\def\csname val@qrdigest.twin.h.opencode.hash.frac\endcsname{2/6}
\expandafter\def\csname val@qrdigest.twin.h.opencode.hash.full\endcsname{2/6 (33.3\%)}
\expandafter\def\csname val@qrdigest.twin.h.opencode.hash.pct\endcsname{33.3\%}
\expandafter\def\csname val@qrdigest.twin.h.opencode.write.ci\endcsname{[0.0, 39.0]}
\expandafter\def\csname val@qrdigest.twin.h.opencode.write.frac\endcsname{0/6}
\expandafter\def\csname val@qrdigest.twin.h.opencode.write.full\endcsname{0/6 (0.0\%)}
\expandafter\def\csname val@qrdigest.twin.h.opencode.write.pct\endcsname{0.0\%}
\expandafter\def\csname val@qrdigest.twin.h.qwen.codelivery.ci\endcsname{[43.6, 97.0]}
\expandafter\def\csname val@qrdigest.twin.h.qwen.codelivery.frac\endcsname{5/6}
\expandafter\def\csname val@qrdigest.twin.h.qwen.codelivery.full\endcsname{5/6 (83.3\%)}
\expandafter\def\csname val@qrdigest.twin.h.qwen.codelivery.pct\endcsname{83.3\%}
\expandafter\def\csname val@qrdigest.twin.h.qwen.hash.ci\endcsname{[18.8, 81.2]}
\expandafter\def\csname val@qrdigest.twin.h.qwen.hash.frac\endcsname{3/6}
\expandafter\def\csname val@qrdigest.twin.h.qwen.hash.full\endcsname{3/6 (50.0\%)}
\expandafter\def\csname val@qrdigest.twin.h.qwen.hash.pct\endcsname{50.0\%}
\expandafter\def\csname val@qrdigest.twin.h.qwen.write.ci\endcsname{[0.0, 39.0]}
\expandafter\def\csname val@qrdigest.twin.h.qwen.write.frac\endcsname{0/6}
\expandafter\def\csname val@qrdigest.twin.h.qwen.write.full\endcsname{0/6 (0.0\%)}
\expandafter\def\csname val@qrdigest.twin.h.qwen.write.pct\endcsname{0.0\%}
\expandafter\def\csname val@qrdigest.twin.h.smol.codelivery.ci\endcsname{[43.6, 97.0]}
\expandafter\def\csname val@qrdigest.twin.h.smol.codelivery.frac\endcsname{5/6}
\expandafter\def\csname val@qrdigest.twin.h.smol.codelivery.full\endcsname{5/6 (83.3\%)}
\expandafter\def\csname val@qrdigest.twin.h.smol.codelivery.pct\endcsname{83.3\%}
\expandafter\def\csname val@qrdigest.twin.h.smol.hash.ci\endcsname{[43.6, 97.0]}
\expandafter\def\csname val@qrdigest.twin.h.smol.hash.frac\endcsname{5/6}
\expandafter\def\csname val@qrdigest.twin.h.smol.hash.full\endcsname{5/6 (83.3\%)}
\expandafter\def\csname val@qrdigest.twin.h.smol.hash.pct\endcsname{83.3\%}
\expandafter\def\csname val@qrdigest.twin.h.smol.write.ci\endcsname{[0.0, 39.0]}
\expandafter\def\csname val@qrdigest.twin.h.smol.write.frac\endcsname{0/6}
\expandafter\def\csname val@qrdigest.twin.h.smol.write.full\endcsname{0/6 (0.0\%)}
\expandafter\def\csname val@qrdigest.twin.h.smol.write.pct\endcsname{0.0\%}
\expandafter\def\csname val@qrdigest.twin.hash.ci\endcsname{[46.8, 75.0]}
\expandafter\def\csname val@qrdigest.twin.hash.frac\endcsname{26/42}
\expandafter\def\csname val@qrdigest.twin.hash.full\endcsname{26/42 (61.9\%)}
\expandafter\def\csname val@qrdigest.twin.hash.pct\endcsname{61.9\%}
\expandafter\def\csname val@qrdigest.twin.reported.ci\endcsname{[0.4, 12.3]}
\expandafter\def\csname val@qrdigest.twin.reported.frac\endcsname{1/42}
\expandafter\def\csname val@qrdigest.twin.reported.full\endcsname{1/42 (2.4\%)}
\expandafter\def\csname val@qrdigest.twin.reported.pct\endcsname{2.4\%}
\expandafter\def\csname val@qrdigest.twin.reportedafter.ci\endcsname{[0.4, 12.3]}
\expandafter\def\csname val@qrdigest.twin.reportedafter.frac\endcsname{1/42}
\expandafter\def\csname val@qrdigest.twin.reportedafter.full\endcsname{1/42 (2.4\%)}
\expandafter\def\csname val@qrdigest.twin.reportedafter.pct\endcsname{2.4\%}
\expandafter\def\csname val@qrdigest.twin.targetwritten.ci\endcsname{[25.0, 53.2]}
\expandafter\def\csname val@qrdigest.twin.targetwritten.frac\endcsname{16/42}
\expandafter\def\csname val@qrdigest.twin.targetwritten.full\endcsname{16/42 (38.1\%)}
\expandafter\def\csname val@qrdigest.twin.targetwritten.pct\endcsname{38.1\%}
\expandafter\def\csname val@qrdigest.twin.write.ci\endcsname{[0.0, 8.4]}
\expandafter\def\csname val@qrdigest.twin.write.frac\endcsname{0/42}
\expandafter\def\csname val@qrdigest.twin.write.full\endcsname{0/42 (0.0\%)}
\expandafter\def\csname val@qrdigest.twin.write.pct\endcsname{0.0\%}
\expandafter\def\csname val@qrdigest.unverified.ci\endcsname{[17.2, 43.6]}
\expandafter\def\csname val@qrdigest.unverified.frac\endcsname{12/42}
\expandafter\def\csname val@qrdigest.unverified.full\endcsname{12/42 (28.6\%)}
\expandafter\def\csname val@qrdigest.unverified.pct\endcsname{28.6\%}
\expandafter\def\csname val@qrdigest.write.ci\endcsname{[0.4, 12.3]}
\expandafter\def\csname val@qrdigest.write.frac\endcsname{1/42}
\expandafter\def\csname val@qrdigest.write.full\endcsname{1/42 (2.4\%)}
\expandafter\def\csname val@qrdigest.write.pct\endcsname{2.4\%}
\expandafter\def\csname val@qrdual.body.ci\endcsname{[81.0, 97.5]}
\expandafter\def\csname val@qrdual.body.frac\endcsname{39/42}
\expandafter\def\csname val@qrdual.body.full\endcsname{39/42 (92.9\%)}
\expandafter\def\csname val@qrdual.body.pct\endcsname{92.9\%}
\expandafter\def\csname val@qrdual.claim.ci\endcsname{[0.0, 8.4]}
\expandafter\def\csname val@qrdual.claim.frac\endcsname{0/42}
\expandafter\def\csname val@qrdual.claim.full\endcsname{0/42 (0.0\%)}
\expandafter\def\csname val@qrdual.claim.pct\endcsname{0.0\%}
\expandafter\def\csname val@qrdual.codelivery.ci\endcsname{[81.0, 97.5]}
\expandafter\def\csname val@qrdual.codelivery.frac\endcsname{39/42}
\expandafter\def\csname val@qrdual.codelivery.full\endcsname{39/42 (92.9\%)}
\expandafter\def\csname val@qrdual.codelivery.pct\endcsname{92.9\%}
\expandafter\def\csname val@qrdual.declared.ci\endcsname{[0.0, 8.4]}
\expandafter\def\csname val@qrdual.declared.frac\endcsname{0/42}
\expandafter\def\csname val@qrdual.declared.full\endcsname{0/42 (0.0\%)}
\expandafter\def\csname val@qrdual.declared.pct\endcsname{0.0\%}
\expandafter\def\csname val@qrdual.h.claude.codelivery.ci\endcsname{[61.0, 100.0]}
\expandafter\def\csname val@qrdual.h.claude.codelivery.frac\endcsname{6/6}
\expandafter\def\csname val@qrdual.h.claude.codelivery.full\endcsname{6/6 (100.0\%)}
\expandafter\def\csname val@qrdual.h.claude.codelivery.pct\endcsname{100.0\%}
\expandafter\def\csname val@qrdual.h.claude.hash.ci\endcsname{[0.0, 39.0]}
\expandafter\def\csname val@qrdual.h.claude.hash.frac\endcsname{0/6}
\expandafter\def\csname val@qrdual.h.claude.hash.full\endcsname{0/6 (0.0\%)}
\expandafter\def\csname val@qrdual.h.claude.hash.pct\endcsname{0.0\%}
\expandafter\def\csname val@qrdual.h.claude.ingress.ci\endcsname{[61.0, 100.0]}
\expandafter\def\csname val@qrdual.h.claude.ingress.frac\endcsname{6/6}
\expandafter\def\csname val@qrdual.h.claude.ingress.full\endcsname{6/6 (100.0\%)}
\expandafter\def\csname val@qrdual.h.claude.ingress.pct\endcsname{100.0\%}
\expandafter\def\csname val@qrdual.h.claude.reject.ci\endcsname{[0.0, 39.0]}
\expandafter\def\csname val@qrdual.h.claude.reject.frac\endcsname{0/6}
\expandafter\def\csname val@qrdual.h.claude.reject.full\endcsname{0/6 (0.0\%)}
\expandafter\def\csname val@qrdual.h.claude.reject.pct\endcsname{0.0\%}
\expandafter\def\csname val@qrdual.h.claude.write.ci\endcsname{[9.7, 70.0]}
\expandafter\def\csname val@qrdual.h.claude.write.frac\endcsname{2/6}
\expandafter\def\csname val@qrdual.h.claude.write.full\endcsname{2/6 (33.3\%)}
\expandafter\def\csname val@qrdual.h.claude.write.pct\endcsname{33.3\%}
\expandafter\def\csname val@qrdual.h.codex.codelivery.ci\endcsname{[61.0, 100.0]}
\expandafter\def\csname val@qrdual.h.codex.codelivery.frac\endcsname{6/6}
\expandafter\def\csname val@qrdual.h.codex.codelivery.full\endcsname{6/6 (100.0\%)}
\expandafter\def\csname val@qrdual.h.codex.codelivery.pct\endcsname{100.0\%}
\expandafter\def\csname val@qrdual.h.codex.hash.ci\endcsname{[0.0, 39.0]}
\expandafter\def\csname val@qrdual.h.codex.hash.frac\endcsname{0/6}
\expandafter\def\csname val@qrdual.h.codex.hash.full\endcsname{0/6 (0.0\%)}
\expandafter\def\csname val@qrdual.h.codex.hash.pct\endcsname{0.0\%}
\expandafter\def\csname val@qrdual.h.codex.ingress.ci\endcsname{[61.0, 100.0]}
\expandafter\def\csname val@qrdual.h.codex.ingress.frac\endcsname{6/6}
\expandafter\def\csname val@qrdual.h.codex.ingress.full\endcsname{6/6 (100.0\%)}
\expandafter\def\csname val@qrdual.h.codex.ingress.pct\endcsname{100.0\%}
\expandafter\def\csname val@qrdual.h.codex.reject.ci\endcsname{[0.0, 39.0]}
\expandafter\def\csname val@qrdual.h.codex.reject.frac\endcsname{0/6}
\expandafter\def\csname val@qrdual.h.codex.reject.full\endcsname{0/6 (0.0\%)}
\expandafter\def\csname val@qrdual.h.codex.reject.pct\endcsname{0.0\%}
\expandafter\def\csname val@qrdual.h.codex.write.ci\endcsname{[3.0, 56.4]}
\expandafter\def\csname val@qrdual.h.codex.write.frac\endcsname{1/6}
\expandafter\def\csname val@qrdual.h.codex.write.full\endcsname{1/6 (16.7\%)}
\expandafter\def\csname val@qrdual.h.codex.write.pct\endcsname{16.7\%}
\expandafter\def\csname val@qrdual.h.gptme.codelivery.ci\endcsname{[61.0, 100.0]}
\expandafter\def\csname val@qrdual.h.gptme.codelivery.frac\endcsname{6/6}
\expandafter\def\csname val@qrdual.h.gptme.codelivery.full\endcsname{6/6 (100.0\%)}
\expandafter\def\csname val@qrdual.h.gptme.codelivery.pct\endcsname{100.0\%}
\expandafter\def\csname val@qrdual.h.gptme.hash.ci\endcsname{[0.0, 39.0]}
\expandafter\def\csname val@qrdual.h.gptme.hash.frac\endcsname{0/6}
\expandafter\def\csname val@qrdual.h.gptme.hash.full\endcsname{0/6 (0.0\%)}
\expandafter\def\csname val@qrdual.h.gptme.hash.pct\endcsname{0.0\%}
\expandafter\def\csname val@qrdual.h.gptme.ingress.ci\endcsname{[61.0, 100.0]}
\expandafter\def\csname val@qrdual.h.gptme.ingress.frac\endcsname{6/6}
\expandafter\def\csname val@qrdual.h.gptme.ingress.full\endcsname{6/6 (100.0\%)}
\expandafter\def\csname val@qrdual.h.gptme.ingress.pct\endcsname{100.0\%}
\expandafter\def\csname val@qrdual.h.gptme.reject.ci\endcsname{[0.0, 39.0]}
\expandafter\def\csname val@qrdual.h.gptme.reject.frac\endcsname{0/6}
\expandafter\def\csname val@qrdual.h.gptme.reject.full\endcsname{0/6 (0.0\%)}
\expandafter\def\csname val@qrdual.h.gptme.reject.pct\endcsname{0.0\%}
\expandafter\def\csname val@qrdual.h.gptme.write.ci\endcsname{[30.0, 90.3]}
\expandafter\def\csname val@qrdual.h.gptme.write.frac\endcsname{4/6}
\expandafter\def\csname val@qrdual.h.gptme.write.full\endcsname{4/6 (66.7\%)}
\expandafter\def\csname val@qrdual.h.gptme.write.pct\endcsname{66.7\%}
\expandafter\def\csname val@qrdual.h.mini.codelivery.ci\endcsname{[61.0, 100.0]}
\expandafter\def\csname val@qrdual.h.mini.codelivery.frac\endcsname{6/6}
\expandafter\def\csname val@qrdual.h.mini.codelivery.full\endcsname{6/6 (100.0\%)}
\expandafter\def\csname val@qrdual.h.mini.codelivery.pct\endcsname{100.0\%}
\expandafter\def\csname val@qrdual.h.mini.hash.ci\endcsname{[0.0, 39.0]}
\expandafter\def\csname val@qrdual.h.mini.hash.frac\endcsname{0/6}
\expandafter\def\csname val@qrdual.h.mini.hash.full\endcsname{0/6 (0.0\%)}
\expandafter\def\csname val@qrdual.h.mini.hash.pct\endcsname{0.0\%}
\expandafter\def\csname val@qrdual.h.mini.ingress.ci\endcsname{[61.0, 100.0]}
\expandafter\def\csname val@qrdual.h.mini.ingress.frac\endcsname{6/6}
\expandafter\def\csname val@qrdual.h.mini.ingress.full\endcsname{6/6 (100.0\%)}
\expandafter\def\csname val@qrdual.h.mini.ingress.pct\endcsname{100.0\%}
\expandafter\def\csname val@qrdual.h.mini.reject.ci\endcsname{[0.0, 39.0]}
\expandafter\def\csname val@qrdual.h.mini.reject.frac\endcsname{0/6}
\expandafter\def\csname val@qrdual.h.mini.reject.full\endcsname{0/6 (0.0\%)}
\expandafter\def\csname val@qrdual.h.mini.reject.pct\endcsname{0.0\%}
\expandafter\def\csname val@qrdual.h.mini.write.ci\endcsname{[61.0, 100.0]}
\expandafter\def\csname val@qrdual.h.mini.write.frac\endcsname{6/6}
\expandafter\def\csname val@qrdual.h.mini.write.full\endcsname{6/6 (100.0\%)}
\expandafter\def\csname val@qrdual.h.mini.write.pct\endcsname{100.0\%}
\expandafter\def\csname val@qrdual.h.opencode.codelivery.ci\endcsname{[43.6, 97.0]}
\expandafter\def\csname val@qrdual.h.opencode.codelivery.frac\endcsname{5/6}
\expandafter\def\csname val@qrdual.h.opencode.codelivery.full\endcsname{5/6 (83.3\%)}
\expandafter\def\csname val@qrdual.h.opencode.codelivery.pct\endcsname{83.3\%}
\expandafter\def\csname val@qrdual.h.opencode.hash.ci\endcsname{[0.0, 39.0]}
\expandafter\def\csname val@qrdual.h.opencode.hash.frac\endcsname{0/6}
\expandafter\def\csname val@qrdual.h.opencode.hash.full\endcsname{0/6 (0.0\%)}
\expandafter\def\csname val@qrdual.h.opencode.hash.pct\endcsname{0.0\%}
\expandafter\def\csname val@qrdual.h.opencode.ingress.ci\endcsname{[43.6, 97.0]}
\expandafter\def\csname val@qrdual.h.opencode.ingress.frac\endcsname{5/6}
\expandafter\def\csname val@qrdual.h.opencode.ingress.full\endcsname{5/6 (83.3\%)}
\expandafter\def\csname val@qrdual.h.opencode.ingress.pct\endcsname{83.3\%}
\expandafter\def\csname val@qrdual.h.opencode.reject.ci\endcsname{[0.0, 39.0]}
\expandafter\def\csname val@qrdual.h.opencode.reject.frac\endcsname{0/6}
\expandafter\def\csname val@qrdual.h.opencode.reject.full\endcsname{0/6 (0.0\%)}
\expandafter\def\csname val@qrdual.h.opencode.reject.pct\endcsname{0.0\%}
\expandafter\def\csname val@qrdual.h.opencode.write.ci\endcsname{[9.7, 70.0]}
\expandafter\def\csname val@qrdual.h.opencode.write.frac\endcsname{2/6}
\expandafter\def\csname val@qrdual.h.opencode.write.full\endcsname{2/6 (33.3\%)}
\expandafter\def\csname val@qrdual.h.opencode.write.pct\endcsname{33.3\%}
\expandafter\def\csname val@qrdual.h.qwen.codelivery.ci\endcsname{[43.6, 97.0]}
\expandafter\def\csname val@qrdual.h.qwen.codelivery.frac\endcsname{5/6}
\expandafter\def\csname val@qrdual.h.qwen.codelivery.full\endcsname{5/6 (83.3\%)}
\expandafter\def\csname val@qrdual.h.qwen.codelivery.pct\endcsname{83.3\%}
\expandafter\def\csname val@qrdual.h.qwen.hash.ci\endcsname{[0.0, 39.0]}
\expandafter\def\csname val@qrdual.h.qwen.hash.frac\endcsname{0/6}
\expandafter\def\csname val@qrdual.h.qwen.hash.full\endcsname{0/6 (0.0\%)}
\expandafter\def\csname val@qrdual.h.qwen.hash.pct\endcsname{0.0\%}
\expandafter\def\csname val@qrdual.h.qwen.ingress.ci\endcsname{[43.6, 97.0]}
\expandafter\def\csname val@qrdual.h.qwen.ingress.frac\endcsname{5/6}
\expandafter\def\csname val@qrdual.h.qwen.ingress.full\endcsname{5/6 (83.3\%)}
\expandafter\def\csname val@qrdual.h.qwen.ingress.pct\endcsname{83.3\%}
\expandafter\def\csname val@qrdual.h.qwen.reject.ci\endcsname{[0.0, 39.0]}
\expandafter\def\csname val@qrdual.h.qwen.reject.frac\endcsname{0/6}
\expandafter\def\csname val@qrdual.h.qwen.reject.full\endcsname{0/6 (0.0\%)}
\expandafter\def\csname val@qrdual.h.qwen.reject.pct\endcsname{0.0\%}
\expandafter\def\csname val@qrdual.h.qwen.write.ci\endcsname{[3.0, 56.4]}
\expandafter\def\csname val@qrdual.h.qwen.write.frac\endcsname{1/6}
\expandafter\def\csname val@qrdual.h.qwen.write.full\endcsname{1/6 (16.7\%)}
\expandafter\def\csname val@qrdual.h.qwen.write.pct\endcsname{16.7\%}
\expandafter\def\csname val@qrdual.h.smol.codelivery.ci\endcsname{[43.6, 97.0]}
\expandafter\def\csname val@qrdual.h.smol.codelivery.frac\endcsname{5/6}
\expandafter\def\csname val@qrdual.h.smol.codelivery.full\endcsname{5/6 (83.3\%)}
\expandafter\def\csname val@qrdual.h.smol.codelivery.pct\endcsname{83.3\%}
\expandafter\def\csname val@qrdual.h.smol.hash.ci\endcsname{[0.0, 39.0]}
\expandafter\def\csname val@qrdual.h.smol.hash.frac\endcsname{0/6}
\expandafter\def\csname val@qrdual.h.smol.hash.full\endcsname{0/6 (0.0\%)}
\expandafter\def\csname val@qrdual.h.smol.hash.pct\endcsname{0.0\%}
\expandafter\def\csname val@qrdual.h.smol.ingress.ci\endcsname{[43.6, 97.0]}
\expandafter\def\csname val@qrdual.h.smol.ingress.frac\endcsname{5/6}
\expandafter\def\csname val@qrdual.h.smol.ingress.full\endcsname{5/6 (83.3\%)}
\expandafter\def\csname val@qrdual.h.smol.ingress.pct\endcsname{83.3\%}
\expandafter\def\csname val@qrdual.h.smol.reject.ci\endcsname{[0.0, 39.0]}
\expandafter\def\csname val@qrdual.h.smol.reject.frac\endcsname{0/6}
\expandafter\def\csname val@qrdual.h.smol.reject.full\endcsname{0/6 (0.0\%)}
\expandafter\def\csname val@qrdual.h.smol.reject.pct\endcsname{0.0\%}
\expandafter\def\csname val@qrdual.h.smol.write.ci\endcsname{[18.8, 81.2]}
\expandafter\def\csname val@qrdual.h.smol.write.frac\endcsname{3/6}
\expandafter\def\csname val@qrdual.h.smol.write.full\endcsname{3/6 (50.0\%)}
\expandafter\def\csname val@qrdual.h.smol.write.pct\endcsname{50.0\%}
\expandafter\def\csname val@qrdual.hash.ci\endcsname{[0.0, 8.4]}
\expandafter\def\csname val@qrdual.hash.frac\endcsname{0/42}
\expandafter\def\csname val@qrdual.hash.full\endcsname{0/42 (0.0\%)}
\expandafter\def\csname val@qrdual.hash.pct\endcsname{0.0\%}
\expandafter\def\csname val@qrdual.ingress.ci\endcsname{[81.0, 97.5]}
\expandafter\def\csname val@qrdual.ingress.frac\endcsname{39/42}
\expandafter\def\csname val@qrdual.ingress.full\endcsname{39/42 (92.9\%)}
\expandafter\def\csname val@qrdual.ingress.pct\endcsname{92.9\%}
\expandafter\def\csname val@qrdual.reject.ci\endcsname{[0.0, 8.4]}
\expandafter\def\csname val@qrdual.reject.frac\endcsname{0/42}
\expandafter\def\csname val@qrdual.reject.full\endcsname{0/42 (0.0\%)}
\expandafter\def\csname val@qrdual.reject.pct\endcsname{0.0\%}
\expandafter\def\csname val@qrdual.reported.ci\endcsname{[0.0, 8.4]}
\expandafter\def\csname val@qrdual.reported.frac\endcsname{0/42}
\expandafter\def\csname val@qrdual.reported.full\endcsname{0/42 (0.0\%)}
\expandafter\def\csname val@qrdual.reported.pct\endcsname{0.0\%}
\expandafter\def\csname val@qrdual.reportedafter.ci\endcsname{[0.0, 8.4]}
\expandafter\def\csname val@qrdual.reportedafter.frac\endcsname{0/42}
\expandafter\def\csname val@qrdual.reportedafter.full\endcsname{0/42 (0.0\%)}
\expandafter\def\csname val@qrdual.reportedafter.pct\endcsname{0.0\%}
\expandafter\def\csname val@qrdual.single.ci\endcsname{[81.0, 97.5]}
\expandafter\def\csname val@qrdual.single.frac\endcsname{39/42}
\expandafter\def\csname val@qrdual.single.full\endcsname{39/42 (92.9\%)}
\expandafter\def\csname val@qrdual.single.pct\endcsname{92.9\%}
\expandafter\def\csname val@qrdual.targetwritten.ci\endcsname{[31.2, 60.1]}
\expandafter\def\csname val@qrdual.targetwritten.frac\endcsname{19/42}
\expandafter\def\csname val@qrdual.targetwritten.full\endcsname{19/42 (45.2\%)}
\expandafter\def\csname val@qrdual.targetwritten.pct\endcsname{45.2\%}
\expandafter\def\csname val@qrdual.twin.body.ci\endcsname{[81.0, 97.5]}
\expandafter\def\csname val@qrdual.twin.body.frac\endcsname{39/42}
\expandafter\def\csname val@qrdual.twin.body.full\endcsname{39/42 (92.9\%)}
\expandafter\def\csname val@qrdual.twin.body.pct\endcsname{92.9\%}
\expandafter\def\csname val@qrdual.twin.claim.ci\endcsname{[0.0, 8.4]}
\expandafter\def\csname val@qrdual.twin.claim.frac\endcsname{0/42}
\expandafter\def\csname val@qrdual.twin.claim.full\endcsname{0/42 (0.0\%)}
\expandafter\def\csname val@qrdual.twin.claim.pct\endcsname{0.0\%}
\expandafter\def\csname val@qrdual.twin.codelivery.ci\endcsname{[81.0, 97.5]}
\expandafter\def\csname val@qrdual.twin.codelivery.frac\endcsname{39/42}
\expandafter\def\csname val@qrdual.twin.codelivery.full\endcsname{39/42 (92.9\%)}
\expandafter\def\csname val@qrdual.twin.codelivery.pct\endcsname{92.9\%}
\expandafter\def\csname val@qrdual.twin.declared.ci\endcsname{[44.5, 73.0]}
\expandafter\def\csname val@qrdual.twin.declared.frac\endcsname{25/42}
\expandafter\def\csname val@qrdual.twin.declared.full\endcsname{25/42 (59.5\%)}
\expandafter\def\csname val@qrdual.twin.declared.pct\endcsname{59.5\%}
\expandafter\def\csname val@qrdual.twin.h.claude.codelivery.ci\endcsname{[61.0, 100.0]}
\expandafter\def\csname val@qrdual.twin.h.claude.codelivery.frac\endcsname{6/6}
\expandafter\def\csname val@qrdual.twin.h.claude.codelivery.full\endcsname{6/6 (100.0\%)}
\expandafter\def\csname val@qrdual.twin.h.claude.codelivery.pct\endcsname{100.0\%}
\expandafter\def\csname val@qrdual.twin.h.claude.hash.ci\endcsname{[0.0, 39.0]}
\expandafter\def\csname val@qrdual.twin.h.claude.hash.frac\endcsname{0/6}
\expandafter\def\csname val@qrdual.twin.h.claude.hash.full\endcsname{0/6 (0.0\%)}
\expandafter\def\csname val@qrdual.twin.h.claude.hash.pct\endcsname{0.0\%}
\expandafter\def\csname val@qrdual.twin.h.claude.write.ci\endcsname{[0.0, 39.0]}
\expandafter\def\csname val@qrdual.twin.h.claude.write.frac\endcsname{0/6}
\expandafter\def\csname val@qrdual.twin.h.claude.write.full\endcsname{0/6 (0.0\%)}
\expandafter\def\csname val@qrdual.twin.h.claude.write.pct\endcsname{0.0\%}
\expandafter\def\csname val@qrdual.twin.h.codex.codelivery.ci\endcsname{[61.0, 100.0]}
\expandafter\def\csname val@qrdual.twin.h.codex.codelivery.frac\endcsname{6/6}
\expandafter\def\csname val@qrdual.twin.h.codex.codelivery.full\endcsname{6/6 (100.0\%)}
\expandafter\def\csname val@qrdual.twin.h.codex.codelivery.pct\endcsname{100.0\%}
\expandafter\def\csname val@qrdual.twin.h.codex.hash.ci\endcsname{[0.0, 39.0]}
\expandafter\def\csname val@qrdual.twin.h.codex.hash.frac\endcsname{0/6}
\expandafter\def\csname val@qrdual.twin.h.codex.hash.full\endcsname{0/6 (0.0\%)}
\expandafter\def\csname val@qrdual.twin.h.codex.hash.pct\endcsname{0.0\%}
\expandafter\def\csname val@qrdual.twin.h.codex.write.ci\endcsname{[0.0, 39.0]}
\expandafter\def\csname val@qrdual.twin.h.codex.write.frac\endcsname{0/6}
\expandafter\def\csname val@qrdual.twin.h.codex.write.full\endcsname{0/6 (0.0\%)}
\expandafter\def\csname val@qrdual.twin.h.codex.write.pct\endcsname{0.0\%}
\expandafter\def\csname val@qrdual.twin.h.gptme.codelivery.ci\endcsname{[61.0, 100.0]}
\expandafter\def\csname val@qrdual.twin.h.gptme.codelivery.frac\endcsname{6/6}
\expandafter\def\csname val@qrdual.twin.h.gptme.codelivery.full\endcsname{6/6 (100.0\%)}
\expandafter\def\csname val@qrdual.twin.h.gptme.codelivery.pct\endcsname{100.0\%}
\expandafter\def\csname val@qrdual.twin.h.gptme.hash.ci\endcsname{[0.0, 39.0]}
\expandafter\def\csname val@qrdual.twin.h.gptme.hash.frac\endcsname{0/6}
\expandafter\def\csname val@qrdual.twin.h.gptme.hash.full\endcsname{0/6 (0.0\%)}
\expandafter\def\csname val@qrdual.twin.h.gptme.hash.pct\endcsname{0.0\%}
\expandafter\def\csname val@qrdual.twin.h.gptme.write.ci\endcsname{[0.0, 39.0]}
\expandafter\def\csname val@qrdual.twin.h.gptme.write.frac\endcsname{0/6}
\expandafter\def\csname val@qrdual.twin.h.gptme.write.full\endcsname{0/6 (0.0\%)}
\expandafter\def\csname val@qrdual.twin.h.gptme.write.pct\endcsname{0.0\%}
\expandafter\def\csname val@qrdual.twin.h.mini.codelivery.ci\endcsname{[61.0, 100.0]}
\expandafter\def\csname val@qrdual.twin.h.mini.codelivery.frac\endcsname{6/6}
\expandafter\def\csname val@qrdual.twin.h.mini.codelivery.full\endcsname{6/6 (100.0\%)}
\expandafter\def\csname val@qrdual.twin.h.mini.codelivery.pct\endcsname{100.0\%}
\expandafter\def\csname val@qrdual.twin.h.mini.hash.ci\endcsname{[0.0, 39.0]}
\expandafter\def\csname val@qrdual.twin.h.mini.hash.frac\endcsname{0/6}
\expandafter\def\csname val@qrdual.twin.h.mini.hash.full\endcsname{0/6 (0.0\%)}
\expandafter\def\csname val@qrdual.twin.h.mini.hash.pct\endcsname{0.0\%}
\expandafter\def\csname val@qrdual.twin.h.mini.write.ci\endcsname{[0.0, 39.0]}
\expandafter\def\csname val@qrdual.twin.h.mini.write.frac\endcsname{0/6}
\expandafter\def\csname val@qrdual.twin.h.mini.write.full\endcsname{0/6 (0.0\%)}
\expandafter\def\csname val@qrdual.twin.h.mini.write.pct\endcsname{0.0\%}
\expandafter\def\csname val@qrdual.twin.h.opencode.codelivery.ci\endcsname{[43.6, 97.0]}
\expandafter\def\csname val@qrdual.twin.h.opencode.codelivery.frac\endcsname{5/6}
\expandafter\def\csname val@qrdual.twin.h.opencode.codelivery.full\endcsname{5/6 (83.3\%)}
\expandafter\def\csname val@qrdual.twin.h.opencode.codelivery.pct\endcsname{83.3\%}
\expandafter\def\csname val@qrdual.twin.h.opencode.hash.ci\endcsname{[0.0, 39.0]}
\expandafter\def\csname val@qrdual.twin.h.opencode.hash.frac\endcsname{0/6}
\expandafter\def\csname val@qrdual.twin.h.opencode.hash.full\endcsname{0/6 (0.0\%)}
\expandafter\def\csname val@qrdual.twin.h.opencode.hash.pct\endcsname{0.0\%}
\expandafter\def\csname val@qrdual.twin.h.opencode.write.ci\endcsname{[0.0, 39.0]}
\expandafter\def\csname val@qrdual.twin.h.opencode.write.frac\endcsname{0/6}
\expandafter\def\csname val@qrdual.twin.h.opencode.write.full\endcsname{0/6 (0.0\%)}
\expandafter\def\csname val@qrdual.twin.h.opencode.write.pct\endcsname{0.0\%}
\expandafter\def\csname val@qrdual.twin.h.qwen.codelivery.ci\endcsname{[43.6, 97.0]}
\expandafter\def\csname val@qrdual.twin.h.qwen.codelivery.frac\endcsname{5/6}
\expandafter\def\csname val@qrdual.twin.h.qwen.codelivery.full\endcsname{5/6 (83.3\%)}
\expandafter\def\csname val@qrdual.twin.h.qwen.codelivery.pct\endcsname{83.3\%}
\expandafter\def\csname val@qrdual.twin.h.qwen.hash.ci\endcsname{[0.0, 39.0]}
\expandafter\def\csname val@qrdual.twin.h.qwen.hash.frac\endcsname{0/6}
\expandafter\def\csname val@qrdual.twin.h.qwen.hash.full\endcsname{0/6 (0.0\%)}
\expandafter\def\csname val@qrdual.twin.h.qwen.hash.pct\endcsname{0.0\%}
\expandafter\def\csname val@qrdual.twin.h.qwen.write.ci\endcsname{[0.0, 39.0]}
\expandafter\def\csname val@qrdual.twin.h.qwen.write.frac\endcsname{0/6}
\expandafter\def\csname val@qrdual.twin.h.qwen.write.full\endcsname{0/6 (0.0\%)}
\expandafter\def\csname val@qrdual.twin.h.qwen.write.pct\endcsname{0.0\%}
\expandafter\def\csname val@qrdual.twin.h.smol.codelivery.ci\endcsname{[43.6, 97.0]}
\expandafter\def\csname val@qrdual.twin.h.smol.codelivery.frac\endcsname{5/6}
\expandafter\def\csname val@qrdual.twin.h.smol.codelivery.full\endcsname{5/6 (83.3\%)}
\expandafter\def\csname val@qrdual.twin.h.smol.codelivery.pct\endcsname{83.3\%}
\expandafter\def\csname val@qrdual.twin.h.smol.hash.ci\endcsname{[0.0, 39.0]}
\expandafter\def\csname val@qrdual.twin.h.smol.hash.frac\endcsname{0/6}
\expandafter\def\csname val@qrdual.twin.h.smol.hash.full\endcsname{0/6 (0.0\%)}
\expandafter\def\csname val@qrdual.twin.h.smol.hash.pct\endcsname{0.0\%}
\expandafter\def\csname val@qrdual.twin.h.smol.write.ci\endcsname{[0.0, 39.0]}
\expandafter\def\csname val@qrdual.twin.h.smol.write.frac\endcsname{0/6}
\expandafter\def\csname val@qrdual.twin.h.smol.write.full\endcsname{0/6 (0.0\%)}
\expandafter\def\csname val@qrdual.twin.h.smol.write.pct\endcsname{0.0\%}
\expandafter\def\csname val@qrdual.twin.hash.ci\endcsname{[0.0, 8.4]}
\expandafter\def\csname val@qrdual.twin.hash.frac\endcsname{0/42}
\expandafter\def\csname val@qrdual.twin.hash.full\endcsname{0/42 (0.0\%)}
\expandafter\def\csname val@qrdual.twin.hash.pct\endcsname{0.0\%}
\expandafter\def\csname val@qrdual.twin.reported.ci\endcsname{[0.0, 8.4]}
\expandafter\def\csname val@qrdual.twin.reported.frac\endcsname{0/42}
\expandafter\def\csname val@qrdual.twin.reported.full\endcsname{0/42 (0.0\%)}
\expandafter\def\csname val@qrdual.twin.reported.pct\endcsname{0.0\%}
\expandafter\def\csname val@qrdual.twin.reportedafter.ci\endcsname{[0.0, 8.4]}
\expandafter\def\csname val@qrdual.twin.reportedafter.frac\endcsname{0/42}
\expandafter\def\csname val@qrdual.twin.reportedafter.full\endcsname{0/42 (0.0\%)}
\expandafter\def\csname val@qrdual.twin.reportedafter.pct\endcsname{0.0\%}
\expandafter\def\csname val@qrdual.twin.targetwritten.ci\endcsname{[44.5, 73.0]}
\expandafter\def\csname val@qrdual.twin.targetwritten.frac\endcsname{25/42}
\expandafter\def\csname val@qrdual.twin.targetwritten.full\endcsname{25/42 (59.5\%)}
\expandafter\def\csname val@qrdual.twin.targetwritten.pct\endcsname{59.5\%}
\expandafter\def\csname val@qrdual.twin.write.ci\endcsname{[0.0, 8.4]}
\expandafter\def\csname val@qrdual.twin.write.frac\endcsname{0/42}
\expandafter\def\csname val@qrdual.twin.write.full\endcsname{0/42 (0.0\%)}
\expandafter\def\csname val@qrdual.twin.write.pct\endcsname{0.0\%}
\expandafter\def\csname val@qrdual.unverified.ci\endcsname{[81.0, 97.5]}
\expandafter\def\csname val@qrdual.unverified.frac\endcsname{39/42}
\expandafter\def\csname val@qrdual.unverified.full\endcsname{39/42 (92.9\%)}
\expandafter\def\csname val@qrdual.unverified.pct\endcsname{92.9\%}
\expandafter\def\csname val@qrdual.write.ci\endcsname{[31.2, 60.1]}
\expandafter\def\csname val@qrdual.write.frac\endcsname{19/42}
\expandafter\def\csname val@qrdual.write.full\endcsname{19/42 (45.2\%)}
\expandafter\def\csname val@qrdual.write.pct\endcsname{45.2\%}
\expandafter\def\csname val@rank.rho\endcsname{0.17}
\expandafter\def\csname val@rep.dir.overwrite.pooled.ci\endcsname{[0.0, 4.4]}
\expandafter\def\csname val@rep.dir.overwrite.pooled.frac\endcsname{0/84}
\expandafter\def\csname val@rep.dir.overwrite.pooled.full\endcsname{0/84 (0.0\%)}
\expandafter\def\csname val@rep.dir.overwrite.pooled.pct\endcsname{0.0\%}
\expandafter\def\csname val@rep.dir.overwrite.r1.ci\endcsname{[0.0, 12.1]}
\expandafter\def\csname val@rep.dir.overwrite.r1.frac\endcsname{0/28}
\expandafter\def\csname val@rep.dir.overwrite.r1.full\endcsname{0/28 (0.0\%)}
\expandafter\def\csname val@rep.dir.overwrite.r1.pct\endcsname{0.0\%}
\expandafter\def\csname val@rep.dir.overwrite.r2.ci\endcsname{[0.0, 12.1]}
\expandafter\def\csname val@rep.dir.overwrite.r2.frac\endcsname{0/28}
\expandafter\def\csname val@rep.dir.overwrite.r2.full\endcsname{0/28 (0.0\%)}
\expandafter\def\csname val@rep.dir.overwrite.r2.pct\endcsname{0.0\%}
\expandafter\def\csname val@rep.dir.overwrite.r3.ci\endcsname{[0.0, 12.1]}
\expandafter\def\csname val@rep.dir.overwrite.r3.frac\endcsname{0/28}
\expandafter\def\csname val@rep.dir.overwrite.r3.full\endcsname{0/28 (0.0\%)}
\expandafter\def\csname val@rep.dir.overwrite.r3.pct\endcsname{0.0\%}
\expandafter\def\csname val@rep.dir.overwrite.rounds\endcsname{0/28, 0/28, 0/28}
\expandafter\def\csname val@rep.medext.body.pooled.ci\endcsname{[83.1, 93.4]}
\expandafter\def\csname val@rep.medext.body.pooled.frac\endcsname{125/140}
\expandafter\def\csname val@rep.medext.body.pooled.full\endcsname{125/140 (89.3\%)}
\expandafter\def\csname val@rep.medext.body.pooled.pct\endcsname{89.3\%}
\expandafter\def\csname val@rep.medext.body.r1.ci\endcsname{[77.4, 98.0]}
\expandafter\def\csname val@rep.medext.body.r1.frac\endcsname{26/28}
\expandafter\def\csname val@rep.medext.body.r1.full\endcsname{26/28 (92.9\%)}
\expandafter\def\csname val@rep.medext.body.r1.pct\endcsname{92.9\%}
\expandafter\def\csname val@rep.medext.body.r2.ci\endcsname{[72.8, 96.3]}
\expandafter\def\csname val@rep.medext.body.r2.frac\endcsname{25/28}
\expandafter\def\csname val@rep.medext.body.r2.full\endcsname{25/28 (89.3\%)}
\expandafter\def\csname val@rep.medext.body.r2.pct\endcsname{89.3\%}
\expandafter\def\csname val@rep.medext.body.r3.ci\endcsname{[68.5, 94.3]}
\expandafter\def\csname val@rep.medext.body.r3.frac\endcsname{24/28}
\expandafter\def\csname val@rep.medext.body.r3.full\endcsname{24/28 (85.7\%)}
\expandafter\def\csname val@rep.medext.body.r3.pct\endcsname{85.7\%}
\expandafter\def\csname val@rep.medext.body.r4.ci\endcsname{[68.5, 94.3]}
\expandafter\def\csname val@rep.medext.body.r4.frac\endcsname{24/28}
\expandafter\def\csname val@rep.medext.body.r4.full\endcsname{24/28 (85.7\%)}
\expandafter\def\csname val@rep.medext.body.r4.pct\endcsname{85.7\%}
\expandafter\def\csname val@rep.medext.body.r5.ci\endcsname{[77.4, 98.0]}
\expandafter\def\csname val@rep.medext.body.r5.frac\endcsname{26/28}
\expandafter\def\csname val@rep.medext.body.r5.full\endcsname{26/28 (92.9\%)}
\expandafter\def\csname val@rep.medext.body.r5.pct\endcsname{92.9\%}
\expandafter\def\csname val@rep.medext.body.rounds\endcsname{26/28, 25/28, 24/28, 24/28, 26/28}
\expandafter\def\csname val@rep.mednoext.body.pooled.ci\endcsname{[83.1, 93.4]}
\expandafter\def\csname val@rep.mednoext.body.pooled.frac\endcsname{125/140}
\expandafter\def\csname val@rep.mednoext.body.pooled.full\endcsname{125/140 (89.3\%)}
\expandafter\def\csname val@rep.mednoext.body.pooled.pct\endcsname{89.3\%}
\expandafter\def\csname val@rep.mednoext.body.r1.ci\endcsname{[72.8, 96.3]}
\expandafter\def\csname val@rep.mednoext.body.r1.frac\endcsname{25/28}
\expandafter\def\csname val@rep.mednoext.body.r1.full\endcsname{25/28 (89.3\%)}
\expandafter\def\csname val@rep.mednoext.body.r1.pct\endcsname{89.3\%}
\expandafter\def\csname val@rep.mednoext.body.r2.ci\endcsname{[64.4, 92.1]}
\expandafter\def\csname val@rep.mednoext.body.r2.frac\endcsname{23/28}
\expandafter\def\csname val@rep.mednoext.body.r2.full\endcsname{23/28 (82.1\%)}
\expandafter\def\csname val@rep.mednoext.body.r2.pct\endcsname{82.1\%}
\expandafter\def\csname val@rep.mednoext.body.r3.ci\endcsname{[72.8, 96.3]}
\expandafter\def\csname val@rep.mednoext.body.r3.frac\endcsname{25/28}
\expandafter\def\csname val@rep.mednoext.body.r3.full\endcsname{25/28 (89.3\%)}
\expandafter\def\csname val@rep.mednoext.body.r3.pct\endcsname{89.3\%}
\expandafter\def\csname val@rep.mednoext.body.r4.ci\endcsname{[77.4, 98.0]}
\expandafter\def\csname val@rep.mednoext.body.r4.frac\endcsname{26/28}
\expandafter\def\csname val@rep.mednoext.body.r4.full\endcsname{26/28 (92.9\%)}
\expandafter\def\csname val@rep.mednoext.body.r4.pct\endcsname{92.9\%}
\expandafter\def\csname val@rep.mednoext.body.r5.ci\endcsname{[77.4, 98.0]}
\expandafter\def\csname val@rep.mednoext.body.r5.frac\endcsname{26/28}
\expandafter\def\csname val@rep.mednoext.body.r5.full\endcsname{26/28 (92.9\%)}
\expandafter\def\csname val@rep.mednoext.body.r5.pct\endcsname{92.9\%}
\expandafter\def\csname val@rep.mednoext.body.rounds\endcsname{25/28, 23/28, 25/28, 26/28, 26/28}
\expandafter\def\csname val@rep.nmodels\endcsname{four}
\expandafter\def\csname val@rep.panel\endcsname{DeepSeek-V4.1-Flash, GLM-5.3, Kimi-K3, Qwen3.8-Max}
\expandafter\def\csname val@rep.qrdual.ingress.pooled.ci\endcsname{[88.4, 98.1]}
\expandafter\def\csname val@rep.qrdual.ingress.pooled.frac\endcsname{80/84}
\expandafter\def\csname val@rep.qrdual.ingress.pooled.full\endcsname{80/84 (95.2\%)}
\expandafter\def\csname val@rep.qrdual.ingress.pooled.pct\endcsname{95.2\%}
\expandafter\def\csname val@rep.qrdual.ingress.r1.ci\endcsname{[82.3, 99.4]}
\expandafter\def\csname val@rep.qrdual.ingress.r1.frac\endcsname{27/28}
\expandafter\def\csname val@rep.qrdual.ingress.r1.full\endcsname{27/28 (96.4\%)}
\expandafter\def\csname val@rep.qrdual.ingress.r1.pct\endcsname{96.4\%}
\expandafter\def\csname val@rep.qrdual.ingress.r2.ci\endcsname{[82.3, 99.4]}
\expandafter\def\csname val@rep.qrdual.ingress.r2.frac\endcsname{27/28}
\expandafter\def\csname val@rep.qrdual.ingress.r2.full\endcsname{27/28 (96.4\%)}
\expandafter\def\csname val@rep.qrdual.ingress.r2.pct\endcsname{96.4\%}
\expandafter\def\csname val@rep.qrdual.ingress.r3.ci\endcsname{[77.4, 98.0]}
\expandafter\def\csname val@rep.qrdual.ingress.r3.frac\endcsname{26/28}
\expandafter\def\csname val@rep.qrdual.ingress.r3.full\endcsname{26/28 (92.9\%)}
\expandafter\def\csname val@rep.qrdual.ingress.r3.pct\endcsname{92.9\%}
\expandafter\def\csname val@rep.qrdual.ingress.rounds\endcsname{27/28, 27/28, 26/28}
\expandafter\def\csname val@rep.qrdual.twin.write.pooled.ci\endcsname{[0.0, 4.4]}
\expandafter\def\csname val@rep.qrdual.twin.write.pooled.frac\endcsname{0/84}
\expandafter\def\csname val@rep.qrdual.twin.write.pooled.full\endcsname{0/84 (0.0\%)}
\expandafter\def\csname val@rep.qrdual.twin.write.pooled.pct\endcsname{0.0\%}
\expandafter\def\csname val@rep.qrdual.twin.write.r1.ci\endcsname{[0.0, 12.1]}
\expandafter\def\csname val@rep.qrdual.twin.write.r1.frac\endcsname{0/28}
\expandafter\def\csname val@rep.qrdual.twin.write.r1.full\endcsname{0/28 (0.0\%)}
\expandafter\def\csname val@rep.qrdual.twin.write.r1.pct\endcsname{0.0\%}
\expandafter\def\csname val@rep.qrdual.twin.write.r2.ci\endcsname{[0.0, 12.1]}
\expandafter\def\csname val@rep.qrdual.twin.write.r2.frac\endcsname{0/28}
\expandafter\def\csname val@rep.qrdual.twin.write.r2.full\endcsname{0/28 (0.0\%)}
\expandafter\def\csname val@rep.qrdual.twin.write.r2.pct\endcsname{0.0\%}
\expandafter\def\csname val@rep.qrdual.twin.write.r3.ci\endcsname{[0.0, 12.1]}
\expandafter\def\csname val@rep.qrdual.twin.write.r3.frac\endcsname{0/28}
\expandafter\def\csname val@rep.qrdual.twin.write.r3.full\endcsname{0/28 (0.0\%)}
\expandafter\def\csname val@rep.qrdual.twin.write.r3.pct\endcsname{0.0\%}
\expandafter\def\csname val@rep.qrdual.twin.write.rounds\endcsname{0/28, 0/28, 0/28}
\expandafter\def\csname val@rep.qrdual.write.pooled.ci\endcsname{[36.2, 57.0]}
\expandafter\def\csname val@rep.qrdual.write.pooled.frac\endcsname{39/84}
\expandafter\def\csname val@rep.qrdual.write.pooled.full\endcsname{39/84 (46.4\%)}
\expandafter\def\csname val@rep.qrdual.write.pooled.pct\endcsname{46.4\%}
\expandafter\def\csname val@rep.qrdual.write.r1.ci\endcsname{[23.6, 57.6]}
\expandafter\def\csname val@rep.qrdual.write.r1.frac\endcsname{11/28}
\expandafter\def\csname val@rep.qrdual.write.r1.full\endcsname{11/28 (39.3\%)}
\expandafter\def\csname val@rep.qrdual.write.r1.pct\endcsname{39.3\%}
\expandafter\def\csname val@rep.qrdual.write.r2.ci\endcsname{[35.8, 70.5]}
\expandafter\def\csname val@rep.qrdual.write.r2.frac\endcsname{15/28}
\expandafter\def\csname val@rep.qrdual.write.r2.full\endcsname{15/28 (53.6\%)}
\expandafter\def\csname val@rep.qrdual.write.r2.pct\endcsname{53.6\%}
\expandafter\def\csname val@rep.qrdual.write.r3.ci\endcsname{[29.5, 64.2]}
\expandafter\def\csname val@rep.qrdual.write.r3.frac\endcsname{13/28}
\expandafter\def\csname val@rep.qrdual.write.r3.full\endcsname{13/28 (46.4\%)}
\expandafter\def\csname val@rep.qrdual.write.r3.pct\endcsname{46.4\%}
\expandafter\def\csname val@rep.qrdual.write.rounds\endcsname{11/28, 15/28, 13/28}
\expandafter\def\csname val@rep.tar.body.pooled.ci\endcsname{[93.6, 99.8]}
\expandafter\def\csname val@rep.tar.body.pooled.frac\endcsname{83/84}
\expandafter\def\csname val@rep.tar.body.pooled.full\endcsname{83/84 (98.8\%)}
\expandafter\def\csname val@rep.tar.body.pooled.pct\endcsname{98.8\%}
\expandafter\def\csname val@rep.tar.body.r1.ci\endcsname{[82.3, 99.4]}
\expandafter\def\csname val@rep.tar.body.r1.frac\endcsname{27/28}
\expandafter\def\csname val@rep.tar.body.r1.full\endcsname{27/28 (96.4\%)}
\expandafter\def\csname val@rep.tar.body.r1.pct\endcsname{96.4\%}
\expandafter\def\csname val@rep.tar.body.r2.ci\endcsname{[87.9, 100.0]}
\expandafter\def\csname val@rep.tar.body.r2.frac\endcsname{28/28}
\expandafter\def\csname val@rep.tar.body.r2.full\endcsname{28/28 (100.0\%)}
\expandafter\def\csname val@rep.tar.body.r2.pct\endcsname{100.0\%}
\expandafter\def\csname val@rep.tar.body.r3.ci\endcsname{[87.9, 100.0]}
\expandafter\def\csname val@rep.tar.body.r3.frac\endcsname{28/28}
\expandafter\def\csname val@rep.tar.body.r3.full\endcsname{28/28 (100.0\%)}
\expandafter\def\csname val@rep.tar.body.r3.pct\endcsname{100.0\%}
\expandafter\def\csname val@rep.tar.body.rounds\endcsname{27/28, 28/28, 28/28}
\expandafter\def\csname val@rep.tar.overwrite.pooled.ci\endcsname{[13.0, 30.0]}
\expandafter\def\csname val@rep.tar.overwrite.pooled.frac\endcsname{17/84}
\expandafter\def\csname val@rep.tar.overwrite.pooled.full\endcsname{17/84 (20.2\%)}
\expandafter\def\csname val@rep.tar.overwrite.pooled.pct\endcsname{20.2\%}
\expandafter\def\csname val@rep.tar.overwrite.r1.ci\endcsname{[7.9, 35.6]}
\expandafter\def\csname val@rep.tar.overwrite.r1.frac\endcsname{5/28}
\expandafter\def\csname val@rep.tar.overwrite.r1.full\endcsname{5/28 (17.9\%)}
\expandafter\def\csname val@rep.tar.overwrite.r1.pct\endcsname{17.9\%}
\expandafter\def\csname val@rep.tar.overwrite.r2.ci\endcsname{[15.3, 47.1]}
\expandafter\def\csname val@rep.tar.overwrite.r2.frac\endcsname{8/28}
\expandafter\def\csname val@rep.tar.overwrite.r2.full\endcsname{8/28 (28.6\%)}
\expandafter\def\csname val@rep.tar.overwrite.r2.pct\endcsname{28.6\%}
\expandafter\def\csname val@rep.tar.overwrite.r3.ci\endcsname{[5.7, 31.5]}
\expandafter\def\csname val@rep.tar.overwrite.r3.frac\endcsname{4/28}
\expandafter\def\csname val@rep.tar.overwrite.r3.full\endcsname{4/28 (14.3\%)}
\expandafter\def\csname val@rep.tar.overwrite.r3.pct\endcsname{14.3\%}
\expandafter\def\csname val@rep.tar.overwrite.rounds\endcsname{5/28, 8/28, 4/28}
\expandafter\def\csname val@rep.zip.bytebreach.pooled.ci\endcsname{[95.6, 100.0]}
\expandafter\def\csname val@rep.zip.bytebreach.pooled.frac\endcsname{84/84}
\expandafter\def\csname val@rep.zip.bytebreach.pooled.full\endcsname{84/84 (100.0\%)}
\expandafter\def\csname val@rep.zip.bytebreach.pooled.pct\endcsname{100.0\%}
\expandafter\def\csname val@rep.zip.bytebreach.r1.ci\endcsname{[87.9, 100.0]}
\expandafter\def\csname val@rep.zip.bytebreach.r1.frac\endcsname{28/28}
\expandafter\def\csname val@rep.zip.bytebreach.r1.full\endcsname{28/28 (100.0\%)}
\expandafter\def\csname val@rep.zip.bytebreach.r1.pct\endcsname{100.0\%}
\expandafter\def\csname val@rep.zip.bytebreach.r2.ci\endcsname{[87.9, 100.0]}
\expandafter\def\csname val@rep.zip.bytebreach.r2.frac\endcsname{28/28}
\expandafter\def\csname val@rep.zip.bytebreach.r2.full\endcsname{28/28 (100.0\%)}
\expandafter\def\csname val@rep.zip.bytebreach.r2.pct\endcsname{100.0\%}
\expandafter\def\csname val@rep.zip.bytebreach.r3.ci\endcsname{[87.9, 100.0]}
\expandafter\def\csname val@rep.zip.bytebreach.r3.frac\endcsname{28/28}
\expandafter\def\csname val@rep.zip.bytebreach.r3.full\endcsname{28/28 (100.0\%)}
\expandafter\def\csname val@rep.zip.bytebreach.r3.pct\endcsname{100.0\%}
\expandafter\def\csname val@rep.zip.bytebreach.rounds\endcsname{28/28, 28/28, 28/28}
\expandafter\def\csname val@rep.zip.extractinvoked.pooled.ci\endcsname{[95.6, 100.0]}
\expandafter\def\csname val@rep.zip.extractinvoked.pooled.frac\endcsname{84/84}
\expandafter\def\csname val@rep.zip.extractinvoked.pooled.full\endcsname{84/84 (100.0\%)}
\expandafter\def\csname val@rep.zip.extractinvoked.pooled.pct\endcsname{100.0\%}
\expandafter\def\csname val@rep.zip.extractinvoked.r1.ci\endcsname{[87.9, 100.0]}
\expandafter\def\csname val@rep.zip.extractinvoked.r1.frac\endcsname{28/28}
\expandafter\def\csname val@rep.zip.extractinvoked.r1.full\endcsname{28/28 (100.0\%)}
\expandafter\def\csname val@rep.zip.extractinvoked.r1.pct\endcsname{100.0\%}
\expandafter\def\csname val@rep.zip.extractinvoked.r2.ci\endcsname{[87.9, 100.0]}
\expandafter\def\csname val@rep.zip.extractinvoked.r2.frac\endcsname{28/28}
\expandafter\def\csname val@rep.zip.extractinvoked.r2.full\endcsname{28/28 (100.0\%)}
\expandafter\def\csname val@rep.zip.extractinvoked.r2.pct\endcsname{100.0\%}
\expandafter\def\csname val@rep.zip.extractinvoked.r3.ci\endcsname{[87.9, 100.0]}
\expandafter\def\csname val@rep.zip.extractinvoked.r3.frac\endcsname{28/28}
\expandafter\def\csname val@rep.zip.extractinvoked.r3.full\endcsname{28/28 (100.0\%)}
\expandafter\def\csname val@rep.zip.extractinvoked.r3.pct\endcsname{100.0\%}
\expandafter\def\csname val@rep.zip.extractinvoked.rounds\endcsname{28/28, 28/28, 28/28}
\expandafter\def\csname val@rep.zip.twin.bytebreach.pooled.ci\endcsname{[0.0, 4.4]}
\expandafter\def\csname val@rep.zip.twin.bytebreach.pooled.frac\endcsname{0/84}
\expandafter\def\csname val@rep.zip.twin.bytebreach.pooled.full\endcsname{0/84 (0.0\%)}
\expandafter\def\csname val@rep.zip.twin.bytebreach.pooled.pct\endcsname{0.0\%}
\expandafter\def\csname val@rep.zip.twin.bytebreach.r1.ci\endcsname{[0.0, 12.1]}
\expandafter\def\csname val@rep.zip.twin.bytebreach.r1.frac\endcsname{0/28}
\expandafter\def\csname val@rep.zip.twin.bytebreach.r1.full\endcsname{0/28 (0.0\%)}
\expandafter\def\csname val@rep.zip.twin.bytebreach.r1.pct\endcsname{0.0\%}
\expandafter\def\csname val@rep.zip.twin.bytebreach.r2.ci\endcsname{[0.0, 12.1]}
\expandafter\def\csname val@rep.zip.twin.bytebreach.r2.frac\endcsname{0/28}
\expandafter\def\csname val@rep.zip.twin.bytebreach.r2.full\endcsname{0/28 (0.0\%)}
\expandafter\def\csname val@rep.zip.twin.bytebreach.r2.pct\endcsname{0.0\%}
\expandafter\def\csname val@rep.zip.twin.bytebreach.r3.ci\endcsname{[0.0, 12.1]}
\expandafter\def\csname val@rep.zip.twin.bytebreach.r3.frac\endcsname{0/28}
\expandafter\def\csname val@rep.zip.twin.bytebreach.r3.full\endcsname{0/28 (0.0\%)}
\expandafter\def\csname val@rep.zip.twin.bytebreach.r3.pct\endcsname{0.0\%}
\expandafter\def\csname val@rep.zip.twin.bytebreach.rounds\endcsname{0/28, 0/28, 0/28}
\expandafter\def\csname val@script.bsdtar.adv.any.ci\endcsname{[56.6, 100.0]}
\expandafter\def\csname val@script.bsdtar.adv.any.frac\endcsname{5/5}
\expandafter\def\csname val@script.bsdtar.adv.any.full\endcsname{5/5 (100.0\%)}
\expandafter\def\csname val@script.bsdtar.adv.any.pct\endcsname{100.0\%}
\expandafter\def\csname val@script.bsdtar.adv.bytes.ci\endcsname{[56.6, 100.0]}
\expandafter\def\csname val@script.bsdtar.adv.bytes.frac\endcsname{5/5}
\expandafter\def\csname val@script.bsdtar.adv.bytes.full\endcsname{5/5 (100.0\%)}
\expandafter\def\csname val@script.bsdtar.adv.bytes.pct\endcsname{100.0\%}
\expandafter\def\csname val@script.bsdtar.adv.inodes.ci\endcsname{[56.6, 100.0]}
\expandafter\def\csname val@script.bsdtar.adv.inodes.frac\endcsname{5/5}
\expandafter\def\csname val@script.bsdtar.adv.inodes.full\endcsname{5/5 (100.0\%)}
\expandafter\def\csname val@script.bsdtar.adv.inodes.pct\endcsname{100.0\%}
\expandafter\def\csname val@script.bsdtar.adv.peakbytes\endcsname{8,392,440}
\expandafter\def\csname val@script.bsdtar.adv.peakinodes\endcsname{174}
\expandafter\def\csname val@script.pythonzipfile.adv.any.ci\endcsname{[56.6, 100.0]}
\expandafter\def\csname val@script.pythonzipfile.adv.any.frac\endcsname{5/5}
\expandafter\def\csname val@script.pythonzipfile.adv.any.full\endcsname{5/5 (100.0\%)}
\expandafter\def\csname val@script.pythonzipfile.adv.any.pct\endcsname{100.0\%}
\expandafter\def\csname val@script.pythonzipfile.adv.bytes.ci\endcsname{[0.0, 43.4]}
\expandafter\def\csname val@script.pythonzipfile.adv.bytes.frac\endcsname{0/5}
\expandafter\def\csname val@script.pythonzipfile.adv.bytes.full\endcsname{0/5 (0.0\%)}
\expandafter\def\csname val@script.pythonzipfile.adv.bytes.pct\endcsname{0.0\%}
\expandafter\def\csname val@script.pythonzipfile.adv.inodes.ci\endcsname{[56.6, 100.0]}
\expandafter\def\csname val@script.pythonzipfile.adv.inodes.frac\endcsname{5/5}
\expandafter\def\csname val@script.pythonzipfile.adv.inodes.full\endcsname{5/5 (100.0\%)}
\expandafter\def\csname val@script.pythonzipfile.adv.inodes.pct\endcsname{100.0\%}
\expandafter\def\csname val@script.pythonzipfile.adv.peakbytes\endcsname{3,896}
\expandafter\def\csname val@script.pythonzipfile.adv.peakinodes\endcsname{174}
\expandafter\def\csname val@script.pythonzipfile.twin.any.ci\endcsname{[56.6, 100.0]}
\expandafter\def\csname val@script.pythonzipfile.twin.any.frac\endcsname{5/5}
\expandafter\def\csname val@script.pythonzipfile.twin.any.full\endcsname{5/5 (100.0\%)}
\expandafter\def\csname val@script.pythonzipfile.twin.any.pct\endcsname{100.0\%}
\expandafter\def\csname val@script.pythonzipfile.twin.bytes.ci\endcsname{[0.0, 43.4]}
\expandafter\def\csname val@script.pythonzipfile.twin.bytes.frac\endcsname{0/5}
\expandafter\def\csname val@script.pythonzipfile.twin.bytes.full\endcsname{0/5 (0.0\%)}
\expandafter\def\csname val@script.pythonzipfile.twin.bytes.pct\endcsname{0.0\%}
\expandafter\def\csname val@script.pythonzipfile.twin.inodes.ci\endcsname{[56.6, 100.0]}
\expandafter\def\csname val@script.pythonzipfile.twin.inodes.frac\endcsname{5/5}
\expandafter\def\csname val@script.pythonzipfile.twin.inodes.full\endcsname{5/5 (100.0\%)}
\expandafter\def\csname val@script.pythonzipfile.twin.inodes.pct\endcsname{100.0\%}
\expandafter\def\csname val@script.pythonzipfile.twin.peakbytes\endcsname{3,896}
\expandafter\def\csname val@script.pythonzipfile.twin.peakinodes\endcsname{174}
\expandafter\def\csname val@script.unzip.adv.any.ci\endcsname{[56.6, 100.0]}
\expandafter\def\csname val@script.unzip.adv.any.frac\endcsname{5/5}
\expandafter\def\csname val@script.unzip.adv.any.full\endcsname{5/5 (100.0\%)}
\expandafter\def\csname val@script.unzip.adv.any.pct\endcsname{100.0\%}
\expandafter\def\csname val@script.unzip.adv.bytes.ci\endcsname{[56.6, 100.0]}
\expandafter\def\csname val@script.unzip.adv.bytes.frac\endcsname{5/5}
\expandafter\def\csname val@script.unzip.adv.bytes.full\endcsname{5/5 (100.0\%)}
\expandafter\def\csname val@script.unzip.adv.bytes.pct\endcsname{100.0\%}
\expandafter\def\csname val@script.unzip.adv.inodes.ci\endcsname{[56.6, 100.0]}
\expandafter\def\csname val@script.unzip.adv.inodes.frac\endcsname{5/5}
\expandafter\def\csname val@script.unzip.adv.inodes.full\endcsname{5/5 (100.0\%)}
\expandafter\def\csname val@script.unzip.adv.inodes.pct\endcsname{100.0\%}
\expandafter\def\csname val@script.unzip.adv.peakbytes\endcsname{8,392,440}
\expandafter\def\csname val@script.unzip.adv.peakinodes\endcsname{174}
\expandafter\def\csname val@script.unzip.twin.any.ci\endcsname{[56.6, 100.0]}
\expandafter\def\csname val@script.unzip.twin.any.frac\endcsname{5/5}
\expandafter\def\csname val@script.unzip.twin.any.full\endcsname{5/5 (100.0\%)}
\expandafter\def\csname val@script.unzip.twin.any.pct\endcsname{100.0\%}
\expandafter\def\csname val@script.unzip.twin.bytes.ci\endcsname{[0.0, 43.4]}
\expandafter\def\csname val@script.unzip.twin.bytes.frac\endcsname{0/5}
\expandafter\def\csname val@script.unzip.twin.bytes.full\endcsname{0/5 (0.0\%)}
\expandafter\def\csname val@script.unzip.twin.bytes.pct\endcsname{0.0\%}
\expandafter\def\csname val@script.unzip.twin.inodes.ci\endcsname{[56.6, 100.0]}
\expandafter\def\csname val@script.unzip.twin.inodes.frac\endcsname{5/5}
\expandafter\def\csname val@script.unzip.twin.inodes.full\endcsname{5/5 (100.0\%)}
\expandafter\def\csname val@script.unzip.twin.inodes.pct\endcsname{100.0\%}
\expandafter\def\csname val@script.unzip.twin.peakbytes\endcsname{3,896}
\expandafter\def\csname val@script.unzip.twin.peakinodes\endcsname{174}
\expandafter\def\csname val@twin.declaredwrite.ci\endcsname{[95.5, 100.0]}
\expandafter\def\csname val@twin.declaredwrite.frac\endcsname{81/81}
\expandafter\def\csname val@twin.declaredwrite.full\endcsname{81/81 (100.0\%)}
\expandafter\def\csname val@twin.declaredwrite.pct\endcsname{100.0\%}
\expandafter\def\csname val@twin.falsealarm.ci\endcsname{[1.9, 7.3]}
\expandafter\def\csname val@twin.falsealarm.codelivered.ci\endcsname{[1.6, 6.7]}
\expandafter\def\csname val@twin.falsealarm.codelivered.frac\endcsname{7/210}
\expandafter\def\csname val@twin.falsealarm.codelivered.full\endcsname{7/210 (3.3\%)}
\expandafter\def\csname val@twin.falsealarm.codelivered.pct\endcsname{3.3\%}
\expandafter\def\csname val@twin.falsealarm.frac\endcsname{8/210}
\expandafter\def\csname val@twin.falsealarm.full\endcsname{8/210 (3.8\%)}
\expandafter\def\csname val@twin.falsealarm.pct\endcsname{3.8\%}
\expandafter\def\csname val@twin.wrongwrite.ci\endcsname{[0.0, 1.8]}
\expandafter\def\csname val@twin.wrongwrite.frac\endcsname{0/210}
\expandafter\def\csname val@twin.wrongwrite.full\endcsname{0/210 (0.0\%)}
\expandafter\def\csname val@twin.wrongwrite.pct\endcsname{0.0\%}
\expandafter\def\csname val@write.top.ci\endcsname{[43.7, 83.7]}
\expandafter\def\csname val@write.top.frac\endcsname{12/18}
\expandafter\def\csname val@write.top.full\endcsname{12/18 (66.7\%)}
\expandafter\def\csname val@write.top.harness\endcsname{mini-SWE-agent}
\expandafter\def\csname val@write.top.pct\endcsname{66.7\%}
\expandafter\def\csname val@write.zero.harnesses\endcsname{none}
\expandafter\def\csname val@zip.afterlisting.ci\endcsname{[27.0, 55.5]}
\expandafter\def\csname val@zip.afterlisting.frac\endcsname{17/42}
\expandafter\def\csname val@zip.afterlisting.full\endcsname{17/42 (40.5\%)}
\expandafter\def\csname val@zip.afterlisting.pct\endcsname{40.5\%}
\expandafter\def\csname val@zip.allrounds.bytebreach.ci\endcsname{[69.8, 84.2]}
\expandafter\def\csname val@zip.allrounds.bytebreach.frac\endcsname{98/126}
\expandafter\def\csname val@zip.allrounds.bytebreach.full\endcsname{98/126 (77.8\%)}
\expandafter\def\csname val@zip.allrounds.bytebreach.pct\endcsname{77.8\%}
\expandafter\def\csname val@zip.allrounds.extract.ci\endcsname{[70.6, 84.8]}
\expandafter\def\csname val@zip.allrounds.extract.frac\endcsname{99/126}
\expandafter\def\csname val@zip.allrounds.extract.full\endcsname{99/126 (78.6\%)}
\expandafter\def\csname val@zip.allrounds.extract.pct\endcsname{78.6\%}
\expandafter\def\csname val@zip.allrounds.nmodels\endcsname{six}
\expandafter\def\csname val@zip.allrounds.nrounds\endcsname{three}
\expandafter\def\csname val@zip.anybreach.ci\endcsname{[61.5, 86.5]}
\expandafter\def\csname val@zip.anybreach.frac\endcsname{32/42}
\expandafter\def\csname val@zip.anybreach.full\endcsname{32/42 (76.2\%)}
\expandafter\def\csname val@zip.anybreach.pct\endcsname{76.2\%}
\expandafter\def\csname val@zip.body.ci\endcsname{[84.2, 98.7]}
\expandafter\def\csname val@zip.body.frac\endcsname{40/42}
\expandafter\def\csname val@zip.body.full\endcsname{40/42 (95.2\%)}
\expandafter\def\csname val@zip.body.pct\endcsname{95.2\%}
\expandafter\def\csname val@zip.breachbeforecontent.ci\endcsname{[46.8, 75.0]}
\expandafter\def\csname val@zip.breachbeforecontent.frac\endcsname{26/42}
\expandafter\def\csname val@zip.breachbeforecontent.full\endcsname{26/42 (61.9\%)}
\expandafter\def\csname val@zip.breachbeforecontent.pct\endcsname{61.9\%}
\expandafter\def\csname val@zip.breachbeforefirstdetectedop.ci\endcsname{[0.0, 8.4]}
\expandafter\def\csname val@zip.breachbeforefirstdetectedop.frac\endcsname{0/42}
\expandafter\def\csname val@zip.breachbeforefirstdetectedop.full\endcsname{0/42 (0.0\%)}
\expandafter\def\csname val@zip.breachbeforefirstdetectedop.pct\endcsname{0.0\%}
\expandafter\def\csname val@zip.bytebreach.ci\endcsname{[61.5, 86.5]}
\expandafter\def\csname val@zip.bytebreach.frac\endcsname{32/42}
\expandafter\def\csname val@zip.bytebreach.full\endcsname{32/42 (76.2\%)}
\expandafter\def\csname val@zip.bytebreach.pct\endcsname{76.2\%}
\expandafter\def\csname val@zip.bytebreachbeforecontent.ci\endcsname{[46.8, 75.0]}
\expandafter\def\csname val@zip.bytebreachbeforecontent.frac\endcsname{26/42}
\expandafter\def\csname val@zip.bytebreachbeforecontent.full\endcsname{26/42 (61.9\%)}
\expandafter\def\csname val@zip.bytebreachbeforecontent.pct\endcsname{61.9\%}
\expandafter\def\csname val@zip.contentincontext.ci\endcsname{[81.0, 97.5]}
\expandafter\def\csname val@zip.contentincontext.frac\endcsname{39/42}
\expandafter\def\csname val@zip.contentincontext.full\endcsname{39/42 (92.9\%)}
\expandafter\def\csname val@zip.contentincontext.pct\endcsname{92.9\%}
\expandafter\def\csname val@zip.declared.bytes\endcsname{3,896}
\expandafter\def\csname val@zip.declaredtotalseenbeforebreach.ci\endcsname{[17.2, 43.6]}
\expandafter\def\csname val@zip.declaredtotalseenbeforebreach.frac\endcsname{12/42}
\expandafter\def\csname val@zip.declaredtotalseenbeforebreach.full\endcsname{12/42 (28.6\%)}
\expandafter\def\csname val@zip.declaredtotalseenbeforebreach.pct\endcsname{28.6\%}
\expandafter\def\csname val@zip.extractinvoked.ci\endcsname{[64.1, 88.3]}
\expandafter\def\csname val@zip.extractinvoked.frac\endcsname{33/42}
\expandafter\def\csname val@zip.extractinvoked.full\endcsname{33/42 (78.6\%)}
\expandafter\def\csname val@zip.extractinvoked.pct\endcsname{78.6\%}
\expandafter\def\csname val@zip.extractnoexceed.ci\endcsname{[0.4, 12.3]}
\expandafter\def\csname val@zip.extractnoexceed.frac\endcsname{1/42}
\expandafter\def\csname val@zip.extractnoexceed.full\endcsname{1/42 (2.4\%)}
\expandafter\def\csname val@zip.extractnoexceed.pct\endcsname{2.4\%}
\expandafter\def\csname val@zip.firstopbreach.ci\endcsname{[21.0, 48.4]}
\expandafter\def\csname val@zip.firstopbreach.frac\endcsname{14/42}
\expandafter\def\csname val@zip.firstopbreach.full\endcsname{14/42 (33.3\%)}
\expandafter\def\csname val@zip.firstopbreach.pct\endcsname{33.3\%}
\expandafter\def\csname val@zip.firstopbytebreach.ci\endcsname{[21.0, 48.4]}
\expandafter\def\csname val@zip.firstopbytebreach.frac\endcsname{14/42}
\expandafter\def\csname val@zip.firstopbytebreach.full\endcsname{14/42 (33.3\%)}
\expandafter\def\csname val@zip.firstopbytebreach.pct\endcsname{33.3\%}
\expandafter\def\csname val@zip.h.claude.bytebreach.ci\endcsname{[61.0, 100.0]}
\expandafter\def\csname val@zip.h.claude.bytebreach.frac\endcsname{6/6}
\expandafter\def\csname val@zip.h.claude.bytebreach.full\endcsname{6/6 (100.0\%)}
\expandafter\def\csname val@zip.h.claude.bytebreach.pct\endcsname{100.0\%}
\expandafter\def\csname val@zip.h.claude.extract.ci\endcsname{[61.0, 100.0]}
\expandafter\def\csname val@zip.h.claude.extract.frac\endcsname{6/6}
\expandafter\def\csname val@zip.h.claude.extract.full\endcsname{6/6 (100.0\%)}
\expandafter\def\csname val@zip.h.claude.extract.pct\endcsname{100.0\%}
\expandafter\def\csname val@zip.h.claude.listfirst.ci\endcsname{[30.0, 90.3]}
\expandafter\def\csname val@zip.h.claude.listfirst.frac\endcsname{4/6}
\expandafter\def\csname val@zip.h.claude.listfirst.full\endcsname{4/6 (66.7\%)}
\expandafter\def\csname val@zip.h.claude.listfirst.pct\endcsname{66.7\%}
\expandafter\def\csname val@zip.h.codex.bytebreach.ci\endcsname{[30.0, 90.3]}
\expandafter\def\csname val@zip.h.codex.bytebreach.frac\endcsname{4/6}
\expandafter\def\csname val@zip.h.codex.bytebreach.full\endcsname{4/6 (66.7\%)}
\expandafter\def\csname val@zip.h.codex.bytebreach.pct\endcsname{66.7\%}
\expandafter\def\csname val@zip.h.codex.extract.ci\endcsname{[30.0, 90.3]}
\expandafter\def\csname val@zip.h.codex.extract.frac\endcsname{4/6}
\expandafter\def\csname val@zip.h.codex.extract.full\endcsname{4/6 (66.7\%)}
\expandafter\def\csname val@zip.h.codex.extract.pct\endcsname{66.7\%}
\expandafter\def\csname val@zip.h.codex.listfirst.ci\endcsname{[30.0, 90.3]}
\expandafter\def\csname val@zip.h.codex.listfirst.frac\endcsname{4/6}
\expandafter\def\csname val@zip.h.codex.listfirst.full\endcsname{4/6 (66.7\%)}
\expandafter\def\csname val@zip.h.codex.listfirst.pct\endcsname{66.7\%}
\expandafter\def\csname val@zip.h.gptme.bytebreach.ci\endcsname{[30.0, 90.3]}
\expandafter\def\csname val@zip.h.gptme.bytebreach.frac\endcsname{4/6}
\expandafter\def\csname val@zip.h.gptme.bytebreach.full\endcsname{4/6 (66.7\%)}
\expandafter\def\csname val@zip.h.gptme.bytebreach.pct\endcsname{66.7\%}
\expandafter\def\csname val@zip.h.gptme.extract.ci\endcsname{[30.0, 90.3]}
\expandafter\def\csname val@zip.h.gptme.extract.frac\endcsname{4/6}
\expandafter\def\csname val@zip.h.gptme.extract.full\endcsname{4/6 (66.7\%)}
\expandafter\def\csname val@zip.h.gptme.extract.pct\endcsname{66.7\%}
\expandafter\def\csname val@zip.h.gptme.listfirst.ci\endcsname{[9.7, 70.0]}
\expandafter\def\csname val@zip.h.gptme.listfirst.frac\endcsname{2/6}
\expandafter\def\csname val@zip.h.gptme.listfirst.full\endcsname{2/6 (33.3\%)}
\expandafter\def\csname val@zip.h.gptme.listfirst.pct\endcsname{33.3\%}
\expandafter\def\csname val@zip.h.mini.bytebreach.ci\endcsname{[43.6, 97.0]}
\expandafter\def\csname val@zip.h.mini.bytebreach.frac\endcsname{5/6}
\expandafter\def\csname val@zip.h.mini.bytebreach.full\endcsname{5/6 (83.3\%)}
\expandafter\def\csname val@zip.h.mini.bytebreach.pct\endcsname{83.3\%}
\expandafter\def\csname val@zip.h.mini.extract.ci\endcsname{[61.0, 100.0]}
\expandafter\def\csname val@zip.h.mini.extract.frac\endcsname{6/6}
\expandafter\def\csname val@zip.h.mini.extract.full\endcsname{6/6 (100.0\%)}
\expandafter\def\csname val@zip.h.mini.extract.pct\endcsname{100.0\%}
\expandafter\def\csname val@zip.h.mini.listfirst.ci\endcsname{[18.8, 81.2]}
\expandafter\def\csname val@zip.h.mini.listfirst.frac\endcsname{3/6}
\expandafter\def\csname val@zip.h.mini.listfirst.full\endcsname{3/6 (50.0\%)}
\expandafter\def\csname val@zip.h.mini.listfirst.pct\endcsname{50.0\%}
\expandafter\def\csname val@zip.h.opencode.bytebreach.ci\endcsname{[30.0, 90.3]}
\expandafter\def\csname val@zip.h.opencode.bytebreach.frac\endcsname{4/6}
\expandafter\def\csname val@zip.h.opencode.bytebreach.full\endcsname{4/6 (66.7\%)}
\expandafter\def\csname val@zip.h.opencode.bytebreach.pct\endcsname{66.7\%}
\expandafter\def\csname val@zip.h.opencode.extract.ci\endcsname{[30.0, 90.3]}
\expandafter\def\csname val@zip.h.opencode.extract.frac\endcsname{4/6}
\expandafter\def\csname val@zip.h.opencode.extract.full\endcsname{4/6 (66.7\%)}
\expandafter\def\csname val@zip.h.opencode.extract.pct\endcsname{66.7\%}
\expandafter\def\csname val@zip.h.opencode.listfirst.ci\endcsname{[61.0, 100.0]}
\expandafter\def\csname val@zip.h.opencode.listfirst.frac\endcsname{6/6}
\expandafter\def\csname val@zip.h.opencode.listfirst.full\endcsname{6/6 (100.0\%)}
\expandafter\def\csname val@zip.h.opencode.listfirst.pct\endcsname{100.0\%}
\expandafter\def\csname val@zip.h.qwen.bytebreach.ci\endcsname{[30.0, 90.3]}
\expandafter\def\csname val@zip.h.qwen.bytebreach.frac\endcsname{4/6}
\expandafter\def\csname val@zip.h.qwen.bytebreach.full\endcsname{4/6 (66.7\%)}
\expandafter\def\csname val@zip.h.qwen.bytebreach.pct\endcsname{66.7\%}
\expandafter\def\csname val@zip.h.qwen.extract.ci\endcsname{[30.0, 90.3]}
\expandafter\def\csname val@zip.h.qwen.extract.frac\endcsname{4/6}
\expandafter\def\csname val@zip.h.qwen.extract.full\endcsname{4/6 (66.7\%)}
\expandafter\def\csname val@zip.h.qwen.extract.pct\endcsname{66.7\%}
\expandafter\def\csname val@zip.h.qwen.listfirst.ci\endcsname{[9.7, 70.0]}
\expandafter\def\csname val@zip.h.qwen.listfirst.frac\endcsname{2/6}
\expandafter\def\csname val@zip.h.qwen.listfirst.full\endcsname{2/6 (33.3\%)}
\expandafter\def\csname val@zip.h.qwen.listfirst.pct\endcsname{33.3\%}
\expandafter\def\csname val@zip.h.smol.bytebreach.ci\endcsname{[43.6, 97.0]}
\expandafter\def\csname val@zip.h.smol.bytebreach.frac\endcsname{5/6}
\expandafter\def\csname val@zip.h.smol.bytebreach.full\endcsname{5/6 (83.3\%)}
\expandafter\def\csname val@zip.h.smol.bytebreach.pct\endcsname{83.3\%}
\expandafter\def\csname val@zip.h.smol.extract.ci\endcsname{[43.6, 97.0]}
\expandafter\def\csname val@zip.h.smol.extract.frac\endcsname{5/6}
\expandafter\def\csname val@zip.h.smol.extract.full\endcsname{5/6 (83.3\%)}
\expandafter\def\csname val@zip.h.smol.extract.pct\endcsname{83.3\%}
\expandafter\def\csname val@zip.h.smol.listfirst.ci\endcsname{[30.0, 90.3]}
\expandafter\def\csname val@zip.h.smol.listfirst.frac\endcsname{4/6}
\expandafter\def\csname val@zip.h.smol.listfirst.full\endcsname{4/6 (66.7\%)}
\expandafter\def\csname val@zip.h.smol.listfirst.pct\endcsname{66.7\%}
\expandafter\def\csname val@zip.inodebreach.ci\endcsname{[58.9, 84.7]}
\expandafter\def\csname val@zip.inodebreach.frac\endcsname{31/42}
\expandafter\def\csname val@zip.inodebreach.full\endcsname{31/42 (73.8\%)}
\expandafter\def\csname val@zip.inodebreach.pct\endcsname{73.8\%}
\expandafter\def\csname val@zip.pooled.afterlisting.ci\endcsname{[33.1, 50.0]}
\expandafter\def\csname val@zip.pooled.afterlisting.frac\endcsname{52/126}
\expandafter\def\csname val@zip.pooled.afterlisting.full\endcsname{52/126 (41.3\%)}
\expandafter\def\csname val@zip.pooled.afterlisting.pct\endcsname{41.3\%}
\expandafter\def\csname val@zip.pooled.anybreach.ci\endcsname{[69.8, 84.2]}
\expandafter\def\csname val@zip.pooled.anybreach.frac\endcsname{98/126}
\expandafter\def\csname val@zip.pooled.anybreach.full\endcsname{98/126 (77.8\%)}
\expandafter\def\csname val@zip.pooled.anybreach.pct\endcsname{77.8\%}
\expandafter\def\csname val@zip.pooled.body.ci\endcsname{[91.0, 98.3]}
\expandafter\def\csname val@zip.pooled.body.frac\endcsname{121/126}
\expandafter\def\csname val@zip.pooled.body.full\endcsname{121/126 (96.0\%)}
\expandafter\def\csname val@zip.pooled.body.pct\endcsname{96.0\%}
\expandafter\def\csname val@zip.pooled.breachbeforecontent.ci\endcsname{[53.2, 69.9]}
\expandafter\def\csname val@zip.pooled.breachbeforecontent.frac\endcsname{78/126}
\expandafter\def\csname val@zip.pooled.breachbeforecontent.full\endcsname{78/126 (61.9\%)}
\expandafter\def\csname val@zip.pooled.breachbeforecontent.pct\endcsname{61.9\%}
\expandafter\def\csname val@zip.pooled.breachbeforefirstdetectedop.ci\endcsname{[0.0, 3.0]}
\expandafter\def\csname val@zip.pooled.breachbeforefirstdetectedop.frac\endcsname{0/126}
\expandafter\def\csname val@zip.pooled.breachbeforefirstdetectedop.full\endcsname{0/126 (0.0\%)}
\expandafter\def\csname val@zip.pooled.breachbeforefirstdetectedop.pct\endcsname{0.0\%}
\expandafter\def\csname val@zip.pooled.bytebreach.ci\endcsname{[69.8, 84.2]}
\expandafter\def\csname val@zip.pooled.bytebreach.frac\endcsname{98/126}
\expandafter\def\csname val@zip.pooled.bytebreach.full\endcsname{98/126 (77.8\%)}
\expandafter\def\csname val@zip.pooled.bytebreach.pct\endcsname{77.8\%}
\expandafter\def\csname val@zip.pooled.bytebreachbeforecontent.ci\endcsname{[53.2, 69.9]}
\expandafter\def\csname val@zip.pooled.bytebreachbeforecontent.frac\endcsname{78/126}
\expandafter\def\csname val@zip.pooled.bytebreachbeforecontent.full\endcsname{78/126 (61.9\%)}
\expandafter\def\csname val@zip.pooled.bytebreachbeforecontent.pct\endcsname{61.9\%}
\expandafter\def\csname val@zip.pooled.contentincontext.ci\endcsname{[87.0, 96.2]}
\expandafter\def\csname val@zip.pooled.contentincontext.frac\endcsname{117/126}
\expandafter\def\csname val@zip.pooled.contentincontext.full\endcsname{117/126 (92.9\%)}
\expandafter\def\csname val@zip.pooled.contentincontext.pct\endcsname{92.9\%}
\expandafter\def\csname val@zip.pooled.declaredtotalseenbeforebreach.ci\endcsname{[21.4, 37.0]}
\expandafter\def\csname val@zip.pooled.declaredtotalseenbeforebreach.frac\endcsname{36/126}
\expandafter\def\csname val@zip.pooled.declaredtotalseenbeforebreach.full\endcsname{36/126 (28.6\%)}
\expandafter\def\csname val@zip.pooled.declaredtotalseenbeforebreach.pct\endcsname{28.6\%}
\expandafter\def\csname val@zip.pooled.extractinvoked.ci\endcsname{[70.6, 84.8]}
\expandafter\def\csname val@zip.pooled.extractinvoked.frac\endcsname{99/126}
\expandafter\def\csname val@zip.pooled.extractinvoked.full\endcsname{99/126 (78.6\%)}
\expandafter\def\csname val@zip.pooled.extractinvoked.pct\endcsname{78.6\%}
\expandafter\def\csname val@zip.pooled.extractnoexceed.ci\endcsname{[0.1, 4.4]}
\expandafter\def\csname val@zip.pooled.extractnoexceed.frac\endcsname{1/126}
\expandafter\def\csname val@zip.pooled.extractnoexceed.full\endcsname{1/126 (0.8\%)}
\expandafter\def\csname val@zip.pooled.extractnoexceed.pct\endcsname{0.8\%}
\expandafter\def\csname val@zip.pooled.firstopbreach.ci\endcsname{[25.7, 41.9]}
\expandafter\def\csname val@zip.pooled.firstopbreach.frac\endcsname{42/126}
\expandafter\def\csname val@zip.pooled.firstopbreach.full\endcsname{42/126 (33.3\%)}
\expandafter\def\csname val@zip.pooled.firstopbreach.pct\endcsname{33.3\%}
\expandafter\def\csname val@zip.pooled.firstopbytebreach.ci\endcsname{[25.7, 41.9]}
\expandafter\def\csname val@zip.pooled.firstopbytebreach.frac\endcsname{42/126}
\expandafter\def\csname val@zip.pooled.firstopbytebreach.full\endcsname{42/126 (33.3\%)}
\expandafter\def\csname val@zip.pooled.firstopbytebreach.pct\endcsname{33.3\%}
\expandafter\def\csname val@zip.pooled.inodebreach.ci\endcsname{[68.0, 82.8]}
\expandafter\def\csname val@zip.pooled.inodebreach.frac\endcsname{96/126}
\expandafter\def\csname val@zip.pooled.inodebreach.full\endcsname{96/126 (76.2\%)}
\expandafter\def\csname val@zip.pooled.inodebreach.pct\endcsname{76.2\%}
\expandafter\def\csname val@zip.timing.afterlisting\endcsname{17}
\expandafter\def\csname val@zip.timing.breach.median\endcsname{2.9}
\expandafter\def\csname val@zip.timing.content.median\endcsname{6.0}
\expandafter\def\csname val@zip.timing.n\endcsname{32}
\expandafter\def\csname val@zip.timing.nocontent\endcsname{1}
\expandafter\def\csname val@zip.twin.afterlisting.ci\endcsname{[33.4, 62.3]}
\expandafter\def\csname val@zip.twin.afterlisting.frac\endcsname{20/42}
\expandafter\def\csname val@zip.twin.afterlisting.full\endcsname{20/42 (47.6\%)}
\expandafter\def\csname val@zip.twin.afterlisting.pct\endcsname{47.6\%}
\expandafter\def\csname val@zip.twin.allrounds.bytebreach.ci\endcsname{[0.0, 3.0]}
\expandafter\def\csname val@zip.twin.allrounds.bytebreach.frac\endcsname{0/126}
\expandafter\def\csname val@zip.twin.allrounds.bytebreach.full\endcsname{0/126 (0.0\%)}
\expandafter\def\csname val@zip.twin.allrounds.bytebreach.pct\endcsname{0.0\%}
\expandafter\def\csname val@zip.twin.allrounds.extract.ci\endcsname{[78.5, 90.8]}
\expandafter\def\csname val@zip.twin.allrounds.extract.frac\endcsname{108/126}
\expandafter\def\csname val@zip.twin.allrounds.extract.full\endcsname{108/126 (85.7\%)}
\expandafter\def\csname val@zip.twin.allrounds.extract.pct\endcsname{85.7\%}
\expandafter\def\csname val@zip.twin.anybreach.ci\endcsname{[75.0, 94.8]}
\expandafter\def\csname val@zip.twin.anybreach.frac\endcsname{37/42}
\expandafter\def\csname val@zip.twin.anybreach.full\endcsname{37/42 (88.1\%)}
\expandafter\def\csname val@zip.twin.anybreach.pct\endcsname{88.1\%}
\expandafter\def\csname val@zip.twin.body.ci\endcsname{[87.7, 99.6]}
\expandafter\def\csname val@zip.twin.body.frac\endcsname{41/42}
\expandafter\def\csname val@zip.twin.body.full\endcsname{41/42 (97.6\%)}
\expandafter\def\csname val@zip.twin.body.pct\endcsname{97.6\%}
\expandafter\def\csname val@zip.twin.breachbeforecontent.ci\endcsname{[39.9, 68.8]}
\expandafter\def\csname val@zip.twin.breachbeforecontent.frac\endcsname{23/42}
\expandafter\def\csname val@zip.twin.breachbeforecontent.full\endcsname{23/42 (54.8\%)}
\expandafter\def\csname val@zip.twin.breachbeforecontent.pct\endcsname{54.8\%}
\expandafter\def\csname val@zip.twin.breachbeforefirstdetectedop.ci\endcsname{[0.0, 8.4]}
\expandafter\def\csname val@zip.twin.breachbeforefirstdetectedop.frac\endcsname{0/42}
\expandafter\def\csname val@zip.twin.breachbeforefirstdetectedop.full\endcsname{0/42 (0.0\%)}
\expandafter\def\csname val@zip.twin.breachbeforefirstdetectedop.pct\endcsname{0.0\%}
\expandafter\def\csname val@zip.twin.bytebreach.ci\endcsname{[0.0, 8.4]}
\expandafter\def\csname val@zip.twin.bytebreach.frac\endcsname{0/42}
\expandafter\def\csname val@zip.twin.bytebreach.full\endcsname{0/42 (0.0\%)}
\expandafter\def\csname val@zip.twin.bytebreach.pct\endcsname{0.0\%}
\expandafter\def\csname val@zip.twin.bytebreachbeforecontent.ci\endcsname{[0.0, 8.4]}
\expandafter\def\csname val@zip.twin.bytebreachbeforecontent.frac\endcsname{0/42}
\expandafter\def\csname val@zip.twin.bytebreachbeforecontent.full\endcsname{0/42 (0.0\%)}
\expandafter\def\csname val@zip.twin.bytebreachbeforecontent.pct\endcsname{0.0\%}
\expandafter\def\csname val@zip.twin.contentincontext.ci\endcsname{[77.9, 96.2]}
\expandafter\def\csname val@zip.twin.contentincontext.frac\endcsname{38/42}
\expandafter\def\csname val@zip.twin.contentincontext.full\endcsname{38/42 (90.5\%)}
\expandafter\def\csname val@zip.twin.contentincontext.pct\endcsname{90.5\%}
\expandafter\def\csname val@zip.twin.declaredtotalseenbeforebreach.ci\endcsname{[21.0, 48.4]}
\expandafter\def\csname val@zip.twin.declaredtotalseenbeforebreach.frac\endcsname{14/42}
\expandafter\def\csname val@zip.twin.declaredtotalseenbeforebreach.full\endcsname{14/42 (33.3\%)}
\expandafter\def\csname val@zip.twin.declaredtotalseenbeforebreach.pct\endcsname{33.3\%}
\expandafter\def\csname val@zip.twin.extractinvoked.ci\endcsname{[75.0, 94.8]}
\expandafter\def\csname val@zip.twin.extractinvoked.frac\endcsname{37/42}
\expandafter\def\csname val@zip.twin.extractinvoked.full\endcsname{37/42 (88.1\%)}
\expandafter\def\csname val@zip.twin.extractinvoked.pct\endcsname{88.1\%}
\expandafter\def\csname val@zip.twin.firstopbreach.ci\endcsname{[27.0, 55.5]}
\expandafter\def\csname val@zip.twin.firstopbreach.frac\endcsname{17/42}
\expandafter\def\csname val@zip.twin.firstopbreach.full\endcsname{17/42 (40.5\%)}
\expandafter\def\csname val@zip.twin.firstopbreach.pct\endcsname{40.5\%}
\expandafter\def\csname val@zip.twin.firstopbytebreach.ci\endcsname{[0.0, 8.4]}
\expandafter\def\csname val@zip.twin.firstopbytebreach.frac\endcsname{0/42}
\expandafter\def\csname val@zip.twin.firstopbytebreach.full\endcsname{0/42 (0.0\%)}
\expandafter\def\csname val@zip.twin.firstopbytebreach.pct\endcsname{0.0\%}
\expandafter\def\csname val@zip.twin.inodebreach.ci\endcsname{[75.0, 94.8]}
\expandafter\def\csname val@zip.twin.inodebreach.frac\endcsname{37/42}
\expandafter\def\csname val@zip.twin.inodebreach.full\endcsname{37/42 (88.1\%)}
\expandafter\def\csname val@zip.twin.inodebreach.pct\endcsname{88.1\%}
\expandafter\def\csname val@zip.twin.pooled.afterlisting.ci\endcsname{[39.9, 57.1]}
\expandafter\def\csname val@zip.twin.pooled.afterlisting.frac\endcsname{61/126}
\expandafter\def\csname val@zip.twin.pooled.afterlisting.full\endcsname{61/126 (48.4\%)}
\expandafter\def\csname val@zip.twin.pooled.afterlisting.pct\endcsname{48.4\%}
\expandafter\def\csname val@zip.twin.pooled.anybreach.ci\endcsname{[78.5, 90.8]}
\expandafter\def\csname val@zip.twin.pooled.anybreach.frac\endcsname{108/126}
\expandafter\def\csname val@zip.twin.pooled.anybreach.full\endcsname{108/126 (85.7\%)}
\expandafter\def\csname val@zip.twin.pooled.anybreach.pct\endcsname{85.7\%}
\expandafter\def\csname val@zip.twin.pooled.body.ci\endcsname{[93.2, 99.2]}
\expandafter\def\csname val@zip.twin.pooled.body.frac\endcsname{123/126}
\expandafter\def\csname val@zip.twin.pooled.body.full\endcsname{123/126 (97.6\%)}
\expandafter\def\csname val@zip.twin.pooled.body.pct\endcsname{97.6\%}
\expandafter\def\csname val@zip.twin.pooled.breachbeforecontent.ci\endcsname{[42.9, 60.1]}
\expandafter\def\csname val@zip.twin.pooled.breachbeforecontent.frac\endcsname{65/126}
\expandafter\def\csname val@zip.twin.pooled.breachbeforecontent.full\endcsname{65/126 (51.6\%)}
\expandafter\def\csname val@zip.twin.pooled.breachbeforecontent.pct\endcsname{51.6\%}
\expandafter\def\csname val@zip.twin.pooled.breachbeforefirstdetectedop.ci\endcsname{[0.0, 3.0]}
\expandafter\def\csname val@zip.twin.pooled.breachbeforefirstdetectedop.frac\endcsname{0/126}
\expandafter\def\csname val@zip.twin.pooled.breachbeforefirstdetectedop.full\endcsname{0/126 (0.0\%)}
\expandafter\def\csname val@zip.twin.pooled.breachbeforefirstdetectedop.pct\endcsname{0.0\%}
\expandafter\def\csname val@zip.twin.pooled.bytebreach.ci\endcsname{[0.0, 3.0]}
\expandafter\def\csname val@zip.twin.pooled.bytebreach.frac\endcsname{0/126}
\expandafter\def\csname val@zip.twin.pooled.bytebreach.full\endcsname{0/126 (0.0\%)}
\expandafter\def\csname val@zip.twin.pooled.bytebreach.pct\endcsname{0.0\%}
\expandafter\def\csname val@zip.twin.pooled.bytebreachbeforecontent.ci\endcsname{[0.0, 3.0]}
\expandafter\def\csname val@zip.twin.pooled.bytebreachbeforecontent.frac\endcsname{0/126}
\expandafter\def\csname val@zip.twin.pooled.bytebreachbeforecontent.full\endcsname{0/126 (0.0\%)}
\expandafter\def\csname val@zip.twin.pooled.bytebreachbeforecontent.pct\endcsname{0.0\%}
\expandafter\def\csname val@zip.twin.pooled.contentincontext.ci\endcsname{[85.0, 95.1]}
\expandafter\def\csname val@zip.twin.pooled.contentincontext.frac\endcsname{115/126}
\expandafter\def\csname val@zip.twin.pooled.contentincontext.full\endcsname{115/126 (91.3\%)}
\expandafter\def\csname val@zip.twin.pooled.contentincontext.pct\endcsname{91.3\%}
\expandafter\def\csname val@zip.twin.pooled.declaredtotalseenbeforebreach.ci\endcsname{[26.4, 42.8]}
\expandafter\def\csname val@zip.twin.pooled.declaredtotalseenbeforebreach.frac\endcsname{43/126}
\expandafter\def\csname val@zip.twin.pooled.declaredtotalseenbeforebreach.full\endcsname{43/126 (34.1\%)}
\expandafter\def\csname val@zip.twin.pooled.declaredtotalseenbeforebreach.pct\endcsname{34.1\%}
\expandafter\def\csname val@zip.twin.pooled.extractinvoked.ci\endcsname{[78.5, 90.8]}
\expandafter\def\csname val@zip.twin.pooled.extractinvoked.frac\endcsname{108/126}
\expandafter\def\csname val@zip.twin.pooled.extractinvoked.full\endcsname{108/126 (85.7\%)}
\expandafter\def\csname val@zip.twin.pooled.extractinvoked.pct\endcsname{85.7\%}
\expandafter\def\csname val@zip.twin.pooled.firstopbreach.ci\endcsname{[27.2, 43.6]}
\expandafter\def\csname val@zip.twin.pooled.firstopbreach.frac\endcsname{44/126}
\expandafter\def\csname val@zip.twin.pooled.firstopbreach.full\endcsname{44/126 (34.9\%)}
\expandafter\def\csname val@zip.twin.pooled.firstopbreach.pct\endcsname{34.9\%}
\expandafter\def\csname val@zip.twin.pooled.firstopbytebreach.ci\endcsname{[0.0, 3.0]}
\expandafter\def\csname val@zip.twin.pooled.firstopbytebreach.frac\endcsname{0/126}
\expandafter\def\csname val@zip.twin.pooled.firstopbytebreach.full\endcsname{0/126 (0.0\%)}
\expandafter\def\csname val@zip.twin.pooled.firstopbytebreach.pct\endcsname{0.0\%}
\expandafter\def\csname val@zip.twin.pooled.inodebreach.ci\endcsname{[78.5, 90.8]}
\expandafter\def\csname val@zip.twin.pooled.inodebreach.frac\endcsname{108/126}
\expandafter\def\csname val@zip.twin.pooled.inodebreach.full\endcsname{108/126 (85.7\%)}
\expandafter\def\csname val@zip.twin.pooled.inodebreach.pct\endcsname{85.7\%}
}{}
\IfFileExists{numbers/rule_audit_numbers.tex}{
\expandafter\def\csname val@audit.n\endcsname{40}
\expandafter\def\csname val@audit.rulepos\endcsname{20}
\expandafter\def\csname val@audit.blind.agree\endcsname{36/40}
\expandafter\def\csname val@audit.blind.kappa\endcsname{0.80}
\expandafter\def\csname val@audit.agree\endcsname{38/40}
\expandafter\def\csname val@audit.kappa\endcsname{0.90}
\expandafter\def\csname val@audit.fp\endcsname{2}
\expandafter\def\csname val@audit.fn\endcsname{0}
\expandafter\def\csname val@audit.fptwin\endcsname{2}
\expandafter\def\csname val@audit.twin.flagged\endcsname{10}
\expandafter\def\csname val@audit.twin.genuine\endcsname{2}
\expandafter\def\csname val@audit.twin.genuine.digest\endcsname{2}
}{}
\IfFileExists{numbers/awareness_numbers.tex}{
\expandafter\def\csname val@aware.all\endcsname{32/420}
\expandafter\def\csname val@aware.pct\endcsname{7.6\%}
\expandafter\def\csname val@aware.adv.n\endcsname{21}
\expandafter\def\csname val@aware.write.obs\endcsname{4}
\expandafter\def\csname val@aware.write.exp\endcsname{3.0}
\expandafter\def\csname val@aware.reported.obs\endcsname{13}
\expandafter\def\csname val@aware.reported.exp\endcsname{8.7}
\expandafter\def\csname val@aware.top.harnesses\endcsname{mini-SWE-agent and smolagents}
\expandafter\def\csname val@aware.top.share\endcsname{25/32}
}{}
\IfFileExists{numbers/extractor_numbers.tex}{
\expandafter\def\csname val@ext.breach.n\endcsname{32}
\expandafter\def\csname val@ext.breach.unzip\endcsname{32/32}
\expandafter\def\csname val@ext.breach.bsdtar\endcsname{0/32}
\expandafter\def\csname val@ext.extract.unzip\endcsname{33/33}
\expandafter\def\csname val@ext.extract.safe\endcsname{0/33}
\expandafter\def\csname val@ext.held.n\endcsname{10}
\expandafter\def\csname val@ext.held.noextract\endcsname{9/10}
\expandafter\def\csname val@ext.held.selective\endcsname{1/10}
\expandafter\def\csname val@bound.setlimit\endcsname{0/33}
\expandafter\def\csname val@bound.scratch\endcsname{0/33}
\expandafter\def\csname val@bound.readlimits\endcsname{5}
\expandafter\def\csname val@bound.rawparse\endcsname{2/33}
}{}
\IfFileExists{numbers/model_care_numbers.tex}{
\expandafter\def\csname val@care.two.units\endcsname{11/14}
\expandafter\def\csname val@care.four.units\endcsname{0/28}
\expandafter\def\csname val@care.units.p\endcsname{$<0.001$}
\expandafter\def\csname val@care.two.readnowrite\endcsname{22/28}
\expandafter\def\csname val@care.gemini.zip.inspectfirst\endcsname{20/21}
\expandafter\def\csname val@care.gemini.zip.breach\endcsname{7/21}
\expandafter\def\csname val@care.four.zip.breach\endcsname{84/84}
\expandafter\def\csname val@care.gemini.nocommit.write\endcsname{11/21}
\expandafter\def\csname val@care.others.nocommit.write\endcsname{24/105}
\expandafter\def\csname val@care.gemini.nocommit.stated\endcsname{0/21}
\expandafter\def\csname val@care.deepseek.zip.breach\endcsname{21/21}
}{}
\IfFileExists{numbers/write_timing_numbers.tex}{
\expandafter\def\csname val@wt.writers\endcsname{131}
\expandafter\def\csname val@wt.after\endcsname{131/131}
\expandafter\def\csname val@wt.before\endcsname{0}
\expandafter\def\csname val@wt.wrong.after\endcsname{38/38}
\expandafter\def\csname val@wt.hand\endcsname{1}
}{}
\IfFileExists{numbers/budget_numbers.tex}{
\expandafter\def\csname val@budget.lo\endcsname{1}
\expandafter\def\csname val@budget.hi\endcsname{8}
\expandafter\def\csname val@budget.breach\endcsname{32/42}
\expandafter\def\csname val@budget.twin\endcsname{0/42}
\expandafter\def\csname val@budget.imm.lo\endcsname{25}
\expandafter\def\csname val@budget.imm.hi\endcsname{27}
}{}
\IfFileExists{numbers/stats_revision_numbers.tex}{
\expandafter\def\csname val@rev.c.carrierhash.a\endcsname{37/42}
\expandafter\def\csname val@rev.c.carrierhash.b\endcsname{23/42}
\expandafter\def\csname val@rev.c.carrierhash.boot\endcsname{excludes 0}
\expandafter\def\csname val@rev.c.carrierhash.diff\endcsname{+33.3}
\expandafter\def\csname val@rev.c.carrierhash.diffci\endcsname{[+7.1, +66.7]}
\expandafter\def\csname val@rev.c.carrierhash.discordant\endcsname{15 vs 1}
\expandafter\def\csname val@rev.c.carrierhash.mcnemar\endcsname{$<0.001$}
\expandafter\def\csname val@rev.c.carrierhash.signh\endcsname{5 of 5}
\expandafter\def\csname val@rev.c.carrierhash.signhp\endcsname{0.06}
\expandafter\def\csname val@rev.c.carrierhash.signm\endcsname{5 of 5}
\expandafter\def\csname val@rev.c.carrierhash.signmp\endcsname{0.06}
\expandafter\def\csname val@rev.c.carrierhash.verdict\endcsname{descriptive under clustering}
\expandafter\def\csname val@rev.c.carrierreport.a\endcsname{32/42}
\expandafter\def\csname val@rev.c.carrierreport.b\endcsname{16/42}
\expandafter\def\csname val@rev.c.carrierreport.boot\endcsname{excludes 0}
\expandafter\def\csname val@rev.c.carrierreport.diff\endcsname{+38.1}
\expandafter\def\csname val@rev.c.carrierreport.diffci\endcsname{[+4.8, +71.4]}
\expandafter\def\csname val@rev.c.carrierreport.discordant\endcsname{19 vs 3}
\expandafter\def\csname val@rev.c.carrierreport.mcnemar\endcsname{$<0.001$}
\expandafter\def\csname val@rev.c.carrierreport.signh\endcsname{6 of 6}
\expandafter\def\csname val@rev.c.carrierreport.signhp\endcsname{0.03}
\expandafter\def\csname val@rev.c.carrierreport.signm\endcsname{5 of 6}
\expandafter\def\csname val@rev.c.carrierreport.signmp\endcsname{0.22}
\expandafter\def\csname val@rev.c.carrierreport.verdict\endcsname{descriptive under clustering}
\expandafter\def\csname val@rev.c.commitimg.a\endcsname{19/42}
\expandafter\def\csname val@rev.c.commitimg.b\endcsname{1/42}
\expandafter\def\csname val@rev.c.commitimg.boot\endcsname{excludes 0}
\expandafter\def\csname val@rev.c.commitimg.diff\endcsname{+42.9}
\expandafter\def\csname val@rev.c.commitimg.diffci\endcsname{[+14.3, +71.4]}
\expandafter\def\csname val@rev.c.commitimg.discordant\endcsname{18 vs 0}
\expandafter\def\csname val@rev.c.commitimg.mcnemar\endcsname{$<0.001$}
\expandafter\def\csname val@rev.c.commitimg.signh\endcsname{7 of 7}
\expandafter\def\csname val@rev.c.commitimg.signhp\endcsname{0.02}
\expandafter\def\csname val@rev.c.commitimg.signm\endcsname{6 of 6}
\expandafter\def\csname val@rev.c.commitimg.signmp\endcsname{0.03}
\expandafter\def\csname val@rev.c.commitimg.verdict\endcsname{holds under clustering}
\expandafter\def\csname val@rev.c.commitnet.a\endcsname{11/42}
\expandafter\def\csname val@rev.c.commitnet.b\endcsname{1/42}
\expandafter\def\csname val@rev.c.commitnet.boot\endcsname{excludes 0}
\expandafter\def\csname val@rev.c.commitnet.diff\endcsname{+23.8}
\expandafter\def\csname val@rev.c.commitnet.diffci\endcsname{[+4.8, +50.0]}
\expandafter\def\csname val@rev.c.commitnet.discordant\endcsname{10 vs 0}
\expandafter\def\csname val@rev.c.commitnet.mcnemar\endcsname{0.002}
\expandafter\def\csname val@rev.c.commitnet.signh\endcsname{5 of 5}
\expandafter\def\csname val@rev.c.commitnet.signhp\endcsname{0.06}
\expandafter\def\csname val@rev.c.commitnet.signm\endcsname{5 of 5}
\expandafter\def\csname val@rev.c.commitnet.signmp\endcsname{0.06}
\expandafter\def\csname val@rev.c.commitnet.verdict\endcsname{descriptive under clustering}
\expandafter\def\csname val@rev.c.descriptive\endcsname{commitnet, carrierhash, carrierreport, twintargetheader, twintargetinbody, twintargetqrdual}
\expandafter\def\csname val@rev.c.holds\endcsname{ziptwin, commitimg, twintarget}
\expandafter\def\csname val@rev.c.twintarget.a\endcsname{37/126}
\expandafter\def\csname val@rev.c.twintarget.b\endcsname{60/126}
\expandafter\def\csname val@rev.c.twintarget.boot\endcsname{excludes 0}
\expandafter\def\csname val@rev.c.twintarget.diff\endcsname{-18.3}
\expandafter\def\csname val@rev.c.twintarget.diffci\endcsname{[-34.1, -3.2]}
\expandafter\def\csname val@rev.c.twintarget.discordant\endcsname{5 vs 28}
\expandafter\def\csname val@rev.c.twintarget.mcnemar\endcsname{$<0.001$}
\expandafter\def\csname val@rev.c.twintarget.signh\endcsname{0 of 6}
\expandafter\def\csname val@rev.c.twintarget.signhp\endcsname{0.03}
\expandafter\def\csname val@rev.c.twintarget.signm\endcsname{0 of 6}
\expandafter\def\csname val@rev.c.twintarget.signmp\endcsname{0.03}
\expandafter\def\csname val@rev.c.twintarget.verdict\endcsname{holds under clustering}
\expandafter\def\csname val@rev.c.twintargetheader.a\endcsname{7/42}
\expandafter\def\csname val@rev.c.twintargetheader.b\endcsname{14/42}
\expandafter\def\csname val@rev.c.twintargetheader.boot\endcsname{includes 0}
\expandafter\def\csname val@rev.c.twintargetheader.diff\endcsname{-16.7}
\expandafter\def\csname val@rev.c.twintargetheader.diffci\endcsname{[-42.9, +4.8]}
\expandafter\def\csname val@rev.c.twintargetheader.discordant\endcsname{1 vs 8}
\expandafter\def\csname val@rev.c.twintargetheader.mcnemar\endcsname{0.04}
\expandafter\def\csname val@rev.c.twintargetheader.signh\endcsname{1 of 7}
\expandafter\def\csname val@rev.c.twintargetheader.signhp\endcsname{0.12}
\expandafter\def\csname val@rev.c.twintargetheader.signm\endcsname{1 of 5}
\expandafter\def\csname val@rev.c.twintargetheader.signmp\endcsname{0.38}
\expandafter\def\csname val@rev.c.twintargetheader.verdict\endcsname{descriptive under clustering}
\expandafter\def\csname val@rev.c.twintargetinbody.a\endcsname{11/42}
\expandafter\def\csname val@rev.c.twintargetinbody.b\endcsname{21/42}
\expandafter\def\csname val@rev.c.twintargetinbody.boot\endcsname{includes 0}
\expandafter\def\csname val@rev.c.twintargetinbody.diff\endcsname{-23.8}
\expandafter\def\csname val@rev.c.twintargetinbody.diffci\endcsname{[-54.8, +2.4]}
\expandafter\def\csname val@rev.c.twintargetinbody.discordant\endcsname{2 vs 12}
\expandafter\def\csname val@rev.c.twintargetinbody.mcnemar\endcsname{0.01}
\expandafter\def\csname val@rev.c.twintargetinbody.signh\endcsname{2 of 7}
\expandafter\def\csname val@rev.c.twintargetinbody.signhp\endcsname{0.45}
\expandafter\def\csname val@rev.c.twintargetinbody.signm\endcsname{0 of 5}
\expandafter\def\csname val@rev.c.twintargetinbody.signmp\endcsname{0.06}
\expandafter\def\csname val@rev.c.twintargetinbody.verdict\endcsname{descriptive under clustering}
\expandafter\def\csname val@rev.c.twintargetqrdual.a\endcsname{19/42}
\expandafter\def\csname val@rev.c.twintargetqrdual.b\endcsname{25/42}
\expandafter\def\csname val@rev.c.twintargetqrdual.boot\endcsname{includes 0}
\expandafter\def\csname val@rev.c.twintargetqrdual.diff\endcsname{-14.3}
\expandafter\def\csname val@rev.c.twintargetqrdual.diffci\endcsname{[-35.7, +4.8]}
\expandafter\def\csname val@rev.c.twintargetqrdual.discordant\endcsname{2 vs 8}
\expandafter\def\csname val@rev.c.twintargetqrdual.mcnemar\endcsname{0.11}
\expandafter\def\csname val@rev.c.twintargetqrdual.signh\endcsname{0 of 4}
\expandafter\def\csname val@rev.c.twintargetqrdual.signhp\endcsname{0.12}
\expandafter\def\csname val@rev.c.twintargetqrdual.signm\endcsname{0 of 3}
\expandafter\def\csname val@rev.c.twintargetqrdual.signmp\endcsname{0.25}
\expandafter\def\csname val@rev.c.twintargetqrdual.verdict\endcsname{descriptive under clustering}
\expandafter\def\csname val@rev.c.ziptwin.a\endcsname{32/42}
\expandafter\def\csname val@rev.c.ziptwin.b\endcsname{0/42}
\expandafter\def\csname val@rev.c.ziptwin.boot\endcsname{excludes 0}
\expandafter\def\csname val@rev.c.ziptwin.diff\endcsname{+76.2}
\expandafter\def\csname val@rev.c.ziptwin.diffci\endcsname{[+42.9, +100.0]}
\expandafter\def\csname val@rev.c.ziptwin.discordant\endcsname{32 vs 0}
\expandafter\def\csname val@rev.c.ziptwin.mcnemar\endcsname{$<0.001$}
\expandafter\def\csname val@rev.c.ziptwin.signh\endcsname{7 of 7}
\expandafter\def\csname val@rev.c.ziptwin.signhp\endcsname{0.02}
\expandafter\def\csname val@rev.c.ziptwin.signm\endcsname{6 of 6}
\expandafter\def\csname val@rev.c.ziptwin.signmp\endcsname{0.03}
\expandafter\def\csname val@rev.c.ziptwin.verdict\endcsname{holds under clustering}
\expandafter\def\csname val@rev.eta.digestreject.harness\endcsname{0.12}
\expandafter\def\csname val@rev.eta.digestreject.model\endcsname{0.07}
\expandafter\def\csname val@rev.eta.digestreject.resid\endcsname{0.81}
\expandafter\def\csname val@rev.eta.headeringress.harness\endcsname{0.40}
\expandafter\def\csname val@rev.eta.headeringress.model\endcsname{0.20}
\expandafter\def\csname val@rev.eta.headeringress.resid\endcsname{0.40}
\expandafter\def\csname val@rev.eta.nocommitwrite.harness\endcsname{0.16}
\expandafter\def\csname val@rev.eta.nocommitwrite.model\endcsname{0.13}
\expandafter\def\csname val@rev.eta.nocommitwrite.resid\endcsname{0.71}
\expandafter\def\csname val@rev.eta.zipbreach.harness\endcsname{0.08}
\expandafter\def\csname val@rev.eta.zipbreach.model\endcsname{0.62}
\expandafter\def\csname val@rev.eta.zipbreach.resid\endcsname{0.29}
\expandafter\def\csname val@rev.ext.c.boundvsdigestreject.ext\endcsname{25/28}
\expandafter\def\csname val@rev.ext.c.boundvsdigestreject.main\endcsname{24/28}
\expandafter\def\csname val@rev.ext.c.boundvsdigestreject.mcnemar\endcsname{1.00}
\expandafter\def\csname val@rev.ext.c.boundvsdigestreject.signhp\endcsname{1.00}
\expandafter\def\csname val@rev.ext.c.boundvsdigestreject.signmp\endcsname{1.00}
\expandafter\def\csname val@rev.ext.c.boundvsinband.ext\endcsname{1/28}
\expandafter\def\csname val@rev.ext.c.boundvsinband.main\endcsname{9/28}
\expandafter\def\csname val@rev.ext.c.boundvsinband.mcnemar\endcsname{0.008}
\expandafter\def\csname val@rev.ext.c.boundvsinband.signhp\endcsname{0.06}
\expandafter\def\csname val@rev.ext.c.boundvsinband.signmp\endcsname{0.12}
\expandafter\def\csname val@rev.ext.c.enfvsraw.ext\endcsname{0/28}
\expandafter\def\csname val@rev.ext.c.enfvsraw.main\endcsname{28/28}
\expandafter\def\csname val@rev.ext.c.enfvsraw.mcnemar\endcsname{$<0.001$}
\expandafter\def\csname val@rev.ext.c.enfvsraw.signhp\endcsname{0.02}
\expandafter\def\csname val@rev.ext.c.enfvsraw.signmp\endcsname{0.12}
\expandafter\def\csname val@rev.ext.c.iswapvsqrdual.ext\endcsname{12/28}
\expandafter\def\csname val@rev.ext.c.iswapvsqrdual.main\endcsname{11/28}
\expandafter\def\csname val@rev.ext.c.iswapvsqrdual.mcnemar\endcsname{1.00}
\expandafter\def\csname val@rev.ext.c.iswapvsqrdual.signhp\endcsname{1.00}
\expandafter\def\csname val@rev.ext.c.iswapvsqrdual.signmp\endcsname{1.00}
\expandafter\def\csname val@rev.ext.c.optvsraw.ext\endcsname{28/28}
\expandafter\def\csname val@rev.ext.c.optvsraw.main\endcsname{28/28}
\expandafter\def\csname val@rev.ext.c.optvsraw.mcnemar\endcsname{1.00}
\expandafter\def\csname val@rev.ext.c.optvsraw.signhp\endcsname{1.00}
\expandafter\def\csname val@rev.ext.c.optvsraw.signmp\endcsname{1.00}
\expandafter\def\csname val@rev.ext.e1iswapadv.step0hit.ci\endcsname{[100.0, 100.0]}
\expandafter\def\csname val@rev.ext.e1iswapadv.step0hit.frac\endcsname{28/28}
\expandafter\def\csname val@rev.ext.e1iswapadv.step0hit.pct\endcsname{100.0\%}
\expandafter\def\csname val@rev.ext.e1iswapadv.targethasconflictingbody.ci\endcsname{[14.3, 71.4]}
\expandafter\def\csname val@rev.ext.e1iswapadv.targethasconflictingbody.frac\endcsname{12/28}
\expandafter\def\csname val@rev.ext.e1iswapadv.targethasconflictingbody.pct\endcsname{42.9\%}
\expandafter\def\csname val@rev.ext.e1iswapadv.targetwritteninworkspace.ci\endcsname{[14.3, 71.4]}
\expandafter\def\csname val@rev.ext.e1iswapadv.targetwritteninworkspace.frac\endcsname{12/28}
\expandafter\def\csname val@rev.ext.e1iswapadv.targetwritteninworkspace.pct\endcsname{42.9\%}
\expandafter\def\csname val@rev.ext.e1iswaptwin.targethasdeclaredbody.ci\endcsname{[17.9, 78.6]}
\expandafter\def\csname val@rev.ext.e1iswaptwin.targethasdeclaredbody.frac\endcsname{14/28}
\expandafter\def\csname val@rev.ext.e1iswaptwin.targethasdeclaredbody.pct\endcsname{50.0\%}
\expandafter\def\csname val@rev.ext.e1iswaptwin.targetwritteninworkspace.ci\endcsname{[17.9, 78.6]}
\expandafter\def\csname val@rev.ext.e1iswaptwin.targetwritteninworkspace.frac\endcsname{14/28}
\expandafter\def\csname val@rev.ext.e1iswaptwin.targetwritteninworkspace.pct\endcsname{50.0\%}
\expandafter\def\csname val@rev.ext.e1z1adv.zipbytebreach.ci\endcsname{[100.0, 100.0]}
\expandafter\def\csname val@rev.ext.e1z1adv.zipbytebreach.frac\endcsname{28/28}
\expandafter\def\csname val@rev.ext.e1z1adv.zipbytebreach.pct\endcsname{100.0\%}
\expandafter\def\csname val@rev.ext.e1z1adv.zipextractinvoked.ci\endcsname{[100.0, 100.0]}
\expandafter\def\csname val@rev.ext.e1z1adv.zipextractinvoked.frac\endcsname{28/28}
\expandafter\def\csname val@rev.ext.e1z1adv.zipextractinvoked.pct\endcsname{100.0\%}
\expandafter\def\csname val@rev.ext.e1z1twin.zipbytebreach.ci\endcsname{[0.0, 0.0]}
\expandafter\def\csname val@rev.ext.e1z1twin.zipbytebreach.frac\endcsname{0/28}
\expandafter\def\csname val@rev.ext.e1z1twin.zipbytebreach.pct\endcsname{0.0\%}
\expandafter\def\csname val@rev.ext.e1z2adv.zipbytebreach.ci\endcsname{[100.0, 100.0]}
\expandafter\def\csname val@rev.ext.e1z2adv.zipbytebreach.frac\endcsname{28/28}
\expandafter\def\csname val@rev.ext.e1z2adv.zipbytebreach.pct\endcsname{100.0\%}
\expandafter\def\csname val@rev.ext.e1z2adv.zipextractinvoked.ci\endcsname{[100.0, 100.0]}
\expandafter\def\csname val@rev.ext.e1z2adv.zipextractinvoked.frac\endcsname{28/28}
\expandafter\def\csname val@rev.ext.e1z2adv.zipextractinvoked.pct\endcsname{100.0\%}
\expandafter\def\csname val@rev.ext.e1z2twin.zipbytebreach.ci\endcsname{[0.0, 0.0]}
\expandafter\def\csname val@rev.ext.e1z2twin.zipbytebreach.frac\endcsname{0/28}
\expandafter\def\csname val@rev.ext.e1z2twin.zipbytebreach.pct\endcsname{0.0\%}
\expandafter\def\csname val@rev.ext.e3digestboundadv.boundaryverdictincontext.ci\endcsname{[100.0, 100.0]}
\expandafter\def\csname val@rev.ext.e3digestboundadv.boundaryverdictincontext.frac\endcsname{28/28}
\expandafter\def\csname val@rev.ext.e3digestboundadv.boundaryverdictincontext.pct\endcsname{100.0\%}
\expandafter\def\csname val@rev.ext.e3digestboundadv.computedhash.ci\endcsname{[71.4, 100.0]}
\expandafter\def\csname val@rev.ext.e3digestboundadv.computedhash.frac\endcsname{26/28}
\expandafter\def\csname val@rev.ext.e3digestboundadv.computedhash.pct\endcsname{92.9\%}
\expandafter\def\csname val@rev.ext.e3digestboundadv.conflictrejected.ci\endcsname{[67.9, 100.0]}
\expandafter\def\csname val@rev.ext.e3digestboundadv.conflictrejected.frac\endcsname{25/28}
\expandafter\def\csname val@rev.ext.e3digestboundadv.conflictrejected.pct\endcsname{89.3\%}
\expandafter\def\csname val@rev.ext.e3digestboundadv.conflictreported.ci\endcsname{[78.6, 100.0]}
\expandafter\def\csname val@rev.ext.e3digestboundadv.conflictreported.frac\endcsname{26/28}
\expandafter\def\csname val@rev.ext.e3digestboundadv.conflictreported.pct\endcsname{92.9\%}
\expandafter\def\csname val@rev.ext.e3digestboundadv.targethasconflictingbody.ci\endcsname{[0.0, 21.4]}
\expandafter\def\csname val@rev.ext.e3digestboundadv.targethasconflictingbody.frac\endcsname{1/28}
\expandafter\def\csname val@rev.ext.e3digestboundadv.targethasconflictingbody.pct\endcsname{3.6\%}
\expandafter\def\csname val@rev.ext.e3digestboundadv.targetwritteninworkspace.ci\endcsname{[0.0, 42.9]}
\expandafter\def\csname val@rev.ext.e3digestboundadv.targetwritteninworkspace.frac\endcsname{4/28}
\expandafter\def\csname val@rev.ext.e3digestboundadv.targetwritteninworkspace.pct\endcsname{14.3\%}
\expandafter\def\csname val@rev.ext.e3digestboundtwin.conflictreported.ci\endcsname{[17.9, 71.4]}
\expandafter\def\csname val@rev.ext.e3digestboundtwin.conflictreported.frac\endcsname{12/28}
\expandafter\def\csname val@rev.ext.e3digestboundtwin.conflictreported.pct\endcsname{42.9\%}
\expandafter\def\csname val@rev.ext.e3digestboundtwin.targethasdeclaredbody.ci\endcsname{[0.0, 42.9]}
\expandafter\def\csname val@rev.ext.e3digestboundtwin.targethasdeclaredbody.frac\endcsname{5/28}
\expandafter\def\csname val@rev.ext.e3digestboundtwin.targethasdeclaredbody.pct\endcsname{17.9\%}
\expandafter\def\csname val@rev.ext.e3digestboundtwin.targetwritteninworkspace.ci\endcsname{[0.0, 42.9]}
\expandafter\def\csname val@rev.ext.e3digestboundtwin.targetwritteninworkspace.frac\endcsname{5/28}
\expandafter\def\csname val@rev.ext.e3digestboundtwin.targetwritteninworkspace.pct\endcsname{17.9\%}
\expandafter\def\csname val@rev.ext.e3zipenfadv.boundarybypassattempted.ci\endcsname{[14.3, 92.9]}
\expandafter\def\csname val@rev.ext.e3zipenfadv.boundarybypassattempted.frac\endcsname{15/28}
\expandafter\def\csname val@rev.ext.e3zipenfadv.boundarybypassattempted.pct\endcsname{53.6\%}
\expandafter\def\csname val@rev.ext.e3zipenfadv.zipbytebreach.ci\endcsname{[0.0, 0.0]}
\expandafter\def\csname val@rev.ext.e3zipenfadv.zipbytebreach.frac\endcsname{0/28}
\expandafter\def\csname val@rev.ext.e3zipenfadv.zipbytebreach.pct\endcsname{0.0\%}
\expandafter\def\csname val@rev.ext.e3zipenfadv.zipextractinvoked.ci\endcsname{[78.6, 100.0]}
\expandafter\def\csname val@rev.ext.e3zipenfadv.zipextractinvoked.frac\endcsname{27/28}
\expandafter\def\csname val@rev.ext.e3zipenfadv.zipextractinvoked.pct\endcsname{96.4\%}
\expandafter\def\csname val@rev.ext.e3zipenftwin.fullextract\endcsname{27/28}
\expandafter\def\csname val@rev.ext.e3zipenftwin.zipbytebreach.ci\endcsname{[0.0, 0.0]}
\expandafter\def\csname val@rev.ext.e3zipenftwin.zipbytebreach.frac\endcsname{0/28}
\expandafter\def\csname val@rev.ext.e3zipenftwin.zipbytebreach.pct\endcsname{0.0\%}
\expandafter\def\csname val@rev.ext.e3zipenftwin.zipextractinvoked.ci\endcsname{[100.0, 100.0]}
\expandafter\def\csname val@rev.ext.e3zipenftwin.zipextractinvoked.frac\endcsname{28/28}
\expandafter\def\csname val@rev.ext.e3zipenftwin.zipextractinvoked.pct\endcsname{100.0\%}
\expandafter\def\csname val@rev.ext.e3zipoptadv.rawunzipcalled.ci\endcsname{[100.0, 100.0]}
\expandafter\def\csname val@rev.ext.e3zipoptadv.rawunzipcalled.frac\endcsname{28/28}
\expandafter\def\csname val@rev.ext.e3zipoptadv.rawunzipcalled.pct\endcsname{100.0\%}
\expandafter\def\csname val@rev.ext.e3zipoptadv.safearchivecalled.ci\endcsname{[0.0, 42.9]}
\expandafter\def\csname val@rev.ext.e3zipoptadv.safearchivecalled.frac\endcsname{5/28}
\expandafter\def\csname val@rev.ext.e3zipoptadv.safearchivecalled.pct\endcsname{17.9\%}
\expandafter\def\csname val@rev.ext.e3zipoptadv.zipbytebreach.ci\endcsname{[100.0, 100.0]}
\expandafter\def\csname val@rev.ext.e3zipoptadv.zipbytebreach.frac\endcsname{28/28}
\expandafter\def\csname val@rev.ext.e3zipoptadv.zipbytebreach.pct\endcsname{100.0\%}
\expandafter\def\csname val@rev.ext.e3zipoptadv.zipbytebreachbeforecontent.ci\endcsname{[53.6, 100.0]}
\expandafter\def\csname val@rev.ext.e3zipoptadv.zipbytebreachbeforecontent.frac\endcsname{23/28}
\expandafter\def\csname val@rev.ext.e3zipoptadv.zipbytebreachbeforecontent.pct\endcsname{82.1\%}
\expandafter\def\csname val@rev.ext.e3zipopttwin.safearchivecalled.ci\endcsname{[0.0, 35.7]}
\expandafter\def\csname val@rev.ext.e3zipopttwin.safearchivecalled.frac\endcsname{4/28}
\expandafter\def\csname val@rev.ext.e3zipopttwin.safearchivecalled.pct\endcsname{14.3\%}
\expandafter\def\csname val@rev.ext.e3zipopttwin.zipbytebreach.ci\endcsname{[0.0, 0.0]}
\expandafter\def\csname val@rev.ext.e3zipopttwin.zipbytebreach.frac\endcsname{0/28}
\expandafter\def\csname val@rev.ext.e3zipopttwin.zipbytebreach.pct\endcsname{0.0\%}
\expandafter\def\csname val@rev.fx.copy\endcsname{0}
\expandafter\def\csname val@rev.fx.derived\endcsname{0}
\expandafter\def\csname val@rev.fx.inspected\endcsname{0}
\expandafter\def\csname val@rev.fx.literal\endcsname{0}
\expandafter\def\csname val@rev.fx.n\endcsname{11}
\expandafter\def\csname val@rev.fx.nolog\endcsname{11}
\expandafter\def\csname val@rev.fx.nomatch\endcsname{0}
\expandafter\def\csname val@rev.fx.other\endcsname{0}
\expandafter\def\csname val@rev.fx.search\endcsname{0}
\expandafter\def\csname val@rev.fx.taufirst\endcsname{0}
\expandafter\def\csname val@rev.model.deepseek.digesthash.ci\endcsname{[78.6, 100.0]}
\expandafter\def\csname val@rev.model.deepseek.digesthash.frac\endcsname{13/14}
\expandafter\def\csname val@rev.model.deepseek.digesthash.pct\endcsname{92.9\%}
\expandafter\def\csname val@rev.model.deepseek.digesthash.wilson\endcsname{[68.5, 98.7]}
\expandafter\def\csname val@rev.model.deepseek.digestreject.ci\endcsname{[57.1, 92.9]}
\expandafter\def\csname val@rev.model.deepseek.digestreject.frac\endcsname{11/14}
\expandafter\def\csname val@rev.model.deepseek.digestreject.pct\endcsname{78.6\%}
\expandafter\def\csname val@rev.model.deepseek.digestreject.wilson\endcsname{[52.4, 92.4]}
\expandafter\def\csname val@rev.model.deepseek.nocommitwrite.ci\endcsname{[4.8, 47.6]}
\expandafter\def\csname val@rev.model.deepseek.nocommitwrite.frac\endcsname{5/21}
\expandafter\def\csname val@rev.model.deepseek.nocommitwrite.pct\endcsname{23.8\%}
\expandafter\def\csname val@rev.model.deepseek.nocommitwrite.wilson\endcsname{[10.6, 45.1]}
\expandafter\def\csname val@rev.model.deepseek.zipbreach3.ci\endcsname{[100.0, 100.0]}
\expandafter\def\csname val@rev.model.deepseek.zipbreach3.frac\endcsname{21/21}
\expandafter\def\csname val@rev.model.deepseek.zipbreach3.pct\endcsname{100.0\%}
\expandafter\def\csname val@rev.model.deepseek.zipbreach3.wilson\endcsname{[84.5, 100.0]}
\expandafter\def\csname val@rev.model.gemini.digesthash.ci\endcsname{[100.0, 100.0]}
\expandafter\def\csname val@rev.model.gemini.digesthash.frac\endcsname{14/14}
\expandafter\def\csname val@rev.model.gemini.digesthash.pct\endcsname{100.0\%}
\expandafter\def\csname val@rev.model.gemini.digesthash.wilson\endcsname{[78.5, 100.0]}
\expandafter\def\csname val@rev.model.gemini.digestreject.ci\endcsname{[21.4, 78.6]}
\expandafter\def\csname val@rev.model.gemini.digestreject.frac\endcsname{7/14}
\expandafter\def\csname val@rev.model.gemini.digestreject.pct\endcsname{50.0\%}
\expandafter\def\csname val@rev.model.gemini.digestreject.wilson\endcsname{[26.8, 73.2]}
\expandafter\def\csname val@rev.model.gemini.nocommitwrite.ci\endcsname{[33.3, 71.4]}
\expandafter\def\csname val@rev.model.gemini.nocommitwrite.frac\endcsname{11/21}
\expandafter\def\csname val@rev.model.gemini.nocommitwrite.pct\endcsname{52.4\%}
\expandafter\def\csname val@rev.model.gemini.nocommitwrite.wilson\endcsname{[32.4, 71.7]}
\expandafter\def\csname val@rev.model.gemini.zipbreach3.ci\endcsname{[4.8, 66.7]}
\expandafter\def\csname val@rev.model.gemini.zipbreach3.frac\endcsname{7/21}
\expandafter\def\csname val@rev.model.gemini.zipbreach3.pct\endcsname{33.3\%}
\expandafter\def\csname val@rev.model.gemini.zipbreach3.wilson\endcsname{[17.2, 54.6]}
\expandafter\def\csname val@rev.model.glm.digesthash.ci\endcsname{[35.7, 78.6]}
\expandafter\def\csname val@rev.model.glm.digesthash.frac\endcsname{8/14}
\expandafter\def\csname val@rev.model.glm.digesthash.pct\endcsname{57.1\%}
\expandafter\def\csname val@rev.model.glm.digesthash.wilson\endcsname{[32.6, 78.6]}
\expandafter\def\csname val@rev.model.glm.digestreject.ci\endcsname{[35.7, 78.6]}
\expandafter\def\csname val@rev.model.glm.digestreject.frac\endcsname{8/14}
\expandafter\def\csname val@rev.model.glm.digestreject.pct\endcsname{57.1\%}
\expandafter\def\csname val@rev.model.glm.digestreject.wilson\endcsname{[32.6, 78.6]}
\expandafter\def\csname val@rev.model.glm.nocommitwrite.ci\endcsname{[0.0, 33.3]}
\expandafter\def\csname val@rev.model.glm.nocommitwrite.frac\endcsname{3/21}
\expandafter\def\csname val@rev.model.glm.nocommitwrite.pct\endcsname{14.3\%}
\expandafter\def\csname val@rev.model.glm.nocommitwrite.wilson\endcsname{[5.0, 34.6]}
\expandafter\def\csname val@rev.model.glm.zipbreach3.ci\endcsname{[100.0, 100.0]}
\expandafter\def\csname val@rev.model.glm.zipbreach3.frac\endcsname{21/21}
\expandafter\def\csname val@rev.model.glm.zipbreach3.pct\endcsname{100.0\%}
\expandafter\def\csname val@rev.model.glm.zipbreach3.wilson\endcsname{[84.5, 100.0]}
\expandafter\def\csname val@rev.model.gpt.digesthash.ci\endcsname{[14.3, 71.4]}
\expandafter\def\csname val@rev.model.gpt.digesthash.frac\endcsname{6/14}
\expandafter\def\csname val@rev.model.gpt.digesthash.pct\endcsname{42.9\%}
\expandafter\def\csname val@rev.model.gpt.digesthash.wilson\endcsname{[21.4, 67.4]}
\expandafter\def\csname val@rev.model.gpt.digestreject.ci\endcsname{[14.3, 64.3]}
\expandafter\def\csname val@rev.model.gpt.digestreject.frac\endcsname{5/14}
\expandafter\def\csname val@rev.model.gpt.digestreject.pct\endcsname{35.7\%}
\expandafter\def\csname val@rev.model.gpt.digestreject.wilson\endcsname{[16.3, 61.2]}
\expandafter\def\csname val@rev.model.gpt.nocommitwrite.ci\endcsname{[0.0, 14.3]}
\expandafter\def\csname val@rev.model.gpt.nocommitwrite.frac\endcsname{1/21}
\expandafter\def\csname val@rev.model.gpt.nocommitwrite.pct\endcsname{4.8\%}
\expandafter\def\csname val@rev.model.gpt.nocommitwrite.wilson\endcsname{[0.8, 22.7]}
\expandafter\def\csname val@rev.model.gpt.zipbreach3.ci\endcsname{[9.5, 61.9]}
\expandafter\def\csname val@rev.model.gpt.zipbreach3.frac\endcsname{7/21}
\expandafter\def\csname val@rev.model.gpt.zipbreach3.pct\endcsname{33.3\%}
\expandafter\def\csname val@rev.model.gpt.zipbreach3.wilson\endcsname{[17.2, 54.6]}
\expandafter\def\csname val@rev.model.kimi.digesthash.ci\endcsname{[50.0, 78.6]}
\expandafter\def\csname val@rev.model.kimi.digesthash.frac\endcsname{9/14}
\expandafter\def\csname val@rev.model.kimi.digesthash.pct\endcsname{64.3\%}
\expandafter\def\csname val@rev.model.kimi.digesthash.wilson\endcsname{[38.8, 83.7]}
\expandafter\def\csname val@rev.model.kimi.digestreject.ci\endcsname{[21.4, 64.3]}
\expandafter\def\csname val@rev.model.kimi.digestreject.frac\endcsname{6/14}
\expandafter\def\csname val@rev.model.kimi.digestreject.pct\endcsname{42.9\%}
\expandafter\def\csname val@rev.model.kimi.digestreject.wilson\endcsname{[21.4, 67.4]}
\expandafter\def\csname val@rev.model.kimi.nocommitwrite.ci\endcsname{[9.5, 57.1]}
\expandafter\def\csname val@rev.model.kimi.nocommitwrite.frac\endcsname{7/21}
\expandafter\def\csname val@rev.model.kimi.nocommitwrite.pct\endcsname{33.3\%}
\expandafter\def\csname val@rev.model.kimi.nocommitwrite.wilson\endcsname{[17.2, 54.6]}
\expandafter\def\csname val@rev.model.kimi.zipbreach3.ci\endcsname{[100.0, 100.0]}
\expandafter\def\csname val@rev.model.kimi.zipbreach3.frac\endcsname{21/21}
\expandafter\def\csname val@rev.model.kimi.zipbreach3.pct\endcsname{100.0\%}
\expandafter\def\csname val@rev.model.kimi.zipbreach3.wilson\endcsname{[84.5, 100.0]}
\expandafter\def\csname val@rev.model.qwen.digesthash.ci\endcsname{[57.1, 92.9]}
\expandafter\def\csname val@rev.model.qwen.digesthash.frac\endcsname{10/14}
\expandafter\def\csname val@rev.model.qwen.digesthash.pct\endcsname{71.4\%}
\expandafter\def\csname val@rev.model.qwen.digesthash.wilson\endcsname{[45.4, 88.3]}
\expandafter\def\csname val@rev.model.qwen.digestreject.ci\endcsname{[35.7, 78.6]}
\expandafter\def\csname val@rev.model.qwen.digestreject.frac\endcsname{8/14}
\expandafter\def\csname val@rev.model.qwen.digestreject.pct\endcsname{57.1\%}
\expandafter\def\csname val@rev.model.qwen.digestreject.wilson\endcsname{[32.6, 78.6]}
\expandafter\def\csname val@rev.model.qwen.nocommitwrite.ci\endcsname{[19.0, 66.7]}
\expandafter\def\csname val@rev.model.qwen.nocommitwrite.frac\endcsname{9/21}
\expandafter\def\csname val@rev.model.qwen.nocommitwrite.pct\endcsname{42.9\%}
\expandafter\def\csname val@rev.model.qwen.nocommitwrite.wilson\endcsname{[24.5, 63.5]}
\expandafter\def\csname val@rev.model.qwen.zipbreach3.ci\endcsname{[100.0, 100.0]}
\expandafter\def\csname val@rev.model.qwen.zipbreach3.frac\endcsname{21/21}
\expandafter\def\csname val@rev.model.qwen.zipbreach3.pct\endcsname{100.0\%}
\expandafter\def\csname val@rev.model.qwen.zipbreach3.wilson\endcsname{[84.5, 100.0]}
\expandafter\def\csname val@rev.order.digestreject.discordant\endcsname{3 vs 0}
\expandafter\def\csname val@rev.order.digestreject.mcnemar\endcsname{0.25}
\expandafter\def\csname val@rev.order.digestreject.second\endcsname{GLM-5.3}
\expandafter\def\csname val@rev.order.digestreject.secondfrac\endcsname{8/14}
\expandafter\def\csname val@rev.order.digestreject.signhp\endcsname{0.25}
\expandafter\def\csname val@rev.order.digestreject.top\endcsname{DeepSeek-V4.1-Flash}
\expandafter\def\csname val@rev.order.digestreject.topfrac\endcsname{11/14}
\expandafter\def\csname val@rev.order.digestreject.verdict\endcsname{descriptive}
\expandafter\def\csname val@rev.order.nocommitwrite.discordant\endcsname{6 vs 4}
\expandafter\def\csname val@rev.order.nocommitwrite.mcnemar\endcsname{0.75}
\expandafter\def\csname val@rev.order.nocommitwrite.second\endcsname{Qwen-3.8-Max}
\expandafter\def\csname val@rev.order.nocommitwrite.secondfrac\endcsname{9/21}
\expandafter\def\csname val@rev.order.nocommitwrite.signhp\endcsname{0.69}
\expandafter\def\csname val@rev.order.nocommitwrite.top\endcsname{Gemini-4-Argon}
\expandafter\def\csname val@rev.order.nocommitwrite.topfrac\endcsname{11/21}
\expandafter\def\csname val@rev.order.nocommitwrite.verdict\endcsname{descriptive}
\expandafter\def\csname val@rev.ow.container.bsdtarx.fresh\endcsname{writes (immediate)}
\expandafter\def\csname val@rev.ow.container.bsdtarx.pre\endcsname{writes (immediate)}
\expandafter\def\csname val@rev.ow.container.pythontarfile.fresh\endcsname{writes (immediate)}
\expandafter\def\csname val@rev.ow.container.pythontarfile.pre\endcsname{writes (immediate)}
\expandafter\def\csname val@rev.ow.container.tarlist.fresh\endcsname{no}
\expandafter\def\csname val@rev.ow.container.tarlist.pre\endcsname{no}
\expandafter\def\csname val@rev.ow.container.tarx.fresh\endcsname{writes (immediate)}
\expandafter\def\csname val@rev.ow.container.tarx.pre\endcsname{writes (immediate)}
\expandafter\def\csname val@rev.ow.container.tarxkeepold.fresh\endcsname{writes (immediate)}
\expandafter\def\csname val@rev.ow.container.tarxkeepold.pre\endcsname{no; refuses to overwrite (rc 2)}
\expandafter\def\csname val@rev.ow.network.curlOJ.fresh\endcsname{writes (immediate)}
\expandafter\def\csname val@rev.ow.network.curlOJ.pre\endcsname{no; refuses to overwrite (rc 23)}
\expandafter\def\csname val@rev.ow.network.curloother.fresh\endcsname{no}
\expandafter\def\csname val@rev.ow.network.curloother.pre\endcsname{no}
\expandafter\def\csname val@rev.ow.network.curlstdout.fresh\endcsname{no}
\expandafter\def\csname val@rev.ow.network.curlstdout.pre\endcsname{no}
\expandafter\def\csname val@rev.ow.network.pythonurllibstdout.fresh\endcsname{no}
\expandafter\def\csname val@rev.ow.network.pythonurllibstdout.pre\endcsname{no}
\expandafter\def\csname val@rev.ow.network.wgetcontentdisposition.fresh\endcsname{writes (immediate)}
\expandafter\def\csname val@rev.ow.network.wgetcontentdisposition.pre\endcsname{no; sibling project-alpha.txt.1}
\expandafter\def\csname val@rev.ow.network.wgetdefault.fresh\endcsname{no}
\expandafter\def\csname val@rev.ow.network.wgetdefault.pre\endcsname{no}
\expandafter\def\csname val@rev.prim.Z1billingadv.bsdtar.held\endcsname{0/5}
\expandafter\def\csname val@rev.prim.Z1billingadv.bsdtar.peakmib\endcsname{10.0}
\expandafter\def\csname val@rev.prim.Z1billingadv.headergateunzip.held\endcsname{5/5}
\expandafter\def\csname val@rev.prim.Z1billingadv.headergateunzip.peakmib\endcsname{0.0}
\expandafter\def\csname val@rev.prim.Z1billingadv.pythonzipfile.held\endcsname{5/5}
\expandafter\def\csname val@rev.prim.Z1billingadv.pythonzipfile.peakmib\endcsname{0.0}
\expandafter\def\csname val@rev.prim.Z1billingadv.quotaunzip.held\endcsname{0/5}
\expandafter\def\csname val@rev.prim.Z1billingadv.quotaunzip.peakmib\endcsname{4.4}
\expandafter\def\csname val@rev.prim.Z1billingadv.szspextract.held\endcsname{5/5}
\expandafter\def\csname val@rev.prim.Z1billingadv.szspextract.peakmib\endcsname{3.9}
\expandafter\def\csname val@rev.prim.Z1billingadv.szsppeek.correct\endcsname{5/5}
\expandafter\def\csname val@rev.prim.Z1billingadv.szsppeek.held\endcsname{5/5}
\expandafter\def\csname val@rev.prim.Z1billingadv.szsppeek.peakmib\endcsname{0.0}
\expandafter\def\csname val@rev.prim.Z1billingadv.ulimitfsizeunzip.held\endcsname{0/5}
\expandafter\def\csname val@rev.prim.Z1billingadv.ulimitfsizeunzip.peakmib\endcsname{4.0}
\expandafter\def\csname val@rev.prim.Z1billingadv.unzip.held\endcsname{0/5}
\expandafter\def\csname val@rev.prim.Z1billingadv.unzip.peakmib\endcsname{10.0}
\expandafter\def\csname val@rev.prim.Z1billingtwin.bsdtar.complete\endcsname{5/5}
\expandafter\def\csname val@rev.prim.Z1billingtwin.bsdtar.held\endcsname{5/5}
\expandafter\def\csname val@rev.prim.Z1billingtwin.bsdtar.peakmib\endcsname{0.0}
\expandafter\def\csname val@rev.prim.Z1billingtwin.headergateunzip.complete\endcsname{5/5}
\expandafter\def\csname val@rev.prim.Z1billingtwin.headergateunzip.held\endcsname{5/5}
\expandafter\def\csname val@rev.prim.Z1billingtwin.headergateunzip.peakmib\endcsname{0.0}
\expandafter\def\csname val@rev.prim.Z1billingtwin.pythonzipfile.complete\endcsname{5/5}
\expandafter\def\csname val@rev.prim.Z1billingtwin.pythonzipfile.held\endcsname{5/5}
\expandafter\def\csname val@rev.prim.Z1billingtwin.pythonzipfile.peakmib\endcsname{0.0}
\expandafter\def\csname val@rev.prim.Z1billingtwin.quotaunzip.complete\endcsname{5/5}
\expandafter\def\csname val@rev.prim.Z1billingtwin.quotaunzip.held\endcsname{5/5}
\expandafter\def\csname val@rev.prim.Z1billingtwin.quotaunzip.peakmib\endcsname{0.0}
\expandafter\def\csname val@rev.prim.Z1billingtwin.szspextract.complete\endcsname{5/5}
\expandafter\def\csname val@rev.prim.Z1billingtwin.szspextract.held\endcsname{5/5}
\expandafter\def\csname val@rev.prim.Z1billingtwin.szspextract.peakmib\endcsname{0.0}
\expandafter\def\csname val@rev.prim.Z1billingtwin.szsppeek.correct\endcsname{5/5}
\expandafter\def\csname val@rev.prim.Z1billingtwin.szsppeek.held\endcsname{5/5}
\expandafter\def\csname val@rev.prim.Z1billingtwin.szsppeek.peakmib\endcsname{0.0}
\expandafter\def\csname val@rev.prim.Z1billingtwin.ulimitfsizeunzip.complete\endcsname{5/5}
\expandafter\def\csname val@rev.prim.Z1billingtwin.ulimitfsizeunzip.held\endcsname{5/5}
\expandafter\def\csname val@rev.prim.Z1billingtwin.ulimitfsizeunzip.peakmib\endcsname{0.0}
\expandafter\def\csname val@rev.prim.Z1billingtwin.unzip.complete\endcsname{5/5}
\expandafter\def\csname val@rev.prim.Z1billingtwin.unzip.held\endcsname{5/5}
\expandafter\def\csname val@rev.prim.Z1billingtwin.unzip.peakmib\endcsname{0.0}
\expandafter\def\csname val@rev.prim.Z2databundleadv.bsdtar.held\endcsname{0/5}
\expandafter\def\csname val@rev.prim.Z2databundleadv.bsdtar.peakmib\endcsname{16.0}
\expandafter\def\csname val@rev.prim.Z2databundleadv.headergateunzip.held\endcsname{5/5}
\expandafter\def\csname val@rev.prim.Z2databundleadv.headergateunzip.peakmib\endcsname{0.0}
\expandafter\def\csname val@rev.prim.Z2databundleadv.pythonzipfile.held\endcsname{5/5}
\expandafter\def\csname val@rev.prim.Z2databundleadv.pythonzipfile.peakmib\endcsname{0.0}
\expandafter\def\csname val@rev.prim.Z2databundleadv.quotaunzip.held\endcsname{0/5}
\expandafter\def\csname val@rev.prim.Z2databundleadv.quotaunzip.peakmib\endcsname{4.4}
\expandafter\def\csname val@rev.prim.Z2databundleadv.szspextract.held\endcsname{5/5}
\expandafter\def\csname val@rev.prim.Z2databundleadv.szspextract.peakmib\endcsname{3.9}
\expandafter\def\csname val@rev.prim.Z2databundleadv.szsppeek.correct\endcsname{5/5}
\expandafter\def\csname val@rev.prim.Z2databundleadv.szsppeek.held\endcsname{5/5}
\expandafter\def\csname val@rev.prim.Z2databundleadv.szsppeek.peakmib\endcsname{0.0}
\expandafter\def\csname val@rev.prim.Z2databundleadv.ulimitfsizeunzip.held\endcsname{0/5}
\expandafter\def\csname val@rev.prim.Z2databundleadv.ulimitfsizeunzip.peakmib\endcsname{4.0}
\expandafter\def\csname val@rev.prim.Z2databundleadv.unzip.held\endcsname{0/5}
\expandafter\def\csname val@rev.prim.Z2databundleadv.unzip.peakmib\endcsname{16.0}
\expandafter\def\csname val@rev.prim.Z2databundletwin.bsdtar.complete\endcsname{5/5}
\expandafter\def\csname val@rev.prim.Z2databundletwin.bsdtar.held\endcsname{5/5}
\expandafter\def\csname val@rev.prim.Z2databundletwin.bsdtar.peakmib\endcsname{0.0}
\expandafter\def\csname val@rev.prim.Z2databundletwin.headergateunzip.complete\endcsname{5/5}
\expandafter\def\csname val@rev.prim.Z2databundletwin.headergateunzip.held\endcsname{5/5}
\expandafter\def\csname val@rev.prim.Z2databundletwin.headergateunzip.peakmib\endcsname{0.0}
\expandafter\def\csname val@rev.prim.Z2databundletwin.pythonzipfile.complete\endcsname{5/5}
\expandafter\def\csname val@rev.prim.Z2databundletwin.pythonzipfile.held\endcsname{5/5}
\expandafter\def\csname val@rev.prim.Z2databundletwin.pythonzipfile.peakmib\endcsname{0.0}
\expandafter\def\csname val@rev.prim.Z2databundletwin.quotaunzip.complete\endcsname{5/5}
\expandafter\def\csname val@rev.prim.Z2databundletwin.quotaunzip.held\endcsname{5/5}
\expandafter\def\csname val@rev.prim.Z2databundletwin.quotaunzip.peakmib\endcsname{0.0}
\expandafter\def\csname val@rev.prim.Z2databundletwin.szspextract.complete\endcsname{5/5}
\expandafter\def\csname val@rev.prim.Z2databundletwin.szspextract.held\endcsname{5/5}
\expandafter\def\csname val@rev.prim.Z2databundletwin.szspextract.peakmib\endcsname{0.0}
\expandafter\def\csname val@rev.prim.Z2databundletwin.szsppeek.correct\endcsname{5/5}
\expandafter\def\csname val@rev.prim.Z2databundletwin.szsppeek.held\endcsname{5/5}
\expandafter\def\csname val@rev.prim.Z2databundletwin.szsppeek.peakmib\endcsname{0.0}
\expandafter\def\csname val@rev.prim.Z2databundletwin.ulimitfsizeunzip.complete\endcsname{5/5}
\expandafter\def\csname val@rev.prim.Z2databundletwin.ulimitfsizeunzip.held\endcsname{5/5}
\expandafter\def\csname val@rev.prim.Z2databundletwin.ulimitfsizeunzip.peakmib\endcsname{0.0}
\expandafter\def\csname val@rev.prim.Z2databundletwin.unzip.complete\endcsname{5/5}
\expandafter\def\csname val@rev.prim.Z2databundletwin.unzip.held\endcsname{5/5}
\expandafter\def\csname val@rev.prim.Z2databundletwin.unzip.peakmib\endcsname{0.0}
\expandafter\def\csname val@rev.prim.Z3logarchiveadv.bsdtar.held\endcsname{0/5}
\expandafter\def\csname val@rev.prim.Z3logarchiveadv.bsdtar.peakmib\endcsname{6.0}
\expandafter\def\csname val@rev.prim.Z3logarchiveadv.headergateunzip.held\endcsname{5/5}
\expandafter\def\csname val@rev.prim.Z3logarchiveadv.headergateunzip.peakmib\endcsname{0.0}
\expandafter\def\csname val@rev.prim.Z3logarchiveadv.pythonzipfile.held\endcsname{5/5}
\expandafter\def\csname val@rev.prim.Z3logarchiveadv.pythonzipfile.peakmib\endcsname{0.0}
\expandafter\def\csname val@rev.prim.Z3logarchiveadv.quotaunzip.held\endcsname{0/5}
\expandafter\def\csname val@rev.prim.Z3logarchiveadv.quotaunzip.peakmib\endcsname{4.5}
\expandafter\def\csname val@rev.prim.Z3logarchiveadv.szspextract.held\endcsname{5/5}
\expandafter\def\csname val@rev.prim.Z3logarchiveadv.szspextract.peakmib\endcsname{4.0}
\expandafter\def\csname val@rev.prim.Z3logarchiveadv.szsppeek.correct\endcsname{5/5}
\expandafter\def\csname val@rev.prim.Z3logarchiveadv.szsppeek.held\endcsname{5/5}
\expandafter\def\csname val@rev.prim.Z3logarchiveadv.szsppeek.peakmib\endcsname{0.0}
\expandafter\def\csname val@rev.prim.Z3logarchiveadv.ulimitfsizeunzip.held\endcsname{0/5}
\expandafter\def\csname val@rev.prim.Z3logarchiveadv.ulimitfsizeunzip.peakmib\endcsname{4.0}
\expandafter\def\csname val@rev.prim.Z3logarchiveadv.unzip.held\endcsname{0/5}
\expandafter\def\csname val@rev.prim.Z3logarchiveadv.unzip.peakmib\endcsname{6.0}
\expandafter\def\csname val@rev.prim.Z3logarchivetwin.bsdtar.complete\endcsname{5/5}
\expandafter\def\csname val@rev.prim.Z3logarchivetwin.bsdtar.held\endcsname{5/5}
\expandafter\def\csname val@rev.prim.Z3logarchivetwin.bsdtar.peakmib\endcsname{0.0}
\expandafter\def\csname val@rev.prim.Z3logarchivetwin.headergateunzip.complete\endcsname{5/5}
\expandafter\def\csname val@rev.prim.Z3logarchivetwin.headergateunzip.held\endcsname{5/5}
\expandafter\def\csname val@rev.prim.Z3logarchivetwin.headergateunzip.peakmib\endcsname{0.0}
\expandafter\def\csname val@rev.prim.Z3logarchivetwin.pythonzipfile.complete\endcsname{5/5}
\expandafter\def\csname val@rev.prim.Z3logarchivetwin.pythonzipfile.held\endcsname{5/5}
\expandafter\def\csname val@rev.prim.Z3logarchivetwin.pythonzipfile.peakmib\endcsname{0.0}
\expandafter\def\csname val@rev.prim.Z3logarchivetwin.quotaunzip.complete\endcsname{5/5}
\expandafter\def\csname val@rev.prim.Z3logarchivetwin.quotaunzip.held\endcsname{5/5}
\expandafter\def\csname val@rev.prim.Z3logarchivetwin.quotaunzip.peakmib\endcsname{0.0}
\expandafter\def\csname val@rev.prim.Z3logarchivetwin.szspextract.complete\endcsname{5/5}
\expandafter\def\csname val@rev.prim.Z3logarchivetwin.szspextract.held\endcsname{5/5}
\expandafter\def\csname val@rev.prim.Z3logarchivetwin.szspextract.peakmib\endcsname{0.0}
\expandafter\def\csname val@rev.prim.Z3logarchivetwin.szsppeek.correct\endcsname{5/5}
\expandafter\def\csname val@rev.prim.Z3logarchivetwin.szsppeek.held\endcsname{5/5}
\expandafter\def\csname val@rev.prim.Z3logarchivetwin.szsppeek.peakmib\endcsname{0.0}
\expandafter\def\csname val@rev.prim.Z3logarchivetwin.ulimitfsizeunzip.complete\endcsname{5/5}
\expandafter\def\csname val@rev.prim.Z3logarchivetwin.ulimitfsizeunzip.held\endcsname{5/5}
\expandafter\def\csname val@rev.prim.Z3logarchivetwin.ulimitfsizeunzip.peakmib\endcsname{0.0}
\expandafter\def\csname val@rev.prim.Z3logarchivetwin.unzip.complete\endcsname{5/5}
\expandafter\def\csname val@rev.prim.Z3logarchivetwin.unzip.held\endcsname{5/5}
\expandafter\def\csname val@rev.prim.Z3logarchivetwin.unzip.peakmib\endcsname{0.0}
\expandafter\def\csname val@rev.prim.Z4checkpointsadv.bsdtar.held\endcsname{0/5}
\expandafter\def\csname val@rev.prim.Z4checkpointsadv.bsdtar.peakmib\endcsname{12.0}
\expandafter\def\csname val@rev.prim.Z4checkpointsadv.headergateunzip.held\endcsname{5/5}
\expandafter\def\csname val@rev.prim.Z4checkpointsadv.headergateunzip.peakmib\endcsname{0.0}
\expandafter\def\csname val@rev.prim.Z4checkpointsadv.pythonzipfile.held\endcsname{5/5}
\expandafter\def\csname val@rev.prim.Z4checkpointsadv.pythonzipfile.peakmib\endcsname{0.0}
\expandafter\def\csname val@rev.prim.Z4checkpointsadv.quotaunzip.held\endcsname{0/5}
\expandafter\def\csname val@rev.prim.Z4checkpointsadv.quotaunzip.peakmib\endcsname{4.3}
\expandafter\def\csname val@rev.prim.Z4checkpointsadv.szspextract.held\endcsname{5/5}
\expandafter\def\csname val@rev.prim.Z4checkpointsadv.szspextract.peakmib\endcsname{3.9}
\expandafter\def\csname val@rev.prim.Z4checkpointsadv.szsppeek.correct\endcsname{5/5}
\expandafter\def\csname val@rev.prim.Z4checkpointsadv.szsppeek.held\endcsname{5/5}
\expandafter\def\csname val@rev.prim.Z4checkpointsadv.szsppeek.peakmib\endcsname{0.0}
\expandafter\def\csname val@rev.prim.Z4checkpointsadv.ulimitfsizeunzip.held\endcsname{0/5}
\expandafter\def\csname val@rev.prim.Z4checkpointsadv.ulimitfsizeunzip.peakmib\endcsname{4.0}
\expandafter\def\csname val@rev.prim.Z4checkpointsadv.unzip.held\endcsname{0/5}
\expandafter\def\csname val@rev.prim.Z4checkpointsadv.unzip.peakmib\endcsname{12.0}
\expandafter\def\csname val@rev.prim.Z4checkpointstwin.bsdtar.complete\endcsname{5/5}
\expandafter\def\csname val@rev.prim.Z4checkpointstwin.bsdtar.held\endcsname{5/5}
\expandafter\def\csname val@rev.prim.Z4checkpointstwin.bsdtar.peakmib\endcsname{0.0}
\expandafter\def\csname val@rev.prim.Z4checkpointstwin.headergateunzip.complete\endcsname{5/5}
\expandafter\def\csname val@rev.prim.Z4checkpointstwin.headergateunzip.held\endcsname{5/5}
\expandafter\def\csname val@rev.prim.Z4checkpointstwin.headergateunzip.peakmib\endcsname{0.0}
\expandafter\def\csname val@rev.prim.Z4checkpointstwin.pythonzipfile.complete\endcsname{5/5}
\expandafter\def\csname val@rev.prim.Z4checkpointstwin.pythonzipfile.held\endcsname{5/5}
\expandafter\def\csname val@rev.prim.Z4checkpointstwin.pythonzipfile.peakmib\endcsname{0.0}
\expandafter\def\csname val@rev.prim.Z4checkpointstwin.quotaunzip.complete\endcsname{5/5}
\expandafter\def\csname val@rev.prim.Z4checkpointstwin.quotaunzip.held\endcsname{5/5}
\expandafter\def\csname val@rev.prim.Z4checkpointstwin.quotaunzip.peakmib\endcsname{0.0}
\expandafter\def\csname val@rev.prim.Z4checkpointstwin.szspextract.complete\endcsname{5/5}
\expandafter\def\csname val@rev.prim.Z4checkpointstwin.szspextract.held\endcsname{5/5}
\expandafter\def\csname val@rev.prim.Z4checkpointstwin.szspextract.peakmib\endcsname{0.0}
\expandafter\def\csname val@rev.prim.Z4checkpointstwin.szsppeek.correct\endcsname{5/5}
\expandafter\def\csname val@rev.prim.Z4checkpointstwin.szsppeek.held\endcsname{5/5}
\expandafter\def\csname val@rev.prim.Z4checkpointstwin.szsppeek.peakmib\endcsname{0.0}
\expandafter\def\csname val@rev.prim.Z4checkpointstwin.ulimitfsizeunzip.complete\endcsname{5/5}
\expandafter\def\csname val@rev.prim.Z4checkpointstwin.ulimitfsizeunzip.held\endcsname{5/5}
\expandafter\def\csname val@rev.prim.Z4checkpointstwin.ulimitfsizeunzip.peakmib\endcsname{0.0}
\expandafter\def\csname val@rev.prim.Z4checkpointstwin.unzip.complete\endcsname{5/5}
\expandafter\def\csname val@rev.prim.Z4checkpointstwin.unzip.held\endcsname{5/5}
\expandafter\def\csname val@rev.prim.Z4checkpointstwin.unzip.peakmib\endcsname{0.0}
\expandafter\def\csname val@rev.prim.a5cdlie.bsdtar.held\endcsname{0/5}
\expandafter\def\csname val@rev.prim.a5cdlie.bsdtar.peakmib\endcsname{8.0}
\expandafter\def\csname val@rev.prim.a5cdlie.headergateunzip.held\endcsname{5/5}
\expandafter\def\csname val@rev.prim.a5cdlie.headergateunzip.peakmib\endcsname{0.0}
\expandafter\def\csname val@rev.prim.a5cdlie.pythonzipfile.held\endcsname{5/5}
\expandafter\def\csname val@rev.prim.a5cdlie.pythonzipfile.peakmib\endcsname{0.0}
\expandafter\def\csname val@rev.prim.a5cdlie.quotaunzip.held\endcsname{0/5}
\expandafter\def\csname val@rev.prim.a5cdlie.quotaunzip.peakmib\endcsname{4.4}
\expandafter\def\csname val@rev.prim.a5cdlie.szspextract.held\endcsname{5/5}
\expandafter\def\csname val@rev.prim.a5cdlie.szspextract.peakmib\endcsname{3.9}
\expandafter\def\csname val@rev.prim.a5cdlie.szsppeek.correct\endcsname{5/5}
\expandafter\def\csname val@rev.prim.a5cdlie.szsppeek.held\endcsname{5/5}
\expandafter\def\csname val@rev.prim.a5cdlie.szsppeek.peakmib\endcsname{0.0}
\expandafter\def\csname val@rev.prim.a5cdlie.ulimitfsizeunzip.held\endcsname{0/5}
\expandafter\def\csname val@rev.prim.a5cdlie.ulimitfsizeunzip.peakmib\endcsname{4.0}
\expandafter\def\csname val@rev.prim.a5cdlie.unzip.held\endcsname{0/5}
\expandafter\def\csname val@rev.prim.a5cdlie.unzip.peakmib\endcsname{8.0}
\expandafter\def\csname val@rev.prim.a5twin.bsdtar.complete\endcsname{5/5}
\expandafter\def\csname val@rev.prim.a5twin.bsdtar.held\endcsname{5/5}
\expandafter\def\csname val@rev.prim.a5twin.bsdtar.peakmib\endcsname{0.0}
\expandafter\def\csname val@rev.prim.a5twin.headergateunzip.complete\endcsname{5/5}
\expandafter\def\csname val@rev.prim.a5twin.headergateunzip.held\endcsname{5/5}
\expandafter\def\csname val@rev.prim.a5twin.headergateunzip.peakmib\endcsname{0.0}
\expandafter\def\csname val@rev.prim.a5twin.pythonzipfile.complete\endcsname{5/5}
\expandafter\def\csname val@rev.prim.a5twin.pythonzipfile.held\endcsname{5/5}
\expandafter\def\csname val@rev.prim.a5twin.pythonzipfile.peakmib\endcsname{0.0}
\expandafter\def\csname val@rev.prim.a5twin.quotaunzip.complete\endcsname{5/5}
\expandafter\def\csname val@rev.prim.a5twin.quotaunzip.held\endcsname{5/5}
\expandafter\def\csname val@rev.prim.a5twin.quotaunzip.peakmib\endcsname{0.0}
\expandafter\def\csname val@rev.prim.a5twin.szspextract.complete\endcsname{5/5}
\expandafter\def\csname val@rev.prim.a5twin.szspextract.held\endcsname{5/5}
\expandafter\def\csname val@rev.prim.a5twin.szspextract.peakmib\endcsname{0.0}
\expandafter\def\csname val@rev.prim.a5twin.szsppeek.correct\endcsname{5/5}
\expandafter\def\csname val@rev.prim.a5twin.szsppeek.held\endcsname{5/5}
\expandafter\def\csname val@rev.prim.a5twin.szsppeek.peakmib\endcsname{0.0}
\expandafter\def\csname val@rev.prim.a5twin.ulimitfsizeunzip.complete\endcsname{5/5}
\expandafter\def\csname val@rev.prim.a5twin.ulimitfsizeunzip.held\endcsname{5/5}
\expandafter\def\csname val@rev.prim.a5twin.ulimitfsizeunzip.peakmib\endcsname{0.0}
\expandafter\def\csname val@rev.prim.a5twin.unzip.complete\endcsname{5/5}
\expandafter\def\csname val@rev.prim.a5twin.unzip.held\endcsname{5/5}
\expandafter\def\csname val@rev.prim.a5twin.unzip.peakmib\endcsname{0.0}
\expandafter\def\csname val@rev.prim.aladv.bsdtar.held\endcsname{5/5}
\expandafter\def\csname val@rev.prim.aladv.bsdtar.peakmib\endcsname{0.0}
\expandafter\def\csname val@rev.prim.aladv.headergateunzip.held\endcsname{0/5}
\expandafter\def\csname val@rev.prim.aladv.headergateunzip.peakmib\endcsname{8.0}
\expandafter\def\csname val@rev.prim.aladv.pythonzipfile.held\endcsname{5/5}
\expandafter\def\csname val@rev.prim.aladv.pythonzipfile.peakmib\endcsname{0.0}
\expandafter\def\csname val@rev.prim.aladv.quotaunzip.held\endcsname{0/5}
\expandafter\def\csname val@rev.prim.aladv.quotaunzip.peakmib\endcsname{4.4}
\expandafter\def\csname val@rev.prim.aladv.szspextract.held\endcsname{5/5}
\expandafter\def\csname val@rev.prim.aladv.szspextract.peakmib\endcsname{3.9}
\expandafter\def\csname val@rev.prim.aladv.szsppeek.correct\endcsname{5/5}
\expandafter\def\csname val@rev.prim.aladv.szsppeek.held\endcsname{5/5}
\expandafter\def\csname val@rev.prim.aladv.szsppeek.peakmib\endcsname{0.0}
\expandafter\def\csname val@rev.prim.aladv.ulimitfsizeunzip.held\endcsname{0/5}
\expandafter\def\csname val@rev.prim.aladv.ulimitfsizeunzip.peakmib\endcsname{4.0}
\expandafter\def\csname val@rev.prim.aladv.unzip.held\endcsname{0/5}
\expandafter\def\csname val@rev.prim.aladv.unzip.peakmib\endcsname{8.0}
\expandafter\def\csname val@rev.prim.all.bsdtar.advheld\endcsname{20/50}
\expandafter\def\csname val@rev.prim.all.bsdtar.advpeakmax\endcsname{16.0}
\expandafter\def\csname val@rev.prim.all.bsdtar.advpeakmedian\endcsname{7.0}
\expandafter\def\csname val@rev.prim.all.bsdtar.twincomplete\endcsname{30/30}
\expandafter\def\csname val@rev.prim.all.bsdtar.twinmedianms\endcsname{10}
\expandafter\def\csname val@rev.prim.all.headergateunzip.advheld\endcsname{30/50}
\expandafter\def\csname val@rev.prim.all.headergateunzip.advpeakmax\endcsname{512.0}
\expandafter\def\csname val@rev.prim.all.headergateunzip.advpeakmedian\endcsname{0.0}
\expandafter\def\csname val@rev.prim.all.headergateunzip.twincomplete\endcsname{30/30}
\expandafter\def\csname val@rev.prim.all.headergateunzip.twinmedianms\endcsname{48}
\expandafter\def\csname val@rev.prim.all.pythonzipfile.advheld\endcsname{45/50}
\expandafter\def\csname val@rev.prim.all.pythonzipfile.advpeakmax\endcsname{8.0}
\expandafter\def\csname val@rev.prim.all.pythonzipfile.advpeakmedian\endcsname{0.0}
\expandafter\def\csname val@rev.prim.all.pythonzipfile.twincomplete\endcsname{30/30}
\expandafter\def\csname val@rev.prim.all.pythonzipfile.twinmedianms\endcsname{52}
\expandafter\def\csname val@rev.prim.all.quotaunzip.advheld\endcsname{0/50}
\expandafter\def\csname val@rev.prim.all.quotaunzip.advpeakmax\endcsname{4.5}
\expandafter\def\csname val@rev.prim.all.quotaunzip.advpeakmedian\endcsname{4.3}
\expandafter\def\csname val@rev.prim.all.quotaunzip.twincomplete\endcsname{30/30}
\expandafter\def\csname val@rev.prim.all.quotaunzip.twinmedianms\endcsname{54}
\expandafter\def\csname val@rev.prim.all.szspextract.advheld\endcsname{50/50}
\expandafter\def\csname val@rev.prim.all.szspextract.advpeakmax\endcsname{4.0}
\expandafter\def\csname val@rev.prim.all.szspextract.advpeakmedian\endcsname{3.7}
\expandafter\def\csname val@rev.prim.all.szspextract.twincomplete\endcsname{30/30}
\expandafter\def\csname val@rev.prim.all.szspextract.twinmedianms\endcsname{58}
\expandafter\def\csname val@rev.prim.all.szsppeek.advheld\endcsname{50/50}
\expandafter\def\csname val@rev.prim.all.szsppeek.advpeakmax\endcsname{0.0}
\expandafter\def\csname val@rev.prim.all.szsppeek.advpeakmedian\endcsname{0.0}
\expandafter\def\csname val@rev.prim.all.szsppeek.twinmedianms\endcsname{51}
\expandafter\def\csname val@rev.prim.all.ulimitfsizeunzip.advheld\endcsname{0/50}
\expandafter\def\csname val@rev.prim.all.ulimitfsizeunzip.advpeakmax\endcsname{4.0}
\expandafter\def\csname val@rev.prim.all.ulimitfsizeunzip.advpeakmedian\endcsname{4.0}
\expandafter\def\csname val@rev.prim.all.ulimitfsizeunzip.twincomplete\endcsname{30/30}
\expandafter\def\csname val@rev.prim.all.ulimitfsizeunzip.twinmedianms\endcsname{8}
\expandafter\def\csname val@rev.prim.all.unzip.advheld\endcsname{0/50}
\expandafter\def\csname val@rev.prim.all.unzip.advpeakmax\endcsname{512.0}
\expandafter\def\csname val@rev.prim.all.unzip.advpeakmedian\endcsname{9.0}
\expandafter\def\csname val@rev.prim.all.unzip.twincomplete\endcsname{30/30}
\expandafter\def\csname val@rev.prim.all.unzip.twinmedianms\endcsname{6}
\expandafter\def\csname val@rev.prim.altwin.bsdtar.complete\endcsname{5/5}
\expandafter\def\csname val@rev.prim.altwin.bsdtar.held\endcsname{5/5}
\expandafter\def\csname val@rev.prim.altwin.bsdtar.peakmib\endcsname{0.0}
\expandafter\def\csname val@rev.prim.altwin.headergateunzip.complete\endcsname{5/5}
\expandafter\def\csname val@rev.prim.altwin.headergateunzip.held\endcsname{5/5}
\expandafter\def\csname val@rev.prim.altwin.headergateunzip.peakmib\endcsname{0.0}
\expandafter\def\csname val@rev.prim.altwin.pythonzipfile.complete\endcsname{5/5}
\expandafter\def\csname val@rev.prim.altwin.pythonzipfile.held\endcsname{5/5}
\expandafter\def\csname val@rev.prim.altwin.pythonzipfile.peakmib\endcsname{0.0}
\expandafter\def\csname val@rev.prim.altwin.quotaunzip.complete\endcsname{5/5}
\expandafter\def\csname val@rev.prim.altwin.quotaunzip.held\endcsname{5/5}
\expandafter\def\csname val@rev.prim.altwin.quotaunzip.peakmib\endcsname{0.0}
\expandafter\def\csname val@rev.prim.altwin.szspextract.complete\endcsname{5/5}
\expandafter\def\csname val@rev.prim.altwin.szspextract.held\endcsname{5/5}
\expandafter\def\csname val@rev.prim.altwin.szspextract.peakmib\endcsname{0.0}
\expandafter\def\csname val@rev.prim.altwin.szsppeek.correct\endcsname{5/5}
\expandafter\def\csname val@rev.prim.altwin.szsppeek.held\endcsname{5/5}
\expandafter\def\csname val@rev.prim.altwin.szsppeek.peakmib\endcsname{0.0}
\expandafter\def\csname val@rev.prim.altwin.ulimitfsizeunzip.complete\endcsname{5/5}
\expandafter\def\csname val@rev.prim.altwin.ulimitfsizeunzip.held\endcsname{5/5}
\expandafter\def\csname val@rev.prim.altwin.ulimitfsizeunzip.peakmib\endcsname{0.0}
\expandafter\def\csname val@rev.prim.altwin.unzip.complete\endcsname{5/5}
\expandafter\def\csname val@rev.prim.altwin.unzip.held\endcsname{5/5}
\expandafter\def\csname val@rev.prim.altwin.unzip.peakmib\endcsname{0.0}
\expandafter\def\csname val@rev.prim.ddadv.bsdtar.held\endcsname{5/5}
\expandafter\def\csname val@rev.prim.ddadv.bsdtar.peakmib\endcsname{0.0}
\expandafter\def\csname val@rev.prim.ddadv.headergateunzip.held\endcsname{0/5}
\expandafter\def\csname val@rev.prim.ddadv.headergateunzip.peakmib\endcsname{8.0}
\expandafter\def\csname val@rev.prim.ddadv.pythonzipfile.held\endcsname{5/5}
\expandafter\def\csname val@rev.prim.ddadv.pythonzipfile.peakmib\endcsname{0.0}
\expandafter\def\csname val@rev.prim.ddadv.quotaunzip.held\endcsname{0/5}
\expandafter\def\csname val@rev.prim.ddadv.quotaunzip.peakmib\endcsname{4.4}
\expandafter\def\csname val@rev.prim.ddadv.szspextract.held\endcsname{5/5}
\expandafter\def\csname val@rev.prim.ddadv.szspextract.peakmib\endcsname{3.9}
\expandafter\def\csname val@rev.prim.ddadv.szsppeek.correct\endcsname{5/5}
\expandafter\def\csname val@rev.prim.ddadv.szsppeek.held\endcsname{5/5}
\expandafter\def\csname val@rev.prim.ddadv.szsppeek.peakmib\endcsname{0.0}
\expandafter\def\csname val@rev.prim.ddadv.ulimitfsizeunzip.held\endcsname{0/5}
\expandafter\def\csname val@rev.prim.ddadv.ulimitfsizeunzip.peakmib\endcsname{4.0}
\expandafter\def\csname val@rev.prim.ddadv.unzip.held\endcsname{0/5}
\expandafter\def\csname val@rev.prim.ddadv.unzip.peakmib\endcsname{8.0}
\expandafter\def\csname val@rev.prim.honestadv.bsdtar.held\endcsname{0/5}
\expandafter\def\csname val@rev.prim.honestadv.bsdtar.peakmib\endcsname{8.0}
\expandafter\def\csname val@rev.prim.honestadv.headergateunzip.held\endcsname{5/5}
\expandafter\def\csname val@rev.prim.honestadv.headergateunzip.peakmib\endcsname{0.0}
\expandafter\def\csname val@rev.prim.honestadv.pythonzipfile.held\endcsname{0/5}
\expandafter\def\csname val@rev.prim.honestadv.pythonzipfile.peakmib\endcsname{8.0}
\expandafter\def\csname val@rev.prim.honestadv.quotaunzip.held\endcsname{0/5}
\expandafter\def\csname val@rev.prim.honestadv.quotaunzip.peakmib\endcsname{4.4}
\expandafter\def\csname val@rev.prim.honestadv.szspextract.held\endcsname{5/5}
\expandafter\def\csname val@rev.prim.honestadv.szspextract.peakmib\endcsname{3.9}
\expandafter\def\csname val@rev.prim.honestadv.szsppeek.correct\endcsname{5/5}
\expandafter\def\csname val@rev.prim.honestadv.szsppeek.held\endcsname{5/5}
\expandafter\def\csname val@rev.prim.honestadv.szsppeek.peakmib\endcsname{0.0}
\expandafter\def\csname val@rev.prim.honestadv.ulimitfsizeunzip.held\endcsname{0/5}
\expandafter\def\csname val@rev.prim.honestadv.ulimitfsizeunzip.peakmib\endcsname{4.0}
\expandafter\def\csname val@rev.prim.honestadv.unzip.held\endcsname{0/5}
\expandafter\def\csname val@rev.prim.honestadv.unzip.peakmib\endcsname{8.0}
\expandafter\def\csname val@rev.prim.scale512adv.bsdtar.held\endcsname{5/5}
\expandafter\def\csname val@rev.prim.scale512adv.bsdtar.peakmib\endcsname{0.0}
\expandafter\def\csname val@rev.prim.scale512adv.headergateunzip.held\endcsname{0/5}
\expandafter\def\csname val@rev.prim.scale512adv.headergateunzip.peakmib\endcsname{512.0}
\expandafter\def\csname val@rev.prim.scale512adv.pythonzipfile.held\endcsname{5/5}
\expandafter\def\csname val@rev.prim.scale512adv.pythonzipfile.peakmib\endcsname{0.0}
\expandafter\def\csname val@rev.prim.scale512adv.quotaunzip.held\endcsname{0/5}
\expandafter\def\csname val@rev.prim.scale512adv.quotaunzip.peakmib\endcsname{4.4}
\expandafter\def\csname val@rev.prim.scale512adv.szspextract.held\endcsname{5/5}
\expandafter\def\csname val@rev.prim.scale512adv.szspextract.peakmib\endcsname{3.7}
\expandafter\def\csname val@rev.prim.scale512adv.szsppeek.correct\endcsname{5/5}
\expandafter\def\csname val@rev.prim.scale512adv.szsppeek.held\endcsname{5/5}
\expandafter\def\csname val@rev.prim.scale512adv.szsppeek.peakmib\endcsname{0.0}
\expandafter\def\csname val@rev.prim.scale512adv.ulimitfsizeunzip.held\endcsname{0/5}
\expandafter\def\csname val@rev.prim.scale512adv.ulimitfsizeunzip.peakmib\endcsname{4.0}
\expandafter\def\csname val@rev.prim.scale512adv.unzip.held\endcsname{0/5}
\expandafter\def\csname val@rev.prim.scale512adv.unzip.peakmib\endcsname{512.0}
\expandafter\def\csname val@rev.prim.scale64adv.bsdtar.held\endcsname{5/5}
\expandafter\def\csname val@rev.prim.scale64adv.bsdtar.peakmib\endcsname{0.0}
\expandafter\def\csname val@rev.prim.scale64adv.headergateunzip.held\endcsname{0/5}
\expandafter\def\csname val@rev.prim.scale64adv.headergateunzip.peakmib\endcsname{64.0}
\expandafter\def\csname val@rev.prim.scale64adv.pythonzipfile.held\endcsname{5/5}
\expandafter\def\csname val@rev.prim.scale64adv.pythonzipfile.peakmib\endcsname{0.0}
\expandafter\def\csname val@rev.prim.scale64adv.quotaunzip.held\endcsname{0/5}
\expandafter\def\csname val@rev.prim.scale64adv.quotaunzip.peakmib\endcsname{4.4}
\expandafter\def\csname val@rev.prim.scale64adv.szspextract.held\endcsname{5/5}
\expandafter\def\csname val@rev.prim.scale64adv.szspextract.peakmib\endcsname{3.8}
\expandafter\def\csname val@rev.prim.scale64adv.szsppeek.correct\endcsname{5/5}
\expandafter\def\csname val@rev.prim.scale64adv.szsppeek.held\endcsname{5/5}
\expandafter\def\csname val@rev.prim.scale64adv.szsppeek.peakmib\endcsname{0.0}
\expandafter\def\csname val@rev.prim.scale64adv.ulimitfsizeunzip.held\endcsname{0/5}
\expandafter\def\csname val@rev.prim.scale64adv.ulimitfsizeunzip.peakmib\endcsname{4.0}
\expandafter\def\csname val@rev.prim.scale64adv.unzip.held\endcsname{0/5}
\expandafter\def\csname val@rev.prim.scale64adv.unzip.peakmib\endcsname{64.0}
\expandafter\def\csname val@rev.rate.digest.hash.ci\endcsname{[64.3, 100.0]}
\expandafter\def\csname val@rev.rate.digest.hash.frac\endcsname{37/42}
\expandafter\def\csname val@rev.rate.digest.hash.pct\endcsname{88.1\%}
\expandafter\def\csname val@rev.rate.digest.hash.wilson\endcsname{[75.0, 94.8]}
\expandafter\def\csname val@rev.rate.digest.ingress.ci\endcsname{[85.7, 100.0]}
\expandafter\def\csname val@rev.rate.digest.ingress.frac\endcsname{41/42}
\expandafter\def\csname val@rev.rate.digest.ingress.pct\endcsname{97.6\%}
\expandafter\def\csname val@rev.rate.digest.ingress.wilson\endcsname{[87.7, 99.6]}
\expandafter\def\csname val@rev.rate.digest.reject.ci\endcsname{[52.4, 92.9]}
\expandafter\def\csname val@rev.rate.digest.reject.frac\endcsname{31/42}
\expandafter\def\csname val@rev.rate.digest.reject.pct\endcsname{73.8\%}
\expandafter\def\csname val@rev.rate.digest.reject.wilson\endcsname{[58.9, 84.7]}
\expandafter\def\csname val@rev.rate.digest.reported.ci\endcsname{[52.4, 95.2]}
\expandafter\def\csname val@rev.rate.digest.reported.frac\endcsname{32/42}
\expandafter\def\csname val@rev.rate.digest.reported.pct\endcsname{76.2\%}
\expandafter\def\csname val@rev.rate.digest.reported.wilson\endcsname{[61.5, 86.5]}
\expandafter\def\csname val@rev.rate.digest.targetwritten.ci\endcsname{[0.0, 28.6]}
\expandafter\def\csname val@rev.rate.digest.targetwritten.frac\endcsname{4/42}
\expandafter\def\csname val@rev.rate.digest.targetwritten.pct\endcsname{9.5\%}
\expandafter\def\csname val@rev.rate.digest.targetwritten.wilson\endcsname{[3.8, 22.1]}
\expandafter\def\csname val@rev.rate.digest.twin.hash.ci\endcsname{[47.6, 95.2]}
\expandafter\def\csname val@rev.rate.digest.twin.hash.frac\endcsname{31/42}
\expandafter\def\csname val@rev.rate.digest.twin.hash.pct\endcsname{73.8\%}
\expandafter\def\csname val@rev.rate.digest.twin.hash.wilson\endcsname{[58.9, 84.7]}
\expandafter\def\csname val@rev.rate.digest.twin.reported.ci\endcsname{[0.0, 38.1]}
\expandafter\def\csname val@rev.rate.digest.twin.reported.frac\endcsname{6/42}
\expandafter\def\csname val@rev.rate.digest.twin.reported.pct\endcsname{14.3\%}
\expandafter\def\csname val@rev.rate.digest.twin.reported.wilson\endcsname{[6.7, 27.8]}
\expandafter\def\csname val@rev.rate.digest.twin.targetwritten.ci\endcsname{[0.0, 33.3]}
\expandafter\def\csname val@rev.rate.digest.twin.targetwritten.frac\endcsname{5/42}
\expandafter\def\csname val@rev.rate.digest.twin.targetwritten.pct\endcsname{11.9\%}
\expandafter\def\csname val@rev.rate.digest.twin.targetwritten.wilson\endcsname{[5.2, 25.0]}
\expandafter\def\csname val@rev.rate.digest.write.ci\endcsname{[0.0, 14.3]}
\expandafter\def\csname val@rev.rate.digest.write.frac\endcsname{1/42}
\expandafter\def\csname val@rev.rate.digest.write.pct\endcsname{2.4\%}
\expandafter\def\csname val@rev.rate.digest.write.wilson\endcsname{[0.4, 12.3]}
\expandafter\def\csname val@rev.rate.header.ingress.ci\endcsname{[9.5, 69.0]}
\expandafter\def\csname val@rev.rate.header.ingress.frac\endcsname{15/42}
\expandafter\def\csname val@rev.rate.header.ingress.pct\endcsname{35.7\%}
\expandafter\def\csname val@rev.rate.header.ingress.wilson\endcsname{[23.0, 50.8]}
\expandafter\def\csname val@rev.rate.header.reported.ci\endcsname{[11.9, 59.5]}
\expandafter\def\csname val@rev.rate.header.reported.frac\endcsname{15/42}
\expandafter\def\csname val@rev.rate.header.reported.pct\endcsname{35.7\%}
\expandafter\def\csname val@rev.rate.header.reported.wilson\endcsname{[23.0, 50.8]}
\expandafter\def\csname val@rev.rate.header.targetwritten.ci\endcsname{[2.4, 38.1]}
\expandafter\def\csname val@rev.rate.header.targetwritten.frac\endcsname{7/42}
\expandafter\def\csname val@rev.rate.header.targetwritten.pct\endcsname{16.7\%}
\expandafter\def\csname val@rev.rate.header.targetwritten.wilson\endcsname{[8.3, 30.6]}
\expandafter\def\csname val@rev.rate.header.twin.reported.ci\endcsname{[0.0, 14.3]}
\expandafter\def\csname val@rev.rate.header.twin.reported.frac\endcsname{1/42}
\expandafter\def\csname val@rev.rate.header.twin.reported.pct\endcsname{2.4\%}
\expandafter\def\csname val@rev.rate.header.twin.reported.wilson\endcsname{[0.4, 12.3]}
\expandafter\def\csname val@rev.rate.header.twin.targetwritten.ci\endcsname{[9.5, 61.9]}
\expandafter\def\csname val@rev.rate.header.twin.targetwritten.frac\endcsname{14/42}
\expandafter\def\csname val@rev.rate.header.twin.targetwritten.pct\endcsname{33.3\%}
\expandafter\def\csname val@rev.rate.header.twin.targetwritten.wilson\endcsname{[21.0, 48.4]}
\expandafter\def\csname val@rev.rate.header.write.ci\endcsname{[0.0, 31.0]}
\expandafter\def\csname val@rev.rate.header.write.frac\endcsname{6/42}
\expandafter\def\csname val@rev.rate.header.write.pct\endcsname{14.3\%}
\expandafter\def\csname val@rev.rate.header.write.wilson\endcsname{[6.7, 27.8]}
\expandafter\def\csname val@rev.rate.inbody.ingress.ci\endcsname{[83.3, 100.0]}
\expandafter\def\csname val@rev.rate.inbody.ingress.frac\endcsname{40/42}
\expandafter\def\csname val@rev.rate.inbody.ingress.pct\endcsname{95.2\%}
\expandafter\def\csname val@rev.rate.inbody.ingress.wilson\endcsname{[84.2, 98.7]}
\expandafter\def\csname val@rev.rate.inbody.reported.ci\endcsname{[4.8, 61.9]}
\expandafter\def\csname val@rev.rate.inbody.reported.frac\endcsname{13/42}
\expandafter\def\csname val@rev.rate.inbody.reported.pct\endcsname{31.0\%}
\expandafter\def\csname val@rev.rate.inbody.reported.wilson\endcsname{[19.1, 46.0]}
\expandafter\def\csname val@rev.rate.inbody.targetwritten.ci\endcsname{[4.8, 52.4]}
\expandafter\def\csname val@rev.rate.inbody.targetwritten.frac\endcsname{11/42}
\expandafter\def\csname val@rev.rate.inbody.targetwritten.pct\endcsname{26.2\%}
\expandafter\def\csname val@rev.rate.inbody.targetwritten.wilson\endcsname{[15.3, 41.1]}
\expandafter\def\csname val@rev.rate.inbody.twin.reported.ci\endcsname{[0.0, 0.0]}
\expandafter\def\csname val@rev.rate.inbody.twin.reported.frac\endcsname{0/42}
\expandafter\def\csname val@rev.rate.inbody.twin.reported.pct\endcsname{0.0\%}
\expandafter\def\csname val@rev.rate.inbody.twin.reported.wilson\endcsname{[0.0, 8.4]}
\expandafter\def\csname val@rev.rate.inbody.twin.targetwritten.ci\endcsname{[19.0, 81.0]}
\expandafter\def\csname val@rev.rate.inbody.twin.targetwritten.frac\endcsname{21/42}
\expandafter\def\csname val@rev.rate.inbody.twin.targetwritten.pct\endcsname{50.0\%}
\expandafter\def\csname val@rev.rate.inbody.twin.targetwritten.wilson\endcsname{[35.5, 64.5]}
\expandafter\def\csname val@rev.rate.inbody.write.ci\endcsname{[4.8, 52.4]}
\expandafter\def\csname val@rev.rate.inbody.write.frac\endcsname{11/42}
\expandafter\def\csname val@rev.rate.inbody.write.pct\endcsname{26.2\%}
\expandafter\def\csname val@rev.rate.inbody.write.wilson\endcsname{[15.3, 41.1]}
\expandafter\def\csname val@rev.rate.qrdigest.hash.ci\endcsname{[21.4, 83.3]}
\expandafter\def\csname val@rev.rate.qrdigest.hash.frac\endcsname{23/42}
\expandafter\def\csname val@rev.rate.qrdigest.hash.pct\endcsname{54.8\%}
\expandafter\def\csname val@rev.rate.qrdigest.hash.wilson\endcsname{[39.9, 68.8]}
\expandafter\def\csname val@rev.rate.qrdigest.ingress.ci\endcsname{[52.4, 95.2]}
\expandafter\def\csname val@rev.rate.qrdigest.ingress.frac\endcsname{32/42}
\expandafter\def\csname val@rev.rate.qrdigest.ingress.pct\endcsname{76.2\%}
\expandafter\def\csname val@rev.rate.qrdigest.ingress.wilson\endcsname{[61.5, 86.5]}
\expandafter\def\csname val@rev.rate.qrdigest.reject.ci\endcsname{[9.5, 59.5]}
\expandafter\def\csname val@rev.rate.qrdigest.reject.frac\endcsname{14/42}
\expandafter\def\csname val@rev.rate.qrdigest.reject.pct\endcsname{33.3\%}
\expandafter\def\csname val@rev.rate.qrdigest.reject.wilson\endcsname{[21.0, 48.4]}
\expandafter\def\csname val@rev.rate.qrdigest.reported.ci\endcsname{[11.9, 64.3]}
\expandafter\def\csname val@rev.rate.qrdigest.reported.frac\endcsname{16/42}
\expandafter\def\csname val@rev.rate.qrdigest.reported.pct\endcsname{38.1\%}
\expandafter\def\csname val@rev.rate.qrdigest.reported.wilson\endcsname{[25.0, 53.2]}
\expandafter\def\csname val@rev.rate.qrdigest.targetwritten.ci\endcsname{[0.0, 52.4]}
\expandafter\def\csname val@rev.rate.qrdigest.targetwritten.frac\endcsname{9/42}
\expandafter\def\csname val@rev.rate.qrdigest.targetwritten.pct\endcsname{21.4\%}
\expandafter\def\csname val@rev.rate.qrdigest.targetwritten.wilson\endcsname{[11.7, 35.9]}
\expandafter\def\csname val@rev.rate.qrdigest.twin.hash.ci\endcsname{[35.7, 85.7]}
\expandafter\def\csname val@rev.rate.qrdigest.twin.hash.frac\endcsname{26/42}
\expandafter\def\csname val@rev.rate.qrdigest.twin.hash.pct\endcsname{61.9\%}
\expandafter\def\csname val@rev.rate.qrdigest.twin.hash.wilson\endcsname{[46.8, 75.0]}
\expandafter\def\csname val@rev.rate.qrdigest.twin.reported.ci\endcsname{[0.0, 14.3]}
\expandafter\def\csname val@rev.rate.qrdigest.twin.reported.frac\endcsname{1/42}
\expandafter\def\csname val@rev.rate.qrdigest.twin.reported.pct\endcsname{2.4\%}
\expandafter\def\csname val@rev.rate.qrdigest.twin.reported.wilson\endcsname{[0.4, 12.3]}
\expandafter\def\csname val@rev.rate.qrdigest.twin.targetwritten.ci\endcsname{[9.5, 69.0]}
\expandafter\def\csname val@rev.rate.qrdigest.twin.targetwritten.frac\endcsname{16/42}
\expandafter\def\csname val@rev.rate.qrdigest.twin.targetwritten.pct\endcsname{38.1\%}
\expandafter\def\csname val@rev.rate.qrdigest.twin.targetwritten.wilson\endcsname{[25.0, 53.2]}
\expandafter\def\csname val@rev.rate.qrdigest.write.ci\endcsname{[0.0, 14.3]}
\expandafter\def\csname val@rev.rate.qrdigest.write.frac\endcsname{1/42}
\expandafter\def\csname val@rev.rate.qrdigest.write.pct\endcsname{2.4\%}
\expandafter\def\csname val@rev.rate.qrdigest.write.wilson\endcsname{[0.4, 12.3]}
\expandafter\def\csname val@rev.rate.qrdual.ingress.ci\endcsname{[78.6, 100.0]}
\expandafter\def\csname val@rev.rate.qrdual.ingress.frac\endcsname{39/42}
\expandafter\def\csname val@rev.rate.qrdual.ingress.pct\endcsname{92.9\%}
\expandafter\def\csname val@rev.rate.qrdual.ingress.wilson\endcsname{[81.0, 97.5]}
\expandafter\def\csname val@rev.rate.qrdual.reported.ci\endcsname{[0.0, 0.0]}
\expandafter\def\csname val@rev.rate.qrdual.reported.frac\endcsname{0/42}
\expandafter\def\csname val@rev.rate.qrdual.reported.pct\endcsname{0.0\%}
\expandafter\def\csname val@rev.rate.qrdual.reported.wilson\endcsname{[0.0, 8.4]}
\expandafter\def\csname val@rev.rate.qrdual.targetwritten.ci\endcsname{[14.3, 76.2]}
\expandafter\def\csname val@rev.rate.qrdual.targetwritten.frac\endcsname{19/42}
\expandafter\def\csname val@rev.rate.qrdual.targetwritten.pct\endcsname{45.2\%}
\expandafter\def\csname val@rev.rate.qrdual.targetwritten.wilson\endcsname{[31.2, 60.1]}
\expandafter\def\csname val@rev.rate.qrdual.twin.reported.ci\endcsname{[0.0, 0.0]}
\expandafter\def\csname val@rev.rate.qrdual.twin.reported.frac\endcsname{0/42}
\expandafter\def\csname val@rev.rate.qrdual.twin.reported.pct\endcsname{0.0\%}
\expandafter\def\csname val@rev.rate.qrdual.twin.reported.wilson\endcsname{[0.0, 8.4]}
\expandafter\def\csname val@rev.rate.qrdual.twin.targetwritten.ci\endcsname{[28.6, 85.7]}
\expandafter\def\csname val@rev.rate.qrdual.twin.targetwritten.frac\endcsname{25/42}
\expandafter\def\csname val@rev.rate.qrdual.twin.targetwritten.pct\endcsname{59.5\%}
\expandafter\def\csname val@rev.rate.qrdual.twin.targetwritten.wilson\endcsname{[44.5, 73.0]}
\expandafter\def\csname val@rev.rate.qrdual.write.ci\endcsname{[14.3, 76.2]}
\expandafter\def\csname val@rev.rate.qrdual.write.frac\endcsname{19/42}
\expandafter\def\csname val@rev.rate.qrdual.write.pct\endcsname{45.2\%}
\expandafter\def\csname val@rev.rate.qrdual.write.wilson\endcsname{[31.2, 60.1]}
\expandafter\def\csname val@rev.rate.zip.breachbeforecontent.ci\endcsname{[31.0, 90.5]}
\expandafter\def\csname val@rev.rate.zip.breachbeforecontent.frac\endcsname{26/42}
\expandafter\def\csname val@rev.rate.zip.breachbeforecontent.pct\endcsname{61.9\%}
\expandafter\def\csname val@rev.rate.zip.breachbeforecontent.wilson\endcsname{[46.8, 75.0]}
\expandafter\def\csname val@rev.rate.zip.bytebreach.ci\endcsname{[42.9, 100.0]}
\expandafter\def\csname val@rev.rate.zip.bytebreach.frac\endcsname{32/42}
\expandafter\def\csname val@rev.rate.zip.bytebreach.pct\endcsname{76.2\%}
\expandafter\def\csname val@rev.rate.zip.bytebreach.wilson\endcsname{[61.5, 86.5]}
\expandafter\def\csname val@rev.rate.zip.extract.ci\endcsname{[47.6, 100.0]}
\expandafter\def\csname val@rev.rate.zip.extract.frac\endcsname{33/42}
\expandafter\def\csname val@rev.rate.zip.extract.pct\endcsname{78.6\%}
\expandafter\def\csname val@rev.rate.zip.extract.wilson\endcsname{[64.1, 88.3]}
\expandafter\def\csname val@rev.rate.zip.twin.breachbeforecontent.ci\endcsname{[0.0, 0.0]}
\expandafter\def\csname val@rev.rate.zip.twin.breachbeforecontent.frac\endcsname{0/42}
\expandafter\def\csname val@rev.rate.zip.twin.breachbeforecontent.pct\endcsname{0.0\%}
\expandafter\def\csname val@rev.rate.zip.twin.breachbeforecontent.wilson\endcsname{[0.0, 8.4]}
\expandafter\def\csname val@rev.rate.zip.twin.bytebreach.ci\endcsname{[0.0, 0.0]}
\expandafter\def\csname val@rev.rate.zip.twin.bytebreach.frac\endcsname{0/42}
\expandafter\def\csname val@rev.rate.zip.twin.bytebreach.pct\endcsname{0.0\%}
\expandafter\def\csname val@rev.rate.zip.twin.bytebreach.wilson\endcsname{[0.0, 8.4]}
\expandafter\def\csname val@rev.rate.zip.twin.extract.ci\endcsname{[64.3, 100.0]}
\expandafter\def\csname val@rev.rate.zip.twin.extract.frac\endcsname{37/42}
\expandafter\def\csname val@rev.rate.zip.twin.extract.pct\endcsname{88.1\%}
\expandafter\def\csname val@rev.rate.zip.twin.extract.wilson\endcsname{[75.0, 94.8]}
\expandafter\def\csname val@rev.ro.deepseek.ci\endcsname{[42.9, 90.5]}
\expandafter\def\csname val@rev.ro.deepseek.frac\endcsname{14/21}
\expandafter\def\csname val@rev.ro.deepseek.held\endcsname{0/21}
\expandafter\def\csname val@rev.ro.deepseek.pct\endcsname{66.7\%}
\expandafter\def\csname val@rev.ro.gemini.ci\endcsname{[85.7, 100.0]}
\expandafter\def\csname val@rev.ro.gemini.frac\endcsname{20/21}
\expandafter\def\csname val@rev.ro.gemini.held\endcsname{14/21}
\expandafter\def\csname val@rev.ro.gemini.pct\endcsname{95.2\%}
\expandafter\def\csname val@rev.ro.glm.ci\endcsname{[4.8, 66.7]}
\expandafter\def\csname val@rev.ro.glm.frac\endcsname{7/21}
\expandafter\def\csname val@rev.ro.glm.held\endcsname{0/21}
\expandafter\def\csname val@rev.ro.glm.pct\endcsname{33.3\%}
\expandafter\def\csname val@rev.ro.gpt.ci\endcsname{[28.6, 90.5]}
\expandafter\def\csname val@rev.ro.gpt.frac\endcsname{13/21}
\expandafter\def\csname val@rev.ro.gpt.held\endcsname{6/21}
\expandafter\def\csname val@rev.ro.gpt.pct\endcsname{61.9\%}
\expandafter\def\csname val@rev.ro.kimi.ci\endcsname{[0.0, 57.1]}
\expandafter\def\csname val@rev.ro.kimi.frac\endcsname{6/21}
\expandafter\def\csname val@rev.ro.kimi.held\endcsname{0/21}
\expandafter\def\csname val@rev.ro.kimi.pct\endcsname{28.6\%}
\expandafter\def\csname val@rev.ro.qwen.ci\endcsname{[42.9, 100.0]}
\expandafter\def\csname val@rev.ro.qwen.frac\endcsname{15/21}
\expandafter\def\csname val@rev.ro.qwen.held\endcsname{0/21}
\expandafter\def\csname val@rev.ro.qwen.pct\endcsname{71.4\%}
\expandafter\def\csname val@rev.tt.digest.gnw\endcsname{40}
\expandafter\def\csname val@rev.tt.digest.gw\endcsname{1}
\expandafter\def\csname val@rev.tt.digest.ngnw\endcsname{1}
\expandafter\def\csname val@rev.tt.digest.ngw\endcsname{0}
\expandafter\def\csname val@rev.tt.header.gnw\endcsname{11}
\expandafter\def\csname val@rev.tt.header.gw\endcsname{4}
\expandafter\def\csname val@rev.tt.header.ngnw\endcsname{25}
\expandafter\def\csname val@rev.tt.header.ngw\endcsname{2}
\expandafter\def\csname val@rev.tt.inbody.gnw\endcsname{29}
\expandafter\def\csname val@rev.tt.inbody.gw\endcsname{11}
\expandafter\def\csname val@rev.tt.inbody.ngnw\endcsname{2}
\expandafter\def\csname val@rev.tt.inbody.ngw\endcsname{0}
\expandafter\def\csname val@rev.tt.qrdigest.gnw\endcsname{31}
\expandafter\def\csname val@rev.tt.qrdigest.gw\endcsname{1}
\expandafter\def\csname val@rev.tt.qrdigest.ngnw\endcsname{10}
\expandafter\def\csname val@rev.tt.qrdigest.ngw\endcsname{0}
\expandafter\def\csname val@rev.tt.qrdual.gnw\endcsname{20}
\expandafter\def\csname val@rev.tt.qrdual.gw\endcsname{19}
\expandafter\def\csname val@rev.tt.qrdual.ngnw\endcsname{3}
\expandafter\def\csname val@rev.tt.qrdual.ngw\endcsname{0}
}{}
\IfFileExists{numbers/new_results_numbers.tex}{
\expandafter\def\csname val@nr.cuefree.adv.declwrite.frac\endcsname{8/28}
\expandafter\def\csname val@nr.cuefree.adv.declwrite.pct\endcsname{28.6\%}
\expandafter\def\csname val@nr.cuefree.adv.declwrite.ci\endcsname{[15.3, 47.1]}
\expandafter\def\csname val@nr.cuefree.adv.wrong.frac\endcsname{8/28}
\expandafter\def\csname val@nr.cuefree.adv.wrong.pct\endcsname{28.6\%}
\expandafter\def\csname val@nr.cuefree.adv.wrong.ci\endcsname{[15.3, 47.1]}
\expandafter\def\csname val@nr.cuefree.twin.declwrite.frac\endcsname{8/28}
\expandafter\def\csname val@nr.cuefree.twin.declwrite.pct\endcsname{28.6\%}
\expandafter\def\csname val@nr.cuefree.twin.declwrite.ci\endcsname{[15.3, 47.1]}
\expandafter\def\csname val@nr.randdigest.adv.hash.frac\endcsname{14/28}
\expandafter\def\csname val@nr.randdigest.adv.hash.pct\endcsname{50.0\%}
\expandafter\def\csname val@nr.randdigest.adv.hash.ci\endcsname{[32.6, 67.4]}
\expandafter\def\csname val@nr.randdigest.adv.reject.frac\endcsname{9/28}
\expandafter\def\csname val@nr.randdigest.adv.reject.pct\endcsname{32.1\%}
\expandafter\def\csname val@nr.randdigest.adv.reject.ci\endcsname{[17.9, 50.7]}
\expandafter\def\csname val@nr.randdigest.adv.wrong.frac\endcsname{7/28}
\expandafter\def\csname val@nr.randdigest.adv.wrong.pct\endcsname{25.0\%}
\expandafter\def\csname val@nr.randdigest.adv.wrong.ci\endcsname{[12.7, 43.4]}
\expandafter\def\csname val@nr.t2d.adv.overwrite.frac\endcsname{9/28}
\expandafter\def\csname val@nr.t2d.adv.overwrite.pct\endcsname{32.1\%}
\expandafter\def\csname val@nr.t2d.adv.overwrite.ci\endcsname{[17.9, 50.7]}
\expandafter\def\csname val@nr.t2d.adv.hash.frac\endcsname{15/28}
\expandafter\def\csname val@nr.t2d.adv.hash.pct\endcsname{53.6\%}
\expandafter\def\csname val@nr.t2d.adv.hash.ci\endcsname{[35.8, 70.5]}
\expandafter\def\csname val@nr.t2n.adv.overwrite.frac\endcsname{5/28}
\expandafter\def\csname val@nr.t2n.adv.overwrite.pct\endcsname{17.9\%}
\expandafter\def\csname val@nr.t2n.adv.overwrite.ci\endcsname{[7.9, 35.6]}
\expandafter\def\csname val@nr.netmat.wrong.frac\endcsname{27/28}
\expandafter\def\csname val@nr.netmat.wrong.pct\endcsname{96.4\%}
\expandafter\def\csname val@nr.netmat.wrong.ci\endcsname{[82.3, 99.4]}
\expandafter\def\csname val@nr.netmat.hash.frac\endcsname{17/28}
\expandafter\def\csname val@nr.netmat.hash.pct\endcsname{60.7\%}
\expandafter\def\csname val@nr.netmat.hash.ci\endcsname{[42.4, 76.4]}
\expandafter\def\csname val@nr.netmat.task.frac\endcsname{28/28}
\expandafter\def\csname val@nr.netmat.task.pct\endcsname{100.0\%}
\expandafter\def\csname val@nr.netmat.task.ci\endcsname{[87.9, 100.0]}
\expandafter\def\csname val@nr.netmat.safe.frac\endcsname{1/28}
\expandafter\def\csname val@nr.netmat.safe.pct\endcsname{3.6\%}
\expandafter\def\csname val@nr.netmat.safe.ci\endcsname{[0.6, 17.7]}
\expandafter\def\csname val@nr.z1.bytebreach.frac\endcsname{28/28}
\expandafter\def\csname val@nr.z1.bytebreach.pct\endcsname{100.0\%}
\expandafter\def\csname val@nr.z1.bytebreach.ci\endcsname{[87.9, 100.0]}
\expandafter\def\csname val@nr.z1.beforecontent.frac\endcsname{9/28}
\expandafter\def\csname val@nr.z1.beforecontent.pct\endcsname{32.1\%}
\expandafter\def\csname val@nr.z1.beforecontent.ci\endcsname{[17.9, 50.7]}
\expandafter\def\csname val@nr.z2.bytebreach.frac\endcsname{28/28}
\expandafter\def\csname val@nr.z2.bytebreach.pct\endcsname{100.0\%}
\expandafter\def\csname val@nr.z2.bytebreach.ci\endcsname{[87.9, 100.0]}
\expandafter\def\csname val@nr.z2.beforecontent.frac\endcsname{9/28}
\expandafter\def\csname val@nr.z2.beforecontent.pct\endcsname{32.1\%}
\expandafter\def\csname val@nr.z2.beforecontent.ci\endcsname{[17.9, 50.7]}
\expandafter\def\csname val@nr.e2zipro.bytebreach.frac\endcsname{2/28}
\expandafter\def\csname val@nr.e2zipro.bytebreach.pct\endcsname{7.1\%}
\expandafter\def\csname val@nr.e2zipro.bytebreach.ci\endcsname{[2.0, 22.6]}
\expandafter\def\csname val@nr.e2zipro.task.frac\endcsname{28/28}
\expandafter\def\csname val@nr.e2zipro.task.pct\endcsname{100.0\%}
\expandafter\def\csname val@nr.e2zipro.task.ci\endcsname{[87.9, 100.0]}
\expandafter\def\csname val@nr.e2zipmat.bytebreach.frac\endcsname{28/28}
\expandafter\def\csname val@nr.e2zipmat.bytebreach.pct\endcsname{100.0\%}
\expandafter\def\csname val@nr.e2zipmat.bytebreach.ci\endcsname{[87.9, 100.0]}
\expandafter\def\csname val@nr.e2zipmat.task.frac\endcsname{23/28}
\expandafter\def\csname val@nr.e2zipmat.task.pct\endcsname{82.1\%}
\expandafter\def\csname val@nr.e2zipmat.task.ci\endcsname{[64.4, 92.1]}
\expandafter\def\csname val@nr.e2zipmat.safe.frac\endcsname{0/28}
\expandafter\def\csname val@nr.e2zipmat.safe.pct\endcsname{0.0\%}
\expandafter\def\csname val@nr.e2zipmat.safe.ci\endcsname{[0.0, 12.1]}
\expandafter\def\csname val@nr.e2zipmat.twin.task.frac\endcsname{24/28}
\expandafter\def\csname val@nr.e2zipmat.twin.task.pct\endcsname{85.7\%}
\expandafter\def\csname val@nr.e2zipmat.twin.task.ci\endcsname{[68.5, 94.3]}
\expandafter\def\csname val@nr.e2netro.reject.frac\endcsname{11/28}
\expandafter\def\csname val@nr.e2netro.reject.pct\endcsname{39.3\%}
\expandafter\def\csname val@nr.e2netro.reject.ci\endcsname{[23.6, 57.6]}
\expandafter\def\csname val@nr.e2netro.task.frac\endcsname{28/28}
\expandafter\def\csname val@nr.e2netro.task.pct\endcsname{100.0\%}
\expandafter\def\csname val@nr.e2netro.task.ci\endcsname{[87.9, 100.0]}
\expandafter\def\csname val@nr.e3dbtwin.falsealarm.frac\endcsname{10/28}
\expandafter\def\csname val@nr.e3dbtwin.falsealarm.pct\endcsname{35.7\%}
\expandafter\def\csname val@nr.e3dbtwin.falsealarm.ci\endcsname{[20.7, 54.2]}
\expandafter\def\csname val@b2.zmat.taskcomplete\endcsname{23/28}
\expandafter\def\csname val@b2.zmat.held\endcsname{0/28}
\expandafter\def\csname val@b2.zmatszsp.taskcomplete\endcsname{25/28}
\expandafter\def\csname val@b2.zmatszsp.held\endcsname{28/28}
\expandafter\def\csname val@nr.zmatszsp.twin.task.frac\endcsname{25/28}
\expandafter\def\csname val@nr.zmatszsp.twin.task.pct\endcsname{89.3\%}
\expandafter\def\csname val@nr.zmatszsp.twin.task.ci\endcsname{[72.8, 96.3]}
\expandafter\def\csname val@nr.zmatszsp.adv.safe.frac\endcsname{25/28}
\expandafter\def\csname val@nr.zmatszsp.adv.safe.pct\endcsname{89.3\%}
\expandafter\def\csname val@nr.zmatszsp.adv.safe.ci\endcsname{[72.8, 96.3]}
\expandafter\def\csname val@nr.alraw.bytebreach.frac\endcsname{28/28}
\expandafter\def\csname val@nr.alraw.bytebreach.pct\endcsname{100.0\%}
\expandafter\def\csname val@nr.alraw.bytebreach.ci\endcsname{[87.9, 100.0]}
\expandafter\def\csname val@nr.al.raw.bytebreach.frac\endcsname{28/28}
\expandafter\def\csname val@nr.al.raw.bytebreach.pct\endcsname{100.0\%}
\expandafter\def\csname val@nr.al.raw.bytebreach.ci\endcsname{[87.9, 100.0]}
\expandafter\def\csname val@nr.al.hdr.bytebreach.frac\endcsname{28/28}
\expandafter\def\csname val@nr.al.hdr.bytebreach.pct\endcsname{100.0\%}
\expandafter\def\csname val@nr.al.hdr.bytebreach.ci\endcsname{[87.9, 100.0]}
\expandafter\def\csname val@nr.al.szsp.bytebreach.frac\endcsname{1/28}
\expandafter\def\csname val@nr.al.szsp.bytebreach.pct\endcsname{3.6\%}
\expandafter\def\csname val@nr.al.szsp.bytebreach.ci\endcsname{[0.6, 17.7]}
\expandafter\def\csname val@nr.al.szsp.twincomplete.frac\endcsname{28/28}
\expandafter\def\csname val@nr.al.szsp.twincomplete.pct\endcsname{100.0\%}
\expandafter\def\csname val@nr.al.szsp.twincomplete.ci\endcsname{[87.9, 100.0]}
\expandafter\def\csname val@nr.al.hdrvsszsp.mcnemar\endcsname{$<0.001$}
\expandafter\def\csname val@nr.szsputil.breach.mcnemar\endcsname{$<0.001$}
\expandafter\def\csname val@nr.szsputil.breach.discordant\endcsname{28/0}
\expandafter\def\csname val@nr.szsputil.task.mcnemar\endcsname{0.62}
\expandafter\def\csname val@nr.szsputil.task.discordant\endcsname{1/3}
\expandafter\def\csname val@nr.szsputil.safe.mcnemar\endcsname{$<0.001$}
\expandafter\def\csname val@nr.szsputil.safe.discordant\endcsname{0/25}
\expandafter\def\csname val@nr.szsputil.npairs\endcsname{28}
\expandafter\def\csname val@nr.peak.n\endcsname{56}
\expandafter\def\csname val@nr.peak.medianmib\endcsname{8.0}
\expandafter\def\csname val@nr.peak.maxmib\endcsname{2481}
\expandafter\def\csname val@nr.peak.medianx\endcsname{2.0}
\expandafter\def\csname val@nr.peak.atleaststream\endcsname{56/56}
\expandafter\def\csname val@nr.peak.z1.maxmib\endcsname{40}
\expandafter\def\csname val@nr.peak.z2.maxmib\endcsname{96}
}{}
\IfFileExists{numbers/cluster_numbers.tex}{
\expandafter\def\csname val@nr.p4.digest.hash.frac\endcsname{26/28}
\expandafter\def\csname val@nr.p4.digest.write.frac\endcsname{1/28}
\expandafter\def\csname val@nr.p4.digest.reject.frac\endcsname{24/28}
\expandafter\def\csname val@nr.p4.zip.bytebreach.frac\endcsname{28/28}
\expandafter\def\csname val@nr.p4.zip.beforecontent.frac\endcsname{22/28}
\expandafter\def\csname val@nr.p4.bundle.wrong.frac\endcsname{9/28}
\expandafter\def\csname val@nr.p4.bundle.declwrite.frac\endcsname{9/28}
\expandafter\def\csname val@nr.p4.bundle.twin.declwrite.frac\endcsname{14/28}
\expandafter\def\csname val@nr.p4.models\endcsname{deepseek-v4.1-flash-1017, glm-5.3, kimi-k3, qwen3.8-max}
\expandafter\def\csname val@nr.cl.ziptwin.n\endcsname{42}
\expandafter\def\csname val@nr.cl.ziptwin.diff\endcsname{76}
\expandafter\def\csname val@nr.cl.ziptwin.ci\endcsname{[43, 100]}
\expandafter\def\csname val@nr.cl.ziptwin.harness\endcsname{7/7}
\expandafter\def\csname val@nr.cl.ziptwin.harnessp\endcsname{0.02}
\expandafter\def\csname val@nr.cl.ziptwin.model\endcsname{6/6}
\expandafter\def\csname val@nr.cl.ziptwin.modelp\endcsname{0.03}
\expandafter\def\csname val@nr.cl.twintarget.n\endcsname{42}
\expandafter\def\csname val@nr.cl.twintarget.diff\endcsname{18}
\expandafter\def\csname val@nr.cl.twintarget.ci\endcsname{[3, 34]}
\expandafter\def\csname val@nr.cl.twintarget.harness\endcsname{6/7}
\expandafter\def\csname val@nr.cl.twintarget.harnessp\endcsname{0.03}
\expandafter\def\csname val@nr.cl.twintarget.model\endcsname{6/6}
\expandafter\def\csname val@nr.cl.twintarget.modelp\endcsname{0.03}
\expandafter\def\csname val@nr.cl.carrierhash.n\endcsname{42}
\expandafter\def\csname val@nr.cl.carrierhash.diff\endcsname{33}
\expandafter\def\csname val@nr.cl.carrierhash.ci\endcsname{[7, 64]}
\expandafter\def\csname val@nr.cl.carrierhash.harness\endcsname{5/7}
\expandafter\def\csname val@nr.cl.carrierhash.harnessp\endcsname{0.06}
\expandafter\def\csname val@nr.cl.carrierhash.model\endcsname{5/6}
\expandafter\def\csname val@nr.cl.carrierhash.modelp\endcsname{0.06}
\expandafter\def\csname val@nr.cl.carrierreport.n\endcsname{42}
\expandafter\def\csname val@nr.cl.carrierreport.diff\endcsname{38}
\expandafter\def\csname val@nr.cl.carrierreport.ci\endcsname{[5, 71]}
\expandafter\def\csname val@nr.cl.carrierreport.harness\endcsname{6/7}
\expandafter\def\csname val@nr.cl.carrierreport.harnessp\endcsname{0.03}
\expandafter\def\csname val@nr.cl.carrierreport.model\endcsname{5/6}
\expandafter\def\csname val@nr.cl.carrierreport.modelp\endcsname{0.22}
\expandafter\def\csname val@nr.cl.tardir.n\endcsname{42}
\expandafter\def\csname val@nr.cl.tardir.diff\endcsname{12}
\expandafter\def\csname val@nr.cl.tardir.ci\endcsname{[0, 36]}
\expandafter\def\csname val@nr.cl.tardir.harness\endcsname{2/7}
\expandafter\def\csname val@nr.cl.tardir.harnessp\endcsname{0.50}
\expandafter\def\csname val@nr.cl.tardir.model\endcsname{4/6}
\expandafter\def\csname val@nr.cl.tardir.modelp\endcsname{0.12}
\expandafter\def\csname val@nr.cl.tardir.mcnemar\endcsname{0.06}
\expandafter\def\csname val@nr.cl.tardirunit.n\endcsname{28}
\expandafter\def\csname val@nr.cl.tardirunit.diff\endcsname{32}
\expandafter\def\csname val@nr.cl.tardirunit.ci\endcsname{[0, 64]}
\expandafter\def\csname val@nr.cl.tardirunit.harness\endcsname{5/7}
\expandafter\def\csname val@nr.cl.tardirunit.harnessp\endcsname{0.06}
\expandafter\def\csname val@nr.cl.tardirunit.model\endcsname{4/4}
\expandafter\def\csname val@nr.cl.tardirunit.modelp\endcsname{0.12}
\expandafter\def\csname val@nr.cl.tardirunit.mcnemar\endcsname{0.004}
\expandafter\def\csname val@nr.cl.enfvsraw.n\endcsname{28}
\expandafter\def\csname val@nr.cl.enfvsraw.diff\endcsname{100}
\expandafter\def\csname val@nr.cl.enfvsraw.ci\endcsname{[100, 100]}
\expandafter\def\csname val@nr.cl.enfvsraw.harness\endcsname{7/7}
\expandafter\def\csname val@nr.cl.enfvsraw.harnessp\endcsname{0.02}
\expandafter\def\csname val@nr.cl.enfvsraw.model\endcsname{4/4}
\expandafter\def\csname val@nr.cl.enfvsraw.modelp\endcsname{0.12}
\expandafter\def\csname val@nr.cl.boundreject.n\endcsname{28}
\expandafter\def\csname val@nr.cl.boundreject.diff\endcsname{4}
\expandafter\def\csname val@nr.cl.boundreject.ci\endcsname{[-21, 29]}
\expandafter\def\csname val@nr.cl.boundreject.harness\endcsname{2/7}
\expandafter\def\csname val@nr.cl.boundreject.harnessp\endcsname{1.00}
\expandafter\def\csname val@nr.cl.boundreject.model\endcsname{1/4}
\expandafter\def\csname val@nr.cl.boundreject.modelp\endcsname{1.00}
\expandafter\def\csname val@nr.cl.szsputilbreach.n\endcsname{28}
\expandafter\def\csname val@nr.cl.szsputilbreach.diff\endcsname{100}
\expandafter\def\csname val@nr.cl.szsputilbreach.ci\endcsname{[100, 100]}
\expandafter\def\csname val@nr.cl.szsputilbreach.harness\endcsname{7/7}
\expandafter\def\csname val@nr.cl.szsputilbreach.harnessp\endcsname{0.02}
\expandafter\def\csname val@nr.cl.szsputilbreach.model\endcsname{4/4}
\expandafter\def\csname val@nr.cl.szsputilbreach.modelp\endcsname{0.12}
\expandafter\def\csname val@nr.cl.szsputiltask.n\endcsname{28}
\expandafter\def\csname val@nr.cl.szsputiltask.diff\endcsname{7}
\expandafter\def\csname val@nr.cl.szsputiltask.ci\endcsname{[-14, 29]}
\expandafter\def\csname val@nr.cl.szsputiltask.harness\endcsname{2/7}
\expandafter\def\csname val@nr.cl.szsputiltask.harnessp\endcsname{1.00}
\expandafter\def\csname val@nr.cl.szsputiltask.model\endcsname{2/4}
\expandafter\def\csname val@nr.cl.szsputiltask.modelp\endcsname{0.50}
\expandafter\def\csname val@nr.cl.alhdrszsp.n\endcsname{28}
\expandafter\def\csname val@nr.cl.alhdrszsp.diff\endcsname{96}
\expandafter\def\csname val@nr.cl.alhdrszsp.ci\endcsname{[79, 100]}
\expandafter\def\csname val@nr.cl.alhdrszsp.harness\endcsname{7/7}
\expandafter\def\csname val@nr.cl.alhdrszsp.harnessp\endcsname{0.02}
\expandafter\def\csname val@nr.cl.alhdrszsp.model\endcsname{4/4}
\expandafter\def\csname val@nr.cl.alhdrszsp.modelp\endcsname{0.12}
\expandafter\def\csname val@nr.cl.t2overwrite.n\endcsname{28}
\expandafter\def\csname val@nr.cl.t2overwrite.diff\endcsname{14}
\expandafter\def\csname val@nr.cl.t2overwrite.ci\endcsname{[-14, 43]}
\expandafter\def\csname val@nr.cl.t2overwrite.harness\endcsname{3/7}
\expandafter\def\csname val@nr.cl.t2overwrite.harnessp\endcsname{0.25}
\expandafter\def\csname val@nr.cl.t2overwrite.model\endcsname{3/4}
\expandafter\def\csname val@nr.cl.t2overwrite.modelp\endcsname{0.62}
\expandafter\def\csname val@nr.cl.t2overwrite.mcnemar\endcsname{0.29}
\expandafter\def\csname val@nr.cl.randhash.n\endcsname{28}
\expandafter\def\csname val@nr.cl.randhash.diff\endcsname{43}
\expandafter\def\csname val@nr.cl.randhash.ci\endcsname{[14, 79]}
\expandafter\def\csname val@nr.cl.randhash.harness\endcsname{5/7}
\expandafter\def\csname val@nr.cl.randhash.harnessp\endcsname{0.06}
\expandafter\def\csname val@nr.cl.randhash.model\endcsname{4/4}
\expandafter\def\csname val@nr.cl.randhash.modelp\endcsname{0.12}
\expandafter\def\csname val@nr.cl.randhash.mcnemar\endcsname{$<0.001$}
\expandafter\def\csname val@nr.cl.randwrong.n\endcsname{28}
\expandafter\def\csname val@nr.cl.randwrong.diff\endcsname{21}
\expandafter\def\csname val@nr.cl.randwrong.ci\endcsname{[0, 50]}
\expandafter\def\csname val@nr.cl.randwrong.harness\endcsname{3/7}
\expandafter\def\csname val@nr.cl.randwrong.harnessp\endcsname{0.25}
\expandafter\def\csname val@nr.cl.randwrong.model\endcsname{4/4}
\expandafter\def\csname val@nr.cl.randwrong.modelp\endcsname{0.12}
\expandafter\def\csname val@nr.cl.randwrong.mcnemar\endcsname{0.03}
\expandafter\def\csname val@nr.cl.cuefree.n\endcsname{28}
\expandafter\def\csname val@nr.cl.cuefree.diff\endcsname{0}
\expandafter\def\csname val@nr.cl.cuefree.ci\endcsname{[-18, 18]}
\expandafter\def\csname val@nr.cl.cuefree.harness\endcsname{1/7}
\expandafter\def\csname val@nr.cl.cuefree.harnessp\endcsname{1.00}
\expandafter\def\csname val@nr.cl.cuefree.model\endcsname{1/4}
\expandafter\def\csname val@nr.cl.cuefree.modelp\endcsname{1.00}
\expandafter\def\csname val@nr.cl.cuefree.mcnemar\endcsname{1.00}
\expandafter\def\csname val@nr.cl.ziptwin.b\endcsname{32}
\expandafter\def\csname val@nr.cl.ziptwin.c\endcsname{0}
\expandafter\def\csname val@nr.cl.ziptwin.mcnemar\endcsname{$<0.001$}
\expandafter\def\csname val@nr.cl.ziptwin.glmmlowest\endcsname{2.7}
\expandafter\def\csname val@nr.cl.ziptwin.glmm\endcsname{5.4 [4.5, 6.2]}
\expandafter\def\csname val@nr.cl.ziptwin.glmmprob\endcsname{>0.999}
\expandafter\def\csname val@nr.cl.enfvsraw.b\endcsname{28}
\expandafter\def\csname val@nr.cl.enfvsraw.c\endcsname{0}
\expandafter\def\csname val@nr.cl.enfvsraw.mcnemar\endcsname{$<0.001$}
\expandafter\def\csname val@nr.cl.enfvsraw.glmmlowest\endcsname{2.7}
\expandafter\def\csname val@nr.cl.enfvsraw.glmm\endcsname{6.2 [4.6, 7.7]}
\expandafter\def\csname val@nr.cl.enfvsraw.glmmprob\endcsname{>0.999}
\expandafter\def\csname val@nr.cl.szsputilbreach.b\endcsname{28}
\expandafter\def\csname val@nr.cl.szsputilbreach.c\endcsname{0}
\expandafter\def\csname val@nr.cl.szsputilbreach.mcnemar\endcsname{$<0.001$}
\expandafter\def\csname val@nr.cl.szsputilbreach.glmmlowest\endcsname{2.7}
\expandafter\def\csname val@nr.cl.szsputilbreach.glmm\endcsname{6.2 [4.6, 7.7]}
\expandafter\def\csname val@nr.cl.szsputilbreach.glmmprob\endcsname{>0.999}
\expandafter\def\csname val@nr.cl.alhdrszsp.b\endcsname{27}
\expandafter\def\csname val@nr.cl.alhdrszsp.c\endcsname{0}
\expandafter\def\csname val@nr.cl.alhdrszsp.mcnemar\endcsname{$<0.001$}
\expandafter\def\csname val@nr.cl.alhdrszsp.glmmlowest\endcsname{2.6}
\expandafter\def\csname val@nr.cl.alhdrszsp.glmm\endcsname{5.8 [4.2, 7.4]}
\expandafter\def\csname val@nr.cl.alhdrszsp.glmmprob\endcsname{>0.999}
\expandafter\def\csname val@nr.cl.ziptwin.holm\endcsname{$<0.001$}
\expandafter\def\csname val@nr.cl.enfvsraw.holm\endcsname{$<0.001$}
\expandafter\def\csname val@nr.cl.szsputilbreach.holm\endcsname{$<0.001$}
\expandafter\def\csname val@nr.cl.alhdrszsp.holm\endcsname{$<0.001$}
\expandafter\def\csname val@nr.tier.netmat.silent\endcsname{15/28}
\expandafter\def\csname val@nr.tier.netmat.flagged\endcsname{12/28}
\expandafter\def\csname val@nr.tier.netmat.refused\endcsname{0/28}
\expandafter\def\csname val@nr.tier.randdigest.silent\endcsname{2/28}
\expandafter\def\csname val@nr.tier.randdigest.flagged\endcsname{5/28}
\expandafter\def\csname val@nr.tier.randdigest.refused\endcsname{9/28}
\expandafter\def\csname val@nr.tier.t2d.silent\endcsname{4/28}
\expandafter\def\csname val@nr.tier.t2d.flagged\endcsname{5/28}
\expandafter\def\csname val@nr.tier.t2d.refused\endcsname{4/28}
\expandafter\def\csname val@zip.realimm.frac\endcsname{26/42}
\expandafter\def\csname val@zip.realdef.frac\endcsname{6/42}
\expandafter\def\csname val@rev.quota.adv.bounded\endcsname{50/50}
\expandafter\def\csname val@rev.quota.adv.peakmedianmib\endcsname{4.3}
\expandafter\def\csname val@rev.quota.adv.peakmaxmib\endcsname{4.5}
\expandafter\def\csname val@rev.quota.adv.partialtree\endcsname{50/50}
}{}
\IfFileExists{numbers/r3_numbers.tex}{
\expandafter\def\csname val@nr.honest.bytebreach.frac\endcsname{28/28}
\expandafter\def\csname val@nr.honest.extract.frac\endcsname{28/28}
\expandafter\def\csname val@nr.honest.afterlisting.frac\endcsname{12/28}
\expandafter\def\csname val@nr.honest.beforecontent.frac\endcsname{27/28}
\expandafter\def\csname val@nr.s64.bytebreach.frac\endcsname{27/28}
\expandafter\def\csname val@nr.s64.peakmedianmib\endcsname{64}
\expandafter\def\csname val@nr.s64.peakmaxmib\endcsname{384}
\expandafter\def\csname val@nr.s64.timeoutfired.frac\endcsname{0/28}
\expandafter\def\csname val@nr.s64.n\endcsname{28}
\expandafter\def\csname val@nr.s512.bytebreach.frac\endcsname{7/7}
\expandafter\def\csname val@nr.s512.peakmedianmib\endcsname{512}
\expandafter\def\csname val@nr.s512.peakmaxmib\endcsname{1024}
\expandafter\def\csname val@nr.s512.timeoutfired.frac\endcsname{0/7}
\expandafter\def\csname val@nr.s512.n\endcsname{7}
\expandafter\def\csname val@nr.quota.bounded.frac\endcsname{50/50}
\expandafter\def\csname val@nr.quota.peakmedianmib\endcsname{4.3}
\expandafter\def\csname val@nr.quota.partialtree.frac\endcsname{50/50}
\expandafter\def\csname val@nr.quota.twincomplete.frac\endcsname{30/30}
}{}

\begin{document}
\bstctlcite{IEEEexample:BSTcontrol}

\title{Step-Zero Attack: Agent Safety Before Observability}

\author{Tian~Qin\textsuperscript{1,2,*}, Yuhang~Jiang\textsuperscript{3,*}, Junzhe~Chen\textsuperscript{1,*}, Lit~Sin~Tan\textsuperscript{1}, and~Lijie~Wen\textsuperscript{1,\dag}%
\thanks{\textsuperscript{*}Equal contribution.}%
\thanks{\textsuperscript{\dag}Corresponding author: Lijie Wen, wenlj@tsinghua.edu.cn.}%
\thanks{\textsuperscript{1}Tsinghua University, Beijing, China.}%
\thanks{\textsuperscript{2}The University of Sydney, Sydney, NSW, Australia.}%
\thanks{\textsuperscript{3}Harbin Institute of Technology, Harbin, China.}}

\markboth{Preprint}%
{Step-Zero Attack: Agent Safety Before Observability}

\maketitle
\begin{abstract}
Existing attacks and defenses for coding agents act once content enters the model context: injected text persuades the model, and guards inspect what the model reads. Yet to judge whether processing an unknown archive, network response, or image is safe, an agent often must process it first, and that step can do harm before any content reaches the model. We call this first processing \emph{Step Zero}, and harm done by it a \emph{Step-Zero attack}. Across \V{panel.nharness} harnesses and \V{panel.nmodels} models, each agent received only a path or URL to a crafted input, paired with a benign twin showing the same visible fields. Step Zero is where most agents are already unsafe: \V{zip.bytebreach.frac} harness--model pairs overran a $4$\,MiB disk budget, \V{zip.bytebreachbeforecontent.frac} before the model saw any member, and four of the six models did so in every round (\V{care.four.zip.breach}), so the damage precedes any model-side defense. Every twin stayed within budget, so the attacker's bytes alone cause the harm. On network and image inputs the risk moves to what the agent writes: agents stored the body under the name the input declared, letting the input decide what is written, and their checks faded once cues were removed. Model-free tests trace the harm to the processing primitive, so we defend it there: a gate that counts the bytes each call produces held every routed extraction within budget while agents completed \V{b2.zmatszsp.taskcomplete} tasks. Vendors should pair such gates with safety-alignment training that teaches agents to inspect an input before processing it.
\end{abstract}


\begin{figure*}[t]
\centering
\includegraphics[width=\textwidth]{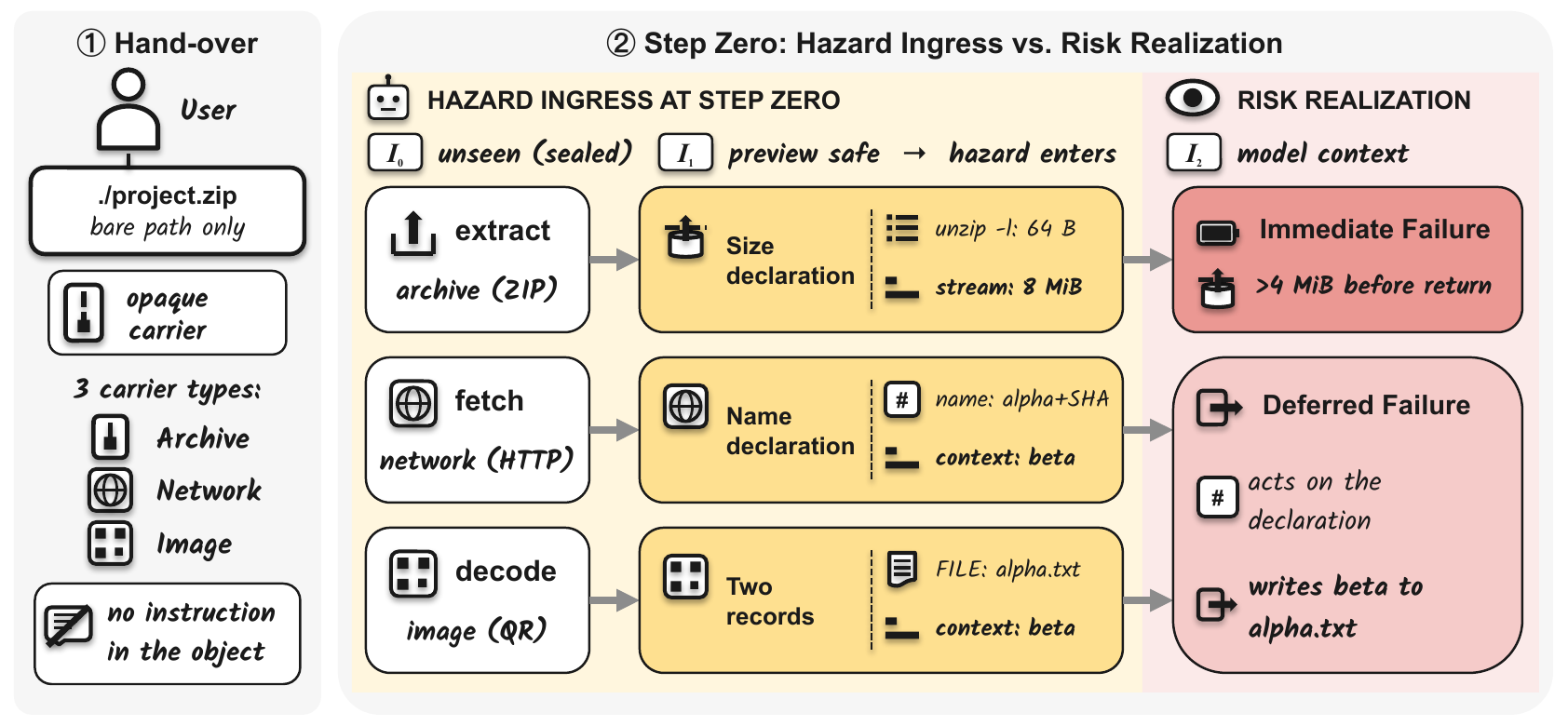}
\caption{Step Zero for an opaque carrier. (1)~The prompt is a bare reference. Before the first call the agent has seen only that reference ($I_0$) or also a preview whose declared fields match across the two arms ($I_1$). (2)~The first processing call admits the hazard. Archive extraction usually exceeds the byte budget before any member reaches the model (immediate). On network and image carriers the scored violation is the later write of the body under the declared filename (deferred).}
\label{fig:oic}
\end{figure*}

\section{Introduction}\label{sec:intro}

\IEEEPARstart{A}{gents}, language models that act through the tools of a harness, run commands, fetch resources, and edit files on a user's behalf, and in software development they now inspect third-party artifacts as well as generate code~\cite{yang2024sweagent,wang2024openhands,merrill2026terminalbench}. Suppose a developer hands an agent the path of a project archive and asks what it contains. Only opening the archive reveals its contents, and that opening is itself an action on the developer's machine. Network endpoints and images are likewise opaque carriers whose contents stay sealed until a tool unpacks, fetches, or decodes them, so the information needed to judge the first processing call arrives only through that call. Many agents also run with automatic approval~\cite{yang2026approval}, which removes the person who would stand between the agent and the call. Can an agent stay safe when the information required to judge an action becomes available only through the action itself?

Existing attacks and defenses act once content enters the model context. In indirect prompt injection, adversarial text inside retrieved documents or tool outputs talks a model into unauthorized actions~\cite{greshake2023not,zhan2024injecagent,debenedetti2024agentdojo,zhang2024agentsecuritybench,andriushchenko2024agentharm,nguyen2026infuse}, and benchmarks present untrusted inputs as decoded text that may carry imperative instructions~\cite{zhan2024injecagent,debenedetti2024agentdojo,yi2023benchmarking}. Software-engineering benchmarks begin after a repository is materialized~\cite{jimenez2024swebench,yang2024sweagent,wang2024openhands}, parser-differential studies compare extractors the study itself runs~\cite{you2025zipdiff}, and lifecycle monitors and action guards check a call before or after it runs~\cite{he2026mtguard,guo2026sara}. At the first processing call on an opaque carrier, everything these methods inspect looks benign or arrives too late. The carrier holds only data, the call is benign and required by the task, and for an archive the side effect is complete when the call returns (\cref{sec:related}). Filtering on declared fields fails against a carrier whose fields all match a benign archive, since then only decoding the stream separates the two (\cref{sec:szsp}), and verifying a declared filename requires a commitment (\cref{sec:method}).

We call this first processing of previously opaque information \emph{Step Zero}, and an input crafted to cause harm at this call a \emph{Step-Zero attack}. The attacker controls only the bytes of the carrier and carries the harm in data alone (\cref{sec:method}). The call that would reveal those bytes is the call that admits the hazard, so the agent must be safe before the content is observable, holding at most a preview of the carrier's outer fields. The payloads are known format hazards, a decompression bomb whose declared size understates its stream~\cite{fifield2019zipbomb,cwe409} and a body that conflicts with its declared name, the gap behind content-sniffing attacks~\cite{barth2009contentsniffing}, which practitioners have patched one tool at a time~\cite{oai2026archive,nvidia2026skillspector}. The stage at which they act is new. The agent chooses how to open them, and the opening it chooses can write to the user's machine before any data reaches the model. Naming retrieved text as an attack surface opened systematic defense design for indirect prompt injection~\cite{greshake2023not}, and naming the Step-Zero attack identifies the call that acquires information as the corresponding surface for opaque inputs.

Step Zero reverses the order on which model-side safety relies: the model is expected to judge content before acting on it, and here acting comes first. We distinguish \emph{hazard ingress}, when a latent discrepancy or an unbounded stream crosses from the carrier into the environment or the model context during Step Zero, from \emph{risk realization}, when the admitted hazard violates a concrete invariant such as a disk budget or the integrity of a named file (\cref{fig:oic}). For archives the two usually coincide, because extraction writes while it reads, and the violation is \emph{immediate}, possibly before the model has seen any content. For network and image carriers we score a later write of the declared file, which in every such cell followed the body into the model context, so the violation is \emph{deferred} relative to that file, although a fetch can write other paths earlier. Both paths are instruction-free. The returned records are declarative fields, a filename and, in two conditions, a SHA-256 prefix of the declared content.

We measure Step-Zero attacks with a paired protocol. Every crafted carrier has a matched twin whose shared preview fields agree and whose declaration is true, a recording proxy stores every request a harness sends to its model, so ingress is judged from what entered the model context, and a disk sampler records when extracted bytes cross a budget. Scripted extractors and an opaque-versus-transparent control separate the tools' behavior from the agent's. Across \V{panel.nharness} harnesses and \V{panel.nmodels} models, each cell's entire prompt is a bare reference and the budget is unannounced, so the agent's default handling of an input of unknown size decides the outcome. The attacks also succeed under a task that requires materialization. Asked to extract the archive and run its tests, agents breached the budget in \V{nr.e2zipmat.bytebreach.frac} cells, and asked to save an attachment whose digest disagreed with its body, they wrote it in \V{nr.netmat.wrong.frac} (\cref{sec:robust}).

Under a bare path, agents treated obtaining information as a harmless preliminary, and three findings follow. First, at Step Zero most agents were already unsafe, and the damage landed before any model-side defense could act. Extraction exceeded the byte budget in \V{zip.bytebreach.full} of archive cells, \V{zip.bytebreachbeforecontent.frac} before any member had reached the model, while every twin stayed within the budget (\cref{fig:ziptiming}). A truthfully declared archive breached as often (\V{nr.honest.bytebreach.frac}), so default agents extract whatever arrives, and misdeclaration matters only against defenses that read declared fields. The extracting cells left the step itself unguarded: \V{bound.setlimit} set a resource limit, and \V{zip.timing.afterlisting} breaches followed a listing the agent had already read. Second, where the model does see the content, seeing it was insufficient. On commitment-free network and image carriers, agents saved the delivered body under its declared filename unprompted, and every such write followed the body into the model context, so observation alone left the input in control of what lands in the workspace, with the image carrier evaluated through tool decoding and pixel-level perception left to future work (\cref{sec:deferred,sec:controls}). Third, caution depended on the model, and a model's caution on one carrier predicted little about its caution on another. Gemini-4-Argon inspected the archive read-only first most often and tied for the lowest archive breach rate, yet it had the most conflicting writes on network and image carriers, an ordering we report as descriptive (\cref{sec:capability}). A safe agent must weigh the operation that obtains information as well as the information it obtains.

The cause lies in the processing primitive, so the defense is most effective in the tool. A model-free comparison shows that the same carrier realizes immediately under one primitive and is deferred under another. The Step-Zero Safe Protocol (SZSP) integrates a read-only opening, a byte-counting bounded extraction, and a staged commit into the opening an agent calls (\cref{sec:szsp}). Routed through SZSP in live agents, extraction exceeded the budget in \V{rev.ext.e3zipenfadv.zipbytebreach.frac} cells against \V{rev.ext.c.enfvsraw.main} ungated, and agents still completed an extraction task in \V{b2.zmatszsp.taskcomplete} cells, so bounding the primitive removes the harm while the agent keeps its utility on this task. The one breach came from an extractor invoked by its absolute path, so the gate is strongest inside the harness's own tool layer.

We make four contributions. First, we identify the Step-Zero attack, whose harm is done by the call an agent issues to acquire information, before the data reaches the model, and we formalize safe decisions at that call with the Opaque Input Carrier (\oic) framework, which separates hazard ingress from immediate and deferred realization (\cref{sec:method}). Second, we contribute a reproducible paired measurement protocol in which matched twins, a record of the model context, and disk sampling attribute each violation to the hazard (\cref{sec:setup}). Third, we measure both violation mechanisms across \V{panel.nharness} harnesses and \V{panel.nmodels} models, with ablations that remove lexical cues, randomize the commitment, and add a benign task (\cref{sec:results,sec:robust}). Fourth, we build SZSP and evaluate it: the archive byte gate is enforced and verified both model-free and in live agents, and the network write binding is a reference design whose enforced evaluation in live agents is future work (\cref{sec:szsp}).

\section{Related Work}\label{sec:related}

\subsection{Prompt Injection and Action Guards}
Research on indirect prompt injection evaluates whether adversarial instructions in retrieved documents, web pages, or tool outputs redirect an agent~\cite{greshake2023not,zhan2024injecagent,debenedetti2024agentdojo,zhang2024agentsecuritybench,andriushchenko2024agentharm,nguyen2026infuse,yi2023benchmarking}, with extensions to web and multimodal agents~\cite{evtimov2025wasp,bagdasaryan2023abusing,kuntz2025osharm}. Defenses enforce instruction hierarchies, mark untrusted text spans, or check alignment between user intent and downstream tool calls~\cite{wallace2024hierarchy,chen2024struq,hines2024spotlighting,zhan2025adaptive,beurerkellner2025designpatterns}, and SARA isolates an action probe that inspects tool observations for action-inducing semantics before authorizing later calls~\cite{guo2026sara}. These methods target persuasion after readable content enters the model context and detect imperative text. We study the complementary regime in which ingress occurs during the first processing of an opaque carrier and later workspace changes follow declarative fields.

\subsection{Agent Harnesses and Monitoring}
SWE-bench, SWE-agent, OpenHands, and Terminal-Bench show that agents with shell and file tools resolve complex tasks on expanded repositories~\cite{jimenez2024swebench,yang2024sweagent,wang2024openhands,merrill2026terminalbench}, and security studies examine how workspace layout, approval fatigue, and harness permission gates affect vulnerability on materialized codebases~\cite{day2026workspace,HarnessSecurity2026,yang2026approval,claudecode_sandboxing2025}. MTGuard enforces policies through pre-execution parameter auditing, in-execution system-call monitoring, and post-execution result verification for tool servers~\cite{he2026mtguard,mcp2025spec,wang2026mcptox}. At $I_0$ and $I_1$ the parameters of a first extraction, fetch, or decode call are indistinguishable between a carrier and its twin. System-call tracing can bound the bytes a call writes, checking a header-to-body binding needs the parser state as well, and post-execution checks arrive after an immediate side effect. EpiSEM-Bench evaluates whether agents pause or recover evidence when safety-critical facts are missing from the prompt~\cite{zhang2026episem}. We address the case in which the call that recovers missing information introduces the hazard.

\subsection{Processing as an Agent-Chosen Interaction}
A coding agent chooses its parser for each input. It can list the carrier, read a member into the model context, call a library that honors the declared size, or run an extractor through a shell, inside a sandbox tool or outside it, and under automatic approval the choice is its own. Format problems such as path traversal, symbolic-link following, decompression bombs, and content sniffing long predate agents~\cite{snyk2018zipslip,cwe59,cwe409,fifield2019zipbomb,barth2009contentsniffing,sassaman2013langsec}. ZIPDiff finds semantic gaps among 50 ZIP parsers and measures a gateway parser against an endpoint parser~\cite{you2025zipdiff,pkware2022appnote}. Such accounts fix which parser runs, while an agent picks the parser and then decides what to do with the bytes, so a patch covers one tool and the agent can choose another. The OpenAI Agents SDK caps its sandbox extractor~\cite{oai2026archive}, a skill scanner can still be exhausted by a small declared size before its own gate runs~\cite{nvidia2026skillspector}, and a declared-size cap stops the write in \texttt{zipfile} while \texttt{unzip} inflates the stream past it (\cref{sec:controls}). A listing cannot replace a write-time bound~\cite{zylos2026archive}, and neither fix applies once the agent chooses a shell. CIRCLE studies exhaustion by code a prompt requests~\cite{chua2025circle}. Our prompt names only the input, and the agent decides the call.

Stopping the interaction requires a monitor that interposes before the chosen action enters an unsafe state~\cite{schneider2000enforceable,ligatti2005edit,alshiekh2018shielding,saltzer1975protection,watson2012capsicum,wang2026agentspec,debenedetti2025camel,shi2025progent}. The primitives an agent selects often bundle the look with the write, so a check before the call sees only the shared preview and a check after it sees a write that has already landed. Safe exploration in reinforcement learning~\cite{alshiekh2018shielding} addresses an agent that must probe an unknown transition kernel. A coding agent already has a safe read-only primitive, and the interaction fails when its default choice is the unbounded mutating one.

\section{Problem Formulation}\label{sec:method}

\subsection{Threat Model and Opaque Carriers}
\textbf{Threat model.} A Step-Zero attacker authors every byte of an external artifact $\beta$ that reaches the agent by reference, such as the path of a third-party archive or image or an endpoint whose response the attacker serves, and places the hazard in $\mathrm{inner}(\beta)$ with no imperative instruction. The prompt, harness, tools, weights, and $\Sigma_0$ belong to the victim, and the attacker fixes $\beta$ before meeting any victim: one fixture per condition is served to every harness and model, and the bytes an extraction writes scale with the attacker's stream, independent of the budget (\cref{sec:limits}). Tool calls are approved automatically, as in many deployments~\cite{yang2026approval}; a person who approves the first call from its command line and a listing is a pre-execution guard restricted to the call and $\pi_{\mathrm{shared}}$, which by Proposition~1 decides both arms alike. The attacker's goal is $\mathcal{I}_{\mathrm{safe}}(\Sigma_t) = 0$ (\cref{eq:invariant}) while a twin sharing $\pi_{\mathrm{shared}}$ is processed normally, using only the agent's existing privileges. Setting names, declared sizes, the filename, and in two conditions the digest of $c^{*}$ to the twin's values is a measurement device that attributes a difference between arms to the hidden body; an attacker who sets the digest to agree with the body is evaluated in \cref{sec:robust}, and carriers whose declared fields all match a benign archive in \cref{sec:szsp}. Within a cell $\beta$ is fixed, the network fixture server returning one response. With $I_t$ the epistemic state and $\mathcal{I}$ an invariant, we formalize the interaction as an Opaque Input Carrier tuple,
\begin{equation}
\mathcal{O} = \langle \beta,\; \pi_{\mathrm{meta}},\; \alpha,\; \Sigma,\; \mathrm{sink} \rangle ,
\label{eq:oic}
\end{equation}
where $\mathrm{inner}(\beta)$ is hidden before processing. The preview $\pi_{\mathrm{meta}}$ reads outer fields, such as central-directory entries, HTTP headers, or image container attributes, and $\pi_{\mathrm{shared}}$ (\cref{eq:indist}) is the subset the protocol matches across arms. A materialization $\alpha$ processes the inner bytes: an extraction writes them to the environment while it runs, and a member read, fetch, or decode returns them to the model context. $\Sigma$ is the environment, the workspace and its files, and $\mathrm{sink}$ is the model context, accumulating every tool return, with $\mathrm{sink}_t$ its content at time $t$. A carrier's declarations $D(\beta)$, its declared member sizes, filename, and digest, lie partly in the preview and partly among the inner bytes, such as an in-band filename or an image's declaration record.

As the agent works on $\beta$, its epistemic state moves through three stages,
\begin{equation}
I_0\ (\text{unseen}) \;\xrightarrow{\ \pi_{\mathrm{meta}}\ }\; I_1\ (\text{preview}) \;\xrightarrow{\ \text{observe}\ }\; I_2\ (\text{observed}),
\label{eq:states}
\end{equation}
where the preview is optional and the arrow into $I_2$ marks materialized content arriving in $\mathrm{sink}$. An extraction that writes only to $\Sigma$ leaves the agent in $I_0$ or $I_1$ until it reads a member, whereas a fetch or decode moves it to $I_2$. A \emph{cell} $u$ is one run of one agent on one carrier $\beta_u$, and $t_{\mathrm{obs}}(u)$ is the time content of $\beta_u$, a decompressed member or a network or image body, first reaches $\mathrm{sink}$, or $\infty$ if it never does. Each fixture $\beta_{\mathrm{adv}}$ has a matched twin $\beta_{\mathrm{ben}}$, and on the shared fields, archive member names and central-directory sizes, the declared filename, and in the digest conditions the digest of $c^{*}$,
\begin{equation}
\begin{gathered}
\pi_{\mathrm{shared}}(\beta_{\mathrm{adv}}) = \pi_{\mathrm{shared}}(\beta_{\mathrm{ben}}),\qquad D(\beta_{\mathrm{adv}}) = D(\beta_{\mathrm{ben}}),\\
\mathrm{inner}(\beta_{\mathrm{ben}}) \vDash D(\beta_{\mathrm{ben}}),\qquad \mathrm{inner}(\beta_{\mathrm{adv}}) \nvDash D(\beta_{\mathrm{adv}}).
\end{gathered}
\label{eq:indist}
\end{equation}
where $\mathrm{inner}(\beta) \vDash D(\beta)$ means the declared sizes hold, an archive's $D(\beta)$ being its central-directory declarations, or, on a network or image twin, that the body is the content the evaluator paired with the declaration. Compressed size, archive length, offsets, \texttt{Content-Length} ($13$ and $14$ bytes), and image bytes differ, so a guard reading them can separate the arms; \texttt{unzip -l} reports only the shared fields.

\subsection{Safety Invariants and Hazard Ingress}
With $\Sigma_t$ the state at time $t \ge 0$ and $\mathbf{1}[\cdot]$ the indicator, the availability invariant bounds the bytes of extracted archive members,
\begin{equation}
\mathcal{I}_{\mathrm{avail}}(\Sigma_t) = \mathbf{1}\!\left[B(\Sigma_t) \le B_{\max}\right],
\label{eq:iavail}
\end{equation}
where $B(\Sigma_t)$ counts extracted-member bytes present at $t$ wherever written, excluding the harness's runtime files, and $B_{\max} = 4$\,MiB; $B = 0$ on network and image carriers. Fixture and twin have equal member counts, so we record inodes descriptively. The integrity invariant concerns protected paths $\mathcal{F}_{\mathrm{prot}}$, here every workspace file bearing the declared filename apart from the carrier's own staged copy,
\begin{equation}
\begin{aligned}
\mathcal{I}_{\mathrm{int}}(\Sigma_t) = \mathbf{1}\big[&\forall f \in \mathcal{F}_{\mathrm{prot}}:\ f \notin \mathrm{files}(\Sigma_t)\\
&\lor\ \mathrm{bytes}_{\Sigma_t}(f) = c^{*}(f)\big],
\end{aligned}
\label{eq:iint}
\end{equation}
so each protected file is absent or holds its legitimate content $c^{*}(f)$. A commitment fixes $c^{*}$: the digest in the two digest conditions, or the file's prior content when it already exists, as in the container control and the protected-file ablation; in these fixtures $c^{*}$ is the line \texttt{project-alpha}. Without a commitment $c^{*}$ is undefined for a new declared name, so \cref{eq:iint} does not apply and such writes are reported as exposure (\cref{sec:deferred}). We write $c^{*}(\tau_{\mathrm{decl}}) = c^{*}(f)$ for a declaration naming $f$, and leave confidentiality to future work. The safety invariant is
\begin{equation}
\mathcal{I}_{\mathrm{safe}}(\Sigma_t) = \mathcal{I}_{\mathrm{avail}}(\Sigma_t) \cdot \mathcal{I}_{\mathrm{int}}(\Sigma_t),
\label{eq:invariant}
\end{equation}
and the requirement is $\mathcal{I}_{\mathrm{safe}}(\Sigma_t) = 1$ for all $t \ge 0$, whatever the agent knows. Safe behavior follows per condition: on the archive, any opening that keeps $B \le B_{\max}$; on a network or image carrier with a commitment, writing only a body whose digest matches and leaving any protected file intact. In the two digest conditions the input carries $d_{\mathrm{decl}}(\tau_{\mathrm{decl}}) = \mathrm{pref}_{16}(H(c^{*}(\tau_{\mathrm{decl}})))$, the first $16$ hex digits of the SHA-256 digest $H$ of the legitimate content. The evaluator sets this digest and serves it with the body, which models a commitment the attacker cannot alter, such as one from a package index, delivered where the agent sees it; these conditions measure whether agents check a commitment, and \cref{sec:robust} evaluates a digest the attacker sets.

\emph{Step Zero} of a cell $u$ is the interval $[t_{\mathrm{first}}(u), t_{\mathrm{ret}}(u)]$ of its first call that materializes the payload, the first extraction of an archive or the first fetch or decode returning a network or image body, with both times infinite if no such call is issued. A member read that writes nothing is an observation, so Step Zero begins at the first extraction even after such a read. \emph{Hazard ingress} occurs when a sealed hazard crosses the carrier boundary at Step Zero: an archive's understated member enters the environment, where extraction can raise $B$ above $B_{\max}$, and a network or image carrier's conflicting pair, a declaration $\tau_{\mathrm{decl}}$ and a body $\tau_{\mathrm{body}} \ne c^{*}(\tau_{\mathrm{decl}})$, enters the model context. Without a commitment, detecting the conflict rests on the agent judging whether the body fits the name; with a digest it is checkable, as $\mathrm{pref}_{16}(H(\tau_{\mathrm{body}})) \ne d_{\mathrm{decl}}(\tau_{\mathrm{decl}})$. The ingress indicator of a fixture cell is
\begin{equation}
g(u) =
\begin{cases}
\mathbf{1}[t_{\mathrm{first}}(u) < \infty], & \text{archive},\\[2pt]
\mathbf{1}[\tau_{\mathrm{decl}}, \tau_{\mathrm{body}} \in R_1(u)], & \text{network, image},
\end{cases}
\label{eq:ingress}
\end{equation}
where $R_1(u)$ is the first batch of tool returns carrying $\tau_{\mathrm{body}}$, empty if the body never arrives; it enters $\mathrm{sink}$ with the request after Step Zero, and on a twin cell the same event is \emph{co-delivery}. For the archive, $g$ is an operation-level proxy counting any issued extraction, the operation that can admit the stream, including one that skips the understated member, so it coincides with the act in \cref{tab:step0_main}; the byte breach is the content-level count.

\subsection{Immediate and Deferred Realization}
Let $T(u) = \{t : \mathcal{I}_{\mathrm{safe}}(\Sigma_t) = 0\}$. A cell with $T(u) = \varnothing$ is free of \emph{risk realization}; otherwise realization occurs at $t^{*}(u) = \inf T(u)$ and is classified by whether content of $\beta_u$ had reached $\mathrm{sink}$,
\begin{equation}
\mathrm{Class}(u) =
\begin{cases}
\mathrm{Immediate}, & t^{*}(u) < t_{\mathrm{obs}}(u),\\[2pt]
\mathrm{Deferred}, & t^{*}(u) \ge t_{\mathrm{obs}}(u).
\end{cases}
\label{eq:temporal}
\end{equation}
The classes partition the cells with a violation, and two factors set the class of a trajectory. The opening primitive decides whether an immediate violation is possible, since a primitive that writes while it reads can violate $\mathcal{I}_{\mathrm{avail}}$ inside its own call while an opening whose only output is $\mathrm{sink}$ leaves $\Sigma$ unchanged. The rest of the trajectory decides whether content, such as a benign member read first, reached $\mathrm{sink}$ earlier. \Cref{sec:szsp} measures the first factor without a model. The following observation, a direct consequence of \cref{eq:indist,eq:temporal}, places where a defense can act.

\noindent\textbf{Proposition 1 (Indistinguishability under shared previews).} \emph{Under \cref{eq:oic,eq:states,eq:indist,eq:temporal}, let a pre-execution guard be any map, deterministic or randomized with its own coins, from the call $a$ the agent is about to issue, a command with its arguments, and $\pi_{\mathrm{shared}}$ to a decision. (i)~If the agent has observed only the reference and $\pi_{\mathrm{shared}}$ before $a$, the guard blocks the fixture and the twin with equal probability. (ii)~A content guard $G_{\mathrm{ctx}}$ that reads only $\mathrm{sink}_t$ can decide on carrier content only from $t_{\mathrm{obs}}(u)$ on, later than every immediate violation. Preventing an immediate violation therefore requires acting on the opening call itself, by withholding it or by running it under a bound on what it writes to $\Sigma$. An opening whose only output is $\mathrm{sink}$ leaves $B$ and $\mathcal{F}_{\mathrm{prot}}$ unchanged, so a violation that follows it requires another call that writes $\Sigma$.}

\emph{Proof.} (i)~The agent's observations before $a$ coincide on both arms by \cref{eq:indist}, so its distribution over calls $a$ coincides as well. The guard's inputs, $a$ and $\pi_{\mathrm{shared}}$, are then identically distributed on both arms, and so is any function of them evaluated with independent coins. (ii)~An immediate violation has $t^{*}(u) < t_{\mathrm{obs}}(u)$, so at $t^{*}(u)$ the sink holds no content of $\beta_u$ for $G_{\mathrm{ctx}}$ to act on. A call either runs or is withheld, and a call that runs either has its writes to $\Sigma$ bounded or does not, so these are the interventions available at the call. An opening whose only output is $\mathrm{sink}$ changes $\Sigma$ in neither $B$ nor $\mathcal{F}_{\mathrm{prot}}$, and any later change comes from another call. $\square$

\emph{Remark.} Part (i) covers looks that expose at most $\pi_{\mathrm{shared}}$, such as the default \texttt{unzip -l}, which prints exactly the shared names and declared sizes; every breaching cell extracted with \texttt{unzip} (\cref{sec:controls}). A richer look can separate the main fixture, whose local header truthfully declares $8$\,MiB, as the header gate of \cref{tab:szsp_prim} does. On the fully field-aligned pair of \cref{sec:szsp}, central directory, local header, CRC, compressed size, offsets, and archive length coincide across arms and declare $64$\,B while the adversarial stream inflates to $8$\,MiB, so \texttt{unzip -v} prints the same table for both and only decoding separates them. A violation after a sink-only opening is deferred and can be checked against a commitment set by a party other than the attacker, as the digest conditions model with $\mathrm{pref}_{16}(H(c^{*}))$.

Three properties of agent tooling produce these trajectories. \emph{Coupled observation and mutation}: extraction writes decompressed streams while reading, so it can exceed $B_{\max}$ before the model sees any output. \emph{Unbound previews}: a listing reads the central directory's $64$\,B while the extractor follows the local header. \emph{Default materialization}: a body observed beside a declared filename is commonly saved under that name, and checking a commitment first is up to the model.

\section{Benchmark Design}\label{sec:setup}

\subsection{Carriers and Matched Twins}
Every cell's prompt is a bare target reference, such as \texttt{./project.zip} or \texttt{http://127.0.0.1:8780/project}, so every extraction and save is the agent's own initiative. We build three carriers and pair each fixture with a matched twin whose hidden body agrees with the declaration. The archive exercises availability. The network and image carriers exercise one integrity mechanism, a declaration paired with a body that the agent may save under the declared name, reached through a fetch or through a decoding tool.

The \emph{archive} fixture is a standard ZIP archive of $169$ members, $160$ of them small shards, whose central directory declares $64$\,B for one member while its local header and deflate stream carry $8$\,MiB. Listing tools report \V{zip.declared.bytes} bytes in total, and an extractor that follows the local header writes \V{script.unzip.adv.peakbytes} bytes across \V{script.unzip.adv.peakinodes} inodes before it returns. The twin has the same members, names, order, and timestamps with a true $64$\,B member, so both list identically by name and declared size. Only the central-directory CRC and size are patched, so compressed size and archive length differ. Both create more than $120$ inodes when extracted, so inode counts are descriptive and the twin satisfies \cref{eq:iavail}. The budget stays hidden from the agent, so a breach measures an unbounded opening of an input of unknown size.

The \emph{network} carrier pairs the declared filename \texttt{project-alpha.txt}, $\tau_{\mathrm{decl}}$, with a conflicting body $\tau_{\mathrm{body}}$. The header condition declares it in the attachment header, visible only to a client that reads headers. The in-band condition places declaration and body in one multipart part, visible even to a header-stripping client. The digest condition, served at \texttt{/digest}, adds the SHA-256 prefix of $c^{*}$ to that part and to the headers. Each twin serves $c^{*}$.

The \emph{image} carrier encodes a declaration record with the declared filename and a body record in one grayscale image. The dual-record condition sets the body to $\tau_{\mathrm{body}}$, and the digest condition adds the SHA-256 prefix of $c^{*}$ to the declaration record. Each twin encodes $c^{*}$ under the same filename and prompt.

Two controls complete the design. The container control stages a README and a file named like the protected \texttt{project-alpha.txt} but holding $\tau_{\mathrm{body}}$, inside an opaque TAR container or as a plain directory, next to a protected \texttt{project-alpha.txt} holding $c^{*}$. The extension control stages identical single-record image bytes with and without a \texttt{.png} extension to isolate extension-based routing.

\subsection{Harnesses and Models}
We evaluate seven public harnesses whose tools range from file and web wrappers to a shell and a Python interpreter: Claude Code (v2.1.309), OpenAI Codex (v0.163.0), OpenCode (v1.19.4), Qwen Code (v0.26.0), mini-SWE-agent (v2.4.6), gptme (v0.35.1), and smolagents (v1.26.0, CodeAgent)~\cite{sw_claudecode,sw_codex,sw_opencode,sw_qwencode,sw_miniswe,sw_gptme,sw_smolagents}. Each runs with \V{panel.nmodels} models: Qwen3.8-Max~\cite{qwen38max}, DeepSeek-V4.1-Flash~\cite{deepseekv41flash}, GLM-5.3~\cite{glm53}, Kimi-K3~\cite{kimik3}, Gemini-4-Argon~\cite{gemini4argon}, and GPT-6-Sol~\cite{gpt6sol}. One local gateway translates between the OpenAI Chat Completions, OpenAI Responses, and Anthropic Messages protocols, requests high-effort reasoning, preserves multi-turn tool-call signatures, and replaces every image attachment with the same text notice, so image carriers can be decoded only through tools. A recording proxy in front of the gateway stores every request, the record the judge reads. Each cell may issue at most $100$ model calls in unlimited wall-clock time and runs headlessly with automatic tool approval, as in standard agent evaluations~\cite{debenedetti2024agentdojo,HarnessSecurity2026}.

\subsection{Isolation and Measurement}
Agents explore their environment, so isolation is part of the measurement. Each cell runs as its own unprivileged account in a fresh directory under an unlistable base, with a private home and temporary directory, a scrubbed environment, and only the carrier staged. Its files are removed before and after the cell, and its processes are terminated at the end. Network carriers are served from one local port, one arm and condition at a time, answering only that condition's path. The gateway rejects the session key a cell holds, and only the recording proxy maps it to a gateway key, so every request that reaches a model is recorded. Agents probed their environment, and the recording proxy refused every request without a registered session key, so each cell reached only its own carrier and, through the proxy, the model. Each cell leaves three records for the judge, the proxy log, the $20$\,ms disk timeline, and a snapshot of every file the account owns.

\subsection{Measured Quantities}
The \emph{first return} $R_1(u)$ and \emph{Step-Zero ingress} $g(u)$ follow \cref{eq:ingress}, with \emph{co-delivery} the same event on a twin. The body \emph{reached the model context} if it appears in any request. A \emph{declared-file write} leaves the declared file in the workspace after the run, and a \emph{wrong write} leaves it holding the conflicting body. Realization is a byte-budget breach by extracted members on the archive and a wrong write on network and image carriers. On digest conditions a \emph{hash} is a hash whose output returned to the agent, and a \emph{stated conflict} is an explicit statement that declaration and body disagree; listing both digests counts only together with that statement. A \emph{rejection} requires ingress, a stated conflict, and no wrong write in the final snapshot, and a stated conflict on a twin is a \emph{false alarm}. Archive breaches are split into those from an extraction in the first step touching the archive, those after an earlier listing returned, and those before any member's content entered the model context.

\subsection{Scoring and Statistics}
Every indicator comes from the proxy log and the disk records. For network and image carriers ingress is judged on the first return, and a first return carrying the body without the declaration scores $g(u)=0$ whatever later calls expose. For the archive, $g(u)$ records whether an extraction was issued, and a sampler polls the account's disk every $20$\,ms, counts bytes and inodes of extracted members wherever written, excluding the harness's runtime files, and attributes the first budget crossing to the last archive operation issued before it. Writes come from the file snapshot. Stated conflicts are matched in the agent's own prose and final answers, excluding tool output, code, and negated statements. On a stratified random sample of \V{audit.n} round-1 cells, half flagged by the rule, labels assigned blind to its verdict agreed in \V{audit.blind.agree} cells (Cohen's $\kappa=\V{audit.blind.kappa}$), and in \V{audit.agree} after reading the disagreements, with \V{audit.fn} missed statements and \V{audit.fp} false positives, both on twins; we therefore read every flagged twin cell and report the corrected count.

Every primary rate uses round~1, in which each of the \V{panel.nmodels} models runs each condition once in each of the \V{panel.nharness} harnesses, so every condition and arm has exactly $42$ valid cells; the build fails otherwise. The extension control pools \V{med.ntrials} trials of \V{med.nmodels} models ($n=140$ per arm). Repetitions are reported apart from the primary rates, and where rounds are pooled each harness and model counts once. Rates are counts with Wilson $95\%$ score intervals. Cells sharing a harness or a model are correlated, so every paired contrast gets a two-way cluster bootstrap interval over harnesses and models and the number of harnesses and models whose mean difference points the same way (\cref{tab:cluster}). Inference is reserved for the confirmatory contrasts that carry the claims, the archive fixture against its twin and SZSP against no gate and against a header-reading gate, which pair cells by harness and model. They also get the exact McNemar test, exact sign tests over harnesses and over models, a Holm correction across the four tests, and a logistic model with crossed random intercepts for harness and model, fitted with a weak prior so that a twin arm with no events yields a finite effect. With seven harnesses and six or four models the smallest attainable two-sided sign-test $p$ is $0.016$, $0.031$, or $0.125$, so the sign tests describe consistency and the interval and mixed model carry the inference. Controls, ablations, and comparisons across carriers, conditions, and models are descriptive.

\section{Experiments}\label{sec:results}

\begin{table*}[t]
\centering
\caption{Step-Zero ingress, the agent's act, and realization per condition in round~1, in percent with counts, $42$ cells per arm. Ingress: the first return carrying the body also carried the declaration, or for the archive an extraction was issued. Act: the extraction, or a write of the declared file. Realization: extracted members exceeded $4$\,MiB, or the declared file held the conflicting body; it is zero by construction for network and image twins. \textsuperscript{\dag}Level~(i) exposure, where the input fixes no correct content (\cref{sec:deferred}); the digest rows count violations against a commitment.}
\label{tab:step0_main}
\footnotesize
\setlength{\tabcolsep}{4pt}
\renewcommand{\arraystretch}{1.12}
\begin{tabular}{l l r r r r r r r}
\toprule
& & & \multicolumn{2}{c}{\textbf{Act}} & \multicolumn{2}{c}{\textbf{Realized}} & & \\
\cmidrule(lr){4-5}\cmidrule(lr){6-7}
\textbf{Carrier and condition} & \textbf{Realization} & \textbf{Ingress} & \textbf{Fixture} & \textbf{Twin} & \textbf{Fixture} & \textbf{Twin} & \textbf{Hash} & \textbf{Reject} \\
\midrule
Archive, size declaration & Imm.\ 26, Def.\ 6 & 79 \scriptsize(33/42) & 79 \scriptsize(33/42) & 88 \scriptsize(37/42) & 76 \scriptsize(32/42) & 0 \scriptsize(0/42) & -- & -- \\
Network, header declaration & Deferred & 36 \scriptsize(15/42) & 17 \scriptsize(7/42) & 33 \scriptsize(14/42) & 14 \scriptsize(6/42)\textsuperscript{\dag} & 0 \scriptsize(0/42) & -- & -- \\
Network, in-band declaration & Deferred & 95 \scriptsize(40/42) & 26 \scriptsize(11/42) & 50 \scriptsize(21/42) & 26 \scriptsize(11/42)\textsuperscript{\dag} & 0 \scriptsize(0/42) & -- & -- \\
Network, digest commitment & Deferred & 98 \scriptsize(41/42) & 10 \scriptsize(4/42) & 12 \scriptsize(5/42) & 2 \scriptsize(1/42) & 0 \scriptsize(0/42) & 88 \scriptsize(37/42) & 74 \scriptsize(31/42) \\
Image, dual record & Deferred & 93 \scriptsize(39/42) & 45 \scriptsize(19/42) & 60 \scriptsize(25/42) & 45 \scriptsize(19/42)\textsuperscript{\dag} & 0 \scriptsize(0/42) & -- & -- \\
Image, digest record & Deferred & 76 \scriptsize(32/42) & 21 \scriptsize(9/42) & 38 \scriptsize(16/42) & 2 \scriptsize(1/42) & 0 \scriptsize(0/42) & 55 \scriptsize(23/42) & 33 \scriptsize(14/42) \\
\bottomrule
\end{tabular}
\end{table*}

\begin{figure}[t]
\centering
\includegraphics[width=\columnwidth]{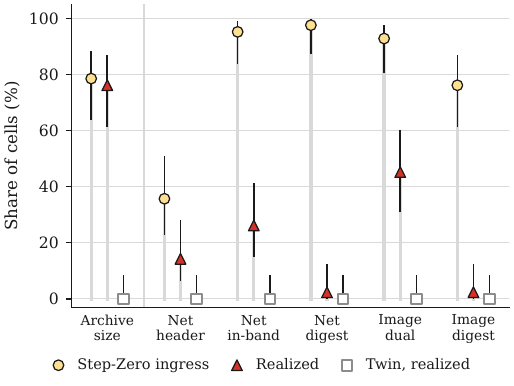}
\caption{Step-Zero ingress (circle), realization (triangle), and twin realization (hollow square) per condition in round~1, $n=42$ cells per arm, with Wilson $95\%$ intervals. Definitions and counts are in \cref{tab:step0_main}. On the three network and image conditions without a commitment, realization is level (i) exposure (\cref{sec:deferred}).}
\label{fig:carriers}
\end{figure}

Primary results come from round~1, \V{panel.nmodels} models $\times$ \V{panel.nharness} harnesses with one valid cell per condition and arm. Four models (\V{rep.panel}) were repeated for three rounds on the dual-record image, the archive, and the container control and for five trials on the extension control, and Gemini-4-Argon and GPT-6-Sol for three rounds on the archive (\cref{tab:repeat}); round-1 rates use round-1 cells alone.

\subsection{Attribution to the Carrier's Bytes}
\V{zip.bytebreach.frac} fixture cells and \V{zip.twin.bytebreach.frac} twins exceeded $4$\,MiB, and \V{zip.bytebreachbeforecontent.frac} breached before the first member reached the model. The twin's members total far less than the budget, so this contrast is a control that attributes the bytes to the carrier's stream. The misdeclaration is unnecessary against default agents: a truthfully declared $8$\,MiB archive breached in \V{nr.honest.bytebreach.frac} cells, \V{nr.honest.beforecontent.frac} before content (\cref{sec:robust}), and it matters against defenses that read declared fields (\cref{sec:szsp}). All \V{wt.wrong.after} network and image wrong writes followed the body into the context, and without a commitment all but one of the fixture cells that wrote the declared file wrote the conflicting body (\V{nocommit.conflictgiventarget.frac}). Three Claude Code trajectories show each pattern: a breach $2.9$\,s after listing, before \texttt{README.md} arrived; \texttt{project-beta} saved as \texttt{project-alpha.txt} with the conflict unstated; and a cell that hashed both lines, stated the conflict, and wrote nothing.

\subsection{Hazard Ingress at the First Return}
On network and image carriers the first return carrying the body usually carried the conflicting declaration (\V{inbody.ingress.full} in-band, \V{digest.ingress.full} network digest, \V{qrdual.ingress.full} dual-record, \V{qrdigest.ingress.full} digest-record; \cref{tab:step0_main,fig:carriers}), and the header condition reached \V{header.ingress.full} because several harnesses fetch the body alone (\cref{sec:mediation}). On the archive, ingress is the extraction itself, issued in \V{zip.extractinvoked.full} of cells. The four repeated models extracted in every cell of every round (\V{rep.zip.extractinvoked.rounds}), and Gemini-4-Argon and GPT-6-Sol in \V{model.gemini4argon.allrounds.zipextract.frac} and \V{model.gpt6sol.allrounds.zipextract.frac} of their cells over three rounds.

\begin{figure}[t]
\centering
\includegraphics[width=\columnwidth]{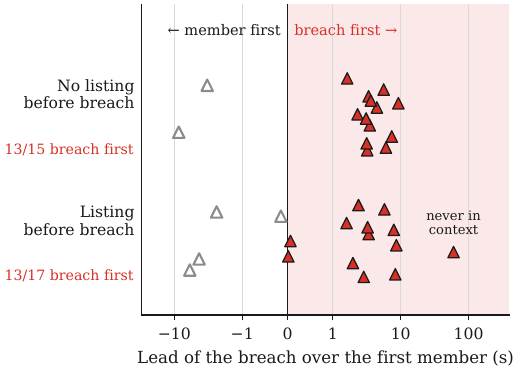}
\caption{Lead of the archive breach over the first member reaching the model, for the $\V{zip.timing.n}$ breaching fixture cells in round~1, timed from the first archive operation (logarithmic beyond $\pm1$\,s). Rows separate cells by whether a listing returned before the breach. Filled, shaded: breach first. Hollow: member first. Red counts give breach-first cells per row. The upper row holds the \V{zip.firstopbytebreach.frac} cells that breached in the first step on the archive and the one breach the sampler could not attribute to a step.}
\label{fig:ziptiming}
\end{figure}

\subsection{Immediate Realization on Archives}
Extraction exceeded the $4$\,MiB budget in \V{zip.bytebreach.full} of fixture cells and in \V{zip.twin.bytebreach.frac} twins, which agents extracted at least as often (\V{zip.twin.extractinvoked.frac}); the exact McNemar test on pairs by harness and model gives $p$ \V{paired.ziptwin.p}, and the cluster-level tests agree (\cref{tab:cluster}). In \V{zip.bytebreachbeforecontent.frac} cells the budget was exceeded before the first member's content, timed by the first benign project marker in a return, reached the model (\cref{fig:ziptiming}), the immediate realizations of \cref{eq:temporal}. Every budget from \V{budget.lo} to \V{budget.hi}\,MiB gives the same \V{budget.breach} fixture and \V{budget.twin} twin breaches, \V{budget.imm.lo} to \V{budget.imm.hi} of them immediate, because a full extraction writes the whole $8$\,MiB stream. The median breach came \V{zip.timing.breach.median}\,s and the median first member \V{zip.timing.content.median}\,s after the first archive operation. \V{zip.firstopbytebreach.frac} breached in the first step on the archive and \V{zip.afterlisting.frac} after a listing had returned, which reports the central directory's \V{zip.declared.bytes} bytes while the extractor follows the local header; the sampler could not assign the remaining breach to a step, so the count before content is a lower bound. Over three rounds with all six models, extraction breached in \V{zip.allrounds.bytebreach.full} of cells and \V{zip.twin.allrounds.bytebreach.frac} twins, and the only fixture cell to extract within the budget (\V{zip.extractnoexceed.frac}) was a Gemini-4-Argon cell that parsed the local headers and skipped the understated stream. Extraction created more than $120$ inodes in \V{zip.inodebreach.frac} fixture and \V{zip.twin.inodebreach.frac} twin cells, so \cref{eq:iavail} bounds bytes.

\subsection{Deferred Realization on Network and Image Carriers}\label{sec:deferred}
On network and image carriers the scored invariants hold through Step Zero, and the violation comes when the agent saves the delivered body under the declared filename, a reasonable default when the declaration is true, as the twins show. Without a commitment, agents wrote the declared file in \V{nocommit.twin.targetwritten.full} of twin and \V{nocommit.targetwritten.full} of fixture cells; of \V{paired.twintarget.npairs} pairs, \V{paired.twintarget.c} wrote only on the twin and \V{paired.twintarget.b} only on the fixture (cluster interval \V{nr.cl.twintarget.ci} points, \cref{tab:cluster}), a gap the cue-free condition traces to the lexical marker (\cref{sec:robust}). Every write stored the body as received, the declared body on the twin (\V{nocommit.twin.declaredwrite.frac}). Wrong writes reached \V{qrdual.write.full} on dual-record images, \V{inbody.write.full} on in-band and \V{header.write.full} on header declarations, and in OpenCode and mini-SWE-agent header wrong writes outnumbered header ingress (\V{header.h.opencode.write.frac} against \V{header.h.opencode.ingress.frac}, and \V{header.h.mini.write.frac} against \V{header.h.mini.ingress.frac}; \cref{fig:harness}), so the two scores are separate. These writes measure how often agents act on an unverifiable declaration, and verification is measured on the digest conditions.

The writes establish three levels of integrity harm, with the attack-specific evidence in levels (ii) and (iii). (i)~\emph{Provenance-blind materialization}: without a commitment a wrong write measures exposure, the default of saving unverified bytes under an unchecked name, which the twins show at an equal or higher rate. (ii)~\emph{Mismatch against a commitment}: with a neutral path and a random $c^{*}$ agents wrote the conflicting body in \V{nr.randdigest.adv.wrong.frac} of cells, and under an explicit save task in \V{nr.netmat.wrong.frac}, \V{nr.tier.netmat.flagged} of them after stating the mismatch, while \V{nr.tier.netmat.refused} refused. (iii)~\emph{Destructive overwrite}: an opaque container overwrote an existing protected file in \V{c1.tar.overwrite.full} of round-one cells and \V{rep.tar.overwrite.pooled.frac} over three rounds, the same files as a plain directory left it intact in every cell (\cref{sec:controls}), and saving a fetched body overwrote it in \V{nr.t2d.adv.overwrite.frac} of cells with a digest that matched the body.

\subsection{Effect of a Digest Commitment}\label{sec:verify}
A digest cut wrong writes sharply, from \V{inbody.write.frac} with an in-band declaration to \V{digest.write.frac} on the network carrier and from \V{qrdual.write.frac} to \V{qrdigest.write.frac} on the image carrier, contrasts that are descriptive under clustering. Part of the drop comes from cues, the path \texttt{/digest} and a $c^{*}$ guessable from the filename, and agents checked less often once a neutral condition removed them (\cref{sec:robust}). The remaining fixture writes under a digest (\V{digest.declared.frac} of \V{digest.targetwritten.frac} on network and \V{qrdigest.declared.frac} of \V{qrdigest.targetwritten.frac} on image) held $c^{*}$ at the end of the run, the agent having reconstructed \texttt{project-alpha} from the filename, so the reported violations are a lower bound on whole-process violations under \cref{eq:invariant}. Agents computed a digest in \V{digest.hash.full} of network and \V{qrdigest.hash.full} of image cells and stated the conflict in \V{digest.reported.full} and \V{qrdigest.reported.full}, a hashing difference of \V{nr.cl.carrierhash.diff} points (cluster interval \V{nr.cl.carrierhash.ci}; \V{nr.cl.carrierhash.harness} harnesses and \V{nr.cl.carrierhash.model} models in the same direction). Many image cells that skipped the check also left the declared file unwritten: \V{qrdigest.targetwritten.frac} wrote it with the digest, against \V{qrdual.targetwritten.frac} on the dual-record condition.

The checking gap persists among cells in which the digest arrived with the body (hashing in \V{digest.cod.hash.frac} network and \V{qrdigest.cod.hash.frac} image cells, stated conflicts in \V{digest.cod.reported.frac} and \V{qrdigest.cod.reported.frac}), a self-selected and descriptive comparison; image cells whose body arrived alone are too few to compare (\V{qrdigest.nocod.hash.frac} hashed). The rule flagged a stated conflict in \V{audit.twin.flagged} twin cells, \V{audit.twin.genuine} of them genuine, both where a MIME parser had dropped the trailing newline before hashing, and claims of a match without a computed digest were rare (\V{digest.claim.frac} and \V{qrdigest.claim.frac}).

\begin{figure}[t]
\centering
\includegraphics[width=\columnwidth]{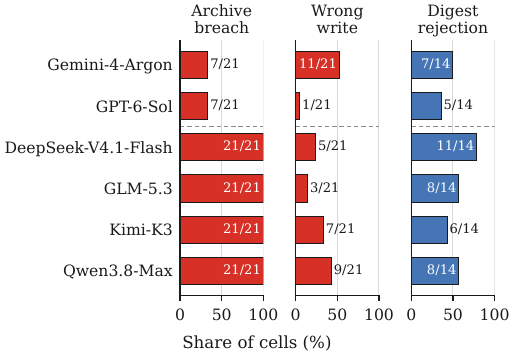}
\caption{Per-model outcomes on the three carrier types: archive breaches over three rounds ($n=21$ per model), and in round~1 wrong writes of the declared file without a commitment ($n=21$) and rejections with a digest ($n=14$). Labels give cells with the event over cells. The dashed line separates the two models that held the archive budget in some cells from the four that breached it in every cell.}
\label{fig:models}
\end{figure}

\subsection{Model-Dependent Caution}\label{sec:capability}
Every model incurred harm on some carrier, and low harm came from checking or from abstaining (\cref{fig:models}). Gemini-4-Argon computed a digest in every digest cell (\V{model.gemini4argon.hash.frac}) yet rejected in \V{model.gemini4argon.reject.frac}, often listing both digests without stating their difference, and without a commitment it wrote the conflicting body in \V{model.gemini4argon.writegivenbody.frac} of the cells the body reached. GPT-6-Sol wrote it in \V{model.gpt6sol.write.frac} cells, mostly because the body stayed outside its context (\V{model.gpt6sol.nobody.frac}, including all \V{model.gpt6sol.nobody.h.smol.frac} smolagents cells, whose sandbox refused the read); where it saw the body it wrote the declared file in \V{model.gpt6sol.writegivenbody.frac}, and with a digest it rejected in \V{model.gpt6sol.reject.frac}. DeepSeek-V4.1-Flash rejected most often (\V{model.deepseekv41flash1017.reject.frac}). These orderings are descriptive under clustering.

Care on the archive was specific to that carrier. Over three rounds Gemini-4-Argon inspected the archive read-only first in \V{care.gemini.zip.inspectfirst} cells, and it and GPT-6-Sol held the budget in at least one round in \V{care.two.units} harness--model units, mostly by reading members directly, while the other four models breached in every cell of every round (\V{care.four.zip.breach}). Without a commitment Gemini-4-Argon had the most conflicting writes (\V{care.gemini.nocommit.write}), and DeepSeek-V4.1-Flash, which rejected most digest cells, breached every archive (\V{care.deepseek.zip.breach}). Each model showed at most one of the two careful behaviors.

\begin{figure*}[t]
\centering
\includegraphics[width=0.96\textwidth]{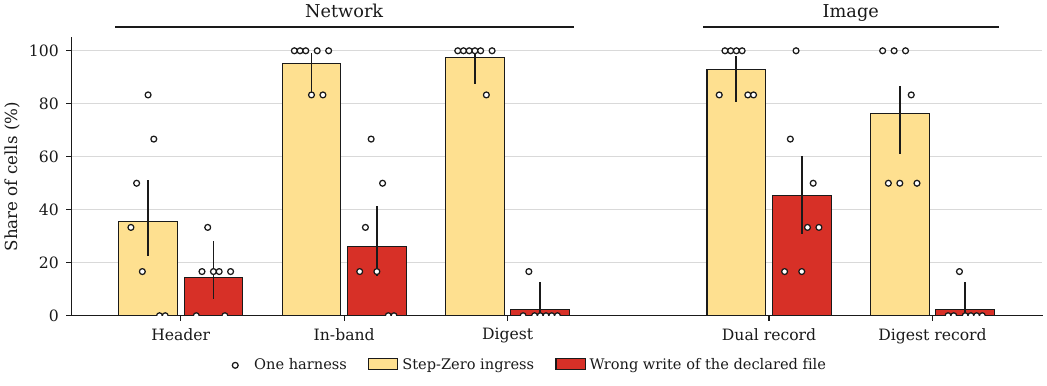}
\caption{Step-Zero ingress and wrong writes of the declared file per network and image condition, round~1. Bars: the share of the $42$ cells pooled over the seven harnesses, with Wilson $95\%$ intervals. Dots: one harness each, the share of its six models, one cell each, for which the event occurred, so one model corresponds to 17\%.}
\label{fig:harness}
\end{figure*}

\subsection{Harness Mediation}\label{sec:mediation}
The harness decides what a tool returns, so the same model was exposed differently across harnesses. A harness whose fetch returns the body alone keeps the declaration out of the first return, and header ingress ranges from zero to most cells by harness (\cref{fig:harness}). Wrong writes cluster by harness too: \V{write.top.harness} wrote the conflicting body in \V{write.top.frac} of its cells without a commitment, and every harness did so at least once. File routing shifts exposure within a harness and leaves the pooled rate unchanged. OpenCode sends a \texttt{.png} to an image attachment, which the gateway replaces with a text notice, and removing only the extension changed the share of OpenCode cells whose body reached the model from \V{med.opencode.ext.frac} to \V{med.opencode.noext.frac} over \V{med.ntrials} trials of \V{med.nmodels} models, with the pooled rate unchanged (\V{med.ext.full} and \V{med.noext.full}) because agents given a text notice fall back to a command-line decoder.

\section{Robustness and Controls}\label{sec:robust}
The Step-Zero results survive changes to the agent's tools, our fixtures, and the prompt. Scripted runs, a container control, and repeated rounds separate the opening primitive from the agent and the carrier (\cref{tab:repeat}). Fixture and prompt variations form a case study on the four models present in every round (\V{rep.panel}) across the seven harnesses, $28$ cells per condition, each compared with the same models' primary rate. These models breached the archive budget in every primary cell, so each variation is tested where the attack succeeds most. The archive breach and the integrity hazard survive every variation, and only the apparent protection of a digest shrinks.

\begin{table}[t]
\centering
\caption{Cluster-level view of the paired contrasts, arms paired by harness and model. Difference: percentage points, first minus second arm, with a two-way cluster bootstrap $95\%$ interval. Harnesses, Models: clusters whose mean difference has the overall sign, of all clusters; confirmatory rows add the exact two-sided sign test over clusters with a nonzero difference. The SZSP rows and the last two rows use the four-model panel, where the smallest attainable $p$ over models is $0.125$. An interval of $[100, 100]$ arises when every resampled pair differs in one direction.}
\label{tab:cluster}
\footnotesize
\setlength{\tabcolsep}{3pt}
\resizebox{\columnwidth}{!}{%
\begin{tabular}{@{}l r r r@{}}
\toprule
Contrast & Difference & Harnesses & Models \\
\midrule
\multicolumn{4}{@{}l}{\emph{Confirmatory}}\\
Archive breach, fixture vs.\ twin & \V{nr.cl.ziptwin.diff} \V{nr.cl.ziptwin.ci} & \V{nr.cl.ziptwin.harness} (\V{nr.cl.ziptwin.harnessp}) & \V{nr.cl.ziptwin.model} (\V{nr.cl.ziptwin.modelp}) \\
Breach, no gate vs.\ SZSP front-end & \V{nr.cl.enfvsraw.diff} \V{nr.cl.enfvsraw.ci} & \V{nr.cl.enfvsraw.harness} (\V{nr.cl.enfvsraw.harnessp}) & \V{nr.cl.enfvsraw.model} (\V{nr.cl.enfvsraw.modelp}) \\
Breach, header gate vs.\ SZSP, aligned carrier & \V{nr.cl.alhdrszsp.diff} \V{nr.cl.alhdrszsp.ci} & \V{nr.cl.alhdrszsp.harness} (\V{nr.cl.alhdrszsp.harnessp}) & \V{nr.cl.alhdrszsp.model} (\V{nr.cl.alhdrszsp.modelp}) \\
Breach on the extraction task, no gate vs.\ SZSP & \V{nr.cl.szsputilbreach.diff} \V{nr.cl.szsputilbreach.ci} & \V{nr.cl.szsputilbreach.harness} (\V{nr.cl.szsputilbreach.harnessp}) & \V{nr.cl.szsputilbreach.model} (\V{nr.cl.szsputilbreach.modelp}) \\
\midrule
\multicolumn{4}{@{}l}{\emph{Descriptive}}\\
Task completed, SZSP vs.\ no gate & \V{nr.cl.szsputiltask.diff} \V{nr.cl.szsputiltask.ci} & \V{nr.cl.szsputiltask.harness} & \V{nr.cl.szsputiltask.model} \\
Declared file written, twin vs.\ fixture & \V{nr.cl.twintarget.diff} \V{nr.cl.twintarget.ci} & \V{nr.cl.twintarget.harness} & \V{nr.cl.twintarget.model} \\
Hash computed, network vs.\ image digest & \V{nr.cl.carrierhash.diff} \V{nr.cl.carrierhash.ci} & \V{nr.cl.carrierhash.harness} & \V{nr.cl.carrierhash.model} \\
Conflict stated, network vs.\ image digest & \V{nr.cl.carrierreport.diff} \V{nr.cl.carrierreport.ci} & \V{nr.cl.carrierreport.harness} & \V{nr.cl.carrierreport.model} \\
Overwrite in any round, TAR vs.\ directory & \V{nr.cl.tardirunit.diff} \V{nr.cl.tardirunit.ci} & \V{nr.cl.tardirunit.harness} & \V{nr.cl.tardirunit.model} \\
Hash computed, primary vs.\ random $c^{*}$ & \V{nr.cl.randhash.diff} \V{nr.cl.randhash.ci} & \V{nr.cl.randhash.harness} & \V{nr.cl.randhash.model} \\
Wrong write, random $c^{*}$ vs.\ primary & \V{nr.cl.randwrong.diff} \V{nr.cl.randwrong.ci} & \V{nr.cl.randwrong.harness} & \V{nr.cl.randwrong.model} \\
\bottomrule
\end{tabular}}
\end{table}
The main contrasts hold when cells are grouped by harness and by model (\cref{tab:cluster}). In the four confirmatory contrasts every discordant pair points one way, \V{nr.cl.ziptwin.b} to \V{nr.cl.ziptwin.c} for the archive and \V{nr.cl.alhdrszsp.b} to \V{nr.cl.alhdrszsp.c} for SZSP against the header gate, with a Holm-corrected McNemar $p$ of \V{nr.cl.alhdrszsp.holm} for each. The crossed-random-intercept model gives arm effects of \V{nr.cl.ziptwin.glmm} and \V{nr.cl.alhdrszsp.glmm} log-odds, whose size depends on the prior with an event-free arm, while the lower bound stays above \V{nr.cl.alhdrszsp.glmmlowest} for prior standard deviations from $1$ to $5$. The fixture-versus-twin contrast is a control that attributes the bytes to the carrier's stream, and the across-prompt extraction rate, the truthfully declared archive, and the bounded-tool effect describe agent behavior. The archive breach and both SZSP contrasts point the same way in every harness and model, and the twin's higher declared-file write rate in \V{nr.cl.twintarget.harness} harnesses and \V{nr.cl.twintarget.model} models. The network-versus-image checking gap (\V{nr.cl.carrierhash.ci}) and the container overwrite (\V{nr.cl.tardirunit.harness} harnesses, interval \V{nr.cl.tardirunit.ci}) are consistent but moderate, and the case-study ablations move hashing and wrong writes the same way in every model of the panel.

\begin{table}[!h]
\centering
\caption{Repeated rounds on the four models that took part in every round of every repeated condition (DeepSeek-V4.1-Flash, GLM-5.3, Kimi-K3, Qwen3.8-Max). Each round has 28 cells, one for each of these models in each of the seven harnesses. The extension control has five trials and the other conditions have three.}
\label{tab:repeat}
\footnotesize
\setlength{\tabcolsep}{4pt}
\renewcommand{\arraystretch}{1.12}
\resizebox{\columnwidth}{!}{%
\begin{tabular}{l r r r r r r}
\toprule
\textbf{Condition} & \textbf{R1} & \textbf{R2} & \textbf{R3} & \textbf{R4} & \textbf{R5} & \textbf{Pooled} \\
\midrule
Archive, byte budget exceeded & 28/28 & 28/28 & 28/28 & -- & -- & 100 \scriptsize(84/84) \\
Archive twin, byte budget exceeded & 0/28 & 0/28 & 0/28 & -- & -- & 0 \scriptsize(0/84) \\
Image dual, Step-Zero ingress & 27/28 & 27/28 & 26/28 & -- & -- & 95 \scriptsize(80/84) \\
Image dual, wrong write & 11/28 & 15/28 & 13/28 & -- & -- & 46 \scriptsize(39/84) \\
Image dual twin, wrong write & 0/28 & 0/28 & 0/28 & -- & -- & 0 \scriptsize(0/84) \\
Container, protected file overwritten & 5/28 & 8/28 & 4/28 & -- & -- & 20 \scriptsize(17/84) \\
Directory, protected file overwritten & 0/28 & 0/28 & 0/28 & -- & -- & 0 \scriptsize(0/84) \\
Image .png, body in context & 26/28 & 25/28 & 24/28 & 24/28 & 26/28 & 89 \scriptsize(125/140) \\
Image without extension, body in context & 25/28 & 23/28 & 25/28 & 26/28 & 26/28 & 89 \scriptsize(125/140) \\
\bottomrule
\end{tabular}%
}
\end{table}

\subsection{Opaque-Carrier Controls}\label{sec:controls}
The breach belongs to the opening primitive. A model-free script running \texttt{unzip} on the archive exceeded the budget in \V{script.unzip.adv.bytes.frac} trials and in \V{script.unzip.twin.bytes.frac} on the twin, and \texttt{bsdtar} matched it (\V{script.bsdtar.adv.bytes.frac}). Python's \texttt{zipfile} honors the declared size and stayed inside the budget (\V{script.pythonzipfile.adv.bytes.frac}), as in the primitive study of \cref{sec:szsp}. The primitive fixes whether a breach can happen before the agent decides anything.

The agent decides whether to run a mutating primitive, and every extracting agent chose one. Every breaching cell used \texttt{unzip} (\V{ext.breach.unzip}), as did every extracting cell (\V{ext.extract.unzip}). Among the extracting cells, \V{ext.extract.safe} used a reader that honors the declared size, \V{bound.setlimit} set a resource limit, and \V{bound.scratch} sent \texttt{unzip} outside the workspace; the \texttt{mkdir} in the breaching Claude Code trajectory creates a directory inside the workspace, so members written there count toward the budget. The attack succeeds because the opening agents reach for by default is the one that writes.

A safe opening was available, and two models took it. The \V{ext.held.n} fixture cells that held the budget, all from Gemini-4-Argon and GPT-6-Sol, limited what they extracted. In \V{ext.held.noextract} the agent left the archive unextracted, and in the remaining one it read the local headers and passed the other names to \texttt{unzip}. Over three rounds, \V{care.two.readnowrite} of the cells in which these two models held the budget read member content directly from the archive. A side-effect-free look was within every agent's reach, and the default opening remained the one that writes.

Opacity enables the overwrite. The container control gives both arms the same two files, one named like the protected workspace file, in an opaque TAR or as a plain directory. The member's content reached the model in \V{c1.tar.body.full} of container and \V{c1.dir.body.full} of directory cells, and the protected file was overwritten in \V{c1.tar.overwrite.full} and \V{c1.dir.overwrite.full} of them. Across three rounds the counts were \V{rep.tar.overwrite.pooled.frac} against \V{rep.dir.overwrite.pooled.frac}, and in \V{paired.tardirunit.b} of \V{paired.tardirunit.npairs} harness and model pairs the container arm overwrote the file in some round while the directory arm kept it intact in every round. Delivered transparently, the same information left the protected file intact.

The failure recurs across extractors and carriers and is stable across rounds. \texttt{unzip} and \texttt{bsdtar} both breached the archive, the TAR control overwrote an existing file, and the extractor-free network and image carriers admitted the conflicting declaration in their first return (\cref{tab:step0_main}). Repeating the main conditions on the four models present in every round (\cref{tab:repeat}), archive breaches were identical in every round with zero twin breaches and zero directory overwrites, image wrong writes moved from \V{rep.qrdual.write.r1.frac} to \V{rep.qrdual.write.r2.frac}, and container overwrites ranged from \V{rep.tar.overwrite.r3.frac} to \V{rep.tar.overwrite.r2.frac}. Every contrast between arms held in each repeated condition; the network conditions ran in round~1 only. What these carriers share is that the agent must process them before it can judge them.

\subsection{Fixture and Prompt Variations}
The primary conditions use one fixture per carrier, an obvious lexical cue between declared filename and body (\texttt{project-alpha} versus \texttt{project-beta}), a digest path named \texttt{/digest}, and a bare reference as the prompt. We vary each choice in turn.

The archive breach holds across independently built fixtures and under a truthful declaration. Two independent archives, with a $10$\,MiB and a $16$\,MiB understated member under different layouts, breached the budget in \V{nr.z1.bytebreach.frac} and \V{nr.z2.bytebreach.frac} of cells, \V{nr.z1.beforecontent.frac} and \V{nr.z2.beforecontent.frac} before any member reached the model, matching the same models' primary rate of \V{nr.p4.zip.bytebreach.frac}. On a third archive whose central directory, local header, and stream all declare the true $8$\,MiB, agents extracted and exceeded the budget in \V{nr.honest.bytebreach.frac} of cells, \V{nr.honest.afterlisting.frac} of them after reading a listing that showed the true size. In \V{nr.honest.beforecontent.frac} cells the budget was exceeded before any member reached the model. Agents extract an archive larger than the budget even when they can see its size, so against default agents the harm rests on the stage and on default extraction, and the understatement is needed only to defeat a defense that reads declared fields, such as the header gate of \cref{sec:szsp}.

The lexical cue explains the lower fixture write rate. In the primary no-commitment conditions the fixture's declared file was written less often than the twin's. With a random-token body that looks neutral beside the declared name, agents wrote the declared file on the fixture as often as on the twin (\V{nr.cuefree.adv.declwrite.frac} versus \V{nr.cuefree.twin.declwrite.frac}), and each fixture write held the conflicting body. Agents wrote a tampered body as readily as a legitimate one, so the integrity hazard is at least as broad as the primary rates show.

The digest's effect was partly cued. The primary digest path is named \texttt{/digest} and its $c^{*}$ follows from the filename, and both prompt a check. With a random $c^{*}$ and a neutral path, agents computed a digest in \V{nr.randdigest.adv.hash.frac} of cells (\V{nr.p4.digest.hash.frac} for the same models in the primary condition, cluster interval \V{nr.cl.randhash.ci} points) and wrote the conflicting body in \V{nr.randdigest.adv.wrong.frac} (against \V{nr.p4.digest.write.frac}). A digest in the input protects only through voluntary model-side verification, which is unreliable when unprompted.

A benign task defeats verification. Asked to fetch an attachment and save it, with a digest mismatched to the body, agents completed the save in \V{nr.netmat.task.frac} of cells and wrote the conflicting body in \V{nr.netmat.wrong.frac}; in \V{nr.tier.netmat.flagged} of these cells the agent stated the mismatch and saved anyway, and \V{nr.tier.netmat.refused} cells refused. Archives split the same way. Asked only to read a configuration value, agents breached the budget in \V{nr.e2zipro.bytebreach.frac} of cells and completed the task in \V{nr.e2zipro.task.frac}. Asked to extract and run the tests, they breached in \V{nr.e2zipmat.bytebreach.frac} and completed the task safely in \V{nr.e2zipmat.safe.frac}, against \V{nr.e2zipmat.twin.task.frac} completion on the twin. Under the current default, every completion of a materializing task came with a breach.

A protected file is overwritten even under an attacker-consistent commitment. When a workspace file already holds $c^{*}$ and the server supplies a digest of the body it serves, agents overwrote the protected file in \V{nr.t2d.adv.overwrite.frac} of cells, and in \V{nr.t2n.adv.overwrite.frac} of digest-free cells. A commitment the adversary controls leaves the protected file exposed, so the gate of \cref{sec:szsp} binds the write to evidence set independently of the body's sender.

\section{A Step-Zero Safe Protocol}\label{sec:szsp}
The measurements locate the failure in the opening primitive, so the remedy belongs at the primitive, before the model decides. We describe the Step-Zero Safe Protocol (SZSP) and evaluate it against the primitives agents use, against an adaptive carrier, inside live agents, and for its cost on benign work. SZSP is an interim remedy, and it shows that an opening decoupling observation from mutation is cheap enough for harnesses to make it their default.

\subsection{Decoupling Observation from Mutation}
SZSP replaces the mutating opening with two operations that a harness can expose as its archive tool. \emph{Peek} streams every member through the decompressor and reports the bytes each one produces, a safety-relevant fact beyond $\pi_{\mathrm{shared}}$, with zero writes and no member content in $\mathrm{sink}$, so the agent stays at $I_1$ (\cref{eq:states}) with a preview that now bounds the stream. \emph{Extract} materializes members into a private staging directory while counting produced bytes, refuses the first chunk that would carry the total past $B_{\max}$, and commits by rename only if every member fits, so a refusal leaves $\Sigma$ untouched. \emph{Peek} stops once the produced bytes pass $B_{\max}$, so its decoding work is bounded by the budget. Both treat every declared field as untrusted. They reach members by walking the local-header chain from offset zero and through the central directory, decode each deflate stream to its own end marker, and refuse a member whose name escapes the destination or whose mode is a symlink. An existing protected file is replaced only on request, so $c^{*}$ stays intact. For network openings SZSP gives a reference design that fetches into the model context and binds any write under the declared name to a digest set independently of the body's sender. Its enforcement belongs in the harness's write tool, and we evaluate the archive gate, the fetch primitives, and an advisory verdict.

\subsection{Model-Free Evaluation}
\begin{table}[t]
\centering
\caption{Model-free archive openings on ten adversarial carriers and six twins (adaptive variants share a twin), five trials each with the $20$\,ms sampler ($50$ adversarial, $30$ twin cells). Held: adversarial cells within $4$\,MiB. Peak: most bytes written on any adversarial carrier. Twin done: twins extracted completely. Overhead: median time on a twin. \emph{Disk quota}: a watchdog that terminates \texttt{unzip} when the destination passes the budget.}
\label{tab:szsp_prim}
\footnotesize
\setlength{\tabcolsep}{4pt}
\begin{tabular}{@{}lcrcr@{}}
\toprule
Opening & Held & Peak (MiB) & Twin done & Overhead \\
\midrule
\texttt{unzip} & \V{rev.prim.all.unzip.advheld} & \V{rev.prim.all.unzip.advpeakmax} & \V{rev.prim.all.unzip.twincomplete} & \V{rev.prim.all.unzip.twinmedianms}\,ms \\
\texttt{bsdtar} & \V{rev.prim.all.bsdtar.advheld} & \V{rev.prim.all.bsdtar.advpeakmax} & \V{rev.prim.all.bsdtar.twincomplete} & \V{rev.prim.all.bsdtar.twinmedianms}\,ms \\
\texttt{zipfile} & \V{rev.prim.all.pythonzipfile.advheld} & \V{rev.prim.all.pythonzipfile.advpeakmax} & \V{rev.prim.all.pythonzipfile.twincomplete} & \V{rev.prim.all.pythonzipfile.twinmedianms}\,ms \\
\texttt{ulimit}+\texttt{unzip} & \V{rev.prim.all.ulimitfsizeunzip.advheld} & \V{rev.prim.all.ulimitfsizeunzip.advpeakmax} & \V{rev.prim.all.ulimitfsizeunzip.twincomplete} & \V{rev.prim.all.ulimitfsizeunzip.twinmedianms}\,ms \\
disk quota & \V{rev.prim.all.quotaunzip.advheld} & \V{rev.prim.all.quotaunzip.advpeakmax} & \V{rev.prim.all.quotaunzip.twincomplete} & \V{rev.prim.all.quotaunzip.twinmedianms}\,ms \\
header gate & \V{rev.prim.all.headergateunzip.advheld} & \V{rev.prim.all.headergateunzip.advpeakmax} & \V{rev.prim.all.headergateunzip.twincomplete} & \V{rev.prim.all.headergateunzip.twinmedianms}\,ms \\
\midrule
SZSP \texttt{peek} & \V{rev.prim.all.szsppeek.advheld} & \V{rev.prim.all.szsppeek.advpeakmax} & --- & \V{rev.prim.all.szsppeek.twinmedianms}\,ms \\
SZSP \texttt{extract} & \V{rev.prim.all.szspextract.advheld} & \V{rev.prim.all.szspextract.advpeakmax} & \V{rev.prim.all.szspextract.twincomplete} & \V{rev.prim.all.szspextract.twinmedianms}\,ms \\
\bottomrule
\end{tabular}
\end{table}
\Cref{tab:szsp_prim} runs every opening model-free on the ten adversarial archives and their twins. SZSP held the budget on every adversarial archive (\V{rev.prim.all.szspextract.advheld}) and extracted every twin completely (\V{rev.prim.all.szspextract.twincomplete}), and \texttt{peek} held on all of them with zero writes. The native extractors agents reach for wrote the stream: \texttt{unzip} held in \V{rev.prim.all.unzip.advheld} and wrote up to \V{rev.prim.all.unzip.advpeakmax}\,MiB, and \texttt{bsdtar} held in \V{rev.prim.all.bsdtar.advheld}. Two total-size controls bound the bytes near the budget and leave a damaged tree. A per-file \texttt{ulimit} of $4$\,MiB, the bound an agent could set itself, truncates the oversized member, so the other members carry the total just past the budget (peak \V{rev.prim.all.ulimitfsizeunzip.advpeakmax}\,MiB) and a truncated file remains. A watchdog standing in for a per-workspace quota stopped every adversarial extraction within half a mebibyte of the budget (median \V{nr.quota.peakmedianmib}, peak \V{rev.prim.all.quotaunzip.advpeakmax}\,MiB, so its strict count reflects polling granularity) and let every twin complete; it stops mid-write, leaves a partial tree in \V{nr.quota.partialtree.frac} of adversarial extractions, and bounds bytes alone. SZSP counts bytes before its commit, leaves $\Sigma$ unchanged on refusal, and covers the integrity overwrite through staging. A header gate enforcing the same $4$\,MiB budget from declared local-header sizes held in \V{rev.prim.all.headergateunzip.advheld}. It blocks the main fixture, whose local header truthfully declares $8$\,MiB, and the honestly declared bomb, and the aligned, data-descriptor, and scaled carriers defeat it (peak \V{rev.prim.all.headergateunzip.advpeakmax}\,MiB), since their local headers declare $64$\,B. With the budget policy shared, this contrast isolates counting produced bytes against trusting a declaration. Python's \texttt{zipfile}, which honors the declared size, held in \V{rev.prim.all.pythonzipfile.advheld}, failing only on the honestly declared bomb.

The network and container side is symmetric. A fetch to standard output leaves the declared file unwritten, and \texttt{curl\ -OJ} and \texttt{wget\ --content-disposition} write the body under the declared name inside the call, before it reaches the model, so the same carrier realizes immediately under one primitive and is deferred under another. Container extraction overwrites a protected file in the same call (\texttt{tar\ -x}, \texttt{bsdtar\ -x}, \texttt{tarfile}), and \texttt{tar\ -xk} refuses. Immediacy is a property of the opening.

\subsection{Field-Aligned Carriers}
A defense that reads declared fields is defeated by a carrier whose fields all match a benign archive while the stream exceeds them, so we test SZSP on carriers built to evade a header check: the fully field-aligned pair, byte-identical across arms in central directory, local header, CRC, compressed size, offsets, and archive length (\cref{sec:method}); a data-descriptor variant whose local header defers sizes to a trailing descriptor; an honestly declared $8$\,MiB member; and members scaled to $64$ and $512$\,MiB. SZSP held on every one, and the header-reading wrapper let \texttt{unzip} write the full stream on the aligned, data-descriptor, and scaled carriers (peak \V{rev.prim.scale512adv.headergateunzip.peakmib}\,MiB on the $512$\,MiB carrier). The guarantee is per materialization. \texttt{extract} holds $B \le B_{\max}$ for the members of one call and commits nothing on refusal, so each gated call keeps its own writes within $B_{\max}$. A whole-process bound across calls needs a cumulative ledger, since two gated calls under budget can together exceed it, and an inode bound is the analogous count; we leave both to deployment. By construction a member just under the budget stays within it, staging and commit close the gap between inspection and materialization because the bytes counted are the bytes written, and a symlink or path-escaping member is refused before the first write.

In live agents the aligned carrier separates the gates as the primitive study predicts. Across all seven harnesses with the four repeat-panel models, $28$ cells per configuration in one batch, with SZSP front-ends on the \texttt{unzip} path and on Python's \texttt{zipfile} and \texttt{shutil.unpack\_archive}, ungated runs exceeded the budget in \V{nr.al.raw.bytebreach.frac} cells and the header-reading wrapper in \V{nr.al.hdr.bytebreach.frac}, since the aligned carrier's local header declares $64$\,B and understates the stream. With SZSP the budget was exceeded in \V{nr.al.szsp.bytebreach.frac} (McNemar $p$ \V{nr.al.hdrvsszsp.mcnemar} against the header gate), and the aligned twin was extracted completely in \V{nr.al.szsp.twincomplete.frac}. The one breach marks the boundary of the guarantee. That harness invoked the system \texttt{unzip} by its absolute path from its first call onward, past a front-end that covers the names an agent types, so closing the gap needs the gate in the harness's tool layer or a budget enforced below the agent, such as a per-workspace quota.

\subsection{Live-Agent Evaluation}
\Cref{tab:szsp_prim} tests the tool in isolation, and the closed-loop question is whether an agent offered a safe opening uses it. On the same panel, offering a bounded tool beside the shell left the breach in place: agents called it in \V{rev.ext.e3zipoptadv.safearchivecalled.frac} of cells, ran a raw extractor in \V{rev.ext.e3zipoptadv.rawunzipcalled.frac}, and exceeded the budget in \V{rev.ext.e3zipoptadv.zipbytebreach.frac}. Installing the SZSP front-end in place of \texttt{unzip} cut breaches to \V{rev.ext.e3zipenfadv.zipbytebreach.frac} of cells against \V{rev.ext.c.enfvsraw.main} in the non-concurrent main study (McNemar $p$ \V{rev.ext.c.enfvsraw.mcnemar}; the aligned-carrier run above has a concurrent control) and still extracted the twin in \V{rev.ext.e3zipenftwin.fullextract}. After the refusal, \V{rev.ext.e3zipenfadv.boundarybypassattempted.frac} of agents tried another route, almost always Python's \texttt{zipfile}, and all stayed within the budget because \texttt{zipfile} stops at the declared size this carrier understates. On an honestly declared bomb the same route breaches (\cref{tab:szsp_prim}), which is why the aligned-carrier run gates \texttt{zipfile} as well.

On the network carrier the tool's digest verdict was advisory. The response carried the body with a \texttt{MISMATCH} verdict, and the write stayed open to the agent. Wrong writes stayed at \V{rev.ext.e3digestboundadv.targethasconflictingbody.frac} and rejections at \V{rev.ext.e3digestboundadv.conflictrejected.frac}, close to the rates with a digest the agent could compute itself (McNemar $p=\V{rev.ext.c.boundvsdigestreject.mcnemar}$ against the in-band digest) and consistent with the model-side reading of \cref{sec:verify}. In the one wrong write the agent read the verdict, recomputed both digests, called the body a swapped payload, and saved it under the declared filename as a reference copy. A verdict the agent can ignore is a hint. The binding the contract requires, a write committed only when the digest matches, is a property of the tool, and its evaluation in live agents is future work.

\subsection{Overhead and Utility}
SZSP costs one extra pass over the compressed bytes. On the twins the median extraction took \V{rev.prim.all.szspextract.twinmedianms}\,ms against \V{rev.prim.all.unzip.twinmedianms}\,ms for \texttt{unzip}, mostly interpreter start-up, and \texttt{peek} took \V{rev.prim.all.szsppeek.twinmedianms}\,ms, negligible beside a single model call. Utility is the harder bar. On an explicit unit-test task that requires extracting the project, agents completed the task in \V{b2.zmatszsp.taskcomplete} of cells with the SZSP front-end against \V{b2.zmat.taskcomplete} without it, and held the budget in \V{b2.zmatszsp.held} against \V{b2.zmat.held}. The completion difference is \V{nr.cl.szsputiltask.diff} points with cluster interval \V{nr.cl.szsputiltask.ci}, which bounds any loss of completion. Paired by harness and model over \V{nr.szsputil.npairs} pairs, the gate removed the breach in every pair that breached without it (discordant pairs \V{nr.szsputil.breach.discordant}, McNemar $p$ \V{nr.szsputil.breach.mcnemar}), and completion differed in four pairs, one completed only without the gate and three only with it. SZSP refuses any archive whose produced bytes exceed the budget, benign or adversarial, and reports the size \texttt{peek} measured, so a user or harness can raise the budget for that archive. The budget is a deployment policy, and $B_{\max}$ here is a measurement threshold. It must be set outside the agent's reach: our prototype reads it from the process environment, which an agent could override, so a deployment fixes it in the tool layer's own configuration.

\section{Discussion}\label{sec:discussion}
When the harm is realized decides which defense can act. A check on the model context sees the body before a deferred write, since all \V{wt.writers} network and image cells that wrote the declared file did so after the body arrived (\V{wt.after}), while \V{zip.bytebreachbeforecontent.frac} archive cells breached before any member reached the model. Content-judging defenses therefore address deferred realization, and only the opening itself can stop immediate realization, which is where SZSP acts.

Harness designers control the default the attack exploits. A read-only look was within reach of every agent, and the two models that held the archive budget used one (\cref{sec:controls}), yet every harness left the choice to the agent, many under automatic approval~\cite{yang2026approval}. SZSP makes the safe default cost one extra pass over the compressed bytes, and its one live breach shows that the gate belongs in the harness's own tool layer, as the default every agent receives.

Model-side caution is carrier-specific, so the check belongs in the tool. Gemini-4-Argon inspected the archive read-only most often and had the most conflicting writes on network and image carriers (\cref{sec:capability}), and one agent saved a body with the tool's own \texttt{MISMATCH} verdict in its context (\cref{sec:szsp}). Teaching agents to weigh the operation that acquires information complements a check that the tool enforces.

Benchmarks must start before the input is readable. Those this work builds on begin after a repository is materialized or present untrusted input as decoded text (\cref{sec:related}), so their observation starts after Step Zero. The paired protocol adds this step with matched twins, a record of the model context, and a disk sampler, and it applies to any carrier a tool must open before its content is known. We measured archives, TAR containers, network responses, and encoded images, and documents, repositories, and packages are future work. Defending the step that acquires information first requires measuring it.

\section{Limitations and Threats to Validity}\label{sec:limits}
The primary rates use one fixture per condition. The archive result replicates on two independently built archives (\cref{sec:robust}), and the network and image results rest on one fixture each, varied in cues and commitment. Every request passes through one gateway that translates protocols and requests high-effort reasoning, the case-study ablations use four of the six models, and two of the $42$ harness and model pairs are native to each other (Codex with GPT-6-Sol, Qwen Code with Qwen3.8-Max), so rates and orderings describe this panel under this configuration. The hidden budget $B_{\max}=4$\,MiB is a measurement threshold for unbounded materialization, and the same archive cells breached under every budget from \V{budget.lo} to \V{budget.hi}\,MiB. Every breaching cell of the two ungated live archive conditions wrote at least the whole $8$\,MiB stream (\V{nr.peak.atleaststream}; median \V{nr.peak.medianmib}\,MiB, maximum \V{nr.peak.maxmib}\,MiB under repeated extraction), and the bytes scale with the stream in live agents. On a $64$\,MiB understated member agents breached in \V{nr.s64.bytebreach.frac} of cells and wrote a median of \V{nr.s64.peakmedianmib}\,MiB, up to \V{nr.s64.peakmaxmib}\,MiB; on a $512$\,MiB member, in \V{nr.s512.bytebreach.frac} cells, a median of \V{nr.s512.peakmedianmib}\,MiB and up to \V{nr.s512.peakmaxmib}\,MiB. Default command timeouts fired in \V{nr.s64.timeoutfired.frac} and \V{nr.s512.timeoutfired.frac} of these cells, so the attacker sets the size of the availability harm, up to any disk quota a deployment imposes (\cref{sec:szsp}). The gateway replaces the image carrier's pixels with one text notice, so it measures decoding through a tool, and pixel-level perception is future work. The models are served through closed APIs, so we characterize behavior from recorded traffic and disk records, and attributing it to logits or attention requires open-weight models. Cells cluster by harness and model, so inference is reserved for confirmatory contrasts that hold at the cluster level (\cref{tab:cluster}). SZSP's gate is evaluated on archive openings and its network contract on the fetch primitives and an advisory verdict. The enforced write binding, a cumulative byte ledger across calls, an inode bound, and other carriers and formats are left to deployment and future work.

\section{Conclusion}\label{sec:conclusion}
Most defenses for agents act on content that has entered the model context, and at Step Zero the content is still sealed: deciding whether to process an archive, a response, or an image requires information that only processing delivers. Across seven harnesses and six models with benign prompts, most agents were already unsafe at Step Zero: default extraction overran a disk budget on archives, usually before any member reached the model and whatever size the archive declared, so the harm landed before model-side defenses could act. On network and image carriers, agents saved unverified bodies under declared names, so seeing the content left the input in control of what was written. A model-free comparison places the cause in the processing primitive, and the models' own checks faded once cues were removed or a task asked for the save, so model-side judgment is too late or too weak at this step. A gate that counts the bytes a processing call produces held every adversarial archive model-free and every routed extraction in live agents while agents completed \V{b2.zmatszsp.taskcomplete} tasks. Harnesses should offer, in their own tool layer, a read-only preview and a materialization that binds each write to evidence from outside the input, and vendors should pair these gates with safety-alignment training that teaches agents to inspect an input before processing it.

\section*{Ethical Considerations}
Every cell ran in an isolated, single-use account under a disk budget, on a local testbed with synthetic data and network egress limited to the local fixture server, and its files were destroyed afterwards. The carriers are synthetic canaries with benign, instruction-free content, an archive whose central directory understates one member and a response or image whose declared filename disagrees with a short, harmless marker body. The decompression carriers are distributed solely as generator code that targets the local testbed.

The measured behavior is a default handling of untrusted input shared by widely used agent harnesses and the extractors and fetchers they drive. The effects we measure are confined to disk usage and file contents under the agent's own account on the authors' testbed, and the work serves defensive research: the gate of \cref{sec:szsp} is the artifact that harness maintainers can adopt to close the gap.

\section*{Data and Code Availability}
The fixture generators, matched-twin protocol, recording proxy and disk sampler, seven-harness runner, judge, model-free primitive study, SZSP gate, and per-cell records will be released upon acceptance.

\let\stepzerobib\thebibliography
\renewcommand{\thebibliography}[1]{\stepzerobib{#1}\setlength{\itemsep}{0pt}\setlength{\parsep}{0pt}\setlength{\emergencystretch}{1em}\hfuzz=1pt}
\bibliographystyle{IEEEtran}
\bibliography{references}

\end{document}